\documentclass[12pt,twoside,a4paper]{article}
\usepackage{graphicx}
\usepackage[dvips]{epsfig}
\usepackage{amssymb}
\usepackage{rotating}
\usepackage{a4wide}

\def\gsim{\mathrel{\rlap{\lower4pt\hbox{\hskip1pt$\sim$}}\raise1pt\hbox{$>$}}}
\def\Att{A_{TT}}

\def\Cer{\v{C}erenkov }
\def\Journal#1#2#3#4{{#1} {\bf #2}, #3 (#4)}

\def\PRL{\em Phys. Rev. Lett.}

\newcommand{\dsdo}{\ensuremath{\frac{{\rm d}\sigma}{{\rm d}\Omega}}}
\newcommand{\dsdt}{\ensuremath{\frac{{\rm d}\sigma}{{\rm d}t}}}
\newcommand{\AN}{\ensuremath{A_{\rm N}}}
\newcommand{\ANN}{\ensuremath{A_{\rm NN}}}
\newcommand{\ASS}{\ensuremath{A_{\rm SS}}}
\newcommand{\ASL}{\ensuremath{A_{\rm SL}}}
\newcommand{\dd}[1]{\ensuremath{{\rm d}#1}}
\newcommand{\DST}{\ensuremath{\Delta\sigma_T}}

\newcommand{\be}{\begin{equation}}
\newcommand{\ee}{\end{equation}}
\newcommand{\ba}{\begin{eqnarray}}
\newcommand{\ea}{\end{eqnarray}}
\newcommand{\JPsi}{\ensuremath{J/\Psi}}
\newcommand\as{\alpha_{\mathrm{S}}}

\def\Ord#1{10^{#1}}
\def\ord#1{\cdot 10^{#1}}
\def\gsim{\mathrel{\rlap{\lower4pt\hbox{\hskip1pt$\sim$}}\raise1pt\hbox{$>$}}}
\def\lsim{\mathrel{\rlap{\lower4pt\hbox{\hskip1pt$\sim$}}\raise1pt\hbox{$<$}}}
\def\ket#1{\left|\,#1\,\right>}

\def\gsim{\mathrel{\rlap{\lower4pt\hbox{\hskip1pt$\sim$}}\raise1pt\hbox{$>$}}}

\begin{document}
\setcounter{secnumdepth}{4}
\setcounter{tocdepth}{4}

\vfill
\begin{center} \LARGE Technical Proposal \\ for \end{center}
\thispagestyle{empty}
\vfill
\begin{center}
  {\noindent\Huge Antiproton--Proton Scattering Experiments with
    Polarization}
\end{center}
\begin{center} \LARGE (${\cal PAX}$ Collaboration) \end{center}
\vfill
\begin{center} {\noindent\Large J\"ulich, May  2005\\} \end{center}
\vfill
\cleardoublepage

\begin{center} \LARGE Technical Proposal \\ for \end{center}
\begin{center}
  {\noindent\Huge Antiproton--Proton Scattering Experiments with
    Polarization}
\end{center}
\begin{center} \LARGE (${\cal PAX}$ Collaboration) \end{center}

\begin{abstract}
  Polarized antiprotons, produced by spin filtering with an internal
  polarized gas target, provide access to a wealth of single-- and
  double--spin observables, thereby opening a new window to physics
  uniquely accessible at the HESR.  This includes a first measurement
  of the transversity distribution of the valence quarks in the
  proton, a test of the predicted opposite sign of the
  Sivers--function, related to the quark distribution inside a
  transversely polarized nucleon, in Drell--Yan (DY) as compared to
  semi--inclusive DIS, and a first measurement of the moduli and the
  relative phase of the time--like electric and magnetic form factors
  $G_{E,M}$ of the proton.  In polarized and unpolarized $p\bar{p}$
  elastic scattering, open questions like the contribution from the
  odd charge--symmetry Landshoff--mechanism at large $|t|$ and
  spin--effects in the extraction of the forward scattering amplitude
  at low $|t|$ can be addressed.  The proposed detector consists of a
  large--angle apparatus optimized for the detection of DY electron
  pairs and a forward dipole spectrometer with excellent particle
  identification.
  
  The design and performance of the new components, required for the
  polarized antiproton program, are outlined. A low--energy Antiproton
  Polarizer Ring (APR) yields an antiproton beam polarization of
  $P_{\bar{p}}$ = 0.3 to 0.4 after about two beam life times, which is
  of the order of 5--10 h.  By using an internal $\rm H^{\uparrow}$
  target and a detector installed in a 3.5 GeV/c Cooler Synchrotron
  Ring (CSR), the Phase--I experimental $\bar{p}^{\uparrow}
  p^{\uparrow}$ program could start in 2014, completely independent of
  the operation of the HESR.  In Phase--II, the CSR serves as an
  injector for the polarized antiprotons into the HESR. A chicane
  system inside the HESR is proposed to guide the high--energy
  $\bar{p}^{\uparrow}$ beam to the PAX detector, located inside the
  CSR straight section. In Phase--II, fixed--target or collider
  $\bar{p}^{\uparrow}p^{\uparrow}$ experiments over a broad energy
  range become possible. In the collider mode, polarized protons
  stored in the CSR up to momenta of 3.5~GeV/c are bombarded head--on
  with 15~GeV/c polarized antiprotons stored in the HESR.  This
  asymmetric double--polarized antiproton--proton collider is ideally
  suited to map e.g.  the transversity distribution in the proton.

The appendices contained in this document were composed only after the
main document had been submitted to the QCD-PAC. Appendix~A discusses
the polari\-zation--transfer technique that PAX will exploit to
produce a beam of polarized antiprotons, and applications of this
technique in the high--energy sector.  The spin--dependence of the
antiproton--proton interaction and the special interest in
double--polarized antiproton-proton scattering at very low energies,
in view of the indications for the protonium state, is elaborated in
Appendix~B.  In Appendix~C, we discuss details of the impact of recent
data from electron--positron collider experiments on the
proton--antiproton physics, accessible in Phase~I of the PAX
experimental program.  A comment on the Next-to-Leading-Order
corrections to the Drell-Yan process is presented in Appendix~D.
Appendix~E describes beam dynamics simulations that have been carried
out recently for the proton--antiproton collider mode of the PAX
experiment making use of the CSR and the HESR. Based on conservative
assumptions about the number of antiprotons accumulated in the HESR,
these calculations indicate that a luminosity of about ${\cal
L}=1.5\times10^{30}$~cm$^{-2}$s$^{-1}$ can be achieved in the PAX
collider mode.  An extensive program of Monte Carlo studies, described
in Appendix~F, has been started to investigate different options for
the PAX detector configuration, aiming at an optimization of the
achievable performance.

\end{abstract}

\cleardoublepage
\section*{Members of the Collaboration}
\subsection*{Alessandria, Italy, Universita$'$ del Piemonte 
Orientale $''$A. Avogadro$''$ and INFN} 
Vincenzo Barone

\subsection*{Beijing, China, School of Physics, Peking University}
Bo--Qiang Ma    

\subsection*{Bochum, Germany, Institut f\"ur 
Theoretische Physik II, Ruhr Universit\"at Bochum}
Klaus Goeke, 
Andreas Metz, and
Peter Schweitzer

\subsection*{Bonn, Germany, Helmholtz--Institut f\"ur Strahlen-- 
und Kernphysik, Universit\"at Bonn}
Jens Bisplinghoff,
Paul--Dieter Eversheim, 
Frank Hinterberger, 
Ulf--G. Mei{\ss}ner,
Heiko Rohd\-je{\ss}, and
Alexander Sibirtsev

\subsection*{Brookhaven, USA, Collider--Accelerator Department, Brookhaven National Laboratory}
Christoph Montag

\subsection*{Brookhaven, USA, RIKEN BNL Research Center, Brookhaven National Laboratory}
Werner Vogelsang

\subsection*{Cagliari, Italy, Dipartimento di Fisica, Universita$'$ di Cagliari and INFN}
Umberto D$'$Alesio, and 
Francesco Murgia

\subsection*{Dublin, Ireland, School  of Mathematics, Trinity College, University of Dublin}
Nigel Buttimore

\subsection*{Dubna, Russia, Bogoliubov Laboratory of Theoretical Physics, 
Joint Institute for Nuclear Research} 
Anatoly Efremov, and 
Oleg Teryaev

\subsection*{Dubna, Russia, Dzhelepov Laboratory of Nuclear Problems, 
Joint Institute for Nuclear Research} 
Sergey Dymov,
Natela Kadagidze, 
Vladimir Komarov, 
Anatoly Kulikov,
Vladimir Kurbatov,
Vladimir Leontiev,
Gogi Macharashvili,
Sergey Merzliakov,
Igor Meshkov,
Valeri Serdjuk,
Anatoly Sidorin,
Alexander Smirnow,
Evgeny Syresin,
Sergey Trusov, 
Yuri Uzikov,
Alexander Volkov, and
Nikolai Zhuravlev

\subsection*{Dubna, Russia, Laboratory of Particle Physics, 
Joint Institute for Nuclear Research} 
Oleg Ivanov,              
Victor Krivokhizhin,
Gleb Meshcheryakov,
Alexander Nagaytsev,
Vladimir Peshekhonov, 
Igor Savin, 
Binur Shaikhatdenov,
Oleg Shevchenko, and
Gennady Yarygin

\subsection*{Erlangen, Germany, Physikalisches Institut, Universit\"at 
Er\-langen--N\"urn\-berg} 
Wolfgang Eyrich,
Andro Kacharava,
Bernhard Krauss,
Albert Lehmann,
Davide Reggiani,
Klaus Rith,
Ralf Seidel,
Erhard Steffens,
Friedrich Stinzing,
Phil Tait, and
Sergey Yaschenko

\subsection*{Ferrara, Italy, Istituto Nazionale di Fisica Nucleare}
Marco Capiluppi, 
Guiseppe Ciullo, 
Marco Contalbrigo, 
Alessandro Drago, 
Paola Ferretti--Dalpiaz, 
Francesca Giordano,
Paolo Lenisa, 
Luciano Pappalardo,
Giulio Stancari,
Michelle Stancari, and
Marco Statera

\subsection*{Frascati, Italy, Istituto Nazionale di Fisica Nucleare}
Eduard Avetisyan,
Nicola Bianchi,
Enzo De Sanctis,
Pasquale Di Nezza,
Alessandra Fantoni,
Cynthia Hadjidakis,
Delia Hasch,
Marco Mirazita,
Valeria Muccifora,
Federico Ronchetti, and
Patrizia Rossi

\subsection*{Gatchina, Russia, Petersburg Nuclear Physics Institute}
Sergey Barsov,
Stanislav Belostotski, 
Oleg Grebenyuk, 
Kirill Grigoriev,
Anton Izotov, 
Anton Jgoun,
Peter Kravtsov, 
Sergey Manaenkov, 
Maxim Mikirtytchiants,
Sergey Mikirtytchiants,
Oleg Miklukho,  
Yuri Naryshkin, 
Alexander Vassiliev, and 
Andrey Zhdanov 

\subsection*{Gent, Belgium,  Department of Subatomic and 
Radiation Physics, University of Gent} 
Dirk Ryckbosch

\subsection*{Hefei, China, Department of Modern Physics, University of Science and Technology
of China}
Yi Jiang,
Hai--jiang Lu,
Wen--gan Ma,
Ji Shen,
Yun--xiu Ye,
Ze--Jie Yin, and
Yong--min Zhang

\subsection*{J\"ulich, Germany, Forschungszentrum J\"ulich, Institut 
f\"ur Kernphysik} 
David  Chiladze, 
Ralf Gebel,
Ralf Engels, 
Olaf Felden,
Johann Haidenbauer,
Christoph Hanhart,  
Michael Hartmann,
Irakli Keshelashvili,
Siegfried Krewald,
Andreas Lehrach,   
Bernd Lorentz,  
Sigfried Martin, 
Ulf--G. Mei{\ss}ner,
Nikolai Nikolaev,  
Dieter Prasuhn, 
Frank Rathmann, 
Ralf Schleichert,
Hellmut Seyfarth,  and
Hans Str\"oher

\subsection*{Kosice, Slovakia, Institute of Experimental Physics, 
Slovak Academy of Sciences and P.J. Safarik University, Faculty of Science} 
Dusan Bruncko, 
Jozef Ferencei, 
J\'an Mu\v sinsk\'y, and  
Jozef Urb\'an

\subsection*{Langenbernsdorf, Germany, Unternehmensberatung und Service--B\"uro (USB), 
Gerlinde Schulteis \& Partner GbR} 
Christian Wiedner (formerly at MPI-K Heidelberg)

\subsection*{Lecce, Italy, Dipartimento di Fisica, Universita$'$ di Lecce and INFN}
Claudio Corian\'o, and 
Marco Guzzi 

\subsection*{Madison, USA, University of Wisconsin} 
Tom Wise

\subsection*{Milano, Italy, Universita' dell'Insubria, Como and INFN sez.} 
Philip Ratcliffe

\subsection*{Moscow, Russia, Institute for Theoretical and 
Experimental Physics}
Vadim Baru,
Ashot Gasparyan,  
Vera Grishina, 
Leonid Kondratyuk, and
Alexander Kudriavtsev

\subsection*{Moscow, Russia, Lebedev Physical Institute}
Alexander Bagulya,
Evgeni Devitsin,
Valentin Kozlov,
Adel Terkulov, and
Mikhail Zavertiaev

\subsection*{Moscow, Russia, Physics Department, Moscow Engineering Physics Institute}
Aleksei Bogdanov,
Sandibek Nurushev,
Vitalii Okorokov,
Mikhail Runtzo, and
Mikhail Strikhanov

\subsection*{Novosibirsk, Russia, Budker Institute for Nuclear Physics}
Yuri Shatunov

\subsection*{Palaiseau, France, Centre de Physique Theorique, Ecole Polytechnique} 
Bernard Pire

\subsection*{Protvino, Russia, Institute of High Energy Physics} 
Nikolai Belikov,     
Boris Chujko,     
Yuri Kharlov,    
Vladislav Korotkov,   
Viktor Medvedev,    
Anatoli Mysnik,     
Aleksey Prudkoglyad,
Pavel Semenov,   
Sergey Troshin, and
Mikhail Ukhanov 

\subsection*{Tbilisi, Georgia, Institute of High Energy Physics and Informatization, 
Tbilisi State University} 
Badri Chiladze,
Nodar Lomidze,
Alexander Machavariani,
Mikheil Nioradze,
Tariel Sakhelashvili, 
Mirian Tabidze, and
Igor Trekov

\subsection*{Tbilisi, Georgia, Nuclear Physics Department, Tbilisi State University} 
Leri Kurdadze, and 
George Tsirekidze

\subsection*{Torino, Italy, Dipartimento di Fisica Teorica, 
Universita di Torino and INFN} 
Mauro Anselmino, 
Mariaelena Boglione, and
Alexei Prokudin

\subsection*{Uppsala, Sweden, Department of Radiation Sciences, Nuclear Physics Division} 
Pia Thorngren--Engblom

\subsection*{Virginia, USA, Department of Physics, University of Virginia} 
Simonetta Liuti

\subsection*{Warsaw, Poland, Soltan Institute for Nuclear Studies} 
Witold Augustyniak, 
Bohdan Marianski, 
Lech Szymanowski, 
Andrzej Trzcinski, and
Pawel Zupranski

\subsection*{Yerevan, Armenia, Yerevan Physics Institute}
Norayr Akopov, 
Robert Avagyan,
Albert Avetisyan,
Garry Elbakyan,
Zaven Hakopov,
Hrachya Marukyan, and
Sargis Taroian

\section*{Spokespersons:} 
Paolo Lenisa, E--Mail: lenisa@mail.desy.de\\
Frank Rathmann, E--Mail: f.rathmann@fz--juelich.de


\cleardoublepage

\tableofcontents

\cleardoublepage

\part{Physics Case\label{partI_phys-case}}
\pagestyle{myheadings} \markboth{Technical Proposal for
  ${\cal PAX}$}{Part I: Physics Case}
\section{Preface}
The polarized antiproton--proton interactions at HESR will allow a
unique access to a number of new fundamental physics observables,
which can be studied neither at other facilities nor at HESR without
transverse polarization of protons and/or antiprotons:\\

\begin{itemize}
\item The transversity distribution is the last leading--twist missing
  piece of the QCD description of the partonic structure of the
  nucleon. It describes the quark transverse polarization inside a
  transversely polarized proton \cite{bdr}. Unlike the more
  conventional unpolarized quark distribution $q(x,Q^2)$ and the
  helicity distribution $\Delta q(x,Q^2)$, the transversity
  $h^q_1(x,Q^2)$ can neither be accessed in deep--inelastic scattering
  of leptons off nucleons nor can it be reconstructed from the
  knowledge of $q(x,Q^2)$ and $\Delta q(x,Q^2)$. It may contribute to
  some single--spin observables, but always coupled to other unknown
  functions.  The transversity distribution is directly accessible
  uniquely via the {\bf double transverse spin asymmetry} $A_{TT}$ in
  the Drell--Yan production of lepton pairs. The theoretical
  expectations for $A_{TT}$ in the Drell--Yan process with
  transversely polarized antiprotons interacting with a transversely
  polarized proton target or beam at HESR are in the 30--40 per cent
  range \cite{abdn,gms}; with the expected antiproton spin--filtering
  rate and luminosity of HESR the PAX experiment is uniquely suited
  for the definitive observation of $h^q_1(x,Q^2)$ of the proton for
  the valence quarks.
\item The PAX measurements can also provide completely new insights
  into the understanding of (transverse) single--spin asymmetries
  (SSA) which have been observed in proton--proton and
  proton--antiproton collisions as well as in lepton--nucleon
  scattering.  For instance through charm production ($\bar p^\uparrow
  \, p\to D\, X$ or $\bar p \, p^\uparrow \to D \, X$) it will be
  possible to disentangle the Sivers \cite{SiversFunction} and the
  Collins mechanisms \cite{CollinsFunction}.  In general, both effects
  contribute to the measured SSA (mostly in $p^\uparrow \, p \to \pi
  \, X$ and $\bar p^\uparrow \, p \to \pi \, X$ ), but in the case of
  charm production the Collins mechanism drops out.  Moreover, in
  conjunction with the data on SSA from the HERMES collaboration
  \cite{hermesSSA,HERMESSiv}, the PAX measurements of the SSA in
  Drell--Yan production on transversely polarized protons can for the
  first time provide a test of the theoretical prediction
  \cite{Collins} of the sign--reversal of the Sivers function from
  semi--inclusive DIS to Drell--Yan processes.  Both studies will
  crucially test and improve our present QCD--description of the
  intriguing phenomenon of SSA.
\item The origin of the unexpected $Q^2$--dependence of the ratio of
  the magnetic and electric form factors of the proton, as observed at
  the Jefferson laboratory \cite{perdrisat}, can be clarified by a
  measurement of their relative phase in the time--like region, which
  discriminates strongly between the models for the form factor. This
  phase can be measured via SSA in the annihilation $ \bar{p}
  p^{\uparrow} \to e^+e^-$ on a transversely polarized target
  \cite{d,brodsky}. The first ever measurement of this phase at PAX
  will also contribute to the understanding of the onset of the pQCD
  asymptotics in the time--like region and will serve as a stringent
  test of dispersion theory approaches to the relationship between the
  space--like and time--like form factors
  \cite{Geshkenbein74,HMDtimelike,egle}.  The double--spin asymmetry
  will fix the relative phase ambiguity and allow independently the
  $G_E-G_M$ separation, which will serve as a check of the Rosenbluth
  separation in the time--like region.
  
\item Arguably, in $p\bar{p}$ elastic scattering the hard scattering
  mechanism can be checked beyond $|t| = {1\over 2}(s-4m_p^2)$
  accessible in the $t$--$u$--symmetric $pp$ scattering, because in
  the $p\bar{p}$ case the $u$--channel exchange contribution can only
  originate from the strongly suppressed exotic dibaryon exchange.
  Consequently, in the $p\bar{p}$ case the hard mechanisms
  \cite{Matveev,BrodskyFarrar,KrollElastic} can be tested at $t$
  almost twice as large as in $pp$ scattering. Even unpolarized large
  angle $p\bar{p}$ scattering data can shed light on the origin of the
  intriguing oscillations around the $s^{-10}$ behavior of the $90^0$
  scattering in the $pp$ channel and put stringent constraints on the
  much disputed charge conjugation--odd independent-scattering
  Landshoff mechanism \cite{Landshoff,RalstonPire,Ramsey,Dutta}.  In
  general, the interplay of different mechanisms is such that single
  and double transverse asymmetries in $p\bar{p}$ scattering are
  expected to be as large as the ones observed in the $pp$ case.
\item The charge conjugation property allows direct monitoring of the
  polarization of antiprotons in HESR and the rate of polarization
  buildup constitutes a direct measurement of the transverse double
  spin asymmetry in the $p\bar{p}$ total cross section. This asymmetry
  has never been measured and its knowledge is crucial for the correct
  extraction of the real part of the forward $p\bar{p}$ scattering
  amplitude from Coulomb--nuclear interference. The PAX results on the
  asymmetry will help to clarify the origin of the discrepancy between
  the dispersion theory calculations \cite{Kroll} and the experimental
  extraction \cite{Armstrong} of the value of the real part of the
  forward scattering amplitude usually made assuming the spin
  independence of forward scattering.
\end{itemize}

\section{Accessing Transversity Distributions}
\subsection{Spin Observables and Transversity} 
There are three leading--twist quantities necessary to achieve a full
understanding of the nucleon quark structure: the unpolarized quark
distribution $q(x,Q^2)$, the helicity distribution $\Delta q(x,Q^2)$
and the transversity distribution $\Delta_{_T} q(x,Q^2)$ [more usually
denoted as $h^q_1(x,Q^2)$] \cite{bdr}. While $\Delta q$ describes the
quark longitudinal polarization inside a longitudinally polarized
proton, the transversity describes the quark transverse polarization
inside a transversely polarized proton at infinite momentum. $h^q_1$
and $\Delta q$ are two independent quantities, which might be equal
only in the non--relativistic, small $Q^2$ limit.  Moreover, the quark
transverse polarization does not mix with the gluon polarization
(gluons carry only longitudinal spin), and thus the QCD evolutions of
$h^q_1$ and $\Delta q$ are quite different. One cannot claim to
understand the spin structure of the nucleon until all three
leading--twist structure functions have been measured.

Whereas the unpolarized distributions are well known, and more and
more information is becoming available on $\Delta q$, nothing is known
experimentally on the nucleon transversity distribution. From the
theoretical side, there exist only a few theoretical models for
$h^q_1$.  An upper bound on its magnitude has been derived: this bound
holds in the naive parton model, and, if true in QCD at some scale, it
is preserved by QCD evolution. Therefore, its verification or disproof
would be by itself a very interesting result.  The reason why $h^q_1$,
despite its fundamental importance, has never been measured is that it
is a chiral--odd function, and consequently it decouples from
inclusive deep--inelastic scattering. Since electroweak and strong
interactions conserve chirality, $h^q_1$ cannot occur alone, but has
to be coupled to a second chiral--odd quantity.

This is possible in polarized Drell--Yan processes, where one measures
the product of two transversity distributions, and in semi--inclusive
Deep Inelastic Scattering (SIDIS), where one couples $h^q_1$ to a new
unknown fragmentation function, the so--called Collins function
\cite{CollinsFunction}. Similarly, one could couple $h^q_1$ and the
Collins function in transverse single--spin asymmetries (SSA) in
inclusive processes like $p^\uparrow \, p \to \pi \, X$.

Both HERMES \cite{HERMESSiv} and COMPASS experiments are now gathering
data on spin asymmetries in SIDIS processes, which should yield
information on some combination of $h^q_1$ and the Collins function.
However, one cannot directly extract information on $h^q_1$ alone: the
measured spin asymmetries can originate also from the Sivers function
\cite{SiversFunction} -- a spin property of quark distributions,
rather than fragmentation -- which does not couple to transversity; in
addition, higher twist effects might still be sizeable at the modest
$Q^2$ of the two experiments, thus making the interpretation of data
less clear.  The transverse SSA experimentally observed in $p^\uparrow
\, p \to \pi \, X$ and $\bar p^\uparrow \, p \to \pi \, X$ processes
\cite{ags,e704,star} can be interpreted in terms of transversity and
Collins functions; however, also here contributions from the Sivers
function are important, or even dominant \cite{ppcol}, and these
processes could hardly be used to extract information on $h_1^q$
alone.
 
\subsection{Transversity in Drell--Yan Processes at PAX} 
The most direct way to obtain information on transversity -- the last
leading--twist missing piece of the QCD nucleon spin structure -- is
the measurement of the double transverse spin asymmetry $A_{TT}$ in
Drell--Yan processes with {\it both transversely polarized beam and
  target}:
\begin{equation}
A_{TT} \equiv \frac{d\sigma^{\uparrow\uparrow} - 
d\sigma^{\uparrow\downarrow}} {d\sigma^{\uparrow\uparrow} 
+ d\sigma^{\uparrow\downarrow}} = \hat{a}_{TT} \> 
\frac{\sum_q e_q^2 \, h_1^q(x_1, M^2) \, h_1^{\bar q}(x_2, M^2)}
{\sum_q e_q^2 \, q(x_1, M^2) \, \bar q(x_2, M^2)}\>, 
\label{att}
\end{equation}
where $q = u, \bar u, d, \bar d, ...$, $M$ is the invariant mass of
the lepton pair and $\hat a_{TT}$ is the double spin asymmetry of the
QED elementary process, $q \bar q \to \ell^+ \ell^-$,
\begin{equation}
\hat a_{TT}= {\sin^2\theta\over 1 + \cos^2\theta} \> \cos 2\phi \,, 
\end{equation}
with $\theta$ the polar angle of the lepton in the $l^+l^-$ rest frame
and $\phi$ the azimuthal angle with respect to the proton
polarization.

The measurement of $A_{TT}$ is planned at RHIC, in Drell--Yan
processes with transversely polarized protons (for a review see
\cite{VogelsangRHIC}). In this case one measures the product of two
transversity distributions, one for a quark and one for an antiquark
(both in a proton).  At RHIC energies one expects measurements at
$\tau = x_1x_2 = M^2/s \simeq 10^{-3}$, which mainly lead to the
exploration of the sea quark proton content, where polarization is
likely to be tiny. Moreover, the QCD evolution of transversity is such
that, in the kinematical regions of RHIC data, $h^q_1(x, Q^2)$ is much
smaller than the corresponding values of $\Delta q(x,Q^2)$ and
$q(x,Q^2)$. All this makes the double spin asymmetry $A_{TT}$ expected
at RHIC very small, of the order of a few percents or less
\cite{bcd,vog}.

The situation with the PAX measurement of the double transverse spin
asymmetry $A_{TT}$ in Drell--Yan processes with {\it polarized
  antiprotons and protons}, $\bar p^\uparrow \, p^\uparrow \to \ell^+
\ell^- \, X$, is entirely different. When combining the fixed target
and the collider operational modes, the PAX experiment will explore
ranges of $s \simeq 30$--200 GeV$^2$ and $M^2 \simeq 4$--100 GeV$^2$,
which are ideal for the measurement of large values of $A_{TT}$. There
are some unique features which strongly suggest to pursue the study of
$h_1^q$ in the $\bar{p}p$ channel with PAX:

\begin{itemize}
\item In $\bar pp$ processes both the quark (from the proton) and the
  antiquark (from the antiproton) contributions are large. For typical
  PAX kinematics in the fixed target mode ($s$ = 30 or 45 GeV$^2$, see
  Sec.~\ref{sec:phaseI}) one has $\tau = x_1x_2 = M^2/s \simeq 0.2$ --
  $0.3$, which means that only quarks and antiquarks with large $x$
  contribute, that is valence quarks for which $h_1^q$ is expected to
  be large. Moreover, at such $x$ and $M^2$ values the QCD evolution
  does not suppress $h^q_1(x, Q^2)$. $A_{TT}/\hat a_{TT}$ is expected
  to be as large as 30\% \cite{abdn}; this is confirmed by direct
  calculations using the available models for transversity
  distributions, some of which predict even larger values, up to
  40--45\% \cite{gms}.  Actually, all these models agree in having
  $|h_1^u| \gg |h_1^d|$ \cite{bdr}, so that Eq. (\ref{att}) for $\bar
  pp$ processes at PAX essentially becomes,
\begin{equation}
A_{TT} \simeq \hat{a}_{TT} \>
\frac{h_1^u(x_1, M^2) \, h_1^u(x_2, M^2)}
{u(x_1, M^2) \, u(x_2, M^2)}\> , \label{atts}
\end{equation}
where all distribution functions refer to protons ($\bar q^{\,\bar p}
= q^p = q$, {\it etc.}). $A_{TT}$ allows then a direct access to
$|h_1(x)|$.
 
\item When running in the collider mode (see Sec.~\ref{sec:phaseII})
  the energy range covered by PAX increases up to $s \simeq 200$
  GeV$^2$ and $M^2 \simeq 100$ GeV$^2$, while the value of $A_{TT}$
  remains safely above 20\%.  The $(x_1, x_2)$ kinematical regions
  covered by the PAX measurements, both in the fixed target and
  collider mode, are described in Fig.~\ref{Figphysics}, left side.
  The plots on the right side show the expected values of the
  asymmetry $A_{TT}$ as a function of Feynman $x_F = x_1 - x_2$, for
  different values of $s$ and $Q^2 = 16$ GeV$^2$.  The collider
  experiment plays, for the transversity distribution $h_1(x,M^2)$,
  the same role polarized inclusive DIS played for the helicity
  distribution $\Delta q(x,Q^2)$, with a kinematical $(x, Q^2)$
  coverage similar to that of the HERMES experiment.
\begin{figure}[hbt]
 \begin{center}
 \includegraphics[width=0.49\linewidth]{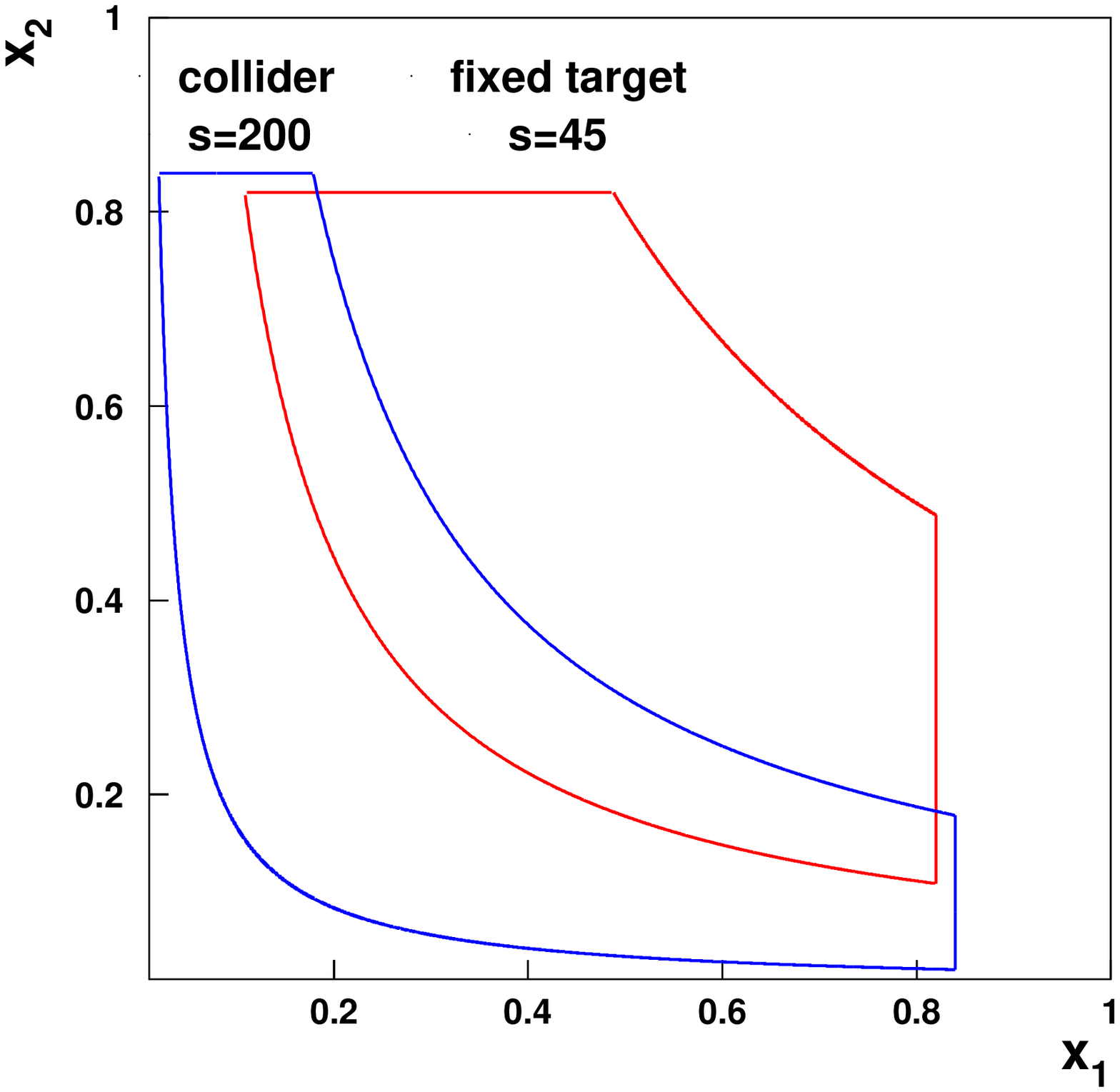}
  \includegraphics[width=0.49\linewidth]{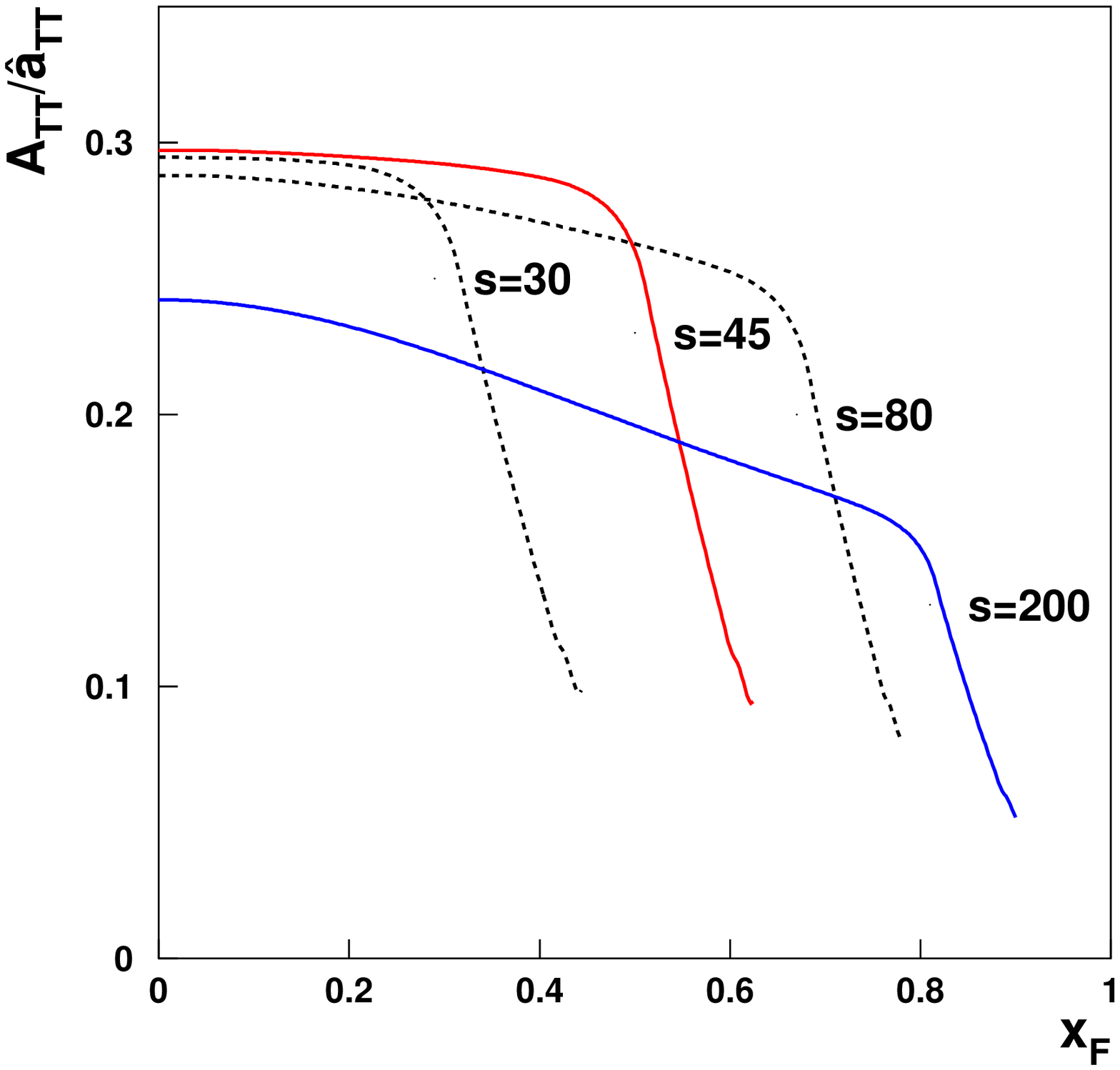}
  \parbox{14cm}{\caption{\label{Figphysics}\small Left: The kinematic region covered by
      the $h_1$ measurement at PAX in phase II. In the asymmetric
      collider scenario (blue) antiprotons of 15~GeV/c impinge on
      protons of 3.5~GeV/c at c.m.  energies of $\sqrt{s}\sim
      \sqrt{200}$~GeV and $Q^2>4$ $\rm GeV^2$.  The fixed target case
      (red) represents antiprotons of 22~GeV/c colliding with a fixed
      polarized target ($\sqrt{s}\sim\sqrt{45}$~GeV).  Right: The
      expected asymmetry as a function of Feynman $x_F$ for different
      values of $s$ and $Q^2=16$ $\rm GeV^2$.}}
\end{center}
\end{figure}
  
\item The counting rates for Drell--Yan processes at PAX are estimated
  in Sec. 4. We notice here that in the quest for $h_1^q$ one should
  not confine to the $M \!>\! 4$ GeV region, which is usually
  considered as the ``safe'' region for the comparison with the pQCD
  computations, as this cut--off eliminates the background from the
  $\JPsi,\Psi'$ production and their subsequent leptonic decay. Also
  the region $1.5 \lsim M \lsim 3$ GeV is free from resonances and can
  be exploited to access $h_1$ via Drell--Yan processes \cite{gms,br}.
  
\item Even the $\JPsi,\Psi'$ resonance region at $M \simeq 3$ GeV
  could be crucial \cite{abdn}. The cross section for dilepton
  production increases by almost 2 orders of magnitude going from $M =
  4$ to $M = 3$ GeV \cite{sps,ana,mmp}: this cross section involves
  unknown quantities related to the $q \bar q-\JPsi$ coupling.
  However, independently of these unknown quantities, the $q \bar
  q-\JPsi$ coupling is a vector one, with the same spinor and Lorentz
  structure as the $q \bar q-\gamma^*$ coupling; similarly for the
  $\JPsi-e^+e^-$ decay. These unknown quantities cancel in the ratio
  giving $A_{TT}$, while the helicity structure remains, so that Eq.
  (\ref{atts}) still holds in the $\JPsi$ resonance region
  \cite{abdn}. This substantially enhances the sensitivity of the PAX
  experiment to $A_{TT}$ and the amount of direct information
  achievable on $h_1^u(x_1,M^2)\,h_1^u(x_2,M^2)$.  The theoretical
  analysis of the NLO corrections to $A_{TT}$ for prompt photon
  production in hadronic collisions has already been accomplished
  \cite{WernerAttNLO}, the full computation of QCD corrections to
  $A_{TT}$, relevant to PAX kinematical values (including the
  $\JPsi,\Psi'$ resonance region), is in progress \cite{bcgr}.

\end{itemize}

\subsection{Transversity in $D$--Meson Production at PAX} 
The double transverse spin asymmetry $A_{TT}$ can be studied also for
other processes; in particular, the open charm production, $\bar
p^\uparrow \, p^\uparrow \to D \, X$ looks like a very promising
channel to extract further information on the transversity
distributions.  At PAX in collider mode ($\sqrt s \simeq \sqrt{210}$
GeV) the production of $D$ mesons with $p_T$ of the order of 2 GeV/$c$
is largely dominated by the $\bar q q \to \bar c c$ elementary process
\cite{umb}; then one has (again, all distribution functions refer to
protons):
\begin{equation}
A_{TT}^{^D}  \simeq 
\frac{\sum_q  h_1^q(x_1) \otimes h_1^q(x_2) 
\otimes \Delta \hat\sigma \otimes D(z)}
{\sum_q q(x_1) \otimes q(x_2) \otimes \hat\sigma \otimes D}
\>, \label{attd}
\end{equation}
which supplies information about the convolution of the transversity
distributions with the fragmentation functions $D(z)$ of $c$ quarks or
antiquarks into $D$ mesons, which are available in the literature;
$\Delta \hat\sigma = \hat\sigma^{\uparrow\uparrow} -
\hat\sigma^{\uparrow\downarrow}$ is the known double spin asymmetry
for the $\bar q q \to \bar c c$ elementary process.  Eq. (\ref{attd})
holds above the resonance region ($M = \sqrt{x_1x_2s}\, > 4$ GeV); the
elementary interaction is a pQCD process, so that the cross section
for $D$--production might even be larger, at the same scale, than the
corresponding one for Drell--Yan processes.  Notice that, once more,
the same channel at RHIC cannot supply information on $h_1$, as at
RHIC energy ($\sqrt s = 200$ GeV), the dominant contribution to $D$
production comes from the $gg \to \bar c c$ elementary channel, rather
than the $\bar q q \to \bar c c$ one \cite{Dpaper}.

\section{Single Spin Asymmetries and Sivers Function}
While the direct access to transversity is the outstanding, unique
possibility offered by the PAX proposal concerning the proton spin
structure, there are several other spin observables related to
partonic correlation functions which should not be forgotten.  These
might be measurable even before the antiproton polarization is
achieved.
   
The perturbative QCD spin dynamics, with the helicity conserving
quark--gluon couplings, is very simple.  However, such a simplicity
does not always show up in the hadronic spin observables.  The
observed single spin asymmetries (SSA) are a symptom of this feature.
By now it is obvious that the non--perturbative, long--distance QCD
physics has many spin properties yet to be explored.  A QCD
phenomenology of SSA seems to be possible, but more data and new
measurements are crucially needed.  A new experiment with antiprotons
scattered off a polarized proton target, in a new kinematical region,
would certainly add valuable information on such spin properties of
QCD.
  
As a first example we consider the transverse SSA
\begin{equation}
A_N = \frac{d\sigma^\uparrow - d\sigma^\downarrow}
{d\sigma^\uparrow + d\sigma^\downarrow} \>,
\end{equation}
measured in $p^{\uparrow} \, p \to \pi \, X$ and $\bar p^{\uparrow} \,
p \to \pi \, X$ processes: the SSA at large values of $x_F$ ($x_F
\gsim 0.4$) and moderate values of $p_T$ ($0.7 < p_T < 2.0$ GeV/$c$)
have been found by several $p^{\uparrow} \, p$ experiments
\cite{ags,e704,star} to be unexpectedly large (up to about 40$\%$),
and similar values and trends of $A_N$ have been observed in
experiments with center of mass energies ranging from 6.6 up to 200
GeV.

The large effects were unexpected because, within the standard
framework of collinear QCD factorization, one has to resort to
subleading twist functions in order to obtain non--zero SSA
\cite{et,qs}.  However, if the factorization approach is extended to
not only include longitudinal but also transverse parton momenta,
non--vanishing SSA emerge already at leading twist.  In such an
approach the above mentioned Sivers parton distribution
\cite{SiversFunction} and Collins fragmentation function
\cite{CollinsFunction} enter.  In order to disentangle both effects
the study of SSA for $D$--meson production ($\bar p^\uparrow \, p \to D
\, X$ or $\bar p \, p^\uparrow \to D \, X$) is very promising.  At the
PAX collider energy, for a final $D$ with $p_T$ of about 2 GeV/c, the
dominant subprocess is $\bar q q \to \bar c c$ \cite{umb}, with the
subsequent fragmentation of a charmed quark into a charmed meson.  In
this elementary annihilation process there is no transverse spin
transfer and the final $c$ and $\bar c$ are not polarized.  Therefore,
there cannot be any contribution to the SSA from the Collins
mechanism.  A SSA could only result from the Sivers mechanism, coupled
to an unpolarized elementary reaction and fragmentation.  A
measurement of a SSA in $\bar p^\uparrow \, p \to D \, X$ or $\bar p
\, p^\uparrow \to D \, X$ would then allow a clean access to the quark
Sivers function, active in an annihilation channel.  This is not the
case at RHIC energies, where the leading subprocess turns out to be
$gg \to \bar c c$, which could lead to information on the gluon Sivers
function \cite{Dpaper}.

The Sivers function (denoted by $f_{1T}^{\perp}$) attracted quite some
interest over the past three years.  It belongs to the class of the
so--called (naive) time--reversal odd (T--odd) parton distributions,
which are in general at the origin of SSA.  Therefore, it was believed
for about one decade that the Sivers function vanishes because of
T--invariance of the strong interaction \cite{CollinsFunction}.
However, in 2002 it was shown that $f_{1T}^{\perp}$ can actually be
non--zero \cite{bhs,Collins}.  In this context it is crucial that the
Wilson line, which ensures color gauge invariance, is taken into
account in the operator definition of the Sivers function.  The Wilson
line encodes initial state interactions in the case of the Drell--Yan
process and final state interactions of the struck quark in the case
of DIS.  The Sivers function, describing the (asymmetric) distribution
of quarks in a transversely polarized nucleon \cite{SiversFunction},
contains a rich amount of information on the partonic structure of the
nucleon.  E.g., it is related to the orbital angular momentum of
partons, and the sign of the Sivers asymmetry of a given quark flavor
is directly connected with the sign of the corresponding anomalous
magnetic moment \cite{burk}.
\begin{figure}[htb]
\begin{center}
  \includegraphics[width=0.70\linewidth]{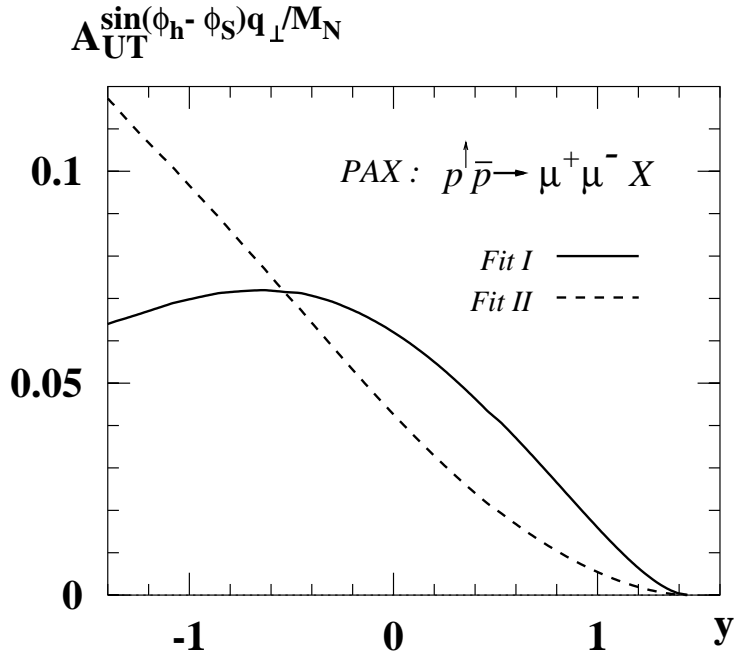}
  \parbox{14cm}{\caption{\small The SSA
      $A_{UT}^{\sin(\phi_h-\phi_S)q_\perp/M_N}$ in Drell-Yan lepton
      pair production, $p^{\uparrow} \bar{p} \to \mu^+ \mu^- X$, as
      function of the rapidity $y$ for typical PAX kinematics ($s = 45
      \, \textrm{GeV}^2$, $M^2 = 2.5 \, \textrm{GeV}^2$).  The
      different curves correspond to equally good fits to the HERMES
      data\cite{HERMESSiv}.\label{SiversDY}}}
\end{center}
\end{figure}

It is now important that the Wilson line can be process dependent.
This property leads to the very interesting prediction that the Sivers
function in Drell--Yan and in semi--inclusive DIS (measured for
instance via the transverse SSA $l \, p^\uparrow \to l \, \pi \, X$)
should have a reversed sign \cite{Collins}, i.e.,
\begin{equation}
f_{1T}^{\perp} \Big|_{DY} = - f_{1T}^{\perp} \Big|_{DIS} \,.
\label{sign}
\end{equation}
In the meantime, the HERMES collaboration has already obtained first
results for the Sivers asymmetry in semi--inclusive DIS
\cite{HERMESSiv}.  Therefore, measuring $f_{1T}^{\perp}$ in Drell--Yan
processes (like $\bar p \, p^\uparrow \to l^+ \, l^- \, X$ or $\bar
p^{\uparrow} \, p \to l^+ \, l^- \, X$) at PAX would check the
clear--cut prediction in Eq.~(\ref{sign}) based on the QCD
factorization approach.  An experimental check of the sign--reversal
would crucially test our present day understanding of T--odd parton
distributions and, consequently, of the very nature of SSA within QCD.
In passing, we note that, within slightly different contexts, recently
several other papers have also stressed the importance of measuring
SSA in Drell--Yan processes \cite{hts,bmt,bl,bq,adm}.

On the basis of the recent HERMES data \cite{HERMESSiv} for
$f_{1T}^{\perp}$ a prediction for the corresponding Sivers asymmetry
in Drell Yan for PAX (for $p^{\uparrow} \bar{p} \to \mu^+ \mu^- X$)
has been reported \cite{egmms}.  The main result of this study is
shown in Fig.~\ref{SiversDY}, where the (weighted) asymmetry
\begin{equation}
 A_{UT}^{\sin(\phi-\phi_S)\frac{q_T}{M_N}}(y,M^2) = 
  2 \; \frac{\sum_a e_a^2 \, x_1f_{1T}^{\perp(1) a/p}(x_1,M^2) \,
                             x_2f_1^{\bar a/\bar{p}}(x_2,M^2)}
            {\sum_a e_a^2 \, x_1f_1^{a/p}(x_1,M^2) \,                 
                             x_2f_1^{\bar a/\bar{p}}(x_2,M^2)} 
\end{equation}
is plotted.  The weighting is performed for technical reasons and is
done with $\sin(\phi - \phi_S)$ ($\phi$ and $\phi_S$ respectively
denoting the azimuthal angle of the virtual photon and the target spin
vector), and with the transverse momentum $\vec{q}_{T}$ of the lepton
pair.  The quantity $f_{1T}^{\perp (1)}$ represents the second moment
of the Sivers function with respect to the transverse quark momentum.
In Fig.~\ref{SiversDY} the asymmetry is displayed as function of the
rapidity $y$ of the lepton pair.  (Note the relation $x_{1/2} =
\sqrt{M^2/s} \, e^{\pm y}$.)  On the basis of this study asymmetries
of the order $5- 10 \%$ can be expected \cite{egmms} --- an effect
which should definitely be measurable at PAX.  This would allow one to
check the predicted sign--flip of the Sivers function in the valence
region, even if the error bars would be large.

In summary, combining information on SSA from $p \, p^\uparrow$ and
$\bar p^{\uparrow} \, p$ processes would greatly help in disentangling
the Sivers and Collins mechanism.  In this context production of
charmed mesons (via $\bar p^\uparrow \, p \to D \, X$ or $\bar p \,
p^\uparrow \to D \, X$) can play a crucial role because these
asymmetries are not sensitive to the Collins function.  We have also
emphasized the importance of measuring the Sivers function in
Drell--Yan.  Through such an experiment, in combination with the
already available information on the Sivers function coming from
semi--inclusive DIS, a crucial check of our current understanding of
the origin of T--odd parton distributions and of SSA within QCD can be
achieved in an unprecedented way.

\section{Electromagnetic Form Factors of the Proton \label{partI_emff}}
The form factors of hadrons as measured both in the space--like and
time--like domains provide fundamental information on their structure
and internal dynamics. Both the analytic structure and phases of the
form factors in the time--like regime are connected by dispersion
relations (DR) to the space--like
regime~\cite{Geshkenbein74,HMDtimelike,egle,baldini,seealso}.  The
recent experiments raised two serious issues: firstly, the Fermilab
E835 measurements of $|G_{M}(q^2)|$ of the proton at time--like $q^2$
= 11.63 and 12.43 GeV$^2$ (\cite{Andreotti} and references therein)
have shown that $|G_{M}(q^2)|$ in the time--like region is twice as
large as in the space--like region (there are some uncertainties
because the direct $G_{E}-G_{M}$ separation was not possible due to
statistics and acceptance); secondly, the studies of the
electron--to--proton polarization transfer in $\overrightarrow e^- \,
p \to e^- \, \overrightarrow p$ scattering at Jefferson Laboratory
\cite{perdrisat} show that the ratio of the Sachs form factors
$G_E(q^2)/ G_M(q^2)$ is monotonically decreasing with increasing
$Q^2=-q^2,$ in strong contradiction with the $G_E/G_M$ scaling assumed
in the traditional Rosenbluth separation method, which may in fact not
be reliable in the space--like region. Notice that the core of the PAX
proposal is precisely the QED electron--to--nucleon polarization
transfer mechanism, employed at Jefferson Laboratory.
\begin{figure}[ht]
\begin{center}
  \includegraphics[width=0.6\linewidth]{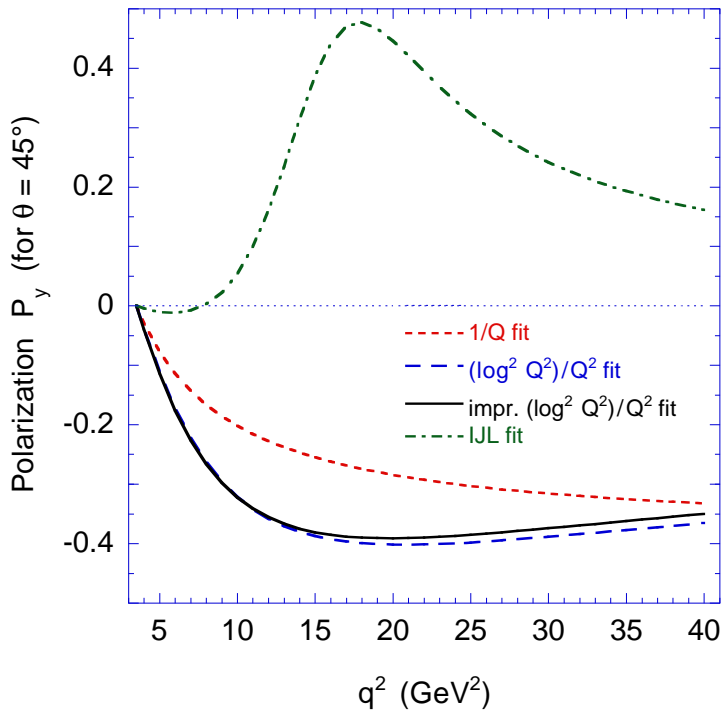}
  \parbox{14cm}{\caption{\label{FFasymmetry}\small Predicted
      single--spin asymmetry ${\cal A}_y={\cal P}_y$ for $\theta =
      45^\circ$ in the time--like region for selected form factor
      fits: $F_2/F_1 \propto 1/Q$ fit \cite{brodsky}, the $(\log^2
      Q^2)/Q^2$ fit of Belitsky {\it et al.} \cite{belitsky02}; an
      improved $(\log^2 Q^2)/Q^2$ fit \cite{BHHK}; and a fit from
      Iachello {\it et al.}, \cite{ijl}.}}
\end{center}
\end{figure}

There is a great theoretical interest in the nucleon time--like form
factors. Although the space--like form factors of a stable hadron are
real, the time--like form factors have a phase structure reflecting
the final--state interactions (FSI) of the outgoing hadrons. Kaidalov
et al. argue that the same FSI effects are responsible for the
enhancement of $|G_{M}(q^2)|$ in the time--like region
\cite{Kaidalov}; their evaluation of the enhancement based on the
variation of Sudakov effects from the space--like to time--like region
is consistent with general requirements from analyticity that FSI
effects vanish at large $q^2$ in the pQCD asymptotics. A recent
discussion can be found in Brodsky et al. \cite{brodsky} ( see also
\cite{egle}). The same property of vanishing FSI at large $q^2$ is
shared by the hybrid pQCD--DR description developed by Hammer,
Meissner and Drechsel \cite{HMDtimelike} and vector--dominance based
models (VDM) \cite{DubnichkaVDM}, which are also able to accommodate
the new results from the Jefferson Laboratory. Iachello et al.
\cite{iachello} stress the need for a better accuracy measurement of
the neutron time--like form factors.
\begin{figure}[hb]
\begin{center}
  \includegraphics[width=0.9\textwidth]{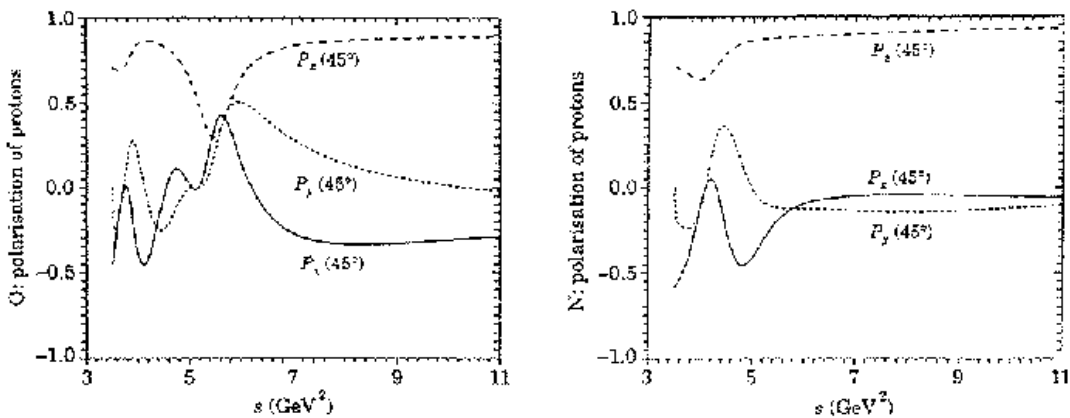}
  \parbox{14cm}{\caption{\label{fig:DubnickaPy}\small Predicted
      single--spin asymmetries (${\cal A}_y={\cal P}_y$) for $\theta =
      45^\circ$ in the time--like region for two versions (O (old) \&
      N (new)) of the analytic and unitary vector--meson dominance
      (VDM) models \cite{d}}}
\end{center}
\end{figure}

Brodsky et al. make a strong point that the new Jefferson Laboratory
results make it critical to carefully identify and separate the
time--like $G_E$ and $G_M$ form factors by measuring the
center--of--mass angular distribution and the polarization of the
proton in $e^+ e^- \to p \bar p$ or the transverse SSA in polarized
$p^{\uparrow} \bar p \to \ell^+ \ell^-$ reactions \cite{brodsky}.  As
noted by Dubnickova, Dubnicka, and Rekalo \cite{d} and by
Rock~\cite{Rock}, the non--zero phase difference between $G_E$ and
$G_M$ entails the normal polarization ${\cal P}_y$ of the final state
(anti)baryons in $e^- e^+\to \overrightarrow p \bar p$ or the
transverse SSA ${\cal A}_y = {\cal P}_y$ in annihilation $p^{\uparrow}
\bar p \to e^- e^+$ on transversely polarized protons: \be \label{py}
{\cal A}_y = { \sin 2\theta \, {\rm Im} G_E^* G_M \over
  [(1+\cos^2\theta)|G_M|^2+\sin^2\theta |G_E|^2/\tau] \sqrt{\tau} }
\ee where $\tau \equiv {q^2 / 4 m_p^2} > 1$ and $\theta$ is the
scattering angle.

As emphasized already by Dubnickova et al. \cite{d} the knowledge of
the phase difference between the $G_E$ and $G_M$ may strongly
constrain the models for the form factors.  More recently there have
been a number of explanations and theoretically motivated fits of the
new data on the proton $F_2/F_1$ ratio
\cite{belitsky02,ralston,miller,BHHK}.  Each of the models predicts a
specific fall--off and phase structure of the form factors from $ s
\leftrightarrow t$ crossing to the time--like domain.  The predicted
single--spin asymmetry is substantial and has a distinct $q^2$
dependence which strongly discriminates between the analytic forms
which fit the proton $G_E/G_M$ data in the space--like region. This is
clearly illustrated in Fig. \ref{FFasymmetry}.  The further
illustration of the discrimination power of $A_y$ comes from the
analytic and unitary vector--meson dominance (VDM) models developed by
Dubnicka et al. \cite{d}, see Fig. \ref{fig:DubnickaPy}, which
indicate a strong model--dependence of $P_y$ and more structure in the
threshold region than suggested by large--$q^2$ parameterizations shown
in Fig. \ref{FFasymmetry}. Finally, as argued in
\cite{DubnickaUnphysical}, the experimental observation of
near--threshold exclusive Drell--Yan reactions $\bar{p}p\to \gamma^*
\pi^o \to e^+e^-\pi^o$ would give unique, albeit a model-dependent,
access to the proton form factors in the unphysical region of $q^2<
4m_p^2$.

\begin{figure}[hb]
  \centering \includegraphics [width=100mm]{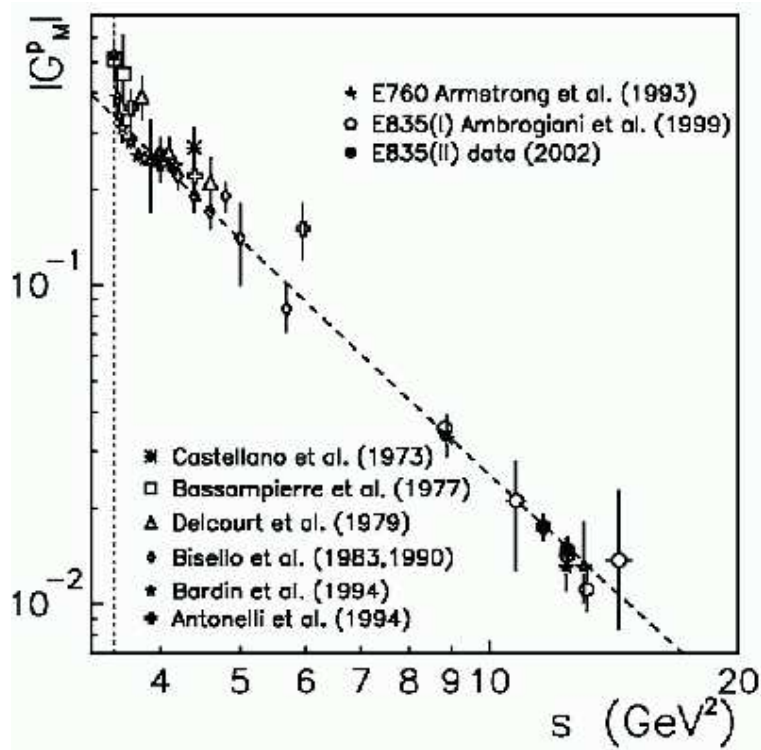}
  \parbox{14cm}{\caption{\label{fig:timelikeFF}\small All existing
      magnetic form factor data of the proton in the time--like region
      obtained with the hypothesis $|G_M|=|G_E|$ versus $s=q^2$, as
      compiled in \cite{Andreotti}; the summary of the earlier data
      can be found in \cite{bib:psip}.}}
\end{figure}

Despite the fundamental implications of the phase for the
understanding of the connection between the space--like and time--like
form factors, such measurements have never been made. The available
data on $|G_M^{(p)}|$ in the time--like region are scarce, as can be
seen from Fig.~\ref{fig:timelikeFF}.

However, these data suggest the existence of additional structures in
the time--like form factor of the proton, especially in the
near--threshold region; as Hammer, Meissner and Drechsel emphasized
\cite{HMDtimelike} that calls for improvements in the
dispersion--theoretical description of form factors.  We also recall
recent indications for the baryonium--like states from BES in the
$\JPsi{\to}\gamma{\bar p}p$ decay \cite{Bai} and from
Belle~\cite{Abe1,Abe2}, which prompted much theoretical activity in
low--energy proton--antiproton interactions (\cite{UlfPbarP} and
references therein). The phase structure of the form factors near
threshold could be much richer than suggested by high--$Q^2$
parameterizations with an oversimplified treatment of the impact of
the unphysical region.

At larger $q^2$ the data from E835 \cite{bib:psip,Andreotti} and E760
\cite{arm} seem to approach the power--law behavior predicted by pQCD.
The PAX experiment would measure the relative phase $\phi_{EM}$ of the
form factors from the SSA data with a transversely polarized proton
target.

The modulus of $G_{E}$ and $G_{M}$ can be deduced from the angular
distribution in an unpolarized measurement for $\bar pp\to e^+e^-$ as
it can be carried out independently at PANDA as well as at PAX.
However, the additional measurement of the transverse double spin
asymmetry in $p^{\uparrow} \bar{p}^{\uparrow} \to \ell^+ \ell^-$, that
is feasible at PAX, could further reduce the systematic uncertainties
of the Rosenbluth separation. We recall that, as emphasized by
Tomasi--Gustaffson and Rekalo \cite{tomasi-gust-rek}, the separation
of magnetic and electric form factors in the time--like region allows
for the most stringent test of the asymptotic regime and QCD
predictions. According to Dubnicka et al.  \cite{d} \be
 \label{Ayy} {\cal A}_{yy} = {  \sin^2\theta
   \, (G_M|^2-|G_E|^2/\tau) /\rm Im \over
   [(1+\cos^2\theta)|G_M|^2+\sin^2\theta |G_E|^2/\tau]}\, . 
\ee
 
Furthermore, in the fixed--target mode, the polarization of the proton
target can readily be changed to the longitudinal direction, and the
in--plane longitudinal--transverse double spin asymmetry would allow
one \cite{d} to measure ${\rm Re} G_E^* G_M$, \be
 \label{Axz} {\cal A}_{xz} = {  \sin2\theta \, {\rm Re} G_E^*  G_M
   \over [(1+\cos^2\theta)|G_M|^2+\sin^2\theta |G_E|^2/\tau]}, \ee
 which would resolve the remaining $\phi_{EM}-(\pi-\phi_{EM})$
 ambiguity from the transverse SSA data.  This will put tight
 constraints on current models of the form factor.

\section{Hard Scattering: Polarized and Unpolarized} 
From the point of view of the theory of elastic and exclusive
two--body reactions, the energy range of HESR corresponds to the
transition from soft mechanisms to hard scattering with the onset of
the power laws for the $s,t,u$--dependence of the differential cross
sections \cite{Matveev,BrodskyFarrar} which have generally been
successful so far (for a review and further references see
\cite{BrodskyHard}).  There remains, though, the open and much debated
issue of the so--called Landshoff inde\-pen\-dent
scattering--mechanism \cite{Landshoff} which gives the odd--charge
symmetry contribution to the $NN$ and $\bar{N}N$ amplitudes and may
dominate at higher energies. The more recent realization of the
importance of the so--called handbag contributions to the amplitudes
of exclusive reactions made possible direct calculations of certain
two--body annihilation cross sections and double--spin asymmetries in
terms of the so--called Generalized Parton Distributions (GPD's)
\cite{Ji,Radyushkin,Ferrara}.  The PAX experiment at HESR is uniquely
poised to address several new aspects of hard exclusive scattering
physics:
\begin{itemize} 
\item The particle identification in the forward spectrometer of PAX
  would allow the measurement of elastic $p\bar{p}$ scattering in the
  small to moderately large $|t|$ in the forward hemisphere and, more
  interestingly, the backward hemisphere at extremely large $t$ not
  accessible in the $t-u$ symmetric $pp$ scattering.
\item The high energy behavior of exotic baryon number, $B=2$,
  exchange in the $u$--channel is interesting in itself. Its
  measurements in the small to moderately large $u$ region of backward
  elastic $\bar{p} p$ scattering will be used for the isolation of
  hard $p\bar{p}$ scattering contribution at large $|u|$.
\item After the isolation of the hard--scattering regime the
  importance of the odd--charge symmetry Landshoff (odderon) mechanism
  can be tested from the onset of the hard scattering regime in
  large--angle elastic $\bar{p}p$ scattering as compared to $pp$
  scattering.
\item The relative importance of odd--charge vs. even--charge
  symmetric mechanisms for the large transverse double spin asymmetry
  $A_{TT}$ in polarized $p^{\uparrow} p^{\uparrow}$ as observed at
  Argonne ZGS and BNL AGS can be clarified by a measurement of
  $A_{TT}$ in polarized $\bar{p}^{\uparrow} p^{\uparrow}$ elastic
  scattering at PAX and the comparison with the earlier data from
  $p^{\uparrow} p^{\uparrow}$ scattering.
\item The future implementation of particle identification in the
  large angle spectrometer of PAX would allow an extension of
  measurements of elastic scattering and two--body annihilation,
  $\bar{p} p \to \gamma\gamma,
  \gamma\pi^0,\pi^+\pi^-,K^+K^-,\Lambda_c\bar{\Lambda}_c,...$ to large
  angles $\theta_{cm} \sim 90^o$.
\item Exclusive Drell--Yan reactions with a lepton pair in the final
  state, accompanied by a photon or meson, may also be studied in the
  framework of the partonic description of baryons. Like in the
  conventional inclusive DY process, the large mass of the lepton pair
  sets the resolution scale of the inner structure of the baryon to
  photon or meson transition processes.
\end{itemize}

The theoretical background behind the high--$t$ or high--$Q^2$ ($Q^2=
M_{e^+e^-}^2$) possibilities of PAX can be summarized as follows:

The scaling power law $s^{-N}$, where $N+2$ is the total number of
elementary constituents in the initial and final state, for exclusive
two--body hard scattering has been in the focus of high--energy
scattering theory ever since the first suggestion in the early 70's of
the constituent counting rules by Matveev et al. \cite{Matveev} and
Brodsky \& Farrar \cite{BrodskyFarrar} and Brodsky \& Hiller
\cite{BrodskyHiller}.  The subsequent hard pQCD approach to the
derivation of the constituent counting rules has been developed in
late 70's--early 80's and has become known as the
Efremov--Radyushkin--Brodsky--Lepage (ERBL) evolution technique
(\cite{Efremov,BrodskyLepage}, see also Chernyak et al.
\cite{Chernyak}). Experimentally, the constituent counting rule proves
to be fairly successful, from the scattering of hadrons on protons to
photoproduction of mesons \cite{Zhu:2002su} to reactions involving
light nuclei, like the photodisintegration of deuterons studied at
Jefferson Lab \cite{Rossi:2004qm,Mirazita:2004rb} A good summary of
the experimental situation is found in Ref. \cite{White:1994tj} and
reviews by Brodsky and Lepage (\cite{BrodskyHard} and references
therein), and is summarized in Table~\ref{tab:E838}, borrowed from the
BNL E838 publication \cite{White:1994tj}.
\begin{table}[htb]
\begin{center}
\renewcommand{\arraystretch}{1.3}
\begin{tabular}{ccccc} \hline\hline
 & & \multicolumn{2}{|c|}{Cross section} & $n-2$ \\
Experiment No. $N$ & Interaction & E838 & E755 & $(\frac{d\sigma}{dt}\sim 1/s^{n-2})$ \\ 
\hline
1 & $\pi^+ p \to p \pi^+$ & $132 \pm 10$ & $4.6 \pm 0.3$ & $6.7 \pm 0.2$ \\
2 & $\pi^- p \to p \pi^-$ & $73 \pm 5$ & $1.7 \pm 0.2$ & $7.5 \pm 0.3$ \\
3 & $K^+ p \to p K^+$ & $219 \pm 30$ & $3.4 \pm 1.4$ & $8.3^{+ 0.6}_{-1.0}$ \\
4 & $K^- p \to p K^-$ & $18 \pm 6$ & $0.9 \pm 0.9$ & $\ge3.9$ \\
5 & $\pi^+ p \to p \rho^+$ & $214 \pm 30$ & $3.4 \pm 0.7$ & $8.3 \pm 0.5$ \\
6 & $\pi^- p \to p \rho^-$ & $99 \pm 13$ & $1.3 \pm 0.6$ & $8.7 \pm 1.0$ \\
7 & $\pi^+ p \to \pi^+\Delta^+$ & $45 \pm 10$ & $2.0 \pm 0.6$ & $6.2 \pm 0.8$ \\
8 & $\pi^- p \to \pi^+\Delta^-$ & $24 \pm 5$ & $\le 0.12$ & $\ge 10.1$ \\
9 & $p p \to pp$ & $3300 \pm 40$ & $48 \pm 5$ & $9.1 \pm 0.2$ \\
10 & $\overline{p} p \to p \overline{p}$ & $75 \pm 8$ & $\le 2.1$ & $\ge 7.5$ \\
\hline \hline
\end{tabular}
\parbox{14cm} {\caption{\label{tab:E838} \small The computation of the
    experiments $N$ of the scaling power in the AFS BNL experiments
    E838 ($E_p=5.9$~GeV/c) and E755 ($E_p=9.9$~GeV/c).  The
    constituent counting predicts $N=8$ for reactions 1 - 8 and $N=10$
    for reactions 9 and 10. (Table from Ref.~\cite{White:1994tj}).}}
\end{center}
\end{table}

The scale for the onset of the genuine pQCD asymptotics can only be
deduced from the experiment, on the theoretical side the new finding
is the importance of the so--called handbag mechanism in the
sub--asymptotic energy range \cite{KrollCompton,BaroneFormFactor}. As
argued by P. Kroll et al., the handbag mechanism prediction for the
sub--asymptotic $s$--dependence of the large--angle elastic $pp$ and
$p\bar{p}$ cross--section \cite{KrollElastic} ,
\begin{eqnarray}
{d\sigma \over dt} \propto {1 \over s^2t^8} \propto {f(\theta) \over s^{10}}
\end{eqnarray}
is similar to that of the constituent quark counting rules of Brodsky
et al. \cite{BrodskyFarrar}.
 
\begin{figure}[htb]
  \centering \includegraphics [width=120mm]{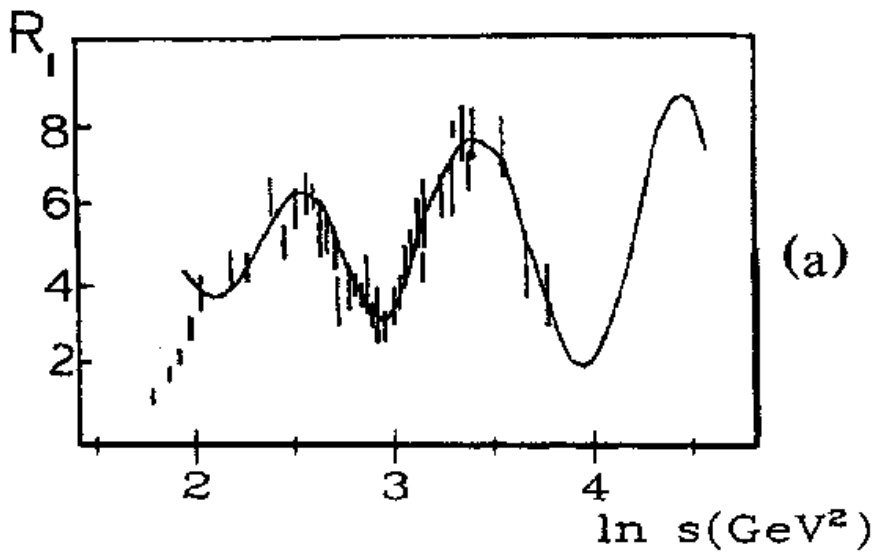}
  \parbox{14cm}{\caption{\label{fig:ralston}\small The energy
      dependence of $R_{1}=s^{10}{d\sigma_{pp}/ dt}|_{90^o}$ for the
      high energy $pp$ elastic scattering at $90^o$ c.m. angle
      compared to the model calculation \cite{RalstonPire} from the
      interference of the Brodsky--Farrar and Landshoff mechanisms.}}
\end{figure}

There remains, though, an open and hot issue of the so--called
Landshoff independent scattering--mechanism \cite{Landshoff}, which
predicts ${d\sigma/ dt} \propto {1/t^8} \propto f_L(\theta)/s^{8}$
and, despite the Sudakov suppression, may dominate at very large $s$.
According to Ralston and Pire \cite{RalstonPire} certain evidence for
the relevance of the Landshoff mechanism in the HESR energy range
comes from the experimentally observed oscillatory $s$--dependence of
$R_{1}=s^{10}{d\sigma/ dt}$, shown in Fig.~\ref{fig:ralston}. Here the
solid curve is the theoretical expectation \cite{RalstonPire} based on
the interference of the Brodsky--Farrar and Landshoff mechanisms.
The Ralston--Pire mechanism has been corroborated to a certain extent
by the experimental finding at BNL of the wash--up of oscillations in
the quasielastic scattering of protons on protons bound in nuclei
(\cite{Aclander:2004zm} and references therein).

\begin{figure}[htb]
  \centering \includegraphics [width=0.46\linewidth]{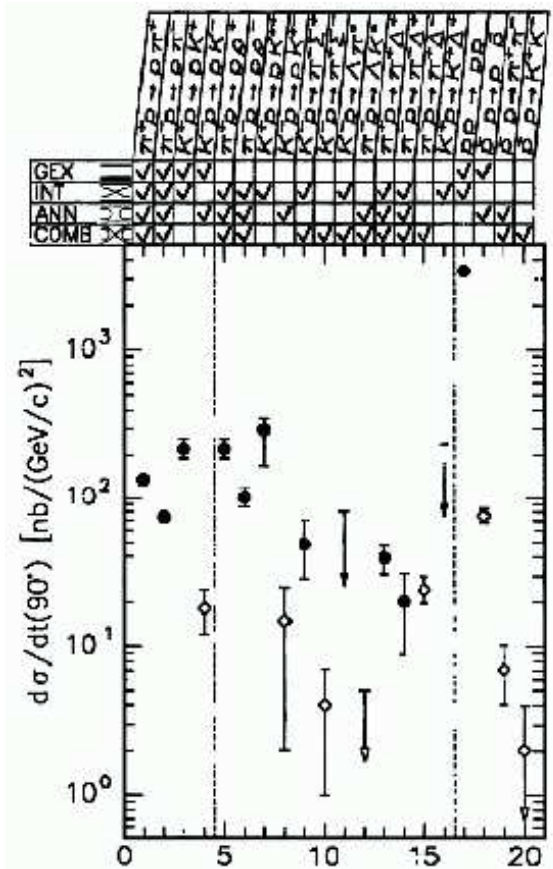}
  \parbox{14cm}{\caption{\label{fig:BNL_E838}\small Differential cross
      sections for the 16 meson--baryon and 4 baryon--baryon reactions
      measured in the BNL AFS experiment E838 \cite{White:1994tj}. The
      four possible quark--gluon diagrams which contribute to each of
      the 20 reactions are given in the chart at the top of the
      figure. The experimental data for those reactions which have a
      contribution from quark interchange mechanism (INT) are shown by
      the solid black points.  }}
\end{figure}

To the lowest order in pQCD the Landshoff amplitude corresponds to the
charge conjugation--odd (odderon) exchange and alters the sign from the
$pp$ to the $p\bar{p}$ case. If the Brodsky--Farrar and/or it's
handbag counterpart were crossing--even, then the Ralston--Pire
scenario for the oscillations would predict the inversion of the sign
of oscillation in $R_1$ from the $pp$ to the $p\bar{p}$ case. Because
the first oscillation in Fig.~\ref{fig:ralston} takes place at $s< 20$
GeV$^2$, this suggests that $\bar{p}p$ elastic scattering at HESR is
ideally suited for testing the oscillation scenarios. Although true in
general, this expectation needs a qualification on the crossing from
the proton--proton to the antiproton--proton channel.  A natural origin
for the constituent counting rules is offered by the quark interchange
mechanism (QIM) which predicts $d\sigma_{el}(pp) \gg
d\sigma_{el}(\bar{p}p)$ in accord with the experimental data from BNL
E838 shown in Fig. \ref{fig:BNL_E838}.

Either the contribution from the independent scattering mechanism is
small or at $E_p=5.9$ GeV in E838 the cancellation of the QIM and the
Landshoff amplitudes is accidentally strong in which case the energy
dependence of $d\sigma_{el}(\bar{p}p)$ could prove exceptionally
non--smooth. On the theoretical side, as early as in 1974, Nielsen and
Neal suggested the version of an independent scattering mechanism
which allows for a substantial crossing--even component
\cite{Nielsen1974}. The Kopeliovich--Zakharov pQCD decameron
(four--gluon) exchange realization \cite{Decameron} of the
Rossi--Veneziano \cite{RossiVeneziano} baryon--junction, much discussed
recently in view of the enhanced yield of baryons in nuclear
collisions at RHIC \cite{Gyulassy}, also is a multiple--scattering
mechanism. The decameron amplitude decreases at large $|t|$ as slowly
as the Landshoff amplitude and contributes only to the $p\bar{p}$
scattering.

The point that polarization observables are sensitive to mechanisms
for the scaling behavior is conspicuous.  As an example we cite the
very recent experimental finding of the onset of pQCD constituent
counting scaling \cite{BrodskyHiller} in photodisintegration of the
deuteron starting from the proton transverse momentum $p_T$ above
about 1.1 GeV \cite{Rossi:2004qm}.  On the other hand, the
experimentally observed non-vanishing polarization transfer from
photons to protons indicates that the observed scaling behavior is
not a result of perturbative QCD \cite{Wijesooriya:2001yu}.

Now we recall that very large double transverse asymmetries have been
observed in hard proton--proton scattering (\cite{Cameron} and
references therein).  The HESR data with polarized antiprotons at PAX
will complement the AGS--ZGS data in a comparable energy range.  In
1974 Nielsen et al. argued \cite{Nielsen1974} that within the
independent scattering models the change from the dominance by
$1\times 1$ parton--parton scattering to the $2\times2$ and $3\times 3$
scattering leads in a natural way to the oscillatory ( and rising with
$t$) behavior of polarization effects. Within this approach Nielsen
et al.  \cite{Nielsen1998} reproduce the gross features of the ZGS
data \cite{Cameron} although they underpredict $A_{TT}$ at largest
$t$.  Within the QCD motivated approach, initiated in
Ref.~\cite{RalstonPire1986}, the helicity properties of different hard
scattering mechanisms have been studied by Ramsey and Sivers
\cite{Ramsey}. These authors tried to extract the normalization of the
Landshoff amplitude from the combined analysis of $pp$ and $p\bar{p}$
elastic scattering and argued it must be small to induce the
oscillations or contribute substantially to the double spin asymmetry
$A_{TT}$. This leaves open the origin of oscillations in $R_1$ but
leads to the conclusion that the double spin asymmetry $A_{TT}$ in
$p^{\uparrow} \bar{p}^{\uparrow}$ at PAX and $p^{\uparrow}
p^{\uparrow}$ as observed at AGS--ZGS must be of comparable magnitude.
The comparison of $A_{TT}$ in the two reactions will also help to
constrain the Landshoff amplitude.  More recently, Dutta and Gao
\cite{Dutta} revisited the Ralston--Pire scenario with allowance for
the helicity--non--conserving $pp$ scattering amplitudes (for the
early discussion of helicity--non--conservation associated with the
Landshoff mechanism, see Ref.  \cite{RalstonPireSCHNC}). They found
good fits to the oscillatory behavior of $R_1$ and the energy
dependence of $A_{TT}$ in $pp$ scattering at 90$^o$ starting from $s
\gsim 8$ GeV$^2$. The extension of predictions from the models by
Nielsen et al. and Dutta et al.  to the crossing antiproton-proton
channel is not yet unique, though.  Brodsky and Teramond make a point
that opening of the $|uuduudc\bar{c}\rangle$ channel at the open charm
threshold would give rise to a broad structure in the $J=L=S=1$
proton-proton partial wave \cite{BrodskyTeramond}. Such a threshold
structure would have a negative parity and affect $p^\uparrow
p^\uparrow$ scattering for parallel spins normal to the scattering
plane. The threshold structure also imitates the "oscillatory" energy
dependence at fixed angle and the model is able to reproduce the gross
features of the $s$ and $t$ dependence of $A_{NN}$.  Arguably, in the
$\bar{p}p$ channel the charm threshold is at much lower energy and the
charm cross section will be much larger, and the Brodsky-Teramond
mechanism would predict $A_{NN}$ quite distinct from that in $pp$
channel.  Still, around the second charm threshold,
$\bar{p}p\to\bar{p}p \bar{c}c$, the $A_{NN}$ for $\bar{p}p$ may repeat
the behavior exhibited in $pp$ scattering.

Finally, the double--spin transverse-longitudinal asymmetry $A_{TL}$ is
readily accessible in the fixed-target mode with the longitudinal
polarization of the target. Its potential must not be overlooked and
needs further theoretical scrutiny.

The differential cross section measured in the BNL E838 experiment is
shown in Fig.~\ref{fig:BNL_E838}. The expected $d\sigma/dt \propto
s^{-10}$ behavior suggests that in the fixed-target Cooler
Synchrotron Ring (CSR) stage (Phase-I) the counting rates will allow
measurements of elastic $p\bar{p}$ scattering, both polarized and
unpolarized, over the whole range of angles. In the fixed-target HESR
stage the measurement of unpolarized scattering can be extended to
energies beyond those of the E838 experiment. The expected counting
rates will also allow the first measurement of the double--spin
observables.

In the comparison of observables for the $pp$ and $\bar{p}p$ elastic
scattering one would encounter manageable complications with the Pauli
principle constraints in the identical particle $pp$ scattering, by
which the spin amplitudes for $pp$ scattering have the
$t$--$u$--(anti)symmetric form $M(\theta)\pm M(\pi-\theta) = M(t)\pm
M(u)$ (\cite{Lehar} and references therein). Regarding the amplitude
structure, the $\bar{p}p$ case is somewhat simpler and offers even
more possibilities for the investigation of hard scattering.  Indeed,
for the hard scattering to be at work, in the general case one demands
that both $|t|$ and $|u|$ are simultaneously large, $|t|\sim |u| \sim
{1\over 2}(s-4m_p^2)$.  Here we notice an important distinction
between the $t$--$u$ asymmetric $\bar{p}p$ from the $t$--$u$ symmetric
identical particle $pp$ elastic scattering.  In the $t$--$u$ symmetric
case the accessible values of $t$ are bound from above by $|t| \leq
|t_{max}| = {1\over 2}(s-4m_p^2)$.  In the $p\bar{p}$ case the
backward scattering corresponds to the strongly suppressed exotic
baryon number two, $B=2$, exchange in the $u$--channel (for a
discussion of the suppression of exotic exchanges see Refs.
\cite{Quigg,Roy} and references therein).  Consequently, the hard
scattering mechanism may dominate well beyond $\theta_{\rm cm}=90^o$
of $p\bar{p}$ elastic scattering. Because of the unambiguous $p$ and
$\bar{p}$ separation in the forward spectrometer, the PAX will for the
first time explore the transition from soft exotic $B=2$ exchange at
$u\sim 0$ to the hard scattering at larger $|u|$: for 15 GeV stored
$\bar{p}$'s the $p-\bar{p}$ separation is possible up to $|u| \leq 4$
GeV$^2$, while $|u| \leq 8$ GeV$^2$ is accessible at 22 GeV. Although
still $|u| \ll s$, these values of $|u|$ are sufficiently large to
suppress the $u$--channel exotic $B=2$ exchange, which allow the
dominance of hard mechanisms, which thus become accessible at values
of $|t| = s-4m_p^2 - |u|$ almost twice as large than in $pp$
scattering at the same value of $s$.  The investigation of the energy
dependence of exotic $B=2$ exchange in the small--$u$ region is
interesting by itself in order to better understand the related
reactions, like the $\pi D$ backward elastic scattering.

Although not all annihilation reactions are readily accessible with
the present detector configuration, they are extremely interesting
from the theoretical standpoint.  Within the modern handbag diagram
description, they probe such fundamental QCD observables as the
Generalized Parton Distributions (GPD's), introduced by Ji and
Radyushkin \cite{Ji,Radyushkin}. These GPD's generalize the
conventional parton-model description of DIS to a broad class of
exclusive and few--body reactions and describe off--forward parton
distributions for polarized as well as unpolarized quarks; the Ferrara
Manifesto, formulated at the recent Conference on the QCD Structure of
the Nucleon (QCD--N'02), lists the determination of GPD's as the major
physics goal of future experiments in the electroweak physics sector
\cite{Ferrara}.  The QCD evolution of GPD's is a combination of the
conventional QCD evolution for DIS parton densities and the ERBL
evolution for the quark distribution amplitudes, GPD's share with the
DIS parton densities and the ERBL hard--scattering amplitudes the hard
factorization theorems: the one and the same set of GPD's at an
appropriate hard scale enters the calculation of amplitudes for a
broad variety of exclusive reactions.

There has been much progress in calculating the electromagnetic form
factors of the nucleon and of the hard Compton scattering amplitudes
in terms of the off--forward extension of the conventional parton
densities
\cite{KrollCompton,BaroneFormFactor,RadyushkinFormFactor,KrollFFnew},
Deeply Virtual Compton Scattering is being studied at all electron
accelerators \cite{DVCSJlab,DVCSHermes} with the purpose to extract
the specific GPD which would allow one to determine the fraction of
the proton's spin carried by the orbital angular momentum of partons
(the Ji sum rule \cite{Ji}).

More recently, the technique of GPD's has been extended by P. Kroll
and collaborators \cite{Kroll2gamma} to the differential cross
sections and spin dependence of annihilation reactions. Here the hard
scale needed for the applicability of the GPD technique is provided by
$|t|$. The theory has been remarkably successful in the simplest case
of $B\bar{B}\to \gamma\gamma$ with two point--like photons (the
inverse reactions $\gamma\gamma \to p\bar{p}$, $\Lambda\bar{\Lambda}$,
and $\Sigma\bar{\Sigma}$ have been studied experimentally by the CLEO
\cite{CLEO} and VENUS \cite{VENUS} collaborations).  A steady progress
is being made by the DESY--Regensburg--Wuppertal group in extending
these techniques to the $p\bar{p}\to \gamma\pi^0$ with a
non--point--like $\pi^0$ in the final state \cite{Kroll-Nonpointlike},
a further generalization to the two--meson final states is expected in
the near future.  As far as the theory of spin dependence of hard
scattering is concerned the theoretical predictions are robust for the
longitudinal double spin asymmetries, and thus their experimental
confirmation will be of great theoretical interest.  To make such
observables accessible experimentally the spin of antiprotons in the
HESR must be rotated by Siberian Snakes.  In addition, the technique
of GPD's should allow one to relate the transverse asymmetries to the
Generalized structure function $h_1^q$ (see above) but such a
relationship has yet to be worked out.

A slightly different application of QCD factorization technique has
been suggested recently by Pire and Szymanowski
\cite{PireSzymanowski1}. They propose to study the exclusive Drell-Yan
annihilation reaction $ p\bar{p}\to \gamma^*\gamma \to e^+e^- \gamma
$. Like in the inclusive DY process, the required hard scale is
provided by the large invariant mass of the lepton pair. Then one can
study such reactions in the forward region which increases the
observed cross section.  The scaling of the cross section at fixed
$Q^2/s$ is then a signal of the applicability of perturbative QCD
techniques. New observables, called proton to photon transition
distribution amplitudes (TDA's), may then be measured which should
shed light on the structure of the baryon wave functions.
Polarization experiments are needed to separate the different TDA's.
The same theoretical framework can also be applied to other reactions
involving mesons in the final state, like $p \bar{p} \to \gamma^* \pi$
or $p \bar{p} \to \gamma^* \rho$ \cite{PireSzymanowski2} (the former
reaction has already been discussed in Sect. 4 as a window to the
time-like form factors of the proton in the unphysical region).
Crossing relates these reactions to backward deep electroproduction
which may be accessed at electron accelerators.

\section{Polarized Antiproton--Proton Soft Scattering}
\subsection{Low--$t$ Physics}
For energies above the resonance region elastic scattering is
dominated by small momentum transfers and therefore total elastic
cross sections are basically sensitive to the small $t$ region only.

Dispersion theory (DT) is based on a generally accepted hypothesis
that scattering amplitudes are analytic in the whole Mandelstam plane
up to singularities derived from unitarity and particle/bound state
poles. This, in combination with unitarity and crossing symmetry,
allows extracting of e.g. the real part of the forward elastic
scattering amplitude from knowledge of the corresponding total cross
sections.  The major unknown in this context is the unphysical region:
a left hand cut that starts at the two pion production threshold and
extends up to the $\bar NN$ threshold, where one is bound to
theoretical models for the discontinuity; the extrapolation to
asymptotic energies is considered to be well understood
\cite{Donnachie} and does not effect the DT predictions in the HESR
energy range.

\begin{figure}[htb]
  \centering \includegraphics [width=0.8\linewidth]{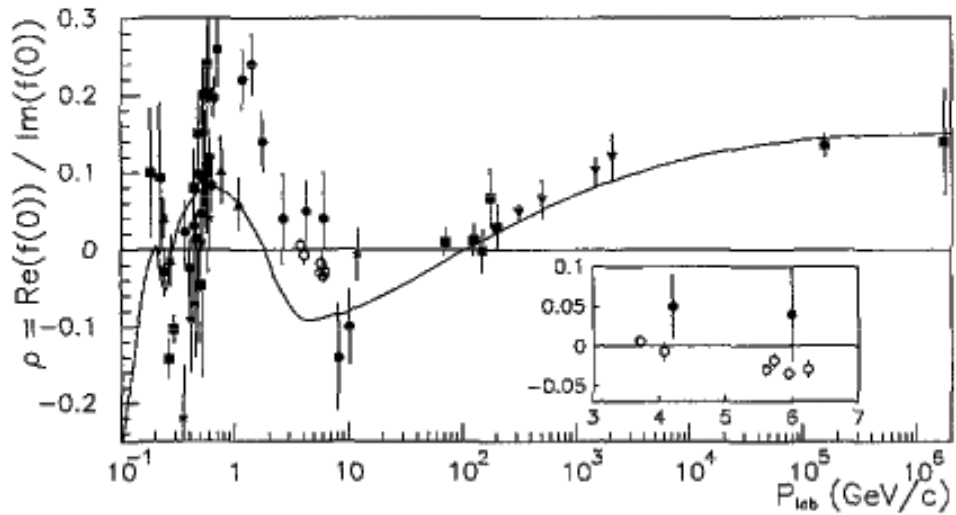}
  \parbox{14cm}{\caption{\label{fig:ReImNew}\small The compilation of
      the experimental data on the ratio of the real to imaginary part
      of the $\bar{p}p$ forward scattering amplitude (from E760
      publication \protect{\cite{Armstrong}}). The insert shows on a
      larger scale the E760 results. The solid line shows the
      predictions from the dispersion relation calculation by P.~Kroll
      et al. \protect{\cite{Kroll}}.}}
\end{figure}

Under certain assumptions, the real part of the forward scattering
amplitude can be extracted from the elastic differential cross section
measured in the Coulomb-nuclear interference (CNI) region
(\cite{KrollCoulombPhase} and references therein). The most recent DT
analysis~\cite{Kroll} reproduces the gross features of the available
data, see Fig. \ref{fig:ReImNew}; still, the experiment suggests more
structure at low energies (which may be related to the near-threshold
structure in the electromagnetic form factor shown in Fig.
\ref{fig:timelikeFF}) and there is a systematic departure of the
theoretical prediction from the experiment in the region between 1 and
10 GeV/c.  In particular the latest precise results from Fermilab E760
Collaboration~\cite{Armstrong} collected in the 3.7 to 6.2 GeV/c
region are in strong disagreement with DT.
There are two explanations possible for this discrepancy.  First one
might doubt the theoretical understanding of the amplitude in the
unphysical region.  In this sense the DT analysis is a strong tool to
explore the unphysical region.  Since the discrepancy of the data to
the result of the DT analysis occurs in a quite confined region, only
a very pronounced structure in the unphysical region could be the
origin. Such a structure can be an additional pole related to a $\bar
p p $ bound state\footnote{Notice that already the present analysis of
  Ref.  \cite{Kroll} contains one pole.}, discussed in Refs.
\cite{Bogdanova,Shapiro,Rossi,Dover}. The appearance of a pole in the
unphysical region might cause a turnover of the real part of the
forward scattering amplitude to small values at momenta above
600~MeV/c~\cite{Grein,Kaseno}.  Indications of such states were seen
recently at BES in the $\JPsi{\to}\gamma{\bar p}p$ decay \cite{Bai}
and Belle~\cite{Abe1,Abe2} and attracted much theoretical attention
(\cite{UlfPbarP} and references therein).

However, there is a second possible reason for the discrepancy of the
DT result and the data, namely that not all assumptions in the
analysis of CNI hold, the strongest one being a negligible spin
dependence in the nuclear interference region \cite{Bourrely}. A
sizable spin dependence of the nuclear amplitude can well change the
analysis used in Ref. \cite{Armstrong}; such a sensitivity to a
possible spin dependence has been discussed earlier \cite{Jenni}. The
quantities to be measured are $\Delta \sigma_T= \sigma (\uparrow
\downarrow) - \sigma (\uparrow \uparrow)$ and $\Delta \sigma_L=\sigma
( \rightleftarrows ) - \sigma ( \rightrightarrows ) $: their knowledge
will eliminate the model--dependent extraction of the real part of the
$p\bar{p}$ scattering amplitude \cite{Jakob}. Please note that a
sizable value of $\Delta \sigma_T$ or $\Delta \sigma_L$ at high
energies is an interesting phenomenon in itself since it contradicts
the generally believed picture that spin effects die out with
increasing energy (see also previous section).

Thus, a measurement of $\Delta \sigma_{L/T}$ in the energy region
accessible at HESR not only allows one to investigate spin effects of
the $\bar p p$ interaction at reasonably high energies but also to pin
down the scattering amplitude in the unphysical region to deepen our
understanding of possible $\bar pp$ bound states.  Especially a
determination of $\Delta \sigma_T$ can be done in a straightforward
way as outlined in the next section. The low-$t$ physics program is
ideally suited for the Phase--I with the polarized fixed target at CSR,
and can further be extended to Phase--II.

\subsection{Total Cross Section Measurement}
The unpolarized total cross section $\sigma_{\rm 0,tot}$ has been
measured at several laboratories over the complete HESR momentum
range; however, the spin dependent total cross section is comprised of
three parts \cite{Bys78}
\begin{equation}
\sigma_{\rm tot} = \sigma_{\rm 0,tot} + 
\sigma_{\rm 1,tot}\vec{P}\cdot\vec{Q} +
\sigma_{\rm 2,tot} (\vec{P}\cdot\hat{k}) (\vec{Q}\cdot\hat{k}).
\label{eq:totCross}
\end{equation}
where $\vec{P},\vec{Q}$ are the beam and target polarizations and
$\hat{k}$ the unit vector along the beam momentum.  Note that the
spin--dependent contributions $\sigma_{1,2}$ are completely unexplored
over the full HESR energy range.  Only one measurement at much higher
energies from E704 at 200~GeV/c \cite{Grosnick:1996sy} has been
reported using polarized antiprotons from parity--non--conserving
$\bar{\Lambda}$--decays.

With the PAX detector the transverse cross section difference $\DST
= -2\sigma_{\rm 1,tot}$ can be accessed by two methods:
\begin{itemize}
\item[(1)] from the rate of polarization buildup for a transversely
  polarized target when only a single hyperfine state is used. The
  contribution from the electrons is known from theory and can be
  subtracted. However, the difference of the time constants for
  polarization buildup with hyperfine states 1 {\em or} 2 (cf.
  Fig.~\ref{fig:breit_rabi_hydrogen}) injected into the target, would
  give direct access to \DST , whereas the contribution from the
  electrons could be extracted from the average.
\item[(2)] from the difference in beam lifetime for a target
  polarization parallel or antiparallel to the beam. A sensitive
  beam--current transformer (BCT) can measure beam lifetimes of the
  antiproton beam after polarization and ramping to the desired
  energy.  An accuracy at the $10^{-4}$ level has been achieved by the
  TRIC experiment at COSY using this method. Access to \DST\ by this
  technique is limited to beam momenta where losses are dominated by
  the nuclear cross section, e.g. above a few GeV/c -- the precise
  limit will be determined by the acceptance of the HESR.
\end{itemize}
Both methods require knowledge of the total polarized target thickness
exposed to the beam. With a calibrated hydrogen source fed into the
storage cell, the target density can be determined to 2--3\% as shown
by the HERMES \cite{HERMESintrep} and FILTEX \cite{rathmann}
experiments.

In principle, $\Delta\sigma_L = -2(\sigma_{\rm 1,tot} + \sigma_{\rm
  2,tot})$ can be measured by the same method; however, a Siberian
snake would be needed in the ring to allow for a stable longitudinal
polarization at the interaction point.

\subsection{Proton--Antiproton Interaction}
The main body of $\bar{N}N$ scattering data has been measured at LEAR
(see \cite{Klempt:2002} for a recent review) and comprises mainly
cross section and analyzing power data, as well as a few data points
on depolarization and polarization transfer.  These data have been
interpreted by phenomenological or meson--exchange potentials by
exploiting the G--parity rule, linking the $\bar{N}N$ and the $NN$
systems.

At the HESR the spin correlation parameters \ANN, \ASS, and \ASL\ can
be accessed for the first time by PAX which would add genuine new
information on the spin dependence of the interaction and help to pin
down parameters of phenomenological $\bar{N}N$ models. This part of
the program will start with the polarized fixed-target experiments
with polarized antiprotons in CSR (Phase--I) and can be extended to
Phase--II.

Besides, available data on the analyzing power from LEAR will be used
for polarimetry to obtain information on the target and beam
polarization, independent from the polarimeter foreseen for the
polarized target (cf. Sec.~\ref{sec:polar}).

\clearpage
\part{Polarized Antiprotons at FAIR\label{sec:APR}}
\pagestyle{myheadings} \markboth{Technical Proposal for
  ${\cal PAX}$}{Part II: Polarized Antiprotons at FAIR}
\section{Overview}
A viable practical scheme which allows us to reach a polarization of
the stored antiprotons at HESR--FAIR of $\simeq 0.3$ has been worked
out and published in Ref.~\cite{ap}. The basic approach to polarizing
and storing antiprotons at HESR--FAIR is based on solid QED
calculations of the spin transfer from electrons to antiprotons
\cite{meyer,horowitz-meyer}, which is being routinely used at
Jefferson Laboratory for the electromagnetic form factor separation
\cite{JlabFF}, and which has been tested and confirmed experimentally
in the FILTEX experiment~\cite{rathmann}.

The PAX Letter--of--Intent was submitted on January 15, 2004. The
physics program of PAX has been reviewed by the QCD Program Advisory
Committee (PAC) on May 14--16, 2004 \cite{paxweb}.  The proposal by
the ASSIA collaboration \cite{ASSIA} to utilize a polarized solid
target and to bombard it with a 45~GeV unpolarized antiproton beam
extracted from the synchrotron SIS100 has been rejected by the GSI
management.  Such measurements would not allow one to determine
$h^q_1(x,Q^2)$, because in single spin measurements $h^q_1(x,Q^2)$
appears always coupled to another unknown fragmentation function.
Following the QCD--PAC report and the recommendation of the Chairman
of the committee on Scientific and Technological Issues (STI) and the
FAIR project coordinator~\cite{paxweb}, the PAX collaboration has
optimized the technique to achieve a sizable antiproton polarization
and is presenting here an updated proposal for experiments at GSI with
polarized antiprotons \cite{ap}.  From various working group meetings
of the PAX collaboration, presented in part in 2004 at several
workshops and conferences \cite{paxweb}, we conclude:
\begin{itemize}
\item Polarization buildup in the HESR ring, operated at the lowest
  possible energy, as discussed in PAX LoI, does not allow one to
  achieve the optimum degree of polarization in the antiproton beam.
  The goal of achieving the highest possible polarization of
  antiprotons and optimization of the figure of merit dictates that
  one polarizes antiprotons in a dedicated low--energy ring (APR). The
  transfer of polarized low--energy antiprotons into the HESR ring
  requires pre--acceleration to about 1.5~GeV/c in a dedicated booster
  ring (CSR).  Simultaneously, the incorporation of this booster ring
  into the HESR complex opens up, quite naturally, the possibility of
  building an asymmetric antiproton--proton collider.
\end{itemize}

The TSR experiment \cite{rathmann} and the analysis of the TSR results
by Meyer and Horowitz \cite{meyer,horowitz-meyer} have shown that
several mechanisms contribute to the buildup of polarization, i.e.,
polarization dependent removal, small--angle scattering
into--the--beam, and interaction with the polarized electrons of the
target atoms. In the case of stored protons, the three mechanisms are
of comparable strength, a comparison of the mechanisms for antiprotons
is discussed in Sec.~\ref{impact}. As a reference point, we discuss
below the electromagnetic transfer of the electron polarization to
scattered antiprotons.

\begin{figure}[h]
\begin{center}
  \includegraphics[width=0.9\linewidth]{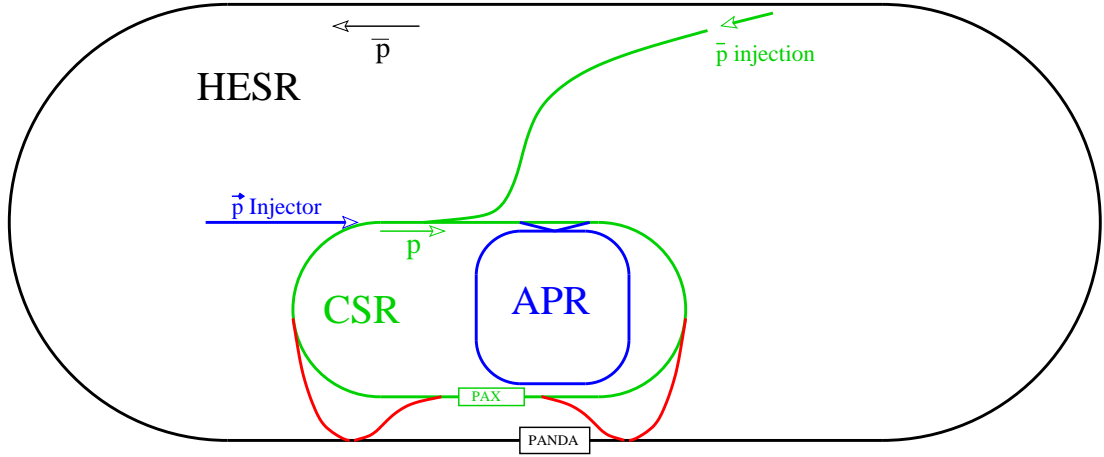}
  \parbox{14cm}{\caption{\label{fig:CSRring}\small The proposed
      accelerator set--up at the HESR (black), with the equipment used
      by the PAX collaboration in Phase--I: CSR (green), APR, beam
      transfer lines and polarized proton injector (all blue).  In
      Phase--II, by adding two transfer lines (red), an asymmetric
      collider is set up. It should be noted that, in this phase, also
      fixed target operation at PAX is possible. (The figure is drawn
      to scale.)}}
\end{center}
\end{figure}

The PAX collaboration proposes an approach that is composed of two
phases. During these the major milestones of the project can be tested
and optimized before the final goal is approached: {\bf An asymmetric
  proton--antiproton collider}, in which polarized protons with
momenta of about 3.5 GeV/c collide with polarized antiprotons with
momenta up to 15 GeV/c. These circulate in the HESR, which has already
been approved and will serve the PANDA experiment. In the following,
we will briefly describe the overall machine setup of the APR, CSR,
and HESR complex, schematically depicted in Fig.~\ref{fig:CSRring}.
The main features of the accelerator setup are:
\begin{enumerate}
\item An Antiproton Polarizer Ring (APR) built inside the HESR area
  with the crucial goal of polarizing antiprotons at kinetic energies
  around $\approx 50$~MeV (see Table~\ref{tab:opt_energies}), to be
  accelerated and injected into the other rings.
\item A second Cooler Synchrotron Ring (CSR, COSY--like) in which
  protons or antiprotons can be stored with a momentum up to 3.5
  GeV/c.  This ring shall have a straight section, where a PAX
  detector could be installed, running parallel to the experimental
  straight section of HESR.
\item By deflection of the HESR beam into the straight section of the
  CSR, both the collider or the fixed--target mode become feasible.
\end{enumerate}
It is worthwhile to stress that, through the employment of the CSR,
effectively a second interaction point is formed with minimum
interference with PANDA. The proposed solution opens the possibility
to run two different experiments at the same time. In order to avoid
unnecessary spin precession, all rings, ARP, CSR and HESR, should be
at the same level such that no vertical deflection is required when
injecting from one ring into the other.\\

In  Sec.~\ref{sec:staging}, we discuss the staging of the physics
program, which should be pursued in two  phases.

\section{Antiproton Polarizer Ring\label{sec_antiprotonpolarizer}}
For more than two decades, physicists have tried to produce beams of
polarized antiprotons \cite{krisch}.  Conventional methods like atomic
beam sources (ABS), appropriate for the production of polarized
protons and heavy ions cannot be applied, since antiprotons annihilate
with matter.  Polarized antiprotons have been produced from the decay
in flight of $\bar{\Lambda}$ hyperons at Fermilab. The achieved
intensities with antiproton polarizations $P>0.35$ never exceeded $1.5
\cdot 10^5$~s$^{-1}$ \cite{grosnick}.  Scattering of antiprotons off a
liquid hydrogen target could yield polarizations of $P\approx 0.2$,
with beam intensities of up to $2 \cdot 10^3$~s$^{-1}$ \cite{spinka}.
Unfortunately, both approaches do not allow efficient accumulation in
a storage ring, which would greatly enhance the luminosity.  Spin
splitting using the Stern--Gerlach separation of the given magnetic
substates in a stored antiproton beam was proposed in 1985
\cite{niinikoski}.  Although the theoretical understanding has much
improved since then \cite{cameron}, spin splitting using a stored beam
has yet to be observed experimentally.
\subsection{\label{process}The Polarizing Process  $\bar{p}+\vec{e} \to \vec{\bar{p}} + e$}
In 1992 an experiment at the Test Storage Ring (TSR) at MPI Heidelberg
showed that an initially unpolarized stored 23~MeV proton beam can be
polarized by spin--dependent interaction with a polarized hydrogen gas
target \cite{rathmann,zapfe,zapfe2}.  In the presence of polarized
protons of magnetic quantum number ${m=\frac{1}{2}}$ in the target,
beam protons with ${m=\frac{1}{2}}$ are scattered less often, than
those with ${m=-\frac{1}{2}}$, which eventually caused the stored beam
to acquire a polarization parallel to the proton spin of the hydrogen
atoms during spin filtering.

\begin{figure}[htb]
\begin{center}
  \includegraphics[width=0.6\linewidth]{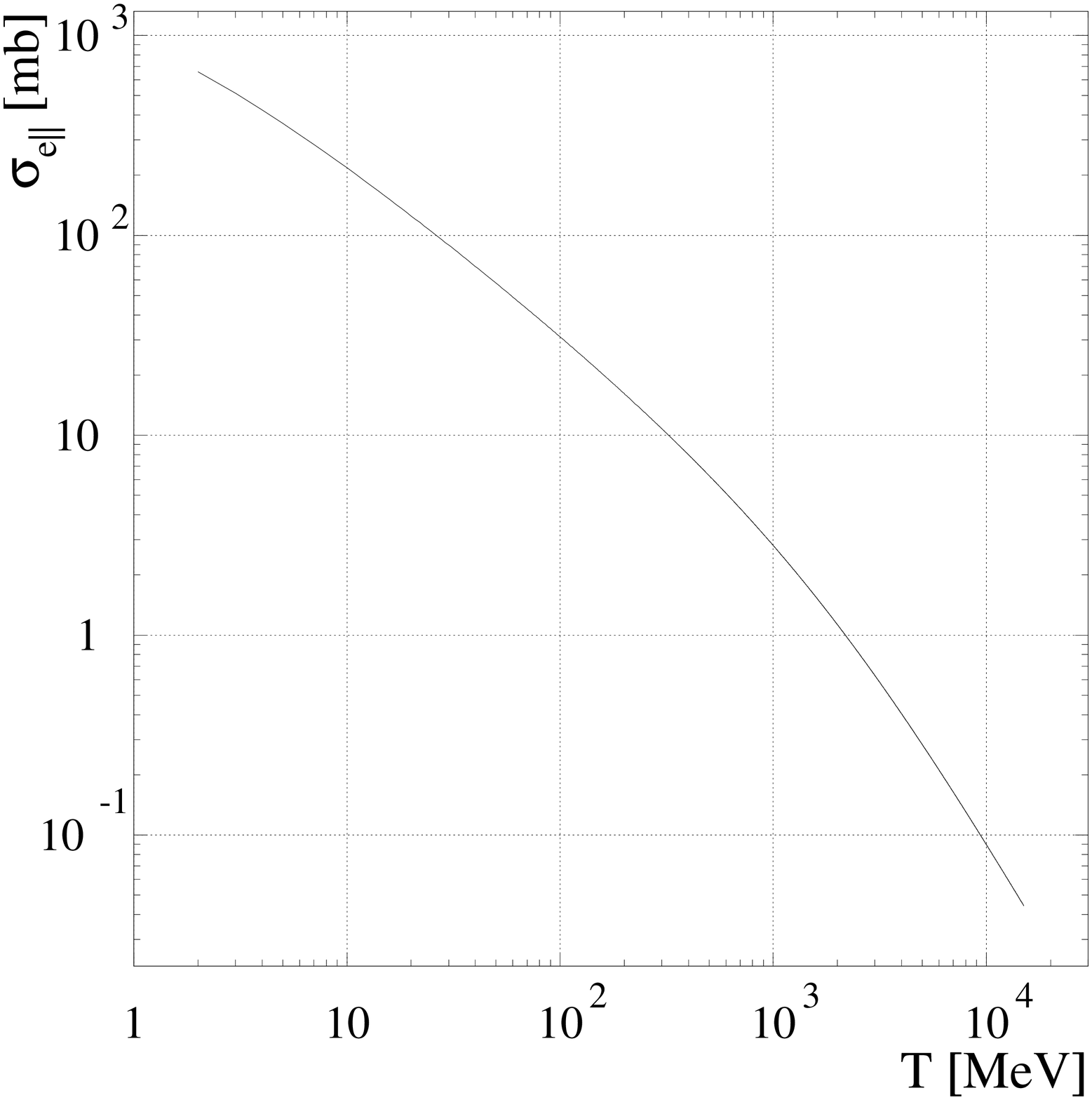}
  \parbox{14cm}{\caption{\label{fig:sigma_long}\small Spin transfer cross
      section ${\sigma_{e_{||}}}$ of antiprotons scattered from
      longitudinally polarized electrons ${(\bar{p} + \vec{e}
        \rightarrow \vec{\bar{p}} + e)}$ as a function of the kinetic
      energy of the antiprotons.}}
\end{center}
\end{figure}

In an analysis by Meyer three different mechanisms were identified,
that add up to the measured result \cite{meyer}.  One of these
mechanisms is spin transfer from the polarized electrons of the
hydrogen gas target to the circulating protons.  Horowitz and Meyer
derived the spin transfer cross section ${p+\vec{e}\rightarrow
  \vec{p}+e}$ (using ${c=\hbar=1}$) \cite{horowitz-meyer},
\begin{equation}
        \sigma_{e_{||}} 
= -\frac{4 \pi \alpha^2 (1+a) 
                          m_e  
                          }{p^2  m_p}\cdot C_0^2 \cdot
                          \frac{v}{2 \alpha}
                          \cdot
                          \sin\left( \frac{2 \alpha}{v} \ln(2pa_0)\right)\;,
        \label{eq:sigma_long}
\end{equation}
where ${\alpha}$ is the fine--structure constant,
$a=\frac{g-2}{2}=1.793$ is the anomalous magnetic moment of the
proton, ${m_e}$ and ${m_p}$ are the rest mass of electron and proton,
${p}$ is the momentum in the CM system, ${a_0=52900 \;{\rm fm}}$ is
the Bohr radius and ${C_0^2=2\pi\eta/[\exp(2\pi\eta)-1]}$ is the
square of the Coulomb wave function at the origin.  The Coulomb
parameter ${\eta}$ is given by ${\eta = -z\alpha/v}$ (for antiprotons,
$\eta$ is positive). ${z}$ is the beam charge number and ${v}$ the
relative velocity of particle and projectile.
In Fig.~\ref{fig:sigma_long} the spin transfer cross section
${\sigma_{e_{||}}}$ of antiprotons scattered from longitudinally
polarized electrons is plotted versus the beam kinetic energy ${T}$. 
\subsection{Design Consideration for the APR}
In the following we evaluate a concept for a dedicated antiproton
polarizer ring (APR).  Antiprotons would be polarized by the
spin--dependent interaction in an electron--polarized hydrogen gas
target. This spin--transfer process is {\it calculable}, whereas, due
to the absence of polarized antiproton beams in the past, a
measurement of the spin--dependent $\bar{p}p$ interaction is still
lacking, and only theoretical models exist \cite{mull}.  The polarized
antiprotons would be subsequently transferred to the HESR for
measurements (Fig.~\ref{fig:CSRring}).

Both the APR and the HESR should be operated as synchrotrons with beam
cooling to counteract emittance growth.  In both rings the beam
polarization should be preserved during acceleration without loss
\cite{pol-conservation}.  The longitudinal spin--transfer cross
section is twice as large as the transverse one \cite{meyer},
$
\sigma_{e_{\parallel}}=2\cdot \sigma_{e_{\bot}}
$, the stable spin direction of the beam at the location of the
polarizing target should therefore be longitudinal as well, which
requires a Siberian snake in a straight section opposite the
polarizing target \cite{snakes}.
\subsubsection{Polarizer Target}
\begin{figure}[htb]
\begin{center}
  \epsfxsize=0.7\linewidth \epsfbox{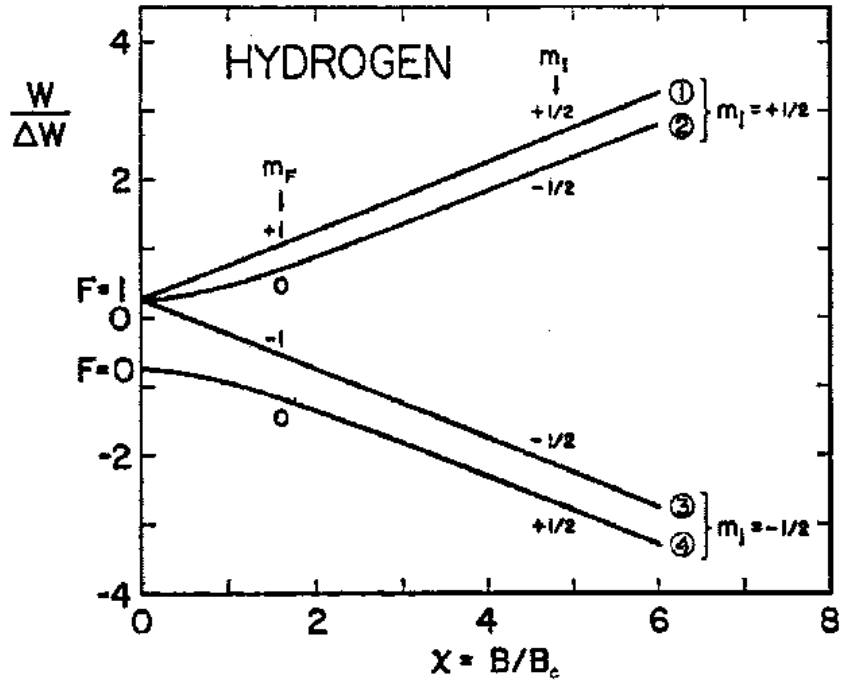}
  \parbox{14cm}{\protect\caption{\small Breit--Rabi diagram of
      hydrogen atoms in units of ${\Delta W=h \times
        1420.4\;\mathrm{MHz} }$ \protect\cite{haeberli}. The magnetic
      field is given in units of ${\chi=B/B_{\mathrm{c}}}$.  The
      critical field for the ground state of hydrogen is
      ${B_c=50.7\;{\rm mT}}$.
      \protect\label{fig:breit_rabi_hydrogen}}}
\end{center}
\end{figure}

A hydrogen gas target of suitable substate population represents a
dense target of quasi--free electrons of high polarization and areal
density. Such a target can be produced by injection of two hyperfine
states with magnetic quantum numbers $|m_J=+\frac{1}{2},
m_I=+\frac{1}{2}\rangle$ and $|+\frac{1}{2}, -\frac{1}{2}\rangle$ into
a strong longitudinal magnetic holding field of about $B_{||}=300$~mT
(Fig.~\ref{fig:breit_rabi_hydrogen}).  
The maximum electron and nuclear target polarizations in such a field
are $Q_e=0.5\cdot (1+\chi/\sqrt{1+\chi^2})= 0.993$ and $Q_z=0.5\cdot
(1-\chi/\sqrt{1+\chi^2})= 0.007$ \cite{haeberli}, where
$\chi=B_{||}/B_c$ and $B_c=50.7$~mT.  Polarized atomic beam sources
presently produce a flux of hydrogen atoms of about
$q=1.2\cdot10^{17}$~atoms/s in two hyperfine states \cite{zelenski}.
Our model calculation for the polarization buildup assumes a moderate
improvement of 20\%, i.e.  a flow of $q=1.5\cdot10^{17}$~atoms/s.
\subsubsection{Beam Lifetime in the APR}
The beam lifetime in the APR can be expressed as function of the
Coulomb--loss cross section $\Delta \sigma_C$ and the total hadronic
$\bar{p}p$ cross section $\sigma_{\mathrm{tot}}$,
\begin{eqnarray}
\tau_{\mathrm{APR}}
 =\frac{1}{(\Delta \sigma_C + 
 \sigma_{\mathrm{tot}}) \cdot d_t 
 \cdot f_{\mathrm{APR}}}\;.
\label{eq:tau_ap}
\end{eqnarray}

The density $d_t$ of a storage cell target depends on the flow of
atoms $q$ into the feeding tube of the cell, its length along the beam
$L_{\mathrm{beam}}$, and the total conductance $C_{\mathrm{tot}}$ of
the storage cell 
$
d_t=\frac{1}{2}\frac{L_{\mathrm{beam}}\cdot q}{C_{\mathrm{tot}}}
$ \cite{steffens}. The conductance of a cylindrical tube
$C_{\mathrm{\circ}}$ for a gas of mass $M$ in the regime of molecular
flow (mean free path large compared to the dimensions of the tube) as
function of its length $L$, diameter $d$, and temperature $T$, is
given by $
C_{\mathrm{\circ}}=3.8\cdot \sqrt{\frac{T}{M}}\cdot
\frac{d^3}{L+\frac{4}{3}\cdot d}.
\label{eq:C_circ}
$ The total conductance $C_{\mathrm{tot}}$ of the storage cell is
given by $
C_{\mathrm{tot}} = C_{\mathrm{\circ}}^{\mathrm{feed}} + 2\cdot
C_{\mathrm{\circ}}^{\mathrm{beam}},
$ where $C_{\mathrm{\circ}}^{\mathrm{feed}}$ denotes the conductance
of the feeding tube and $C_{\mathrm{\circ}}^{\mathrm{beam}}$ the
conductance of one half of the beam tube.  The diameter of the beam
tube of the storage cell should match the ring acceptance angle
$\Psi_{\mathrm{acc}}$ at the target, $
d_{\mathrm{beam}}=2\cdot \Psi_{\mathrm{acc}}\cdot \beta
$, where for the $\beta$--function at the target, we use
$\beta=\frac{1}{2}\,L_{\mathrm{beam}}$.  One can express the target
density in terms of the ring acceptance, $d_t \equiv
d_t(\Psi_{\mathrm{acc}})$, where the other parameters used in the
calculation are listed in Table~\ref{tab:polarizer_section}.
\begin{table}
\begin{center}
\begin{tabular}{l|l|l}
  \hline\hline
  circumference of APR         & $L_{\mathrm{APR}}$      & 150 m  \\
  $\beta$--function at target  & $\beta$                & 0.2 m  \\
  radius of vacuum chamber    & $r$                    & 5 cm   \\
  gap height of magnets       & $2\, g$                & 14 cm \\\hline
  ABS flow into feeding tube  & $q$                    & $1.5 \cdot 10^{17}$ atoms/s \\ 
  storage cell length         & $L_{\mathrm{beam}}$    & 40 cm  \\
  feeding tube diameter       & $d_{\mathrm{feed}}$    & 1 cm   \\
  feeding tube length         & $L_{\mathrm{feed}}$    & 15 cm  \\
  longitudinal holding field  & $B_{||}$               & 300 mT \\
  electron polarization       & $Q_e$                  & 0.9    \\    
  cell temperature            & $T$                    & 100 K  \\\hline\hline 
\end{tabular}
\parbox{14cm}{\caption{\label{tab:polarizer_section}\small Parameters
    of the APR and the polarizing target section. }}
\end{center}
\end{table}

\begin{figure}[htb]
\begin{center}
  \includegraphics[width=0.80\linewidth]{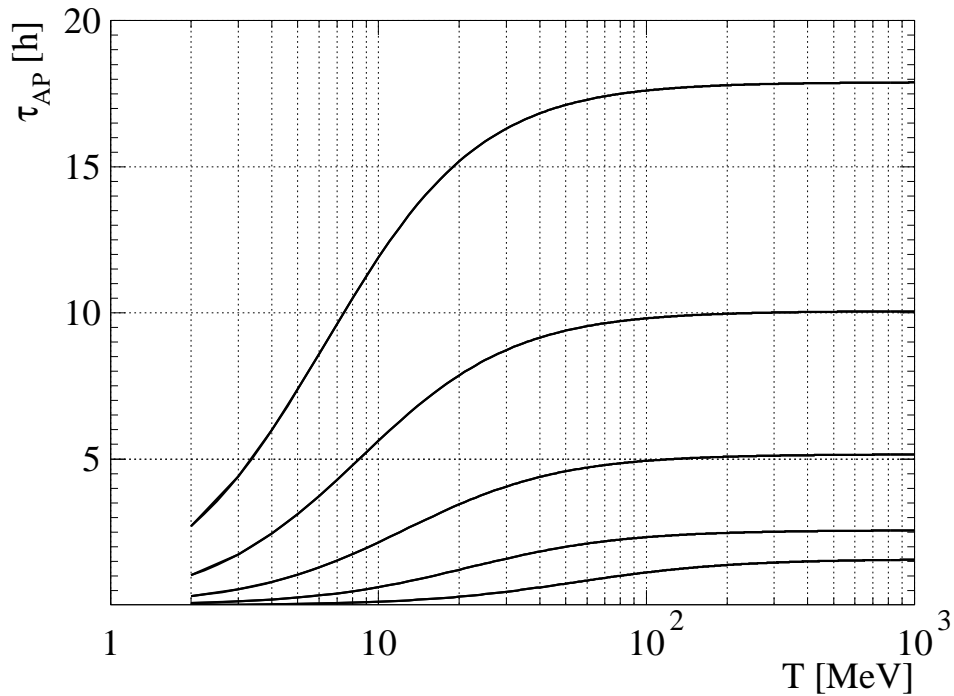}
  \parbox{14cm}{\caption{\label{fig:tau_ap}\small Beam lifetime in the
      APR as function of kinetic energy $T$.  From top to bottom the
      lines denote $\Psi_{\mathrm{acc}}=50$, 40, 30, 20, and
      10~mrad.}}
\end{center}
\end{figure}

The Coulomb--loss cross section $\Delta \sigma_C$ (using
${c=\hbar=1}$) can be derived analytically in terms of the square of
the total energy $s$ by integration of the Rutherford cross section,
taking into account that only those particles are lost that undergo
scattering at angles larger than $\Psi_{\mathrm{acc}}$,
\begin{eqnarray}
\Delta \sigma_C(\Psi_{\mathrm{acc}})=4\pi\alpha^2 
\frac{(s-2m_{\bar{p}}^2)^2 \, 4m_{\bar{p}}^2}
{s^2(s-4m_{\bar{p}}^2)^2}
\,
\left(\frac{1}{\Psi_{\mathrm{acc}}^2} - \frac{s}{4m_{\bar{p}}^2}\right)\;.
\end{eqnarray}
The total hadronic cross section is parameterized using a function
inversely proportional to the Lorentz parameter
$\beta_{\mathrm{lab}}$.  Based on the $\bar{p}p$ data \cite{PDG} the
parameterization
$
\sigma_{\mathrm{tot}}=\frac{75.5}{\beta_{\mathrm{lab}}}\;(\mathrm{mb})
$
yields a description of $\sigma_{\mathrm{tot}}$ with $\approx 15$\%
accuracy up to $T\approx 1000$~MeV.  The APR revolution frequency is
given by
\begin{eqnarray}  
f_{\mathrm{APR}}=\frac{\beta_{\mathrm{lab}} \cdot c}{L_{\mathrm{APR}}}\;.
\label{eq:f_ap}
\end{eqnarray}
The resulting beam lifetime in the APR as function of the kinetic
energy $T$ is depicted in Fig.~\ref{fig:tau_ap} for different
acceptance angles $\Psi_{\mathrm{acc}}$.
\subsection{Polarization Buildup}
\begin{figure}[htb]
\begin{center}
  \includegraphics[width=0.80\linewidth]{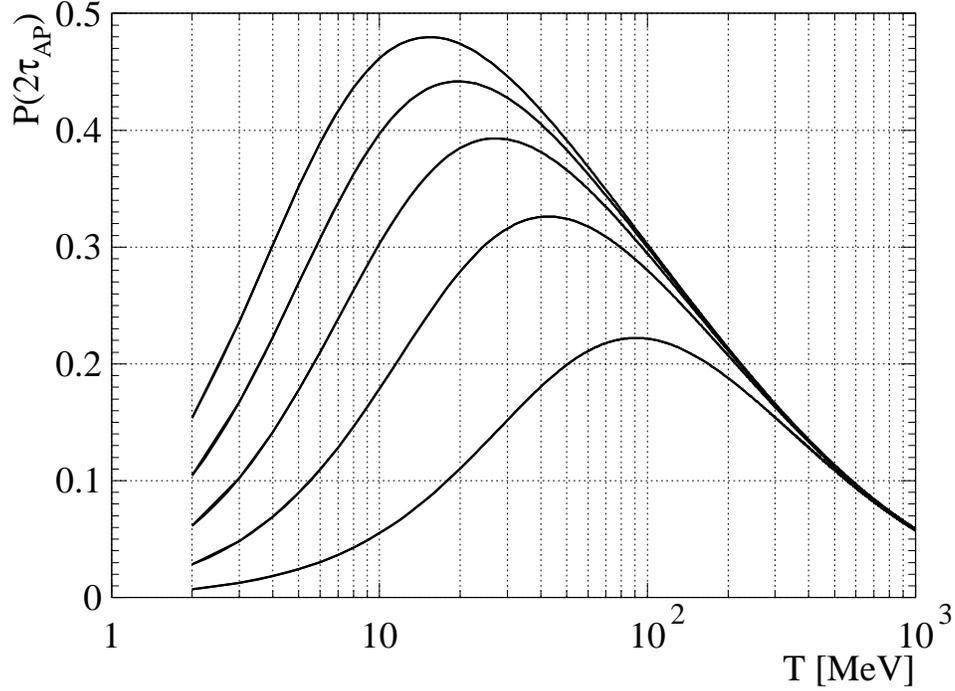}
  \parbox{14cm}{\caption{\label{fig:pol}\small Antiproton beam
      polarization $P(2 \cdot \tau_{\mathrm{APR}})$
      [Eq.~(\ref{eq:p_of_t})] as function of beam energy for different
      acceptance angles $\Psi_{\mathrm{acc}}$.  (Lines are organized
      as in Fig.~\ref{fig:tau_ap}.)}}
\end{center}
\end{figure}

The buildup of polarization due to the spin--dependent $\bar{p} e$
interaction in the target [Eq.~(\ref{eq:sigma_long})] as function of
time $t$ is described by
\begin{eqnarray}
P(t)=\tanh\left(\frac{t}{\tau_p}\right)\;,\, \mathrm{where} \;
\tau_p=\frac{1}{\sigma_{e_{\parallel}} \, d_t \, f_{\mathrm{APR}} \, Q_e}
\label{eq:p_of_t}
\end{eqnarray}
denotes the polarization buildup time. The time dependence of the beam
intensity is described by
\begin{eqnarray}
I(t)=I_0 \cdot \exp{\left(-\frac{t}{\tau_{\mathrm{APR}}}\right)} 
         \cdot \cosh{\left(\frac{t}{\tau_p}\right)}\;,
\label{eq:I_of_t}
\end{eqnarray}
where $I_0=N_{\bar{p}}^{\mathrm{APR}}\cdot f_{\mathrm{APR}}$.  The
quality of the polarized antiproton beam can be expressed in terms of
the figure of merit \cite{ohlsen}
\begin{eqnarray}
\mathrm{FOM}(t)=P(t)^2 \cdot I(t)\;.
\label{eq:fom}
\end{eqnarray}
The optimum interaction time $t_{\mathrm{opt}}$, where
$\mathrm{FOM}(t)$ reaches the maximum, is given by $\frac{\mathrm{d}}
{\mathrm{d}\,t} \mathrm{FOM}(t) =0$.  For the situation discussed
here, $t_{\mathrm{opt}}=2 \cdot \tau_{\mathrm{APR}}$ constitutes a good
approximation that deviates from the true values by at most 3\%. The
magnitude of the antiproton beam polarization $P(t_{\mathrm{opt}})$
based on electron spin transfer [Eq.~(\ref{eq:p_of_t})] is depicted in
Fig.~\ref{fig:pol} as function of beam energy $T$ for different
acceptance angles $\Psi_{\mathrm{acc}}$.
\subsubsection{Space--Charge Limitations}
\begin{figure}[htb]
\begin{center}
  \includegraphics[width=0.8\linewidth]{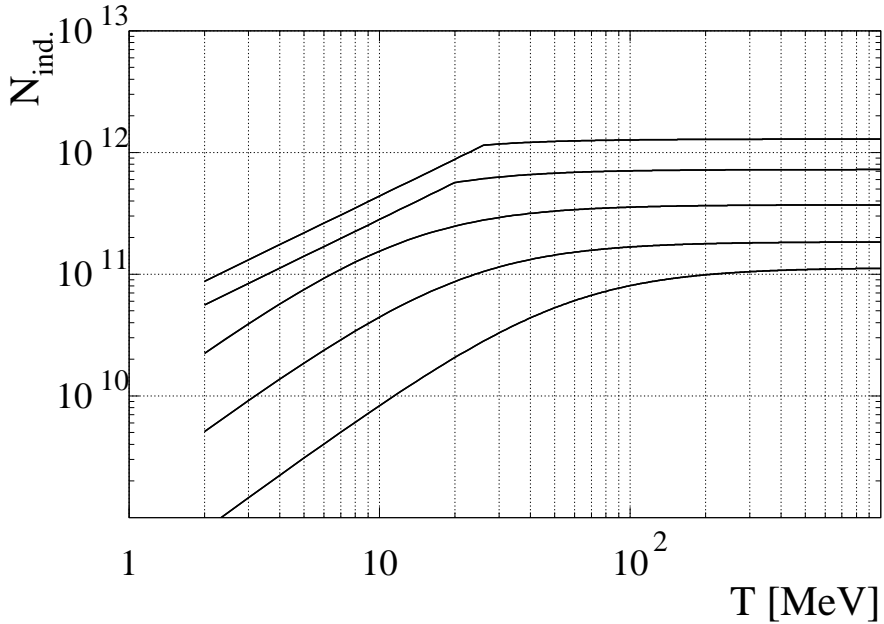}
  \parbox{14cm}{\caption{\label{fig:nri}\small Individual particle
      limit $N_\mathrm{{ind.}}$ for the five different ring acceptance
      angles $\Psi_{\mathrm{acc}}$ (50 -- 10~mrad) as function of beam
      energy.  (Lines are organized as in Fig.~\ref{fig:tau_ap}.)}}
\end{center}
\end{figure}

The number of antiprotons stored in the APR may be limited by
space--charge effects. With an antiproton production rate of
$R=10^7$~$\bar{p}/s$, the number of antiprotons available at the
beginning of the filtering procedure corresponds to
\begin{eqnarray}
N_{\bar{p}}^{\mathrm{APR}}(t=0)=R \cdot 2 \cdot \tau_{\mathrm{APR}}\;.
\label{eq:n_ap}
\end{eqnarray}
The individual particle limit in the APR is given by \cite{bovet}
\begin{eqnarray}
N_{\mathrm{ind.}}=2\, \pi \, \varepsilon \, \beta_{\mathrm{lab}}^2 \,
\gamma_{\mathrm{lab}}^3 \, (r_p \, F)^{-1} \, \Delta Q\;,
\label{eq:n_ind}
\end{eqnarray}
where $\varepsilon=\Psi_{\mathrm{acc}}^2\cdot \beta$ denotes the
vertical and horizontal beam emittance, $\beta_{\mathrm{lab}}$ and
$\gamma_{\mathrm{lab}}$ are the Lorentz parameters, $r_p=1.5347 \cdot
10^{-18}$~m is the classical proton radius, and $\Delta Q=0.01$ is the
allowed incoherent tune spread. The form factor $F$ for a circular vacuum
chamber \cite{bovet} is given by $F=1+\left(a_y \cdot \frac{a_x +
    a_y}{r^2}\right) \cdot \varepsilon_2 \cdot
(\gamma_{\mathrm{lab}}^2-1) \cdot \frac{r^2}{g^2}$, where the mean
semi--minor horizontal $(x)$ and vertical $(y)$ beam axes
$a_{x,y}=\sqrt{\varepsilon \cdot \beta_{x,y}}$ are calculated from the
mean horizontal and vertical $\beta$--functions
$\beta_{x,y}=L_{\mathrm{APR}}\cdot (2\pi\nu)^{-1}$ for a betatron--tune
$\nu=3.6$.  For a circular vacuum chamber and straight magnet pole pieces
the image force coefficient $\varepsilon_2=0.411$.  The parameter $r$
denotes the radius of the vacuum chamber and $g$ half of the height of
the magnet gaps (Table~\ref{tab:polarizer_section}).
In Fig.~\ref{fig:nri} the individual particle limit
is plotted for the different acceptance angles.
\subsubsection{Optimum Beam Energies for the Polarization Buildup}
\begin{figure}[htb]
\begin{center}
  \includegraphics[width=0.80\linewidth]{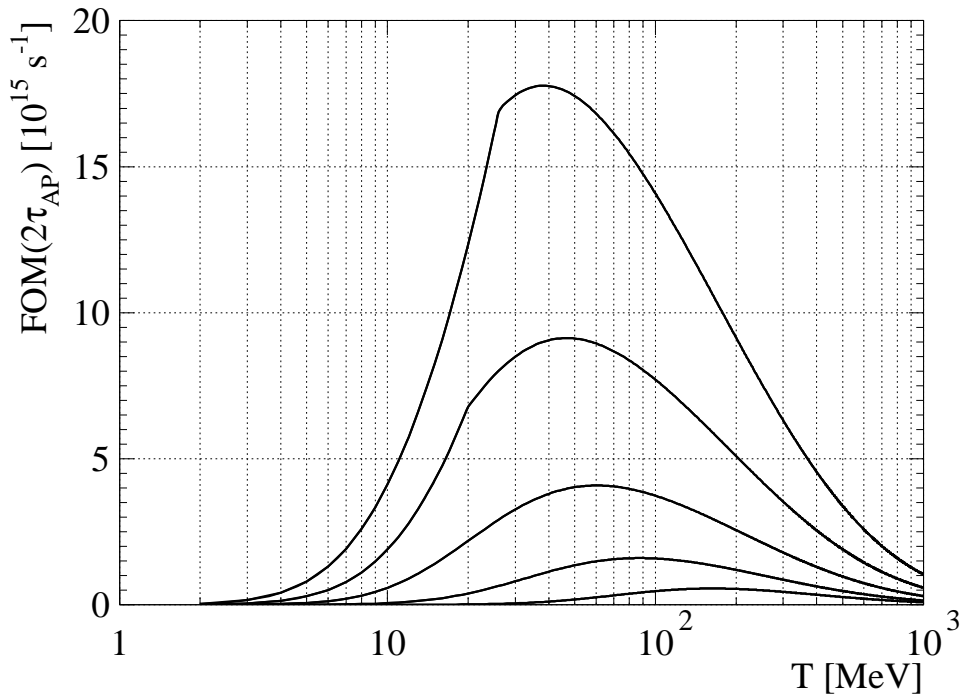}
  \parbox{14cm}{\caption{\label{fig:fom}\small Figure of Merit for the
      polarized antiproton beam for filtering times $t=2\cdot
      \tau_{\mathrm{APR}}$ as function of beam energy. The parameters
      associated with the maxima are summarized in
      Table~\ref{tab:opt_energies}. (Lines are organized as in
      Fig.~\ref{fig:tau_ap}.)}}
\end{center}
\end{figure}

The optimum beam energies for different acceptance angles at which the
polarization buildup works best, however, cannot be obtained from the
maxima in Fig.~\ref{fig:pol}. In order to find these energies, one has
to evaluate at which beam energies the FOM [Eq.~(\ref{eq:fom})],
depicted in Fig.~\ref{fig:fom}, reaches a maximum. The optimum beam
energies for polarization buildup in the APR are listed in
Table~\ref{tab:opt_energies}. The limitations due to space--charge,
$N_{\bar{p}}^{\mathrm{APR}}>N_{\mathrm{ind.}}$ [Eqs.~(\ref{eq:n_ap},
\ref{eq:n_ind})], are visible as kinks in Fig.~\ref{fig:fom} for the
acceptance angles $\Psi_{\mathrm{acc}}=40$ and 50~mrad, however, the
optimum energies are not affected by space--charge.
\begin{table}
\begin{center}
\begin{tabular}{c|c|c|c}
$\Psi_{\mathrm{acc}}$ (mrad) & $T$ (MeV) & $\tau_{\mathrm{APR}}$ (h) & $P(2\,\tau_{\mathrm{APR}})$ \\ \hline\hline
10                           & 167       & 1.2                       & 0.19 \\
20                           & 88        & 2.2                       & 0.29 \\
30                           & 61        & 4.6                       & 0.35 \\
40                           & 47        & 9.2                       & 0.39 \\
50                           & 39        & 16.7                      & 0.42 \\ \hline\hline
\end{tabular}
\parbox{14cm}{\caption{\label{tab:opt_energies}\small Kinetic beam
    energies where the polarized antiproton beam in the APR reaches
    the maximum FOM for different acceptance angles. }}
\end{center}
\end{table}
\subsubsection{\label{sec:pol_el_tgts}Polarized Targets containing only Electrons}
Spin filtering in a {\it pure} electron target greatly reduces the
beam losses, because $\sigma_{\mathrm{tot}}$ disappears and Coulomb
scattering angles in $\bar{p}e$ collisions do not exceed
$\Psi_{\mathrm{acc}}$ of any storage ring.  With stationary electrons
stored in a Penning trap, areal densities of about $10^{12}$
electrons/cm$^2$ may be reached in the future \cite{traps}. A typical
electron cooler operated at 10~kV with polarized electrons of
intensity $\approx 1$~mA ($I_e \approx 6.2\cdot 10^{15}$~electrons/s)
\cite{pol-el-sources}, $A=1$~cm$^2$ cross section, and $l=5$~m length
reaches $d_t=I_e \cdot l \cdot (\beta_{\mathrm{lab}}\, c\,A)^{-1}= 5.2
\cdot 10^{8}$~electrons/cm$^2$, which is six orders of magnitude short
of the electron densities achievable with a neutral hydrogen gas
target.  For a pure electron target the spin transfer cross section is
$\sigma_{e_{||}}=670$~mb (at $T = 6.2$~MeV) \cite{horowitz-meyer},
about a factor $15$ larger than the cross sections associated to the
optimum energies using a gas target (Table~\ref{tab:opt_energies}).
One can therefore conclude that with present day technologies, both
above discussed alternatives are no match for spin filtering using a
polarized gas target.
\subsection{Luminosity Estimate for a Fixed Target in the HESR}
In order to estimate the luminosities, we use the parameters of the
HESR ($L_{\mathrm{HESR}}=440$~m).  After spin filtering in the APR for
$t_{\mathrm{opt}}=2 \cdot \tau_{\mathrm{APR}}$, the number of
polarized antiprotons transfered to HESR is
$N_{\bar{p}}^{\mathrm{APR}}(t=0)/e^2$ [Eq.~(\ref{eq:n_ap})].  The beam
lifetime in the HESR at $T=15$~GeV for an internal polarized hydrogen
gas target of $d_t=7 \cdot 10^{14}$~cm$^{-2}$ is about
$\tau_{\mathrm{HESR}} = 12$~h [Eqs.~(\ref{eq:tau_ap}, \ref{eq:f_ap})],
where the target parameters from Table~\ref{tab:polarizer_section}
were used, a cell diameter $d_{\mathrm{beam}}=0.8$~cm, and
$\sigma_{\mathrm{tot}}= 50$~mb.  Subsequent transfers from the APR to
the HESR can be employed to accumulate antiprotons.  Eventually, since
$\tau_{\mathrm{HESR}}$ is finite, the average number of antiprotons
reaches equilibrium, $
\overline{N_{\bar{p}}^{\mathrm{HESR}}}=R/e^2 \cdot
\tau_{\mathrm{HESR}}= 5.6\cdot 10^{10}
$, independent of $\tau_{\mathrm{APR}}$. An average luminosity of $
\bar{{\cal L}}=
R/(e^2 \cdot \sigma_{\mathrm{tot}})= 2.7 \cdot 10^{31} \;
\mathrm{cm}^{-2}\mathrm{s}^{-1}\
$ can be achieved, with antiproton beam polarizations depending on the
APR acceptance angle $\Psi_{\mathrm{acc}}$
(Table~\ref{tab:opt_energies}).

To summarize, we have shown that with a dedicated large acceptance
antiproton polarizer ring ($\Psi_{\mathrm{acc}}=10$ to 50~mrad), beam
polarizations of $P=0.2$ to 0.4 could be reached. The energies at
which the polarization buildup works best range from $T=40$ to
170~MeV.  In equilibrium, the average luminosity for
double--polarization experiments in an experimental storage ring (e.g.
HESR) after subsequent transfers from the APR could reach $\bar{{\cal
    L}}=2.7 \cdot 10^{31} \; \mathrm{cm}^{-2}\mathrm{s}^{-1}$.

\subsection{Technical Realization of the APR} \label{sec:tech_APR}
Antiprotons are conveniently polarized at an energy of $\approx
50$~MeV ($\beta_c = 0.28$) with an adequate gas target
\cite{steffens}. A storage ring is ideal to efficiently achieve a high
degree of beam polarization due to the repeated beam traversal of the
target. The beam degradation, the geometrical blow--up, and the
subsequent smearing of the beam energy needs to be corrected by
phase--space cooling, preferably by electron cooling. The shaking of
the beam \cite{shaking}, leading to unwanted instabilities caused by
positive ions accumulated around the beam, can be eliminated by a
suitable RF cavity.  Since the antiprotons should be longitudinally
polarized, the ring has to contain a Siberian snake \cite{snakes}.
Finally efficient systems for injection and extraction of the
antiproton beam have to be provided in the ring as well.  The
consequences of these insertions are at first, sufficient space in the
ring and secondly, various specifications of the antiproton beam at
the positions of these insertions, i.e. constraints on the
ion--optical parameters. Obviously, for a high antiproton
polarization, the divergence of the beam at the polarizer target
should be large. The antiproton beam will be injected by stacking in
phase space. The extraction will be done by bunch--to--bunch transfer.
In the empty ring, the antiproton beam lifetime should be a several
tens of hours, which sets also the requirements for the vacuum system.

The antiproton polarizer, discussed here, would provide highly
polarized antiproton beams of unprecedented quality. In particular the
implementation of this option at the Facility for Antiproton and Ion
Research would open new and unique research opportunities for
spin--physics experiments in $\bar{p}p$ interactions at the
HESR and CSR.
\subsubsection{Constraints}
Following the known requests, we describe here a design for such a
ring. Four straight sections are required for the following insertions:

\begin{enumerate}

\item Injection and extraction of the antiproton beam, for which free
  space of 4~m is foreseen.  

\item For the gas target a low $\beta$--section is required in order
  to obtain a small beam spot and a large angular divergence, in this
  straight section a free space of 1~m is provided.

\item The opposite straight section should be reserved for the
  Siberian snake to longitudinally align the antiproton spin.
  
\item For electron cooling, in the straight section opposite to the
  injection straight, a free space of 4~m is reserved.

\item A small RF cavity may be provided in any section.  

\end{enumerate}

Various ion optical conditions have to be met in the four straight
sections: 

\begin{itemize}
\item[A)] In the target, e-cooler, and the Siberian snake sections the
  beam cross section has to be circular and the beams phase space
  ellipse has to be upright.
\item[B)] In all straight sections the dispersion should be zero. 
\item[C)] The antiproton beam in the e-cooling section should be
  parallel and its cross section should be variable in order to match
  the size of the electron beam. 
\item[D)] The radius of the beam spot at the target should be $\approx
  10$~mm.  
\end{itemize}

The present APR is designed for antiprotons of 40~MeV, corresponding
to a momentum of $p = 276$~MeV/c and a magnetic rigidity of $B\rho=
0.924$~Tm.  Finally a large acceptance of the APR is required. We have
anticipated an acceptance of the ring of \be
\epsilon_{x,y}=500\,\pi\mathrm{mm}\,\mathrm{mrad}\,, \ee sufficient to
accommodate a flux of at least $10^{11}$~$\bar{p}$/s.

To provide longitudinal polarized beam at the position of the storage
cell, an integrated field strength of roughly $1.2$~Tm is required in
the opposite straight section of the APR (1.15 (1.04)~Tm for 50
(40)~MeV). Two scenarios are possible in combination with the
solenoids of the electron cooler. The electron cooler is located in a
different straight section than the snake. In this case correcting
solenoids should be utilized to compensate for spin motion in the main
solenoid of the cooling system. One could also apply the electron
cooler solenoids as a snake. The integrated field of a conventional
electron cooler, like the one in use at COSY is about 0.15~T with an
effective length of 2m.  Together with two compensation solenoids,
this would provide a half--snake at APR energies. An additional
solenoid of $0.6~Tm$ would then be sufficient to achieve the required
integrated field.  All solenoids would have to be in the same straight
section, opposite to the storage cell.  Only the additional solenoid
would have to be rampable. For beam extraction and transfer, the beam
should be vertically polarized.  Therefore the snake has to be either
adiabatically turned off (first solution) or ramped to opposite field
strength (second solution) in order to compensate for the cooler
solenoids. In both cases, depolarizing resonances can be crossed
during ramping, since the spin tune moves from half integer for a full
snake to $\gamma G$ = 1.88. If the fractional betatron tune in the APR
is chosen to be larger than 0.83 or smaller than 0.17, no first order
resonances are crossed. If this should not be possible, the ramping
speed of the snake has to be chosen in such a way as to minimize
polarization losses during spin resonance crossing.

\subsubsection{Layout of the APR Lattice}
The optical condition A, outlined above, leads to the utilization of
symmetric quadrupole triplets. Thus for the arcs of the APR ring
design, where the beam is bent by 90 degrees, the following structure
has been chosen:
\begin{eqnarray}
 \mathrm{Triplet\;1}-\mathrm{Bend}(45^\circ)- \mathrm{Triplet\;
    2}-\mathrm{Bend}(45^\circ)-\mathrm{Triplet\;1}
\end{eqnarray}
In detail, denoting with F and D the focusing and defocusing
strengths of the quadrupoles, the 2 triplets are realized by:
\begin{eqnarray}
 \mathrm{Triplet\;1}=\mathrm{D}-\mathrm{F}-\mathrm{D}\; 
\mathrm{and}\;\mathrm{Triplet\;2}=\mathrm{F}-\mathrm{D}-\mathrm{F}
\end{eqnarray}

\begin{figure}[htb]
\begin{center}
  \includegraphics[width=0.70\linewidth]{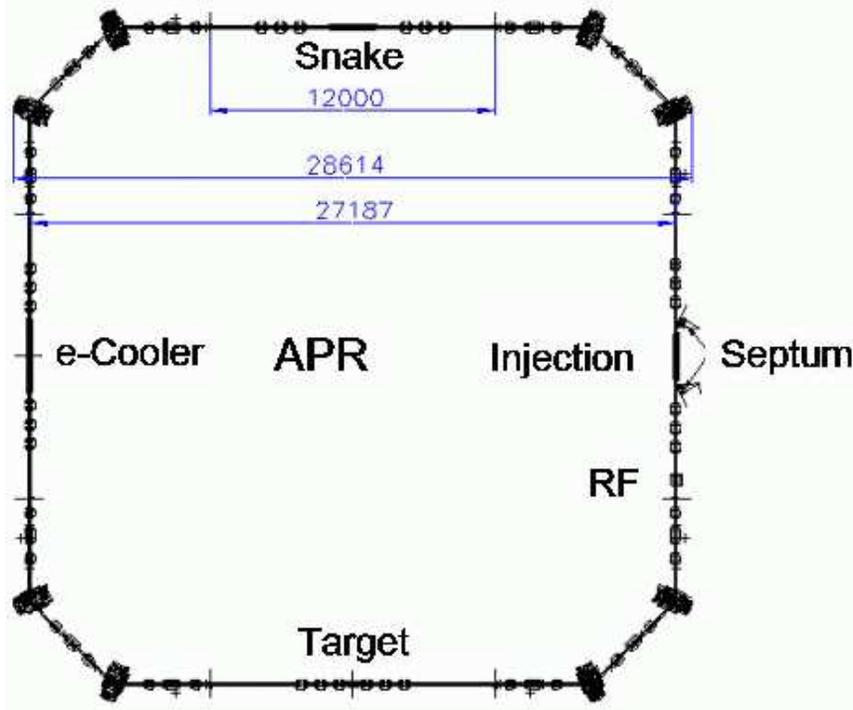}
  \parbox{14cm}{\caption{\label{fig:APR-floor} \small Floor plan of
      the APR lattice.}}
\end{center}
\end{figure}

For each of the four straight sections two triplets are foreseen to
meet the ion--optical conditions mentioned above. For three straight
sections two quadrupole doublets are needed in addition to provide the
requested beam sizes and divergences, e.g. the low $\beta$--section
around the gas target. The ion--optical imaging through each arc is
telescopic, or more accurately, it constitutes a $(-1)$-- telescope.
In Fig.~\ref{fig:APR-floor} a floor plan of the suggested APR is
shown, which could be realized within a floor space of $30 \times
30$~m$^2$.
The main parameters of the ring lattice are contained in
Table~\ref{tab:APRringparameters}. The important specifications for
the dipole and quadrupole magnets are listed in
Table~\ref{tab:dipole_quads} for the case of realizing the magnets
conventionally by water--cooled electromagnets. In
sec.~\ref{sec:perm_mag} an economic solution based on permanent
magnets is discussed.
\begin{table}[hbt]
\begin{center}
\renewcommand{\arraystretch}{1.2}
\begin{tabular}{|p{7cm}|c|r|l|}
\hline
  Periodicity                         &  $P$                &    4               &             \\\hline
  Circumference                       &  $C$                &    100.2           &m            \\\hline
  Floor area                          &  $F$                &    $30\times30$    & m$^2$       \\\hline  
  Magnetic rigidity                   &  $B\rho$            &    0.924           & Tm          \\\hline
  $\gamma$                            &                     &    1.043           &   \\\hline
  $\beta$                             &  $v/c$              &    0.283           &   \\\hline
  Beam emittance                      &  $\epsilon_{x,y}$   &    500             &$\pi$~mm~mrad  \\\hline
  Number of particles                 &                     &    $10^{11}$       &   \\\hline
  Number of straight sections         &                     &    4               &   \\\hline
  Length of  straight sections        & $l_{\mathrm{s.s.}}$ &    12              &m            \\\hline
  Number of arcs                      &                     &    4               &   \\\hline
  Length of arcs                      & $l_a$               &    13.05    &       m            \\\hline
  Number of 45$^\circ$ dipole magnets &                     &    8               &   \\\hline
  Number of quadrupoles               &                     &    72              &   \\\hline
  Length of electron cooler solenoid  & $L_{\mathrm{sol.}}$ &    3.2      & m              \\\hline
  Cavity length                       & $L_{\mathrm{RF}}$   &    0.4      & m           \\\hline
  Injection                           &\multicolumn{3}{l|}{Septum---Kicker}          \\\hline
  Extraction                          &\multicolumn{3}{l|}{Kicker---Septum}           \\\hline
  Cavity type                         &\multicolumn{3}{l|}{Finemet}                   \\\hline
\end{tabular}
\parbox{14cm}{\caption{\label{tab:APRringparameters}\small Main
    parameters of the APR lattice.}}
\end{center}
\end{table} 
\begin{table}[hbt]
\begin{center}
  \renewcommand{\arraystretch}{1.2}
\begin{tabular}{|l|r|l||l|r|l|}
\hline
 \multicolumn{3}{|c||}{Dipole magnets}    & \multicolumn{3}{c|}{Quadrupole magnets}   \\\hline\hline  
 \multicolumn{3}{|c||}{8}                 & \multicolumn{3}{c|}{72}                   \\\hline
 bending angle  & 45     & deg.           & aperture diameter & 135.1/143.3    & mm   \\\hline
 edge angle     & 22.5   & deg.           & effective length  & 25/50          & cm   \\\hline
 arc length     & 0.725  & m              & gradient          & $1.33-1.62$    & T/m  \\\hline
 gap height     & 143    & mm             & pole tip field    & $0.19 - 0.22$  & T    \\\hline
 magnetic field & 1      & T              & weight            & 220/440        & kg   \\\hline
 weight         & 7000   & kg             &                   &                &       \\\hline
\end{tabular}
\parbox{14cm}{\caption{\label{tab:dipole_quads}\small Specifications
    of the electrically powered dipole and quadrupole magnets for the
    APR. Aperture diameter, effective length, and weight are given for
    the singlet and triplet quadrupole magnets.}}
\end{center}
\end{table} 
\subsubsection{Features of the APR Design}
For the layout of the APR, the computer code WinAgile \cite{winagile}
has been used. In the following, we discuss the most important optical
transfer functions. From the $\beta$--functions $\beta_{x,y}$, the
radius of the corresponding beam size $(r_x,r_y)$ is readily obtained
from the relation
\begin{eqnarray}
(r_x,r_y)=\sqrt{\beta_{x,y}\cdot \epsilon_{x,y}}\;,
\end{eqnarray}
where $\epsilon_{x,y}$ denote the horizontal and vertical phase space,
respectively.

\begin{figure}[htb]
\begin{center}
  \includegraphics[width=0.60\linewidth]{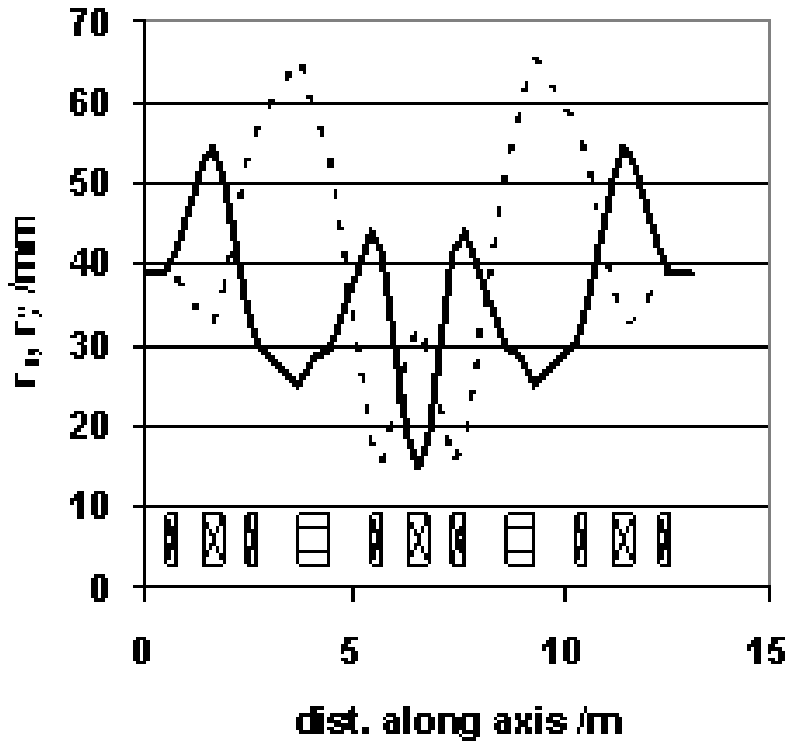}
  \parbox{14cm}{\caption{\label{fig:orbit_arcs} \small Beam radius
      $(r_x,r_y)$ along one arc of the APR lattice ($r_x$ -- solid;
      $r_y$ -- dashed).}}
\end{center}
\end{figure}

\begin{figure}[htb]
\begin{center}
  \includegraphics[angle=270,width=0.60\linewidth]{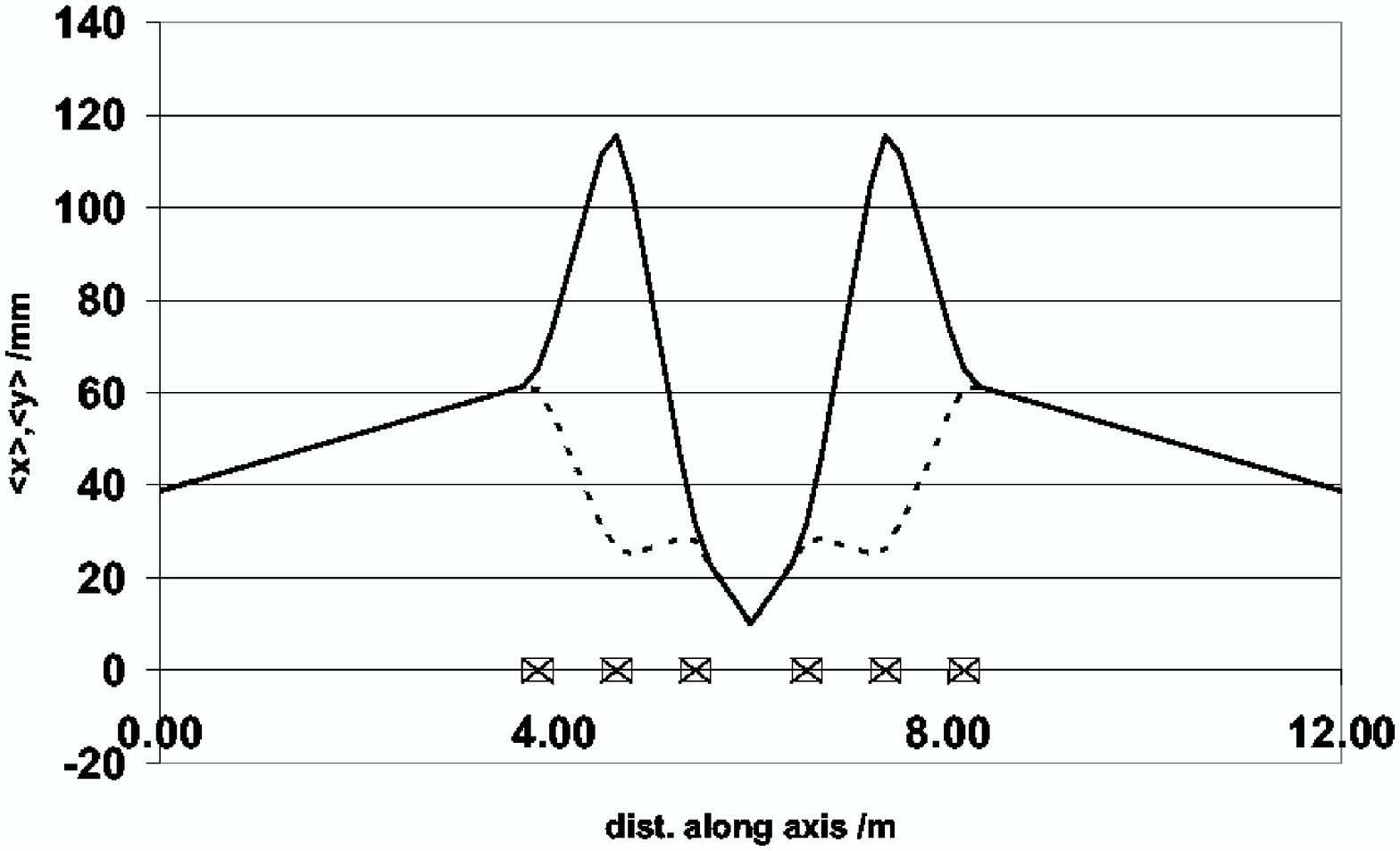}
  \parbox{14cm}{\caption{\label{fig:beamsize} \small Beam radius
      $(r_x,r_y)$ around the gas target ($r_x$ -- solid line, $r_y$ --
      dashed).}}
\end{center}
\end{figure}

In Fig.~\ref{fig:orbit_arcs} the radius of the beam spot is shown for
one of the four arcs of the present APR design. The large phase space
of the beam leads to large sizes of the gaps and diameters of the
dipole magnets and quadrupoles, as listed in Table~\ref{tab:dipole_quads}. 
The beam envelopes in the target section are shown in
Fig~\ref{fig:beamsize}, where a spot size of $r = 10$~mm is required
together with an angular divergence of 50 mrad.
The phase space ellipse of the beam here is upright ($\alpha_{x,y} =
0$).

\begin{figure}[htb]
\begin{center}
  \includegraphics[width=0.60\linewidth]{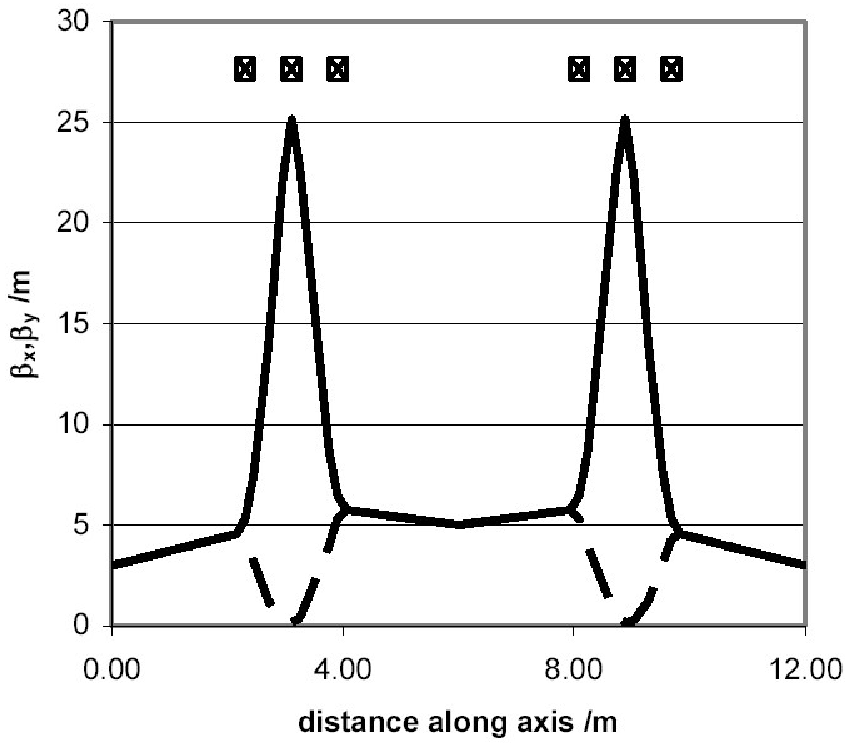}
  \parbox{14cm}{\caption{\label{fig:beta_functions_ecool}\small
      $\beta$--functions in the electron cooler section ($\beta_x$ --
      solid line, $\beta_y$ -- dashed).}}
\end{center}
\end{figure}

In Fig.~\ref{fig:beta_functions_ecool}, the $\beta$--functions for the
straight section provided for the electron cooling solenoid. Here
$\beta_x=\beta_y = 5$~m.  Depending on the cross section of the
electron beam $\beta_{x,y}$ can be adjusted to values between 1 and
10~m.
For the last section with Siberian snake and RF cavity, a wide variety
of optical conditions may be realized with the foreseen quadrupoles.
The same applies to the injection/ejection section.

\subsubsection{Discussion of the APR Design}
The APR lattice design presented here provides telescopic imaging
through each of the 4 arcs of the ring. The quadrupole arrangements in
the straight sections enable tuning of the beam within a wide choice
of optical conditions. Plausible assumptions about the dimensions of
the insertions have been made. The design of these insertions depends on the
final lattice, and can be easily modified.

\subsubsection{Realization of the APR based on Permanent Magnets\label{sec:perm_mag}}
An important aspect concerning the realization of the APR concerns the
employed conventional electromagnets (dipoles and quadrupoles).  Since
all APR magnets are set to a specific magnetic field or field
gradient, the use of permanent magnets is an advisable alternative.
For magnetic quadrupoles, the optical quality achieved so far with
permanent magnetic material is excellent.  Their great advantage would
be: 
\begin{enumerate}
\item Compactness of the quadrupoles, 
\item savings in operational and investment costs, and
\item simple installation (neither power nor cooling water required).
\end{enumerate}
A preliminary estimate for replacing the 72 conventional quadrupoles
in this APR design with permanent ones leads to length reductions from
25 (50) cm to 10 (12) cm for the 2 families with a corresponding
weight reduction from 220 (440) to 11 (31) kg. The reduced weight
would also have profound implications for the support structure,
alignment etc. The permanent quadrupole magnets would save operational
power costs of 150 kW and the investment for water cooling. The design
of dipoles with a specific bending strength employing permanent
magnetic material is complicated, since they consist of an arrangement
of circular slabs with appropriate spacing.  A suitable design is
under preparation and commercially available from the company UGS
\cite{ugs}, which also offers the permanent magnet quadrupoles.
Although details are not available at present, size and weight
reductions for the dipoles are similar. The operation of the
conventional dipoles would require a total power of 309 kW.

The antiproton polarizer, discussed here, would provide highly
polarized antiproton beams of unprecedented quality. In particular the
implementation of this option at the Facility for Antiproton and Ion
Research would open new and unique research opportunities for
spin--physics experiments in $\bar{p}p$ interactions at the
HESR.

\section{Cooler Synchrotron Ring  CSR} \label{sec:CSR}
The CSR has to accelerate polarized protons and antiprotons to momenta
between 600~MeV/c and 3.65~GeV/c and is expected to be very similar to
COSY \cite{cosy}.  A polarized ion source to provide a vector
polarized proton beam is needed which is accelerated to about 50~MeV
before injection into the CSR \cite{gebel,linac}. The polarization is
determined by measuring the asymmetry of $pC$ scattering from a carbon
fiber target \cite{edda}.  Additional polarimeters have to be
installed after the LINAC to optimize the transition units in the
polarized source for a high degree of polarization.
\subsection{Injector LINAC for 50 MeV Polarized Protons}
Low energy superconducting proton LINACs have been discussed recently
as an optimum solution in terms of justifiable resources and available
space.  A layout for such a linac was worked out and proposed as a new
injector for the COSY together with advanced ion sources and two
interchangeable rf--quadrupoles (RFQ).  It was designed to deliver
both polarized and unpolarized pulsed H$^-$/D$^-$ beams at a kinetic
energy of ~50MeV with a maximum repetition rate of 2Hz.  The pulse
length was limited to 500~ms and the beam current to 2~mA (peak).
This compact design leads to a total LINAC length of less than 20~m.
For producing the short pulses of high--intensity polarized
H$^-$/D$^-$ beams, a CIPIOS--type ion source like the one used at IUCF
(Bloomington, Indiana) is suitable \cite{CIPIOS}.

It should be noted that after the termination of the HERA program in
2007, the DESY proton LINAC might be available, which would also be a
perfect CSR injector.

\subsection{Acceleration of Polarized Proton and Antiproton Beam} 
In a strong--focusing synchrotron like the CSR two different types of
strong depolarizing resonances are excited, namely imperfection
resonances caused by magnetic field errors and misalignments of the
magnets, and intrinsic resonances excited by horizontal fields due to
the vertical focusing.  In the momentum range of CSR, five
imperfection resonances have to be crossed. Vertical correction
dipoles or a weak partial snake of a few percentage can be utilized to
overcome all imperfection resonances by exciting adiabatic spin flips
without polarization losses.  The number of intrinsic resonances
depends on the superperiodicity of the lattice. In principle a typical
magnetic structure of a synchrotron ring in this energy range allows
to adjust superperiodicities of $P=2$ or even 6 like in COSY. However,
due to symmetry--breaking modification of the interaction region and
strong magnetic fields of detector magnets and the electron cooling
system, a superperiodicity of $P=1$ is expected, leading to about ten
intrinsic resonances.  A tune--jump system consisting of fast
quadrupoles has especially been developed to handle intrinsic
resonances at COSY and will also be suitable for the CSR.

\subsubsection{Imperfection Resonances} 
The imperfection resonances for polarized protons and antiprotons are
listed in Table \ref{tab.imp}.  They are crossed during acceleration,
if the number of spin precessions per revolution of the particles in
the ring is an integer ($\gamma G = k$, $k$: integer).  The resonance
strength depends on the vertical closed orbit deviation.
\begin{table}[hbt]
\begin{center}
\renewcommand{\arraystretch}{1.2}
\begin{tabular}{lrrrrr}
\hline
  $\gamma G$ & $E_{kin}$ (MeV) & {$P$ (MeV/c)} \\
\hline
2 & 108.4 & 463.8   \\ 
3 & 631.8 & 1258.7   \\ 
4 & 1155.1 & 1871.2  \\
5 & 1678.5 & 2442.6  \\ 
6 & 2201.8 & 2996.4  \\ 
\hline
\end{tabular}
\parbox{14cm}{\caption{\small Proton beam energies and momenta at
    which imperfection resonances occur in the CSR.}}\label{tab.imp}
\end{center}
\end{table} 
A solenoid with low magnetic field acting as a weak partial snake or
vertical correction dipoles can be utilized in the CSR to preserve the
polarization by exciting adiabatic spin flips.  Both methods are
successfully utilized in COSY and have the capability to overcome all
imperfection resonances in the momentum range of the CSR.

\subsubsection{Intrinsic Resonances} 
The number of intrinsic resonances depends on the superperiodicity $P$
of the lattice, which is given by the number of identical periods in
the accelerator.  CSR will be a synchrotron with a racetrack design
consisting of two 180$^{\circ}$ arc sections connected by straight
sections. One straight section will be modified to allow interactions
with the circulation beam in the HESR. In this case the
superperiodicity of the ring will be one.

If the straight sections are tuned as telescopes with 1:1 imaging,
giving a 2$\pi$ betatron phase advance, one then obtains for the
resonance condition $\gamma G = k \cdot P \pm (Q_y -2)$, where $k$ is
an integer and $Q_y$ is the vertical betatron tune.  The corresponding
intrinsic resonances in the momentum range of the CSR are listed in
Table \ref{tab.intr} for different superperiodicities $P$ and a
vertical betatron tune of $Q_y = 3.61$.

\begin{table}[hbt]
\begin{center}
\renewcommand{\arraystretch}{1.2}
\begin{tabular}{lrrrrr}
\hline
  $P$ & $\gamma G$ & $E_{kin}$ (MeV) & {$P$ (MeV/c)} \\
\hline
1,2 & $6-Q_y$ & 312.4 & 826.9 \\
1 & $-1+Q_y$ &  427.5 & 992.4  \\
1 & $7-Q_y$ &   835.6 &  1505.3 \\
1,2 & $0+Q_y$ & 950.7 & 1639.3 \\
1,2,6 & $8-Q_y$ & 1358.8 & 2096.5 \\
1 & $1+Q_y$ &  1473.9 & 2222.0 \\
1 & $9-Q_y$ &   1882.0 &  2659.4 \\
1,2 & $2+Q_y$ & 1997.1 & 2781.2 \\
1,2 & $10-Q_y$ & 2405.2 & 3208.9 \\ 
1 & $3+Q_y$ &   2520.3 & 3328.6 \\ 
\hline
\end{tabular}
\parbox{14cm}{\caption{\small Beam energy and momenta at which
    intrinsic resonances occur in the CSR for a working point
    $Q_y=3.61$, and superperiodicities of $P=1$, 2, and 6.}}
\label{tab.intr}
\end{center}
\end{table} 
Intrinsic resonances in the CSR can be compensated by fast tune jumps
with a similar system like in COSY.  Due to symmetry--breaking
installations like detector magnets and the arrangements for the
interaction zone, the superperiodicity of the lattice is reduced to
one, leading to ten intrinsic resonances.  It has been proved that the
tune--jump system of COSY can handle all ten intrinsic resonances in
this momentum range.  Therefore the same system is proposed for the
CSR.  Polarization measurements during acceleration confirm that the
proposed concept allows the acceleration of a vertically polarized
proton beam with polarization losses of only a few percent up to the
maximum momentum of COSY. Therefore it is the ideal system for the
CSR, which has very similar beam parameters.

\section{PAX Requirements on the HESR Design} \label{sec:HESR}
\subsection{Introduction}
For the Phase--II experimental program, polarized antiprotons stored
in the HESR with energies up to 14.1~GeV are required. The APR--CSR
system will provide antiprotons of up to 2.5~GeV. Therefore, polarized
antiprotons have to be accelerated up to the highest HESR energies.

Instead of employing SIS~100 for acceleration, it appears to be much
more economic to perform acceleration directly within the HESR. An
efficient accumulation of polarized antiprotons in HESR, however,
requires that at each cycle the remaining stack of polarized
antiprotons is decelerated down to the injection energy and, after
injection of additional antiprotons, is accelerated back to high
energies.

Compared with the present HESR design \cite{GSIdesign}, the following
additional features are required:

\begin{itemize}
\item Acceleration using an rf--cavity,
\item Slow ramping of the ring magnets,
\item Stable spin during ramping and flat top by means of Siberian snakes, 
\item Spin manipulation of stored protons and antiprotons by means of
  rf--dipoles and solenoids,
\item Injection of polarized antiprotons from CSR into the HESR, and
\item Guiding the high--energy HESR beam to the PAX experiment in the
  CSR straight section by means of a chicane system.
\end{itemize}

\subsection{Polarization Preservation}
Acceleration and storage of polarized proton and antiproton beams in
medium and high energy circular accelerator is complicated by numerous
depolarizing spin resonances.  In the following we discuss possible
scenarios to accelerate and store polarized beams in the HESR.

The spin motion in an external electromagnetic field is governed by
the so--called Thomas--BMT equation\cite{bmt2}, leading to a spin tune
of $\nu_{sp} = \gamma G$. \footnote{$G$ is the anomalous magnetic
  moment of the particle and $\gamma = E/m$ the Lorentz factor.  The
  $G$--factor is quoted as 1.792847337(29) for protons, 1.800(8) for
  antiprotons \cite{PDG}.}  In a strong--focusing ring like the HESR
imperfection and intrinsic spin resonances can depolarize the beam.

\subsubsection{Depolarizing Resonances}

In total 25 imperfection resonances ranging from $\gamma G = 4$ to
$28$, and 50 intrinsic resonances from $\gamma G = 16-Q_y$ to $16+Q_y$
for a vertical betatron tune of about $Q_y=12.2$ have to be crossed
during acceleration.  The corresponding imperfection and intrinsic
resonances in the momentum range of the HESR for a vertical working
point of $Q_y=12.14$ and superperiodicity $P=1$ are listed in Table
\ref{tab:impintr}
\begin{table}[hbt]
\begin{center}
\renewcommand{\arraystretch}{1.2}
\begin{tabular}{lrrrrr}
\hline
  $\gamma G = ...$ & $ \gamma G = ... \pm Q_y$ & $E_{kin}$ (GeV) & {$P$ (GeV/c)} \\
\hline
  &     16-  &  1.082 & 1.789  \\
4 &          &  1.155 & 1.871  \\
  &     -8+  &  1.228 & 1.953  \\
  &     17-  &  1.605 & 2.364  \\
5 &          &  1.678 & 2.443  \\
  &     -7+  &  1.752 & 2.521  \\
  &     18-  &  2.129 & 2.920  \\
6 &          &  2.202 & 2.997  \\
  &     -6+  &  2.275 & 3.073  \\
  &     19-  &  2.652 & 3.465  \\
7 &          &  2.725 & 3.541  \\
  &     -5+  &  2.798 & 3.617  \\
  &     20-  &  3.175 & 4.005  \\
8 &          &  3.248 & 4.080  \\       
...      & ...       & ...   & ... \\                   
   &     15+  &  13.265 &  14.172  \\
   &     40-  &  13.642 &  14.550  \\
28 &          &  13.715 &  14.624  \\
   &     16+  &  13.789 &  14.697  \\
\hline
\end{tabular}
\parbox{14cm}{\caption{\label{tab:impintr}\small Beam energy and
    momenta at which imperfection and intrinsic resonances occur in
    the HESR for a working point of $Q_y=12.14$.}}
\end{center}
\end{table}  
The strength of the resonances depends on the orbit excursions for
imperfection resonances and focusing structure of the lattice and beam
emittance for intrinsic resonances and is ranging from $10^{-2}$ to
$10^{-6}$ for the expected beam parameter.  Due to coupling introduced
by the 15~Tm solenoid of the Electron Cooler also strong coupling spin
resonances are excited.  The large number of resonances to be overcome
in the HESR makes it very hard to apply techniques of individual
manipulation of single spin resonances \cite{khiari,huang,bai}.
Siberian snakes seem to be to only option to guarantee a setup with
low polarization losses during acceleration.

\subsubsection{Siberian Snake with Combined Fields}
\begin{figure}[hbt]
\begin{center}
  \epsfxsize=0.40\linewidth\epsfbox{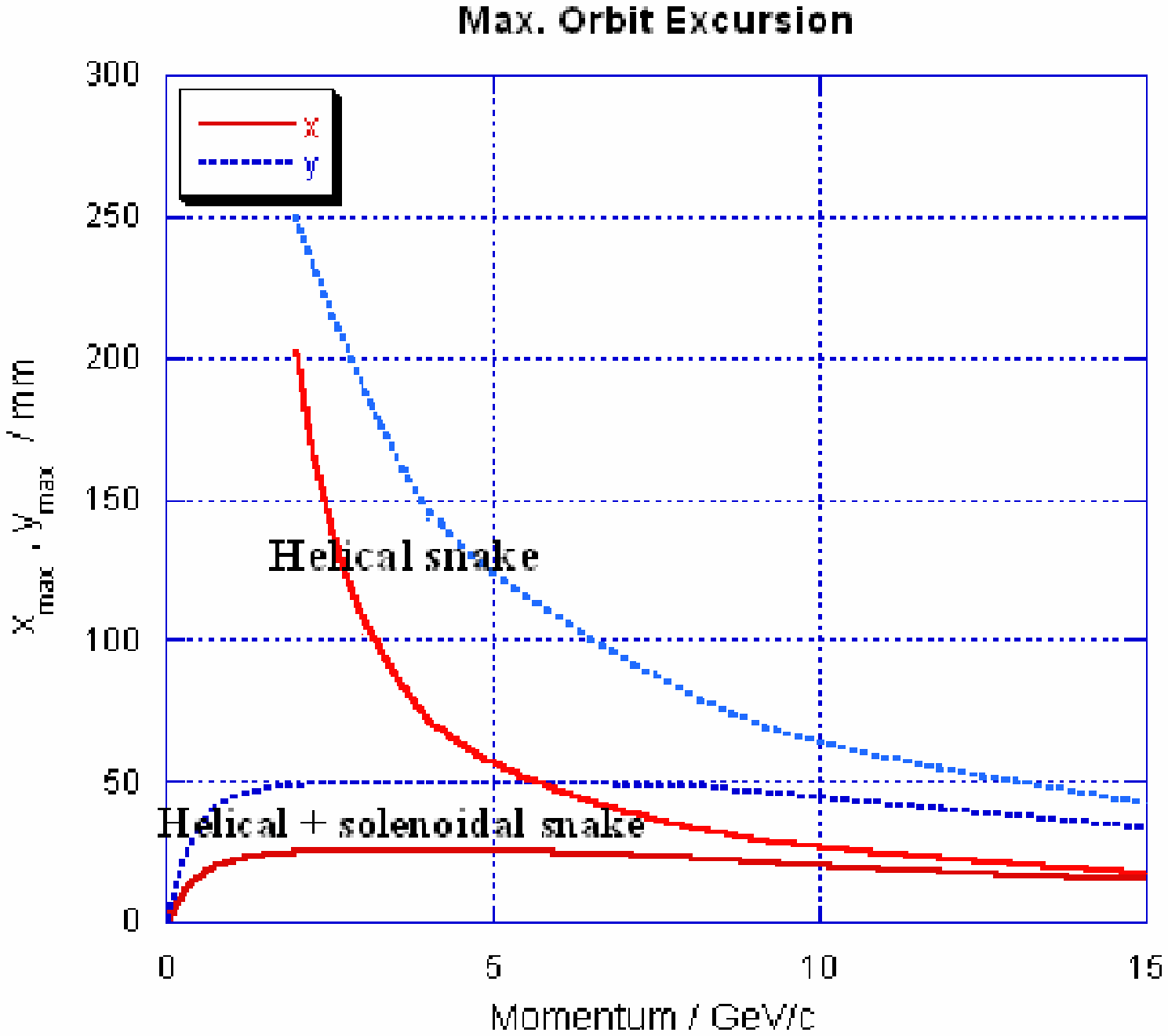}
  \epsfxsize=0.40\linewidth\epsfbox{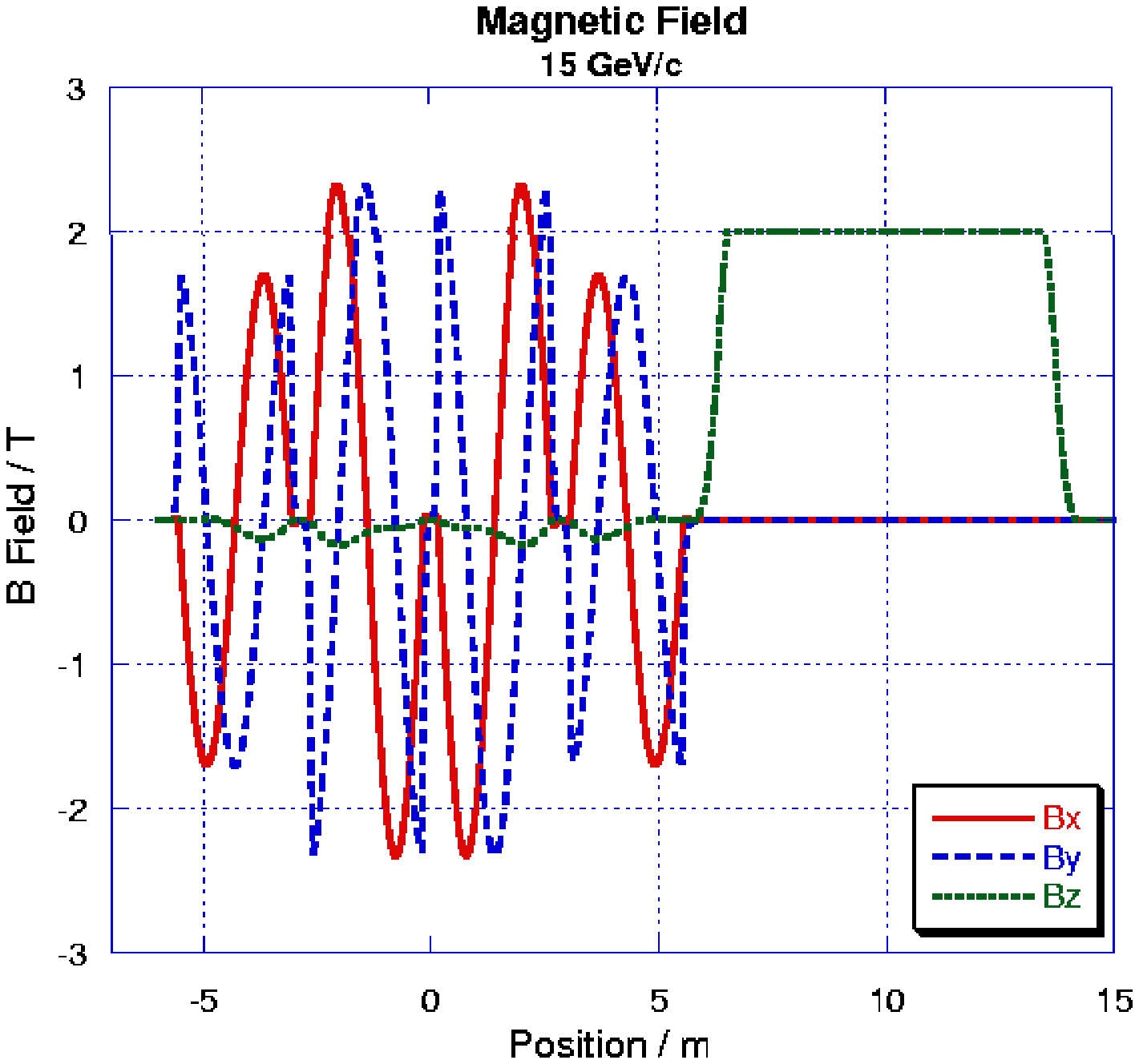}
\end{center}
\begin{center}
  \epsfxsize=0.40\linewidth\epsfbox{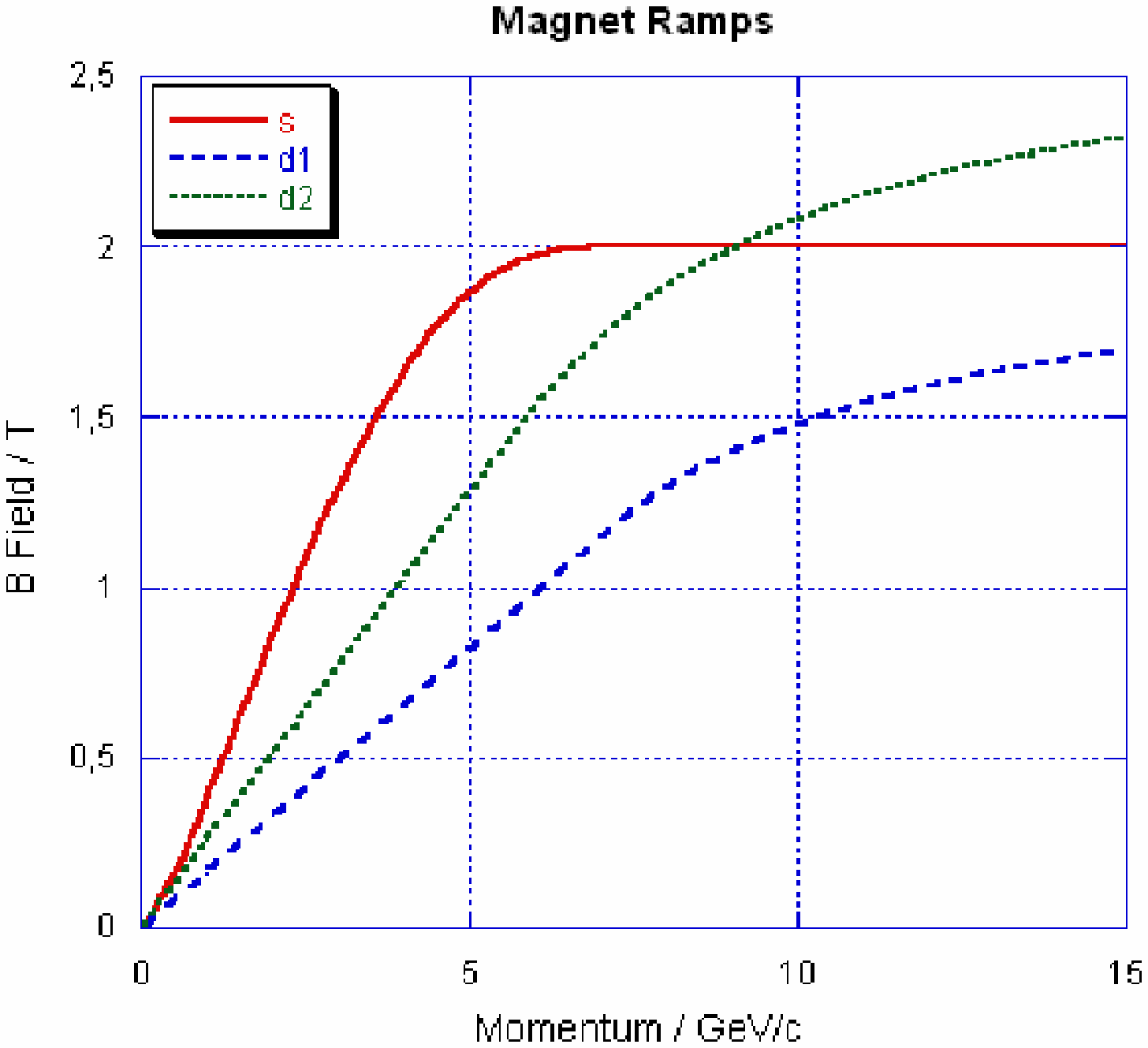}
  \epsfxsize=0.40\linewidth\epsfbox{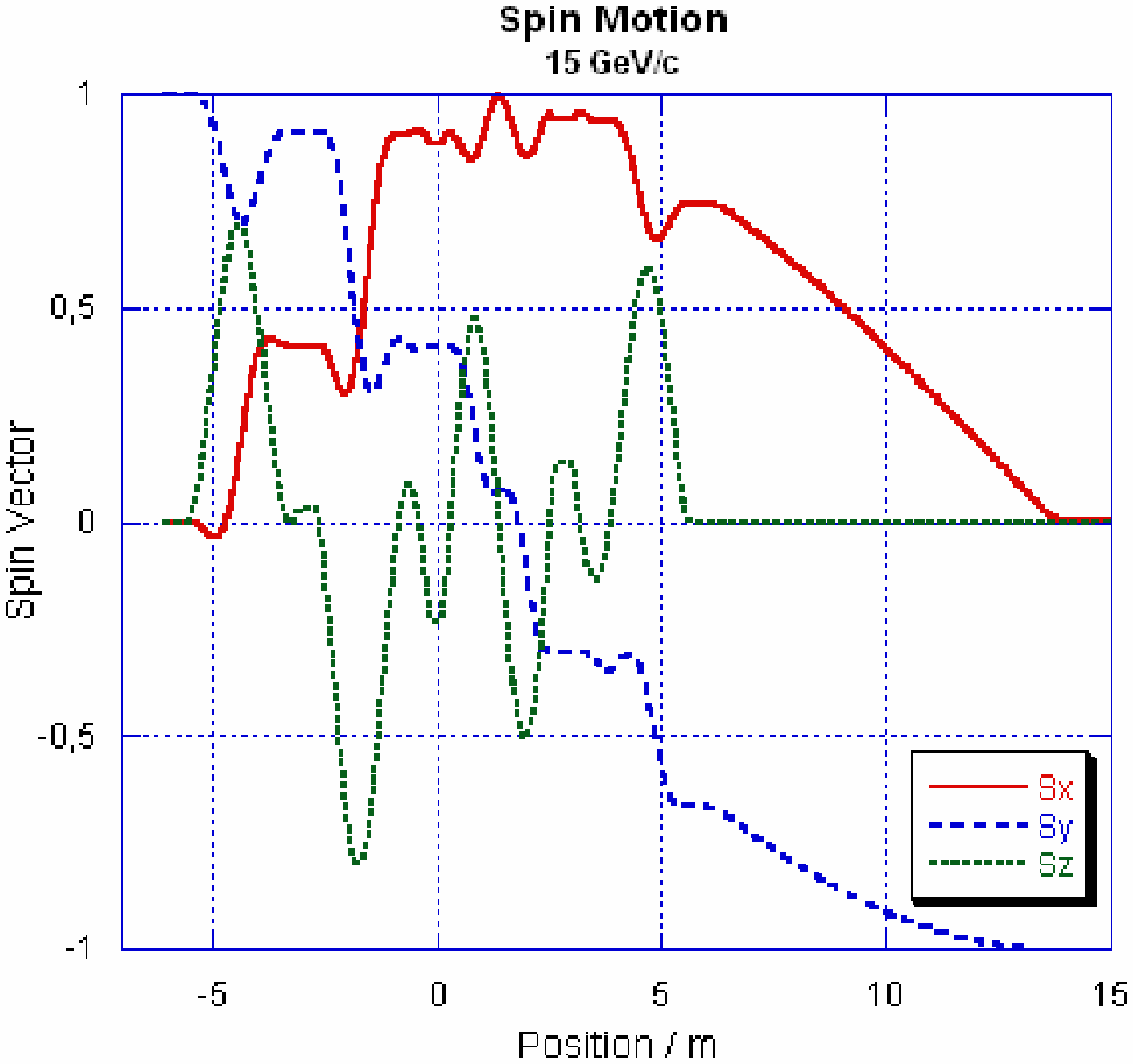}
  \parbox{14cm}{\caption{\label{snake1}\small Layout of a full
      Siberian snake with combined helical dipole and solenoidal
      magnetic fields.}}
\end{center}
\end{figure}

In the HESR momentum range it is difficult use a
RHIC--type~\cite{rsnake} helical dipole snake due to large orbit
excursions as shown in the upper left plot of Fig. \ref{snake1}.
A solenoidal field would require a pretty high integrated field
strength of roughly 60~Tm.  Therefore a magnet system with a
combination of both field types was investigated, consisting of four
RHIC--type helical dipole magnets with a maximum field of 2.5~T and a
15~Tm solenoid (see upper right plot in the same Figure). To provide a
full spin flip in the whole momentum range the snake magnets have to
be ramped according to the values given in the lower left plot, where
$s$ is the solenoid and $d1, d2$ are the two helical dipole field
values.  The resulting spin motion at 15~GeV/c is shown in the lower
right plot.  This magnet system provides a full spin flip in the whole
momentum range by keeping the maximum closed orbit excursion below
5~cm.  Spin rotation induced by the DC Cooler solenoid at any possible
field level can be compensated by the rampable 15~Tm snake solenoid,
if snake and Cooler are installed in the same straight section.

\subsubsection{Siberian Snake with Solenoidal Fields}
\begin{figure}[hbt]
\begin{center}
  \epsfxsize=0.8\linewidth \epsfbox{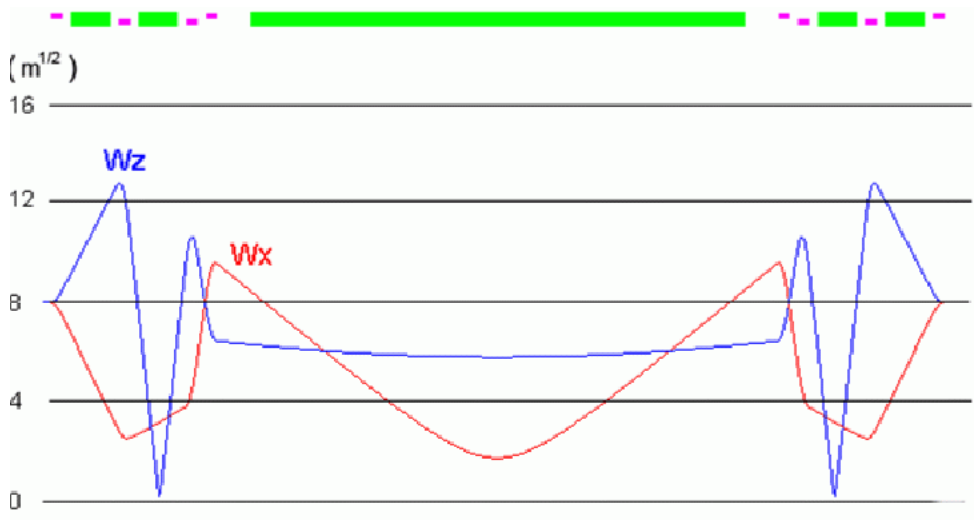}
  \parbox{14cm}{\caption{\label{snake2}\small Layout of a Siberian
      snake with solenoidal fields (top) and optical functions
      (bottom).}}
\end{center}
\end{figure}

The second proposed scheme contains four solenoids grouped on either
side of the Cooler (see upper sketch of Fig. \ref{snake2}) with the
same total integrated field strength of 15~Tm like the Cooler
solenoid.  From injection up to about 7.5~GeV/c all five solenoids
will provide a full snake.  At higher momenta they will work as
partial snake with about $50\%$ partial snake at top momentum.
To compensate for coupling, two groups of four quadrupole magnets are
needed with rotation angles up to $8.6$, $6.3$, $4.3$ and $4.3$
degree.  The rotation angles of the quadrupoles have to be adjusted
for different solenoid fields and beam momenta.  The whole magnet
insertion provides a betatron phase advance of $\pi$ and $2\pi$ in the
two transverse planes and has a total length of 56 m.  To preserve
polarization at first order spin resonances, the fractional part of
the betatron tune has to be kept close to integer in the range $0.75 <
Q_{frac} < 0.25$.  This scheme does not excite orbit excursion and
compensates for transverse phase space coupling.  Applying higher
integrated solenoidal field strength, it could also serve as a full
Siberian snake.\\

The most serious drawback of a combined field scheme is large orbit
excursion in the snake, which could be a major restriction for the
beam quality in the HESR.  Furthermore, ramping of the
super--conducting snake magnets remains to be solved.  Good field
quality of the superconducting ring magnets is essential to apply a
partial Siberian snake in order to keep the strength of higher--order
spin resonances small combined with high flexibility of the lattice
allowing for betatron tunes close to integer.  A decision for one of
the proposed schemes should be taken after intense particle and spin
tracking including field errors and technical layout of the snake
magnets.
\subsection{Spin Manipulation of Polarized Protons and Antiprotons}
Many polarized scattering experiments require frequent spin--direction
reversals (spin--flips) during storage of the polarized beam to reduce
their systematic errors.  For maximum luminosity it is necessary to
reverse the spins of the already stored antiprotons during the
accumulation process, as was done in many IUCF experiments.  Spin
resonances induced by either an rf--solenoid or rf--dipole are well
proven techniques to produce spin--flips in a controlled way.  Spin
flipping and spin manipulation of a stored beam was first studied in
the IUCF Cooler Ring at 270 MeV \cite{krisch1}. In 2002, the SPIN@COSY
collaboration was founded to continue these unique polarized beam
studies in the GeV--regime at COSY \cite{spinatcosy}.  Remarkably high
measured proton spin--flip efficiencies of $99.92 \pm 0.04 \%$ were
achieved by ramping the frequency of a strong ferrite--core
water--cooled RF dipole through an rf--induced spin resonance at
2.1GeV/c \cite{krisch2}.  The weak energy dependence of the
spin--resonance strength induced by transversal rf fields indicates
that only a slightly stronger rf dipole should allow efficient
spin--flips of polarized antiprotons up to the maximum energy of the
HESR.

\subsection{Interaction Region Design for the Asymmetric Collider}
To maximize luminosity, both the proton and the antiproton beam need
to be focused to small spot sizes at the interaction point (IP) of the
facility.  Both beams also have to be separated close to the IP into
their respective storage rings. Though a crossing angle would greatly
simplify the interaction region design, luminosity loss due to the
relatively long bunches makes it less desirable. We have therefore
designed an interaction region for head--on collisions, where beams are
magnetically separated, taking advantage of the unequal beam energies.

Due to the hourglass effect, the finite bunch length of some $30\,{\rm
  cm}$ results in a minimum reasonable $\beta$--function at the IP of
$\beta^{\ast}=0.3\,{\rm m}.$ With normalized emittances of
$\epsilon_{p,n}=1.7\pi\,\mu{\rm m}$ after cooling for the proton beam
and $\epsilon_{\bar{p},n}=20\pi\,\mu{\rm m}$ for the antiproton beam
and the requirement of equal beam sizes of both beams at the
interaction point to avoid emittance blow--up of the larger beam due to
the beam--beam effect, resulting $\beta$--functions at the IP are
$\beta_p^{\ast}=0.3\,{\rm m}$ for the proton beam and
$\beta_{\overline{p}}^{\ast}=1.0\,{\rm m}$ for the antiproton beam.
Table \ref{parameters} lists the design parameters of the interaction
region.
\begin{table}[hbt]
\begin{center}
\begin{tabular}{|l|c|}
\hline
bunch length $\sigma_s$ & 0.3\,m\\
proton normalized emittance $\epsilon_{p.n}$ & $1.7\pi\,\mu{\rm m}$\\ 
antiproton normalized emittance $\epsilon_{\overline{p}.n}$ & 
$20\pi\,\mu{\rm m}$\\
proton $\beta^{\ast}$ & $0.3\,{\rm m}$\\
antiproton $\beta^{\ast}$ & $1.0\,{\rm m}$  \\ 
\hline
\end{tabular}
\parbox{14cm}{\caption{\label{parameters}\small Parameter table.}}
\end{center}
\end{table}

With equal emittances in both transverse planes, the best beam--beam
performance is obtained with round beams at the IP. The required equal
$\beta$--functions in both planes are provided by low--$\beta$
quadrupole triplets near the interaction point. The quadrupoles for
the low--energy proton beam are actually common to both beams; an
additional vertical dipole field separates the two beams by deflecting
the low--energy proton beam to a larger angle than the high--energy
antiprotons. This field configuration is achieved by realization of
those magnets as superconducting quadrupoles with additional dipole
windings.

\begin{figure}
\begin{center}
  \epsfig{file=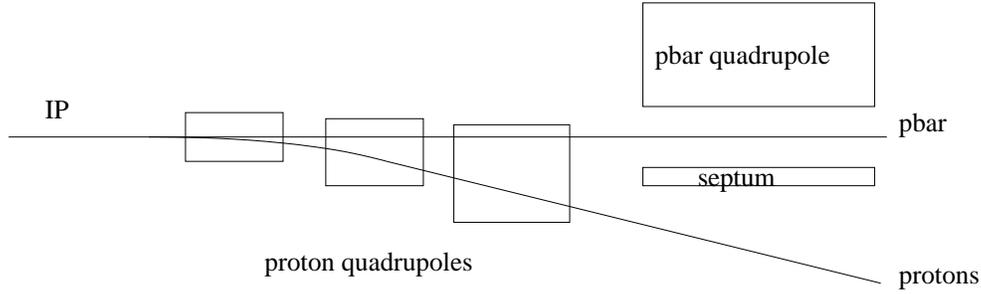,width=0.8\linewidth}
  \parbox{14cm}{\caption{\label{irscheme}\small Schematic drawing of
      the interaction region configuration (top view).}}
\end{center}
\end{figure}

The low--$\beta$ triplet quadrupoles in the antiproton ring are
designed as normal--conducting septum quadrupoles to minimize the
required beam separation at the location of the first magnet. This
configuration is schematically shown in Figure \ref{irscheme}.

\begin{figure}[hbt]
\begin{center}
  \epsfig{file=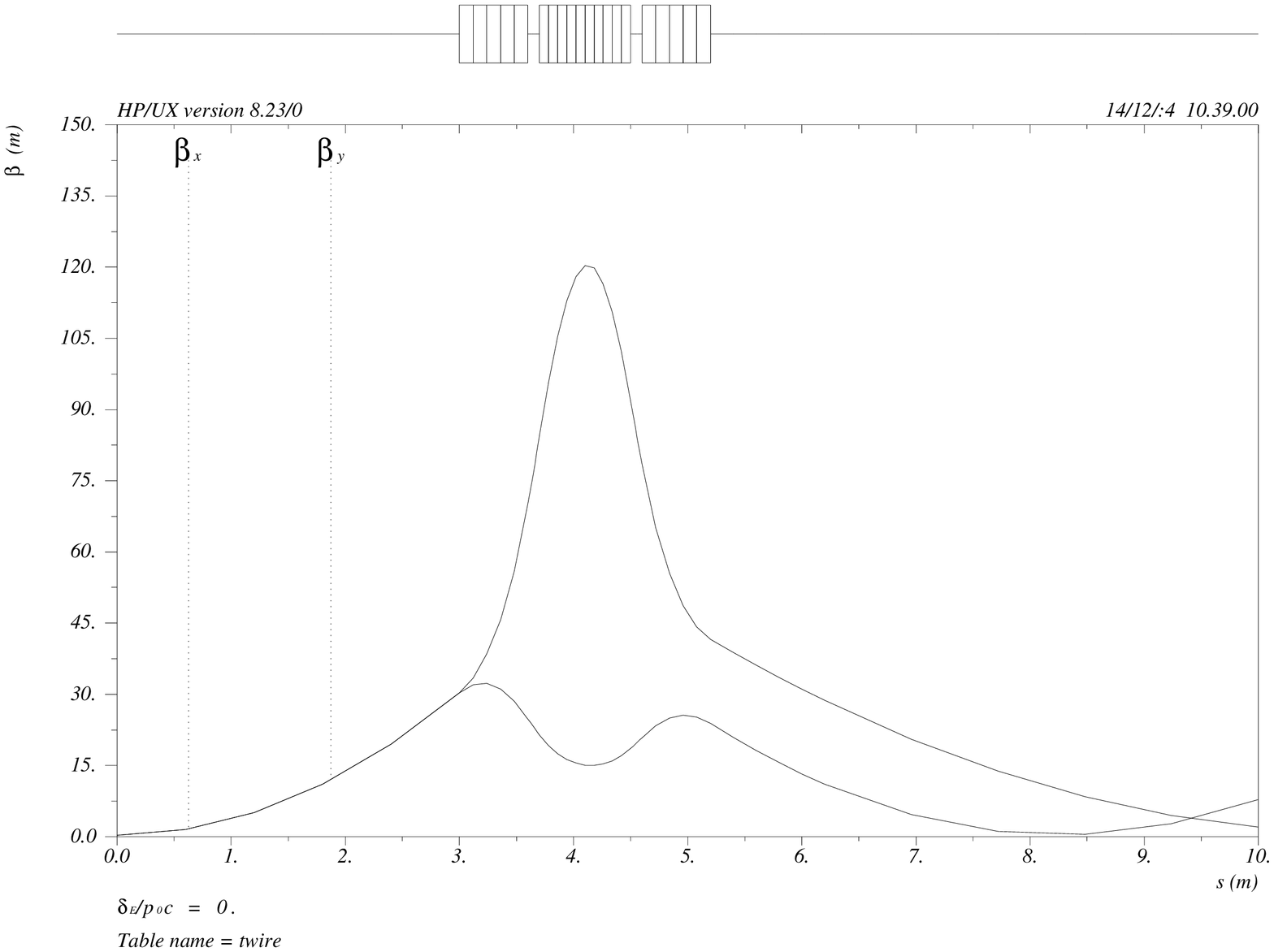, height=0.8\linewidth, angle=270}
  \parbox{14cm}{\caption{\label{protons}\small Proton low--$\beta$ lattice
      (one side only).  Focusing is provided by a superconducting
      quadrupole triplet, starting at a distance of $3.0\,{\rm m}$
      from the interaction point. The first and third magnet of the
      triplet are horizontally focusing, while the center quadrupole
      focuses in the vertical plane.}}
\end{center}
\end{figure}

\begin{figure}[hbt]
\begin{center}
  \epsfig{file=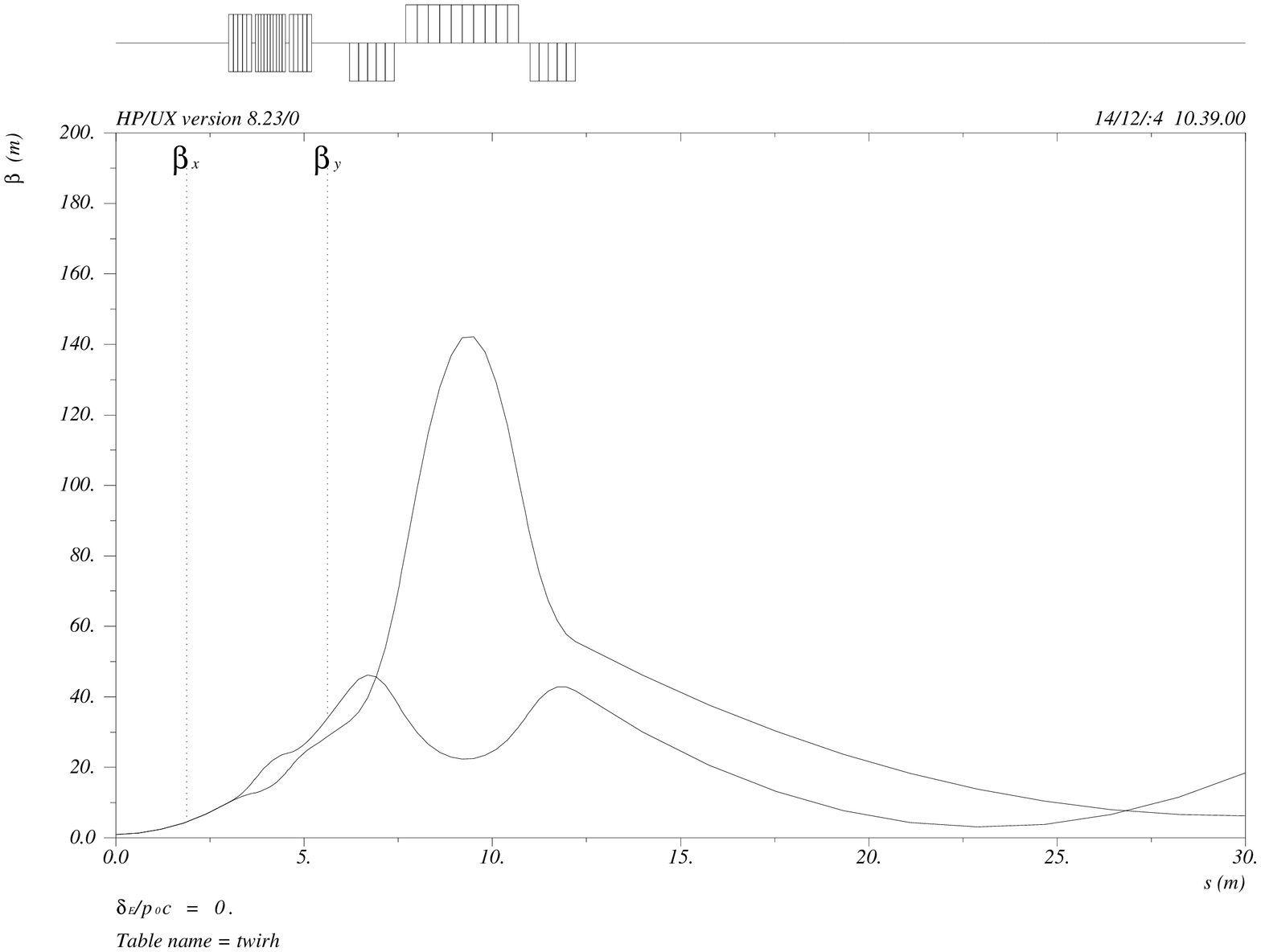,height=0.8\linewidth, angle=270}
  \parbox{14cm}{\caption{\label{antiprotons}\small Antiproton
      $\beta$--functions around the IP (one side only).  Focusing is
      provided by a normal--conducting septum quadrupole triplet.  The
      superconducting proton low--$\beta$ triplet on the left has only
      little effect on the antiproton optics due to the higher beam
      rigidity.}}
\end{center}
\end{figure}

Focusing for both beams is provided by quadrupole triplets. The shared
superconducting low--$\beta$ magnets for the low--energy proton beam
have peak fields of about $1.5\,{\rm T}$ for a beam pipe radius
sufficient to provide a minimum aperture of $12\sigma$ for both beams,
where $\sigma$ denotes the transverse rms beam size.
Figure~\ref{protons} shows the proton low--functions around the IP.
The normal--conducting septum quadrupoles for the high--energy
antiproton beam have a peak field below $1.0\,{\rm T}$ for a minimum
aperture of $12\sigma.$ The resulting $\beta$--functions are depicted
in Figure \ref{antiprotons}.
The septum itself is assumed to have a thickness of $5\,{\rm mm}$ in
the horizontal mid--plane of the magnet. This is achieved by a
triangular cut--out on the outside of the septum plate, which can be
tolerated in terms of saturation since the magnetic field on the
septum plate in the horizontal mid--plane of the magnet vanishes.
Taking into account the wall thickness of the two beam pipes at the
septum, the required separation of the two beams there is
approximately
\begin{eqnarray}
d&=&12\sigma_p+12\sigma_{\overline{p}}+10\,{\rm mm}\\
&=&67\,{\rm mm}.
\end{eqnarray} 
This is achieved by a dipole field of $0.14\,{\rm T}$ over the total
length of the superconducting quadrupoles.

With the parameters given in Table \ref{parameters}, the resulting
beam--beam parameter
\begin{eqnarray}
\xi=\frac{r_p}{4\pi\sigma^2}\cdot\frac{\beta^{\ast}}{\gamma}\cdot N
\end{eqnarray}
can be calculated.  Here, $r_p$ denotes the classical proton radius,
$\gamma$ the Lorentz factor of the beam under consideration, and $N$
the number of protons (antiprotons) per bunch.  For $N=1.0\cdot
10^{11}$ protons and antiprotons, the resulting beam--beam tune shift
is $\xi_p=6.25\cdot 10^{-3}$ for the proton beam and
$\xi_{\overline{p}}=7.5\cdot 10^{-3}$ for the antiproton beam, which
seems to be realistically achievable, based on experience at existing
hadron colliders.

\subsubsection{Luminosity Estimate for the Asymmetric Collider}
The resulting luminosity for a single proton bunch colliding with
three antiproton bunches of equal intensity of
$N_p=N_{\overline{p}}=1.0\cdot 10^{11}$ particles yields
\begin{eqnarray}
{\cal{L}}&=&\frac{N_pN_{\overline{p}}f_c}{4\pi\sigma}\\
&=&1\cdot 10^{30}\,{\rm cm}^{-2}{\rm sec}^{-1},
\end{eqnarray}
where $f_c$ denotes the bunch crossing frequency.

It should be emphasized here that in order to maximize luminosity the
circumference ratio of the two storage rings needs to be reflected in
the number of bunches circulating in each ring; hence the number of
antiproton bunches needs to be three times larger than the number of
proton bunches.
\subsubsection{Intensity dependent Limitations}
Due to the high bunch charge of $N=1.0\cdot 10^{11}$ particles per
bunch in conjunction with low beam energies, intensity dependent
effects need to be studied carefully. Here we list some estimates of
high intensity effects that are to be expected at the proposed
facility.
\paragraph{Touschek Lifetime}
Single scattering of particles within the same bunch leads to momentum
transfer from the transverse into the longitudinal plane, where
particles get lost if their longitudinal momentum exceeds the momentum
acceptance of the machine. The resulting Touschek lifetime is
expressed as \cite{piwi}
\begin{eqnarray}
\tau_{\rm
Touschek}^{-1}&=&\frac{Nr_p^2c}{8\pi\sigma_x\sigma_y\sigma_s}
\cdot\frac{\left(\frac{\Delta p}{p}\right)^{-1}}{\gamma^2}\cdot
D(\xi).
\end{eqnarray}
Here, $r_p$ denotes the classical proton radius, while $c$ is the
velocity of light. The function $D(\xi)\approx 0.3$ for realistic
parameter ranges \cite{piwi}.  Assuming a momentum acceptance of
$\Delta p/p=0.01$ and using average values of
$\sigma_x=\sigma_y=3.5\,{\rm mm}$ and $\sigma_s=0.3\,{\rm m}$ for the
transverse and longitudinal rms beam sizes, respectively, this yields
a Touschek lifetime exceeding a year. This effect is therefore of no
concern for the facility.
\paragraph{Intrabeam Scattering} 
Multiple scattering of particles within the same bunch, also called 
intrabeam scattering, leads to emittance growth in all three dimensions. 
Growth rates are calculated as \cite{piwi}
\begin{eqnarray}
\frac{1}{T_p}&=&\left\langle A\frac{\sigma_h^2}{\sigma_p^2}f(a,b,q)
\right\rangle,\\
\frac{1}{T_x}&=&\left\langle A\left[f\left(\frac{1}{a}, \frac{b}{a}, 
\frac{q}{a}
\right)+\frac{D_x\sigma_h^2}{\sigma_{x\beta}^2}f(a,b,q)\right]\right\rangle,\\
\frac{1}{T_y}&=&\left\langle A\left[f\left(\frac{1}{b}, \frac{a}{b}, 
\frac{q}{b}
\right)+\frac{D_y\sigma_h^2}{\sigma_{y\beta}^2}f(a,b,q)\right]\right\rangle,
\end{eqnarray}
with 
\begin{eqnarray}
A&=&\frac{r_p^2cN}{64\pi^2\beta^3\gamma^4\epsilon_x\epsilon_y\sigma_s\sigma_p},\\
\frac{1}{\sigma_h^2}&=&\frac{1}{\sigma_p^2}+\frac{D_x^2}{\sigma_{x\beta}^2}
+\frac{D_y^2}{\sigma_{y\beta}^2},\\
a&=&\frac{\sigma_h\beta_x}{\gamma\sigma_{x\beta}},\\
b&=&\frac{\sigma_h\beta_y}{\gamma\sigma_{y\beta}},\\
q&=&\sigma_h\beta\sqrt{\frac{2d}{r_p}}.
\end{eqnarray}
$\epsilon_x$ and $\epsilon_y$ denote the horizontal and vertical emittance, 
respectively, while $d$ is the smaller of the horizontal and vertical rms 
beam sizes.

Using the beam emittances given in Table \ref{parameters},
$\sigma_p=1.0\cdot 10^{-3},$ and average $\beta$--functions
$\beta_x=\beta_y=30\,{\rm m},$ the value of the function $f(a,b,q)$
can be obtained from Ref. \cite{piwi} as $f(a,b,q)\approx -100.$ This
yields a longitudinal emittance growth rate of
\begin{eqnarray}
T_p^{-1}\approx 10^{-3}\,{\rm sec}^{-1},
\end{eqnarray}
and results of similar magnitude for the transverse rates $T_x^{-1}$ and $T_y^{-1},$
which is comparable to the design cooling times. This is a major concern for the 
proposed facility and requires a thorough investigation, taking into account the 
actual accelerator lattices.

\section{Polarimetry} \label{sec:polar}
The beam and target polarization will be determined by the following
scheme: First the target polarization using an unpolarized antiproton
beam is established by either one of two methods:
\begin{itemize}
\item[(1)] with reference to a suitable sampling polarimeter of the
  Breit--Rabi \cite{HERMES-tgt-pol} or Lamb--shift \cite{LSP-pol}
  type, which spin--analyzes a small fraction of atomic hydrogen
  extracted from the target cell.
\item[(2)] elastic proton--antiproton scattering data at low energies
  (500--800~MeV) where analyzing power data from PS172 \cite{Kunne}
  are available. Scattering data of lower precision extend up to
  2.5~GeV \cite{albrow}.
\end{itemize}
This allows one to calibrate a suitable detector asymmetry, derived
from elastic scattering, in terms of an effective analyzing power.
Since target and beam analyzing power in $\bar p p$ scattering are
identical, the polarization of the beam can now be measured with an
unpolarized target (e.g. by injecting unpolarized hydrogen gas into
the cell).  When subsequent fills of the HESR are made with different
beam energies, it is straightforward to establish polarization
standards at any energy within the HESR range by exploiting the fact,
that the target polarization is constant with time -- or monitored by
the sampling polarimeter -- and independent of energy
\cite{pollock-calib}.

\cleardoublepage
\part{Staging of Experiments \label{sec:staging}}
\pagestyle{myheadings} 
\markboth{Technical Proposal for ${\cal PAX}$}{Part III: Staging of Experiments}
The PAX collaboration proposes an approach that is composed of
different stages.  During these the major milestones of the project
can be tested and optimized before the final goal is approached: a
polarized proton--antiproton asymmetric collider, in which about
3.5~GeV/c polarized protons will collide head--on with polarized
antiprotons with momenta up to 15 GeV/c.

\section{Preparatory Phase}
\subsection{Accelerator: Design and Construction of the Antiproton Polarizer Ring}
Tuning and commissioning of the APR will require a beam of polarized
protons.  Such a beam and a hall including infrastructure are readily
available at COSY--J\"ulich. This makes Institut f\"ur Kernphysik of
the Forschungszentrum J\"ulich the ideally suited site for the design,
construction and testing of the APR.
\subsection{Physics in the Preparatory Phase}
The calculation of the polarization transfer employed to polarize the
antiprotons is text book QED physics \cite{QED-physics}. The
polarization transfer technique is at the core of an extensive physics
program at JLAB dedicated to experiments on the separation of the
charge and magnetic form factors of the proton \cite{JlabFF}. The
existence of the effect has been verified in the FILTEX experiment at
TSR--Heidelberg in 1992 with a 23 MeV proton beam.  For antiprotons the
optimal energy is around 50 MeV (see Sec. 8.3).  We don't consider
necessary a further demonstration of the validity of this fundamental
QED derivation and, unless specifically requested by the QCD--PAC, we
would not embark on such endeavor. Instead, we would very much prefer
to concentrate on the design, construction, commissioning and direct
proof with the APR that protons can be polarized to a high degree with
a design energy of 50~MeV.  If however, the QCD--PAC will ask us for a
pre--APR test, this can be carried out at COSY.

\begin{itemize}
  
\item A verification of $\sigma_{EM_\perp}$ at 40, 70 and 100 MeV is
  possible using the polarized internal target at the ANKE interaction
  point.
    
  The measurement can be performed by injecting pure states $\ket 1$
  or $\ket 3$ in a weak transverse target guide field (10 G). In this
  situation the electron target polarization $Q_e$ is equal to the
  proton target polarization $Q_p$ and $Q_p$ can be measured by $pp$
  elastic scattering by using the ANKE spectator detector system
  \cite{Schleichert:2003}.
\end{itemize}

\subsection{Development of Polarized Sources}
The polarization mechanism relies on an efficient, high--intensity
source for polarized hydrogen atoms. In the PAX collaboration, most of
the world--expertise on polarized sources is already present
(Erlangen, Ferrara, Gatchina, J\"ulich, Madison).  A program for the
development of a new generation of high--intensity atomic beam sources
has been already started in Ferrara and it will be pursued and
extended during the preparatory phase of the PAX experiment.

\section{Phase--I \label{sec:phaseI}}
\subsection{Accelerator: Transfer of APR and CSR to FAIR}
APR and CSR will be placed inside the HESR.  The straight sections of
CSR and HESR are parallel, whereby an additional IP, independent of
the PANDA IP, is formed.

\subsection{Physics in Phase--I}
A beam of unpolarized or polarized antiprotons with momentum up to 3.5
GeV/c in the CSR ring, colliding on a polarized hydrogen target in the
PAX detector will be available.  This phase is independent of the HESR
performance.  This first phase, at moderately high--energy, will allow
for the first time the measurement of the time--like proton form
factors in single and double polarized reactions from close to
threshold up to 3.5 GeV/c.  It will be possible to determine several
(single and double) spin asymmetries in the elastic $p \bar{p} \to p
\bar{p}$ process.  By detecting back scattered antiprotons one can
also explore hard scattering regions of large $t$: in the
proton--proton scattering reaching the same region of $t$ requires a
twice higher energy.  This would allow us to carry out the
measurements of form factors with a fixed polarized hydrogen target
bombarded by antiprotons orbiting in the CSR with momenta of
3.5~GeV/c.  The CSR would be fed with antiprotons from the APR.

\section{Phase--II\label{sec:phaseII}}  
\subsection{Accelerator: HESR modifications to collider mode or to polarized internal
  target.}
A chicane for CSR and HESR has to be built to bring the proton beam of
the CSR and the antiproton beam of the HESR to a collision point at
the PAX IP.

\subsection{Physics in Phase--II}
This phase will allow the first ever direct measurement of the quark
transversity distribution $h_1$, by measuring the double transverse
spin asymmetry $\Att$ in Drell--Yan processes $p^{\uparrow}
\bar{p}^{\uparrow} \rightarrow e^+ e^- X$ as a function of Bjorken $x$
and $Q^2$ (= $M^2$).  Two possible scenarios might be foreseen to
perform the measurement.
\begin{itemize}
\item[(a)] A beam of polarized antiprotons from 1.5 GeV/c up to 15
  GeV/c circulating in the HESR, colliding on a beam of polarized
  protons of 3.5 GeV/c circulating in the CSR.  This scenario requires
  to demonstrate that a suitable luminosity is reachable.  Deflection
  of the HESR beam to the PAX detector in the CSR is necessary (see
  Fig.~\ref{fig:CSRring}).  By properly varying the energy of the two
  colliding beams, this setup would allow a measurement of the
  transversity distribution $h_1$ in the valence region of $0.1<x<0.8$
  with corresponding $Q^2$ in the range $4<Q^2<100$ $\rm GeV^2$ (see
  Fig.~\ref{Figphysics}). $\Att$ is predicted to be larger than 0.2 in
  the full kinematic range, see Fig.~\ref{Figphysics}, and the cross
  section is large enough to get $\sim 2000$ events per day at a
  luminosity of $5\cdot 10^{30}$ $\rm cm^{-2}s^{-1}$.  Such an
  experiment can be considered, for $h_1$, the analogous of polarized
  DIS for the helicity distribution $\Delta q$; the kinematical
  coverage in $(x,Q^2)$ will be similar to that of the HERMES
  experiment.
  
\item[(b)] Should the requested luminosity for the collider not be
  reachable, a fixed target experiment can be performed.  A beam of 22
  GeV/c (15 GeV/c) polarized antiprotons circulating in the HESR, can
  be used to collide on a polarized internal hydrogen target.  Also
  this scenario requires the deflection of the HESR beam to the PAX
  detector in the CSR (see Fig.~\ref{fig:CSRring}).  A theoretical
  discussion of the significance of the measurement of $\Att$ for a 15
  GeV/c beam impinging on a fixed target is given is Refs.
  \cite{abdn} and \cite{gms} and the recent review paper
  \cite{brodsky}.  This measurement will explore the valence region of
  $0.3<x<0.8$ with corresponding $4<Q^2<16$ ${\rm GeV}^2$, see
  Fig.~\ref{Figphysics}.  In this region $\Att$ is predicted to be
  large (of the order of 0.3, or more) and the expected number of
  events can be of the order of 2000 per day.
\end{itemize}

\cleardoublepage
\part{Detector\label{partIV}}
\pagestyle{myheadings} 
\markboth{Technical Proposal for ${\cal PAX}$}{Part IV: Detector}
\section{Requirements and Design Considerations}
Since a definite choice has not been made yet on the accelerator
configuration (fixed target or collider), it might be premature at
this stage to define the detector details. Still some general
considerations can be anticipated. What we present here is just meant
as a conceptual detector design suitable for the PAX physics program.
Possible alternative scenarios will be discussed at the end in
Sec.~\ref{sec:alt-scen}.

In the following we focus on the most challenging task of the PAX
physics program: the measurement of the Drell--Yan reaction to access
the transversity distribution $h_1$. A detector optimized for this
task can be designed to be suitable to achieve also the other goals of
the PAX experimental program.

In the detector concept, presented below, the following criteria have
been pursued:
\begin{itemize}
\item optimize the acceptance at large angles. The double spin
  asymmetry of the $q \bar q \to \ell^+ \ell^-$ QED elementary
  process,
\begin{equation}
\hat a_{TT}= {\sin^2\theta\over 1 + \cos^2\theta} \> \cos 2\phi \,,
\end{equation}
maximizes the sensitivity of the measured asymmetry $\Att$ for the
transversity distribution $h_1$ for 90--degree scattering in the
center--of--mass of the partonic system, where $\sin^2\theta\sim 1$
(see Eq.~(\ref{att}). We note that also the $\overline{p}p$ elastic
scattering at 90$^\circ$ c.m.  and the form--factor measurements
benefit from a large angle detector;
\item trigger efficiently on the rare Drell--Yan events. At the PAX
  energy the few nb Drell--Yan cross section should be clearly identified from 
  a total $\overline{p}p$ cross--section of about 50 mb;
\item cope with the overwhelming background. The lepton identification
  should provide a rejection factor of the order of $\Ord{4}$ to
  $\Ord{5}$ against hadrons. Secondary leptons, produced in meson
  decays and in secondary interactions in the detector material,
  should be vetoed as well;
\item provide of the order of 1~\% resolution for the invariant mass
  of the lepton pairs, in order to efficiently distinguish the
  contribution of the resonances ($\JPsi$ and $\psi$) from the
  continuum;
\item (if the collider option would be pursued) be compatible with the
  asymmetric collider lattice. In particular any effect of the
  spectrometer magnet should be perfectly compensated in order to not
  degrade the beam polarization;
\item provide a unique facility as complete as possible and flexible
  to allow the study of auxiliary processes and additional physical
  channels which might become interesting during the next 10 years.
\end{itemize}

In order to best match the above requirements, the PAX detector is
designed to measured electron--positron pairs of large invariant mass.

\subsection{Physical Channels}
To reveal rare reactions like the Drell--Yan process and the
$\overline{p}p\rightarrow e^+e^-$ annihilation, the PAX detector has
to be conceived as a large acceptance apparatus capable of
unambiguously identifying electron--positron pairs of large invariant
mass and precisely measuring their momenta. The detector has to be
able to measure electron pairs with large opening angle, in a wide
kinematic range with good angular and energy resolution.  A clear
particle identification is required to separate the electrons of the
wanted processes from the large pion background.

Reactions characterized by two--body hadronic final states like
elastic scattering, present a higher cross--section and put less
stringent constraints on the detector design. They can be identified
by measuring scattering angles and momenta of the hadronic particles
by employing coplanarity and total momentum conservation.  Finally,
the PAX detector can measure the energy of gammas from radiative
processes and $\pi^0$ and $\eta$ decays.

\subsection{Particle Identification}
The Drell--Yan production rate is of the order of $\Ord{-7}$ of the
total $\overline{p}p$ reaction rate and results in a low yield of the
$e^+e^-$ signal per interaction. In order to maximize the dilepton
detection efficiency, the PAX spectrometer must provide a large
geometrical acceptance. At the same time, the high interaction rate
(of the order of 10$^6$~s$^{-1}$) together with the hadron
multiplicity represents a serious challenge to the trigger system
which has to select the events containing the lepton tracks. An
accurate lepton identification can only be achieved by detectors which
are highly insensitive to the large flux of hadrons.  In order to
minimize the background from lepton misidentification at a typical
prevailing $e/\pi$ ratio of $\Ord{-4}$, redundant recognition of
lepton tracks is essential. For the considered range of momenta
(between $0.5$ and $10$ GeV/c) electrons offer the advantage with
respect to muons of that they can be identified in a hadron blind gas
threshold \Cer detector.  This device can be operated on a fast time
scale to meet the stringent trigger requirements in the high--rate
high--multiplicity environment.  Additional discrimination against
pions can be provided both by the cluster lateral profile in the
electromagnetic calorimeter (CAL) and by the $E/p$ ratio between the
energy $E$ measured in the CAL and the momentum $p$ measured in the
spectrometer.  With these constraints, the required rejection factor
of the order of $10^{10}$ against hadronic events (corresponding to
$10^5$ for single track events) is achievable, as demonstrated in
other experiments \cite{E835nbarn, PS170}.

\subsection{Magnetic Field Configuration}

The spectrometer magnet has to be compatible with the \Cer detector
which needs to work in a field free region. A toroid configuration
satisfies this requirement resulting in a negligible fringe field both
along the beam line and external volume where the \Cer detector is
located.  A toroid field is always orthogonal to the particle
momentum, hence the bending effect is optimized regardless of the
scattering angle.  An eight--coil configuration can be designed to
give excellent coverage over the azimuthal angles, facilitating the
detection of the $\cos(2\phi)$ modulation of the transverse asymmetry,
covering the region around $2\phi=n\pi$, where the sensitivity is
largest (see Fig.~\ref{innerPAX}).  The azimuthal acceptance is
optimized by placing service and support structures of the various
detectors in the shadow of the toroid coils.  The detectors are
arranged in an azimuthally eight--fold segmented, frustum--like
geometry.  For compactness, the inner tracking close to the
interaction point is provided by silicon strip detectors. The outer
tracking, behind the magnet, is provided by two sets of conventional
drift--chambers. The \Cer detector is placed outside the magnet, in
between the two sets of drift--chamber, fitting the $0.6-0.8$~m
tracking arm required to obtain the $\sim1$~\% momentum resolution
with a typical spatial resolution of $\sim200$~$\mu$m.

\subsection{Mass Resolution}
A lepton pair invariant mass resolution close to 1~\% is required to
isolate the charmed resonance signals from the continuum.  This value
necessitates that the momentum resolution is of the same order of
magnitude for lepton tracks with a momentum range between 0.5 and 10
GeV/c. This large momentum range leads to a non--focusing geometry
with a transverse momentum kick up to $\Delta p\sim 0.3$ GeV/c.  The
maximum value of $\Delta p$ is required in the forward region where
the particle momenta are larger and is provided by an integrated field
along the particle trajectory of $Bl\sim 1$ Tm. In order to limit the
cost of the external electromagnetic calorimeter, the detector should
be compact: by assuming a value of $l=0.7$ m in the forward direction,
the required magnetic field is of the order of 1.5 T.  A transverse
momentum kick of $\Delta p\sim 0.3$ GeV/c puts constraints on the
position resolution of the tracking detectors to achieve the required
momentum resolution. At a momentum of 10 GeV/c, the deflection angle
equals about $2^\circ$. A model calculation shows that an uncertainty
in the position lower than $30$ $\mu m$ ($200$ $\mu m$) in the inner
(outer) tracking region is needed to obtain a relative momentum
resolution of about 1 \%. This resolution can be provided by
conventional silicon strip detectors (SiD) close to the target and
drift--chamber modules (DCH) outside the magnet. The high spatial
resolution provided by the silicon detector will help to identify the
background leptons from secondary vertex (like $D^\pm$ decays).  The
required momentum resolution can only be achieved by keeping multiple
scattering in the magnetic field region small. If only multiple
scattering is taken into account, the momentum resolution of $\sim 1$
\% at the smallest accepted momenta of $0.5$ GeV/c requires an
effective thickness less than $0.05$ $X_0$.  The amount of material in
the tracking region should be minimized also to reduce the gamma
conversion probability and energy loss by radiation. The first active
tracking layers placed inside the target vacuum chamber can be used to
veto gamma conversions as close as possible to the interaction point.

\subsection{General Remarks}
The present conceptual detector design is based on existing
experiments which have been proven to perform well measuring
final--state particles and energies similar to the ones anticipated at
PAX.

Once the PAX experiment is approved, each of the detector components
will be further optimized depending on the chosen beam configuration.
The performance of the detectors will finally depend (and benefit)
from the development in technology over the course of the next ten
years.  Additional detectors can also be implemented to enhance the
flexibility of the PAX instrument. As an example, hadron
identification at forward scattering angles could be considerably
improved by adding an internally--reflecting ring--imaging \Cer
detector~\cite{BABARdet}.

The PAX detector will be mounted on a platform which can be moved on
rails in and out of the beam line.  For simplicity, in the following
we will refer only to Drell--Yan processes, the argument holds for
$\JPsi$ decays and time--like electromagnetic form factors as well.

\section {Overview of the PAX Spectrometer}  

\begin{figure}[htb]
  \centering \includegraphics*[width=0.8\linewidth]{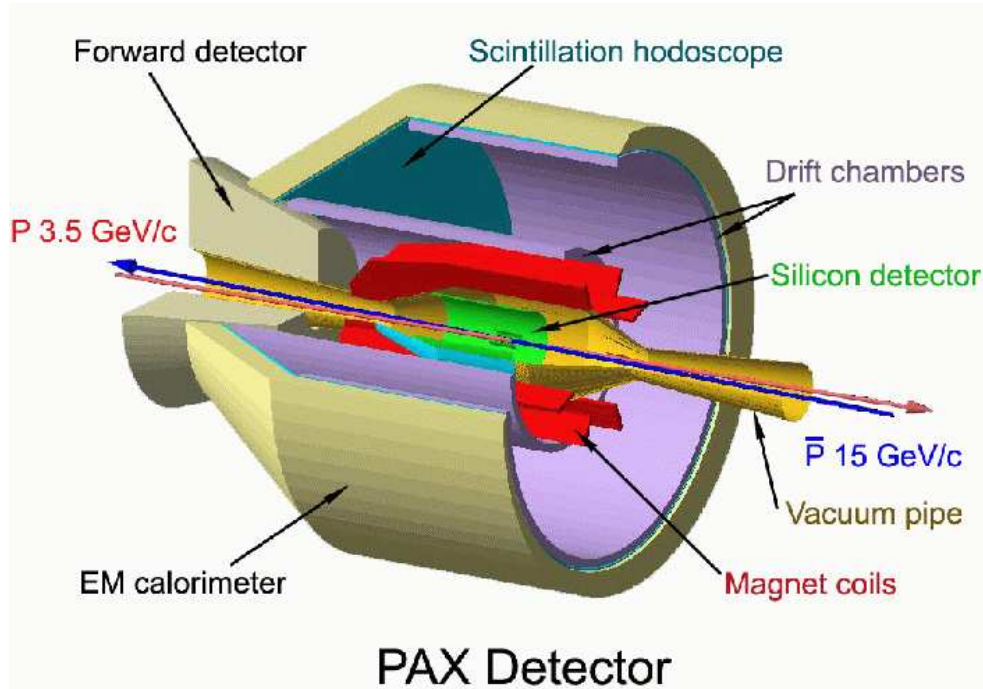}
  \parbox{14cm}{\caption{\small \label{artistic} Artists view of the
      PAX spectrometer.}}
\end{figure}

\begin{figure}[h]
  \centering \includegraphics*[width=0.79\linewidth]{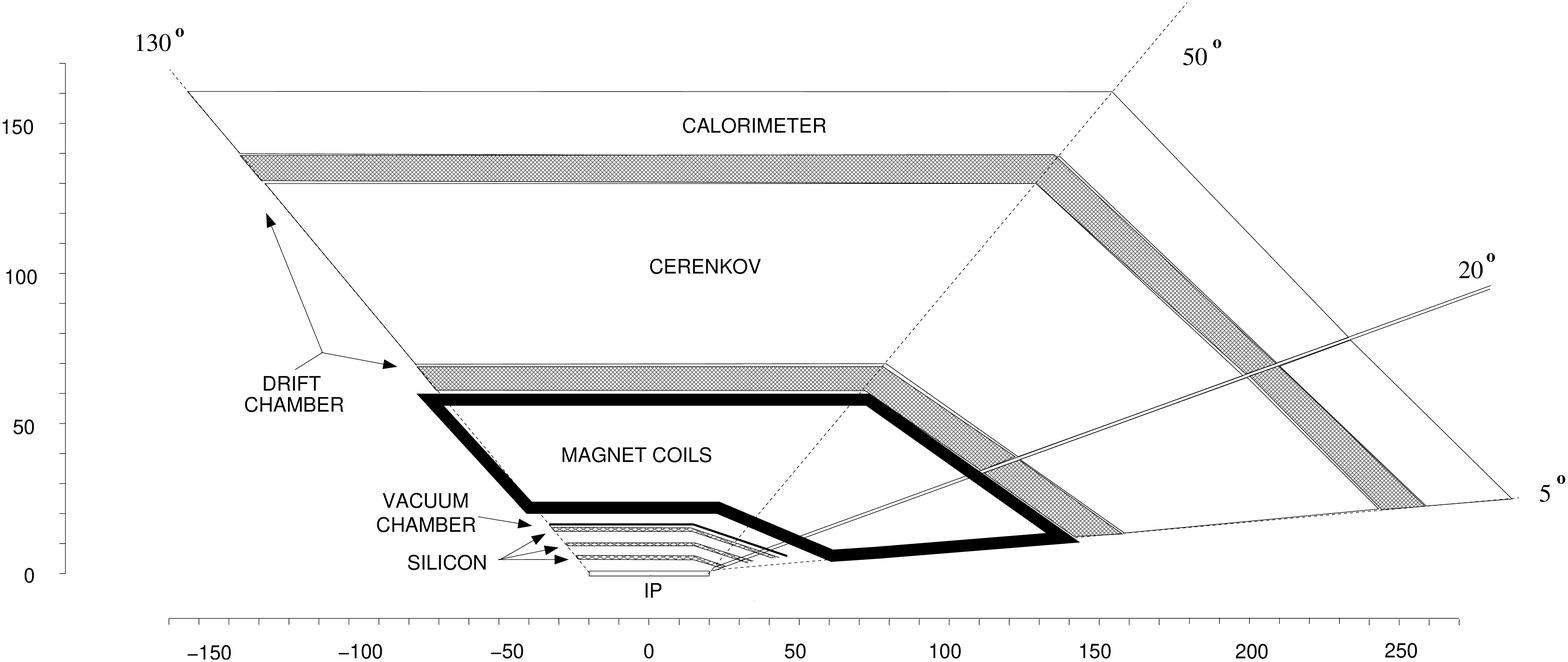}
  \centering \includegraphics*[width=0.79\linewidth]{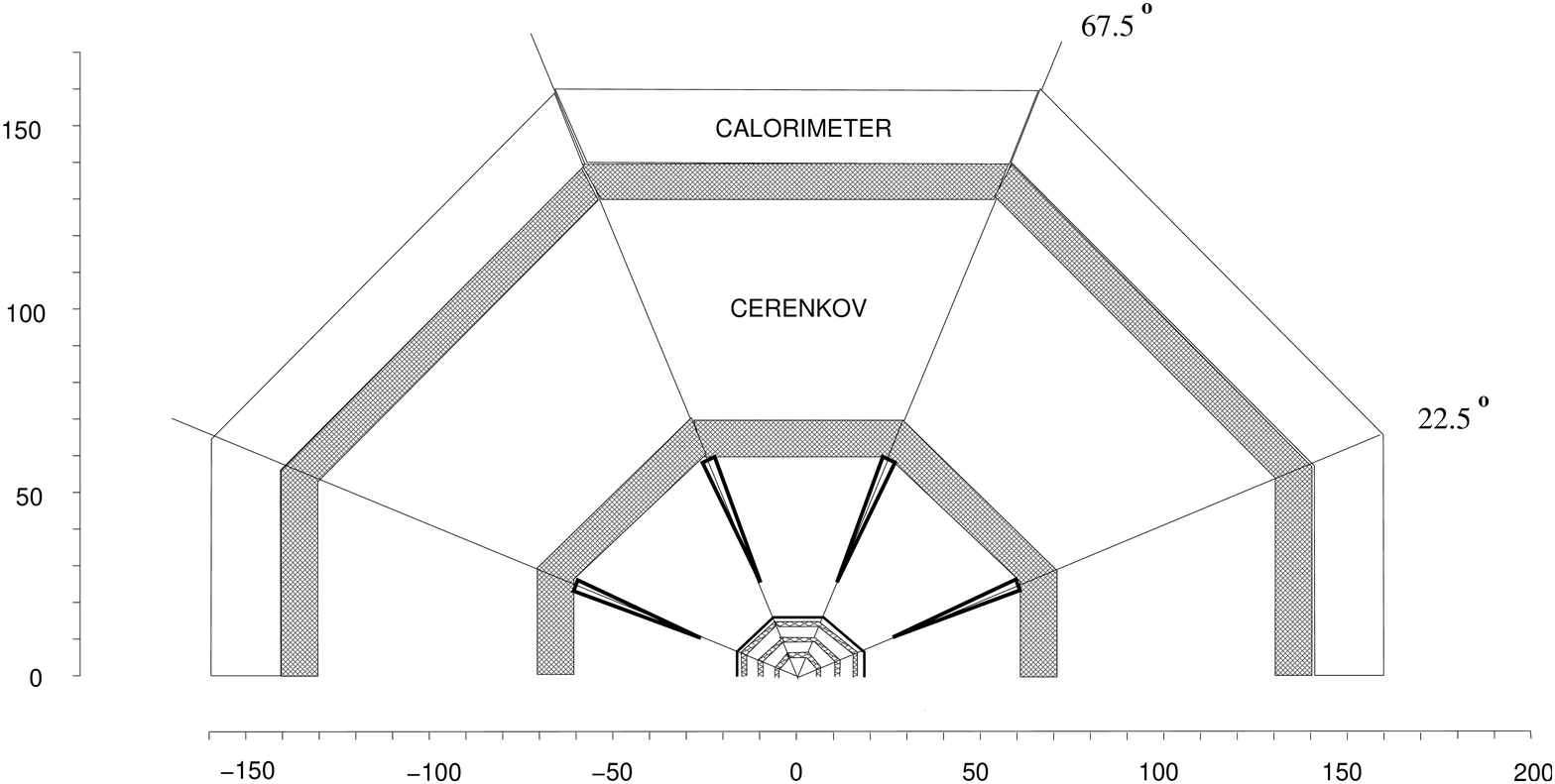}
  \parbox{14cm}{\caption{\label{innerPAX}\small Sketch of the PAX
      detector, showing a side view ([$z,y$], top) and a view in beam
      direction ([$x,y$], bottom).  The optional forward detector,
      sensitive at laboratory polar angles between $5^\circ$ and
      $20^\circ$, is also indicated. }}
\end{figure}

The PAX large--acceptance spectrometer (LAS) (Fig.~\ref{artistic}) is
optimized to detect electromagnetic final states with two charged
tracks of high invariant mass.
A clear identification of electrons is required to separate scattered
electrons of the Drell--Yan mechanism from the large $\pi$ background.
The detector is designed to also detect two--body hadron reactions
using kinematical constraints, i.e.  coplanarity and total momentum
conservation.  Moreover it can measure the energy of gammas from
radiative processes and $\pi^0$ and $\eta$ decays.

The very inner part of the detector is devoted to triggering and
tracking of charged particles.  The scattering angles as well as the
initial trajectory for the determination of the particle's momentum
are measured by a vertex tracking system consisting of three layers of
double--sided silicon strip detectors (SiD). The momentum measurement
is completed by two sets of drift chambers behind the magnet (DC). A
possible additional set of drift chambers inside the magnet (MC) would
improve the matching between the inner and outer tracks and help to
resolve multiple tracks and to identify gamma conversions. This will
allow us to detect low--momentum tracks which do not reach the
external section of the spectrometer.

A threshold \Cer counter (CER) provides trigger capability for
electrons and positrons produced in Drell--Yan processes.  Electron
and photon energies and directions are measured by the CAL.  Both the
CER and the CAL provide fast response and can be employed in the
selection of electromagnetic particles to obtain a $\pi/e$ rejection
factor $\lesssim$ 100 at trigger level and larger than $10^4$ in the
off--line analysis for a single track. A pre--shower detector (PS) can
possibly be added in front of the calorimeter to improve the pion
rejection and the resolution on the impact point of photons.  The
outermost set of drift chambers can be filled with Xe--methane gas and
can be complemented by a radiator to provide an additional $\pi$
rejection through transition radiation detection.

The central detector is designed to assure full acceptance between
$\pm 20^0$ and $\pm 130^0$ for polar angles in the laboratory frame.
It is not active in the small sectors of azimuthal angles in
correspondence of the toroid coils. In these sectors the service
systems for the detectors and the support structure of the system can
be mounted.  In the present design, no real limit exists for the
maximum acceptable polar angle: the above values can be taken as
indicative for the collider option and can be easily adapted to match
different beam configurations. The detectors point toward the central
part of the interaction region in a projective geometry.


An internal reflecting ring--imaging \Cer counter (DIRC) can be
employed to identify charged hadrons (pions, kaons and protons). This
provides flavor separation in single--spin asymmetry investigations,
as well as in the analysis of other semi--inclusive and exclusive
channels. The CAL allows the reconstruction of neutral pions in the
hadronic final state. CER and CAL are included in the trigger together
with a first hodoscope placed in front of the magnet and a second one
before the calorimeter.




\subsection{Toroid Magnet} 
The toroid magnet consists of 8 coils symmetrically placed around the
beam axis. A support ring upstream of the target hosts the supply
lines for electric power and for liquid helium. At the downstream end,
an hexagonal plate compensates the magnetic forces to hold the coils
in place.  The field lines of an ideal toroid magnet are always
perpendicular to the path of the particles originating from the beam
line. Since the field intensity increases inversely proportional to
the radial distance: greater bending power is available for particles
scattered at smaller angles, which have higher momenta.  These
properties help to design a compact spectrometer that keeps the
investment costs for the detector tolerable.  The production of such a
field requires the insertion of the coils into the tracking volume
shadowing part of the azimuthal acceptance.  Preliminary studies show
that the use of superconducting coils, made by a $\rm Nb_3Sn$--Copper
core surrounded by a winding of Aluminum for support and cooling,
allows one to reach an azimuthal detector acceptance in excess of
80\%, while the radius of the inner tracking volume can be kept below
0.7~m.


\subsection{Silicon Detector} 
Three layers of double--sided silicon strip detector provide a precise
vertex reconstruction and tracking of the particles before they reach
the magnet. The design is based on the vertex detector of
BABAR~\cite{BABARdet}, with a smaller number of silicon layers to
minimize the radiation length of the tracking material.  The read--out
electronic can be mounted in the front and back parts of the detector,
outside the acceptance of the spectrometer. With a pitch of $50-100$
$\rm \mu m$ it is possible to reach an intrinsic spatial resolution of
$10-20$ $\rm \mu m$.  Such a spatial resolution should allow one to
partially resolve and reject the secondary decays of $D^\pm$ mesons
into leptons. Due to the associated production of charmed mesons, this
kind of background can not be completely eliminated by subtracting the
dilepton events with the same charge.  The number of channels required
to cover the 40 cm long interaction point is of the order of
$2\ord{5}$, comparable with the BABAR experiment \cite{BABARdet}.

\subsection{Drift Chambers} 

The required position resolution of $200$~$\mu$m after passage through
the magnetic field can be achieved using conventional drift
chambers~\cite{HADESdch, CLASdch}. On the basis of an existing
set--up~\cite{HADESdch}, the chambers are assembled as modules
consisting of four pairs of tracking planes with wires at
$-30^\circ,\,0^\circ,\,0^\circ,+30^\circ$ with respect to the
direction transverse to the plane of the coil, i.e. parallel to the
magnetic field lines.  The wires of the planes oriented at $0^\circ$
are staggered in order to resolve left--right ambiguities.  Uniform
arrays of cathode wires separate the different cell layers. The tilted
angles are optimized to provide a two--dimensional hit--point with
emphasis on the polar (bending) coordinate.  The support structure of
the wires, together with the feedthroughs and the circuit boards for
sector pairs, can fit within the shadow of the coils~\cite{CLASdch}.
With a cell size of 0.5~cm, the total number of channels is about
32000. The expected momentum resolution is of the order of $1$~\% over
the kinematic range of the experiment.

\subsection{\Cer Detector}
A threshold gas \Cer counter is used to trigger and select electrons
in the presence of a large hadronic background.  The required
background event reduction at trigger level is of the order of $10^3$,
and is achieved with a $\pi/e$ rejection factor of $\lesssim 100$ on
the single charged particle.  The counter occupies a $60-80$~cm thick
shell around the inner tracking detector and is divided into eight
identical azimuthal sectors.  Each sector can be further subdivided
into s small number of gas cells. In this way the gas filling of each
cell can be optimized cover the momentum range of interest by
exploiting the correlation between scattering angle and particle
momentum~\cite{E835cer}.  Freon--12 and $\rm CO_2$ are suitable gases
for the energies involved in the PAX
experiment~\cite{E835cer,PHENIXpid}. \Cer photons are reflected by
aluminized carbon--fiber mirrors towards photomultipliers that can be
located outside of the spectrometer acceptance within the shadow of
the toroid coils~\cite{CLAScer}.  The spatial resolution in the polar
angle can be enhanced by segmenting the mirror and focalizing the
light onto a row of photomultipliers~\cite{CLAScer}.  A solution with
multi-wire proportional chambers as photon detector with a solid CsI
cathode pad is under study to obtain a higher spatial
resolution~\cite{HADEScer}.


\subsection{Calorimeter} 
The calorimeter consists of radiation--resistant lead tungstate ($\rm
PbWO_4$) crystals which have been selected for their high density,
short radiation length and small Moliere radius. Unlike the
CMS~\cite{CMScal} and ALICE~\cite{ALICEcal} experiments, the PAX
calorimeter is not immersed inside a high magnetic field. This makes
the use of photomultiplier tubes possible, to provide less noise and
better resolution at low energies.  The crystals are arranged in a
barrel and in a conically--shaped endcap of the experiment in a
quasi--projective geometry.  They are supported from outside to
minimize the material preceding the active region.  The blocks have an
area of $4\times4\,\rm cm^2$ and a variable length matching the mean
momentum of the impinging particles.  The expected energy resolution
can be parameterized as $\sigma(E)/E[\%]=1.8/\sqrt{(E[{\rm
    GeV}])}+0.4$~\cite{BTEVcal}.  For the conditions at PAX this
translates into an energy resolution of $2-3$~\%.  The calorimeter is
used to trigger on Drell--Yan electrons by selecting the events with
energy deposited in two clusters, both with a corresponding positive
\Cer signal, which have large invariant mass.  A back--to--back
topology and a total deposited energy equal to the center--of--mass
energy can be required to trigger on $\overline{p}p\rightarrow e^+e^-$
annihilation events.  Assuming an electron identification efficiency
of 90~\%, a hadron rejection factor of several hundreds can be
achieved by using the lateral profile of the deposited energy and the
$E/p$ ratio between energy $E$ deposited in the calorimeter and the
momentum $p$ measured in the spectrometer.  The estimated number of
crystals in the calorimeter is about 15000.

\subsection{Hodoscopes} 
Two hodoscope planes are used for triggering together with CAL and CER
detectors. A front trigger scintillator is placed directly upstream of
the vacuum chamber. It consists of eight--fold segmented foils of
standard plastic scintillator, 3.2 mm thick (0.7~\% radiation length),
read--out by phototubes. A second scintillator hodoscope is placed in
front of the calorimeter.  The counter is composed of modules of fast
scintillators with a large attenuation length (300--400 cm).  The
modules are longitudinally oriented providing a barrel geometry. The
scintillation light is detected by photomultiplier tubes. To provide
additional particle identification, a passive Pb radiator (2 radiation
lengths) can be placed in front of the external hodoscope to initiate
electromagnetic showers that deposit typically much more energy in the
scintillator than minimum--ionizing particles.

\subsection{Forward Spectrometer}
A forward spectrometer (FS) covering scattering angles below
20$^\circ$, is envisaged to complement the LAS during data--taking at
high energy.  The tracking system of the forward spectrometer has the
same design as the central one, and will benefit from the large field
integral of the toroidal magnet at small angles (where the coils are
closer each other).

The particle identification of the forward spectrometer has to be
adapted to the chosen beam configuration.  The sensitivity on the
transverse distribution $h_1$ in the forward direction is poor,
because $\hat a_{TT}$ is small there. In the asymmetric collider mode,
Drell--Yan events with scattering angles below $20^\circ$ in the
laboratory frame basically do not improve the statistical precision of
the $h_1$ measurement.  The forward region will be crucial for the SSA
measurements at high energy and large values of $x_F$, so that the
forward particle identification has to be optimized for hadrons. One
promising option is to make use of a RICH detector \cite{hermes-tdr}.
For a fixed--target experiment with 22 GeV/c beam momentum, the lower
limit of the useful scattering angle reduces to $5^\circ$ due to the
relativistic boost. The PID of the FS has to be optimized for electron
detection like the LAS.

\subsection{Recoil Detector}
A silicon recoil detector system is needed for the low--$t$
antiproton--proton elastic scattering program and will only be
installed for these measurements, so that radiation damage will be a
minor issue.
\begin{figure}[htb]
\begin{center}
  \includegraphics[width=0.9\textwidth]{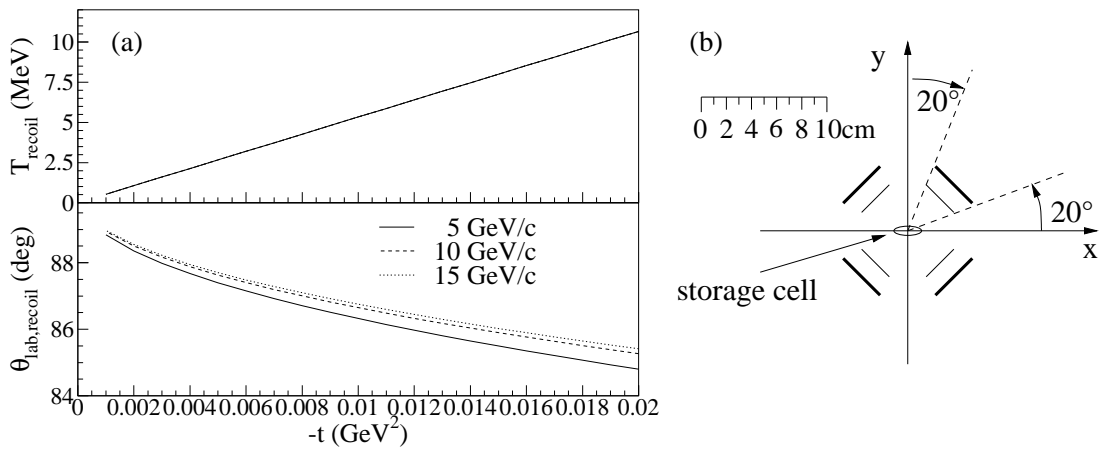}
  \parbox{14cm}{\caption{\label{fig:low_t_kin} \small (a) Laboratory
      kinetic energy (top) and scattering angles (bottom) of the
      recoil proton for three different beam momenta.  (b) Cross
      sectional view of the recoil detector.}  }
\end{center}
\end{figure}

At very low momentum transfer ($|t| = 0.002 \ldots 0.02$~GeV$^2$) the
recoil protons are detected by silicon strip detectors close to
90$^\circ$ laboratory angle. At these angles (cf.
Fig.~\ref{fig:low_t_kin}a) the protons of interest have energies
between 1 and 10~MeV and are stopped in a telescope comprised of a
65~$\mu$m thin surface barrier detector and a 1~mm thick microstrip
detector. Such a telescope \cite{Schleichert:2003} has already been
successfully operated in a similar environment at the ANKE experiment
at COSY/J\"ulich.  In view of the comparably large cross section
(${\rm d}\sigma/{\rm d}t > 150$~mb/GeV$^2$) a precise measurement of
the recoil {\em energy} is sufficient both to determine $t$ and to
cleanly identify elastically scattered protons as in the E760
experiment at FNAL~\cite{Armstrong:1996}.

Four of these detectors cover azimuthal angle intervals of $\Delta
\phi \approx 50^\circ$ in four quadrants, sketched in
Fig.~\ref{fig:low_t_kin} (b). The acceptance is matched to the central
part of the interaction region (collider mode) or of the storage cell
(fixed--target mode). For the fixed--target Phase--I at moderately
high energies, dedicated cell walls can be made as thin as 5 $\mu$m
Teflon -- as demonstrated by the PINTEX--experiment at
IUCF~\cite{PINTEX} -- and allow the detection of recoil protons down
to about 500~keV kinetic energy.  However, the low recoil momenta
prohibit the use of a strong target guide field, such that
measurements must be taken with a single pure hyperfine state and a
weak (some mT) guide field to avoid strong bending of the recoils at
very low $t$. The toroid magnet of the spectrometer provides a field
free region around the interaction point and does not disturb the
recoil trajectory.

The spin--dependent cross section \dsdt\ for vertical beam
polarization $P_y$ and a transverse target polarization $Q_x$, or
$Q_y$ is given by
\begin{equation}
\begin{array}{rcl}
\dsdo(\vec{P},\vec{Q},\theta,\phi) = \left.\dsdo\right|_{\rm unpol.}
\left( 1 \right. &+& \left[(P_y+Q_y)\cos\phi + Q_x\sin\phi\right]\AN\\
&+& P_y Q_y\left[ \ANN \cos^2\phi + \ASS \sin^2\phi \right] \\
&+& \left.P_y Q_x\left[ \ASS - \ANN\right] \sin\phi\cos\phi\right) \\
\end{array}
\label{eq:phimod}
\end{equation}
which relates by integration over $\phi$ and change of variables to
the differential cross section difference
\begin{equation}
\frac{\dd{\DST}}{\dd{t}} = - \dsdt\left(\ANN(t)+\ASS(t)\right)
\label{eq:dst}
\end{equation}
with $\DST = \sigma\left(\uparrow\downarrow\right) -
\sigma\left(\uparrow\uparrow\right)$.  With this experiment spin
correlation parameter $A_{NN}$, $A_{SS}$ as well as the analyzing
power $A_N$ of $\bar{p}p$ elastic scattering are accessible. 

\subsection{Trigger and Data Acquisition}
The actual interaction rate achievable by the PAX experiment will be
precisely estimated only after the final beam configuration is known.
However, an upper limit is anticipated to be of the order of a few
MHz.  A reduction factor of $10^3$ at trigger level is then required
to keep the rate of read--out events below a few kHz.

For the different physics issues to be studied, dedicated trigger
schemes have to be employed. The detection of Drell--Yan electron
pairs will be accomplished by using coincidences comprising
multiplicity information from the hodoscopes, silicon detectors, CAL
and segments of the \Cer counter in the LAS.  For single--spin
asymmetries a single--prong trigger, derived from the forward
scintillator hodoscopes and the calorimeter, can be used in a similar
way as in the HERMES experiment~\cite{HERMES01}. For the low--$t$
elastic antiproton--proton scattering the recoil--detector hodoscope
provides self--triggering capability for low energy hadrons, as
demonstrated at ANKE~\cite{Schleichert:2003}.

High luminosity (above $10^{32}$\,cm$^{-2}$s$^{-1}$ in the case of
unpolarized antiproton beam) and wide solid angle acceptance lead to
high counting rates of the detectors. Under these condition the
architecture of the trigger and the data acquisition systems are
essential in defining the capability of the setup to collect data
without large dead time losses.

The trigger system has to be flexible enough to cover different
physics issues which demand different trigger selection criteria. It
is planned to use a multi--level trigger composed of a fast first
level trigger and hardware and software processors at higher levels.
The experience obtained in running of HERMES~\cite{HERMES01},
ANKE~\cite{anke} and other experiments \cite{DIRACtrig} will be
employed to a considerable extent.  Due to the diversity of the
various detectors, sufficient capability for event--selection at the
trigger level is available, i.e. multiplicity information, energy loss
and total energy, particle identification, tracking and hit--map
correlations.

Pipelines and de--randomizing buffers will be used to store the events
during the processing at the low--level trigger stages.  The event
builder will collect information from all the detector readout
branches.  The event builder protocol has still to be selected in view
of fast developing network and computing technologies.

\section{Detector Phases}

\subsection{Detector Phase--I}
The fixed--target program of Phase--I in the CSR ring (see Sec. 13)
concentrates on the time--like proton form factors and elastic
scattering measurements. The simple and overconstraint kinematics of
these events puts less stringent requirements on the detector
performance. The momentum resolution, for instance, is not a crucial
issue, as demonstrated by the E835 experiment~\cite{E835nbarn}. The
measurements can start even before the detector is completed and can
be used to test and optimize each of the sub--systems, i.e. tracking
system and \Cer PID.  The trigger is provided by two back--to--back
tracks. The \Cer signals above threshold and the total energy
deposited in the calorimeter equal to the center--of--mass energy, can
be employed to trigger the rare electron events.  At the CSR energies,
the outgoing particles have an almost isotropic distribution and a
momentum between $0.5$ and $1.5$ GeV/c.  Hadron identification is
provided by time--of--flight measurement using the
hodoscopes,~\cite{HADESdch, CLASdch}.

\subsubsection{The gaseous Fixed-Target}  
In Phase--I, a storage cell gas target is inserted downstream of the
detector.  The atomic beam source (ABS) and the injection tube of the
cell are placed in the empty solid angle in front of the detector.
The conventional design of a storage cell target is described in
details elsewhere (Sec. 8.2.1).  The polarized gas atoms leave the
target cell through the open ends and are differentially pumped by two
stages along the beam pipe.  This minimizes the degradation of the
vacuum and thus its effect on the stored beam.  The transition from
the cell to the beam pipe could be made smooth using perforated tubes,
to avoid the possible generation of wake fields that could cause
heating and increase the emittance of the beam. Given the importance
of the acceptance in the forward direction, the first pumping system
at the cell position is located upstream, like the ABS.  The vacuum
region extends inside the conic--shaped internal space of the detector
and reaches the second pumping system located just behind the CAL.
Particles scattered into the detector exit the vacuum region through a
0.3 mm stainless--steel foil (corresponding to 0.5~\% of radiation
length) stretched by the toroid coil supports.

\subsubsection{The Target Magnet} 
In order to minimize the material inside the detector acceptance, the
magnet is composed of two superconducting coils surrounding the target
in the horizontal plane and providing a vertical field up to $0.3$ T
in the cell volume.  The coils can be shaped or correcting coils can
be added to improve the homogeneity of the field (if it is required to
avoid depolarization effects from the beam current structure). The
coils run inside a cooling tube where the liquid He is continuously
flowing.  The magnet is inside the vacuum region to provide thermal
insulation.  Four correcting dipoles are added to the beam line to
compensate the effect of the PAX magnets on the orbits of the protons
and/or antiprotons.  For the case where longitudinal target
polarization is required, the transverse field will be ramped down.
If only one hyperfine state is injected, a longitudinal holding field
of some mT is sufficient and can be provided by conventional Helmholtz
coils. (When only one hyperfine state is injected in the target,
spin--relaxation processes like spin--exchange collisions are
practically absent and the condition for a strong holding field is
consequently relaxed.)

\subsection{Detector Phase--II}
The asymmetric--collider program of Phase--II concentrates on the $h_1$
measurement.  The inclusive $\overline{p}p\rightarrow e^+e^-X$
Drell--Yan process has poor kinematic constraints. The intrinsic
transverse momentum of the quarks, for instance, breaks the
coplanarity of the $e^+e^-$ pair.  A rejection factor of $10^3$
against background events is required to reduce the rate from a few
MHz to kHz levels.  The trigger asks for two tracks in opposite
hemispheres above the \Cer threshold. To reduce low--energetic
combinatorial background, a cut on the dilepton invariant mass is
applied using the deposited energy and the impact point at the
calorimeter. The first layer of silicon is used to veto gamma
conversions.

\section{Performance Summary}
\begin{figure}[htb]
  \centering \includegraphics*[width=0.8\linewidth]{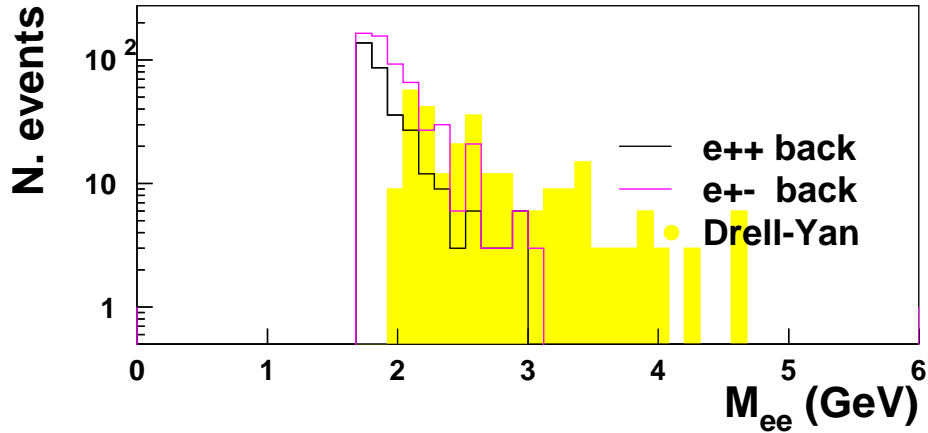}
  \parbox{14cm}{\caption{\small \label{DYback}Background estimation
      for the collider mode. The Drell--Yan signal is generated with a
      minimum invariant mass of $2$ $\rm GeV/c^2$.  The background
      below $1.5$ $\rm GeV/c^2$ is not shown since it is vetoed by the
      trigger.  Only the gamma conversions taking place before the
      second tracking layer are taken into account.  A perfect
      performance of the PID and tracking system is supposed here. The
      generated statistics is $4\ord{9}$ $\overline{p}p$ interactions,
      corresponding to about 4 hours of data taking at a luminosity of
      $5\ord{30}$ $\rm cm^{-2}s^{-1}$. } }
\end{figure}

The major sources of background to the Drell--Yan process are the
$\pi^0$ (and $\eta$) Dalitz--decays, the gamma conversions and the
charmed--meson decays. Dileptons coming from one single light meson
decay or gamma conversion can be identified by the low invariant mass.
Only multiple decays or conversions may generate a dilepton with
invariant mass larger than 2 $\rm GeV/c^2$. An additional
electromagnetic particle in the event can be used to identify the
parent $\pi^0$ or photon and to reject the candidate. The residual
background can be finally subtracted by identification of a lepton
pair with the same charge.  Due to large mass and associated
production, charmed--mesons tend to produce dangerous unlike--sign
candidates at high invariant mass. This kind of background can be
reduced by reconstructing the secondary vertex of the decay with the
silicon detector: the $D^+, D^-$ mesons have a lifetime comparable to
the one of B--mesons whose decay length is routinely measured at the
B--factories.  In the collider mode, after subtraction of the
combinatorial background, a $\Ord{-1}$ contamination from
charmed--meson decays is left (see Fig.~\ref{DYback}). In the
fixed--target mode the center--of--mass energy is too low to generate
a significant contamination from charmed mesons.

For the count rate estimates, we will focus on the Drell--Yan process,
the reaction with the highest demand on luminosity.  Other reaction
channels of interest have larger cross--sections or, like single--spin
asymmetries, may use an unpolarized antiproton beam.

The experimental uncertainty for double--spin asymmetries depends on
the number of observed events $N$ as well as on the degree of
polarization of the two beams. A value of ($P~\gsim~0.80$) can be
assumed for the proton beam polarization, whereas values of
$Q~\approx~0.30$ are anticipated for the antiproton beam
polarization~\cite{ap}, see Sec. 8.  The error is then roughly given
by $(P Q\sqrt{N})^{-1} = 4/\sqrt{N}$.  It should be noted that an
extensive study is foreseen to optimize the spin--filtering process:
any beam polarization acquired in addition by the antiproton beam will
reduce experimental uncertainties linearly.


\begin{figure}[htb]
\begin{center}
  \includegraphics[width=0.43\linewidth]{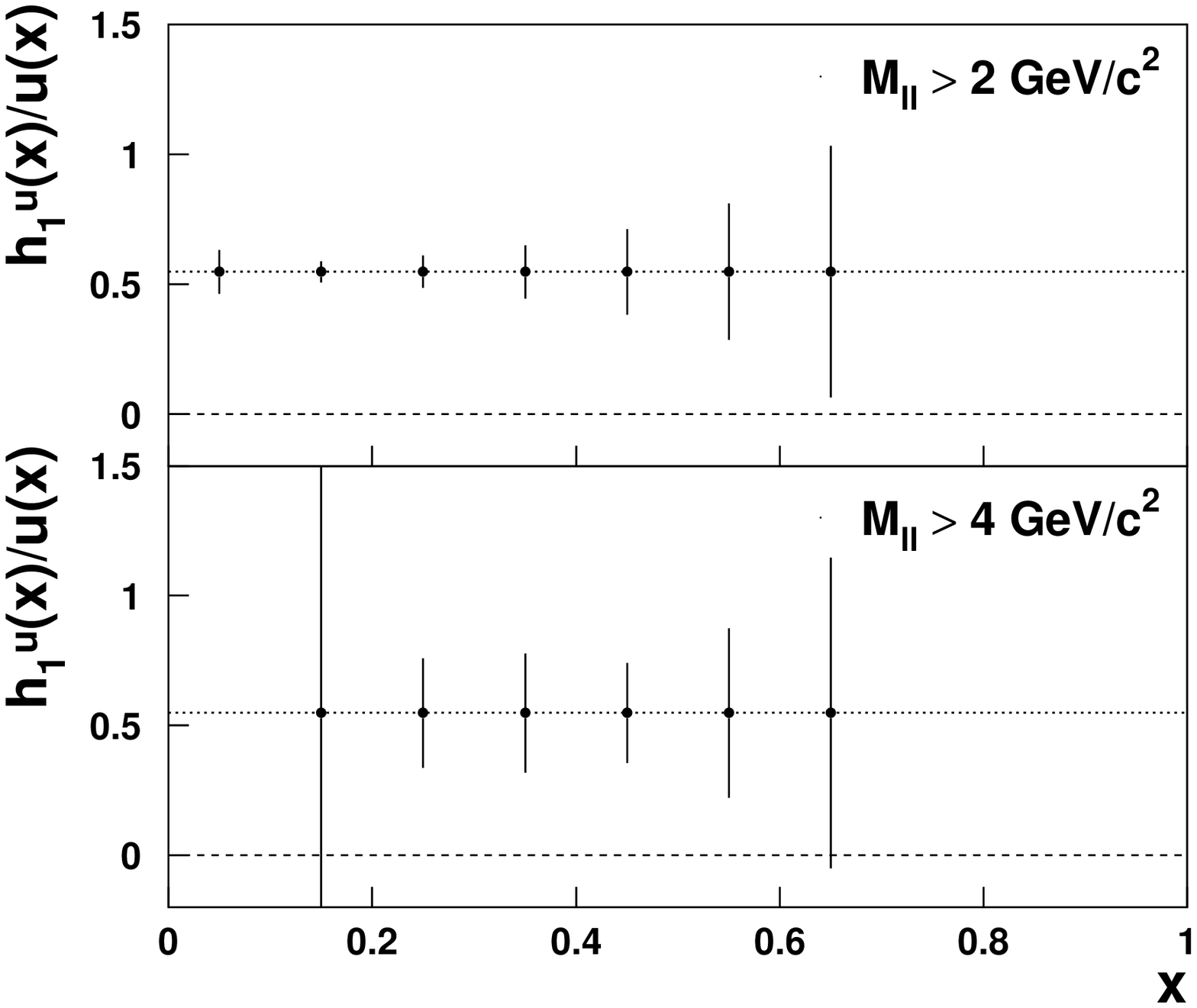}
  \includegraphics[width=0.43\linewidth]{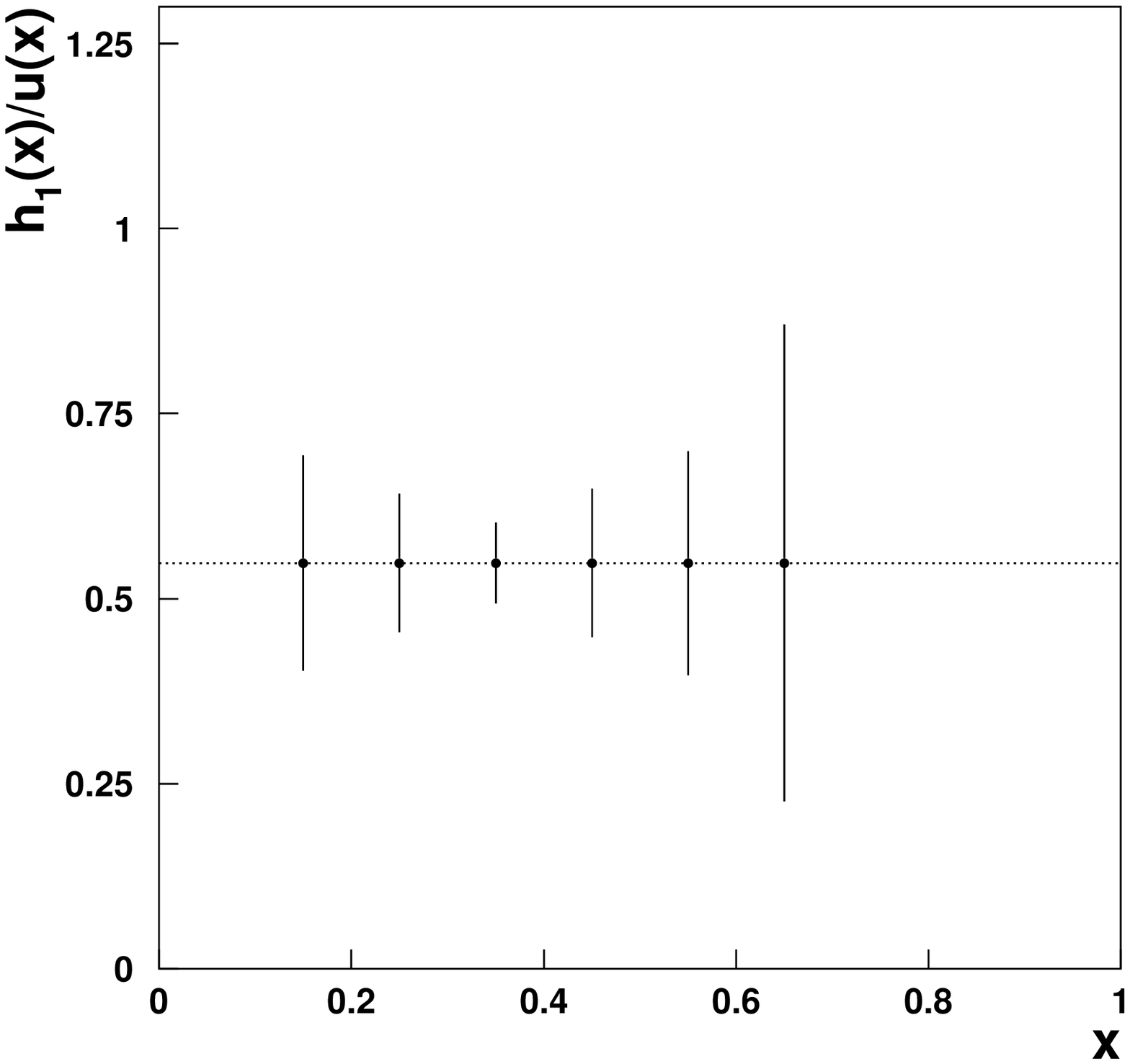}
  \parbox{14cm}{\caption{ \label{h1resu}\small Expected precision of
      the $h_1$ measurement for one year of data taking at 50 \%
      efficiency (180 days). An indicative $A_{TT}/\hat{a}_{TT}=0.3$
      has been considered in the simulation. For the collider mode a
      luminosity of $5\ord{30}$ $\rm cm^{-2}s^{-1}$ and a polar angle
      acceptance between $20^\circ$ and $130^\circ$ was assumed
      (left). For the fixed--target mode at 22 GeV/c beam momentum a
      luminosity of $2.7\ord{31}$ $\rm cm^{-2}s^{-1}$ and a polar
      angle acceptance between $5^\circ$ and $50^\circ$ was assumed
      (right).}}
\end{center}
\end{figure}
For every event, the Bjorken $x$ of the proton and $\overline{x}$ of
the antiproton can be extracted from the measured invariant mass
($M^2=Q^2$) and from the longitudinal momentum ($p_L=s/2 \cdot x_F $)
of the lepton pair.  In the $u$ dominance hypothesis the
$\Att(x,\overline{x})$ asymmetry is related to the convolution of the
transversity distributions $h_1(x)\cdot h_1(\overline{x})$ by the
relation
\begin{eqnarray*}
\Att(x,\overline{x})=\hat{a}_{TT} \; 
\frac{h_1(x)}{u(x)}\frac{h_1(\overline{x})}{u(\overline{x})}\;.
\end{eqnarray*}
During one year of data taking, $h_1(x)$ can be measured in a wide $x$
range, from $0.7$ down to $0.05$, covering the most interesting
valence region and extending to low values of $x$ where theoretical
predictions show the largest deviations. The precision achievable in
180 days (one year of data taking with 50 \% of efficiency) is shown
in Fig.~\ref{h1resu}.  These numbers entail only the non--resonant
contribution to the Drell--Yan process: the \JPsi\ will enhance the
number of events in the $M^2$~=~6--16~GeV$^2$ range considerably.

For low--$t$ proton--antiproton elastic scattering, 
recoil--detectors, with a typical area of 5$\times$4~cm$^2$ each, will
be mounted with an angular acceptance matched to the center of the
storage cell. Count rates of 6$\times10^6$/week per $t$--bin of
0.0005~GeV$^2$ width are expected to be achieved with a polarized
antiproton beam of $P=30\%$ polarization. Assuming a target polarization
of $Q=90\%$ spin correlation parameters can be measured to a precision of
0.01 within a week, so this program can be finished within a few
weeks of beam time.

\section{Alternative Scenarios \label{sec:alt-scen}}
Two alternative scenarios were considered for the detector design. The
first is to select Drell--Yan events with muons.  The second is to use
a conventional solenoid magnet as spectrometer magnet for the collider
mode.

\subsection{Drell--Yan with Muons}
The design of a detector for Drell--Yan with muons at the PAX energies
presents a host of difficulties, which make this a challenging task.
The only known way of separating muons from hadrons is to use their
low interaction probability and consequent high capability of
penetrating large amounts of heavy absorbing material.  The hadronic
background can originate from decays of charged pions and kaons before
reaching the absorber and from hadrons penetrating the material
(punch--through).  At PAX energies, a typical detector with about 1 m
of space before the absorber for charged particle tracking, particle
identification and electron/photon detection, cannot guarantee a
$\pi/\mu$ rejection factor of $\Ord{2}$~\cite{TOPAZ} due to the high
probability of meson decays.  These decays could be only reduced by
moving the absorber closer to the interaction region at the expense of
completely losing all the flexibility of the experimental facility. In
any case, empty space has to be foreseen for the vacuum chamber as a
housing for the vertex detector, so that the secondary--muon rejection
factor can hardly exceed $\Ord{3}$.  The absorber puts also severe
limitations on the achievable resolution of the dilepton invariant
mass.  This approach was extensively adopted in the past for
Drell--Yan measurements, but at a much higher energy where the decay
probability reduces and the absorber filtering is more effective.  To
keep the background at an acceptable level and not being able to
isolate the charmed resonances from the continuum due to the poor
resolution, those experiments are limited to a kinematic range of
dilepton invariant masses larger than 4 GeV~\cite{Drell-Yan-exp}.
Such a constraint would limit both the achievable statistics and the
covered range in Bjorken $x$ at PAX.  Even a large number of heavily
instrumented material segments do not suppress hadrons faking muons
entirely.  Refined studies at the B--meson $e^+e^-$ facilities show
that the maximum rejection factors against hadron punch--through
achievable with active absorbers are of the order of
$\Ord{3}$~\cite{Artuso, CLEO}. The above values of pion rejection
factors are well below the $\Ord{4}$ to $\Ord{5}$ threshold required
for an efficient Drell--Yan measurement.

Since there are several powerful well--established techniques to
distinguish electrons from pions in the energy range of the PAX
experiment, and since these are compatible with a precise measurement
of the particle momenta, the Drell--Yan electron channel was selected
to be a much more effective solution for PAX.

\subsection{Solenoid Magnet}
A solenoid field is a natural option for a collider detector since it
provides an almost homogeneous field in a large empty volume suitable
for tracking.  This choice presents some drawbacks for the PAX physics
case. Since the spin of the beam particles undergoes precession inside
the longitudinal field of the spectrometer, a set of
counteracting--solenoids inside the interaction region is required
such that the longitudinal field integral that protons and antiprotons
experience before a collision vanishes. Although it is in principle
feasible, this solution would complicate the design of the collider
IP.  The correction has to work for any beam momentum employed in
order to vary the center--of--mass energy of the collisions and
explore different kinematic regimes. Moreover, the correction can not
be exact over the about 30 cm long IP bunch--length and the
unavoidable spin precession will be energy dependent.  The solenoid is
not compatible with a transverse polarized target, hence is not
suitable for Phase--I or Phase--II (b) of the PAX physics program (see
Sec. 13 and 14).

Because the threshold \Cer detector can not be efficiently operated in
the strong field of a solenoid magnet, a possible alternative is to
employ a transition radiation detector (TRD).  The transition
radiation, being proportional to the relativistic $\gamma$ factor of
the particle, is effective at high momenta and the required $\Ord{3}$
pion rejection factor might represent a challenge at PAX
energies~\cite{ATLAStrd, ALICEtrd}.  Additionally, TRDs typically add
ten times more material than a \Cer detector (more than 0.1 radiation
length), thus enhancing gamma conversions and energy losses by
radiation of the Drell--Yan electrons.  Triggering on rare
di--electron events in the high--multiplicity PAX environment is not
trivial. A pattern recognition, required to link the electromagnetic
signatures to the helicoidal trajectory of the particle, is possible
only with massive use of computing power.

It is desirable to match the bending power of the spectrometer to the
momenta of the emerging particles. In the forward region, where the
angle between the particle path and the beam direction is small, the
rigidity increases and it is more difficult to provide the necessary
bending power.  A solenoid centered on the beam provides the maximum
bending at large angles in the laboratory frame. Since there is no
transverse force on a particle traveling along a magnet field line and
the solenoid field lines are parallel to the beam axis, bending is
reduced eventually to zero, as the scattering angle is decreased to
zero.

The torus has none of the above drawbacks.  It neither disturbs the
beams nor the operation of the \Cer detector since the generated field
is limited to a well defined region with negligible fringe fields.
The field lines of a toroid field centered on the beam line are always
perpendicular to the paths of particles originating from the axis.
Its field intensity increases inversely with radius from the beam,
such that larger bending power becomes available at small angles. The
drawbacks of the toroid could be a not homogeneous field and the
presence of the coil material inside the tracking volume. The first
issue simply requires to use a detailed field map for tracking.  As a
conservative approach, it is planned to exploit the over--constraint
kinematics of the physical channels of PAX during Phase--I to test and
optimize the spectrometer performance before starting the more
challenging measurements of PAX Phase--II.  The second issue causes a
reduction of the azimuthal acceptance; the loss in acceptance can be
minimized by using superconducting technology to save material and by
placing the support structures and utility lines of the detector in
the blind region of the toroid coils. The approach of PHENIX to use
Helmholtz coils together with large field--driving magnet yokes would
have an even more severe impact on the azimuthal
acceptance~\cite{PHENIXmag}.  Due to the arguments mentioned above a
toroidal field approach has been taken to be more suitable for the PAX
program than a solenoid field.  A feasibility study about the
possibility to use the PANDA detector in the collider--mode must
necessarily be started on the basis of the arguments discussed above.

\cleardoublepage
\part{Organization}
\pagestyle{myheadings} 
\markboth{Technical Proposal for ${\cal PAX}$}{Part V: Organization}
\section{Logistics of the Experiment}
\subsection{Floor Space}
As mentioned in chapter 7, all rings of the HESR complex should be in
one plane.  It is assumed that a hall for the CSR is provided, similar
in space as the COSY hall.  The APR is located inside the CSR.
Furthermore we assume that PAX is the only user of CSR, allowing for a
permanent installation of the PAX detector in one of the straight
sections. For the experiment, a total space of about 300~m$^2$ is
required. If the experiment has to be moved in and out, this space has
to be enlarged to about 450~m$^2$.

Concerning the required height of the experimental hall, we assume a
beam 3~m above the platform and 5~m above the floor. The upper edge of
the detector frames is assumed at 7.5~m above floor, resulting in a
maximum height of the crane hook of 10.5~m above the floor. With 1.5~m
for the crane structure itself, an inner hall volume for the fixed
experiment of about 3600~m$^3$ and for the movable experiment of about
5400~m$^3$ is estimated (Tab.~\ref{tab:hall-volume}).

\begin{table}[htb]
\centering
\renewcommand{\arraystretch}{1.3}
  \begin{tabular}{|p{8cm}|r|}
\hline
Height of the experimental hall & 12~m         \\\hline
Crane hook                      & 10.5~m above floor\\\hline
Assumed beam height             & 5 m above floor \\\hline\hline
Volume of the hall (exp. fixed)   & 300 m$^2 \times 12\,{\rm m} = 3600$~m$^3$\\\hline
Volume of the hall (exp. movable) & 450 m$^2 \times 12\,{\rm m} = 5400$~m$^3$\\\hline
  \end{tabular}
\parbox{14cm}{\caption{\label{tab:hall-volume} \small Requirements of the 
experimental hall for PAX.}}
\end{table}

In addition laboratory space and a control room attached to the
experimental hall are required.

\subsection{Radiation Environment}
The PAX experiment will operate with long beam lifetimes and thus slow
antiproton consumption. Requirements for radiation safety at the
target location will not be enhanced with respect to other areas along
the HESR.

\subsection{Responsibilities and Manpower}
Although the APR needs institutional responsibilities, we assume that
after demonstration and testing the APR will become part of the FAIR
facility and is operated and maintained by the laboratory, i.e.  GSI.
\subsubsection{Institutional Responsibilities (preliminary list)}
\begin{itemize}

\item{\bf{Antiproton Polarizer Ring}}\\
  Ferrara, J\"ulich

\item{\bf{Targets (APR, CSR/HESR)}}\\
  Ferrara, Erlangen, J\"ulich, Gatchina, Madison
\begin{itemize}
\item{{Target Magnet}}\\
  Ferrara, Gatchina
\end{itemize}

\item{\bf{Large--Acceptance Spectrometer}}

\begin{itemize}
\item{Toroid Magnet}\\
  Ansaldo (Italy)
\end{itemize}

\begin{itemize}
\item{Vertex Detector}\\
  J\"ulich
\end{itemize}

\begin{itemize}
\item{Electromagnetic Calorimeter}\\ 
Frascati, Moscow, Protvino,  and Yerevan
\end{itemize}

\begin{itemize}
\item{Drift Chambers}\\
  Gatchina, Dubna
\end{itemize}

\begin{itemize}
\item{Threshold Cherenkov}\\
  Dubna, Gent
\end{itemize}

\item{\bf{Forward Spectrometer (optional)}}
  
  Partially recuperated from the HERMES experiment.

\begin{itemize}
\item{Particle Identification} \\
  Dubna
\end{itemize}

\item{\bf{Data Acquisition and Trigger}}\\
  Dubna, Protvino, Gatchina, J\"ulich
 
\item{\bf{Computing, Technical Software, Simulations}}\\
  Ferrara, Tbilisi, Dubna, Beijing, Hefei
\end{itemize}

\subsubsection{Manpower}
The present status of manpower available for the experiment can be
inferred from the Collaboration list (see p. 5ff). The list comprises
about 170 participants from 13 countries and 35 institutions. Based on
inquiries and discussions, we expect a very significant further growth
of the collaboration in the next couple of years.
\subsection{Schedule}
\subsubsection{Milestones}


\begin{table}[htb]
\centering
\renewcommand{\arraystretch}{1.3}
\begin{tabular}{crp{13cm}}
\hline
\#                 & Date       & Milestone\\\hline\hline
1                  & 12/2006    & Technical Design Reports finished; overall design of all PAX components \\
2                  & 12/2008    & Fabrication of APR components finished \\
3                  & 12/2010    & Spin filtering with protons in the APR tested \\
4                  & 12/2010    & Fabrication of detector components finished \\
5                  & 12/2011    & CSR operational with protons at FAIR \\
6                  & 12/2012    & Installation of PAX detector (Phase--I) finished\\
7                  & 12/2013    & Commissioning of PAX detector (Phase--I) with protons/antiprotons done\\
8                  & 12/2013    & Phase--I polarized antiproton facility (CSR \& APR) ready for experiments\\
9                  & $>$ 2015   & Commissioning of HESR in the (Phase--II) double--polarized collider mode \\\hline\hline
\end{tabular}
\parbox{14cm}{\caption{\label{tab:milestones} \small Milestones of the PAX experiment.}}
\end{table}

\subsubsection{Timelines}
\begin{figure}[htb]
  \centerline{
\epsfxsize=\linewidth                    
\epsfbox{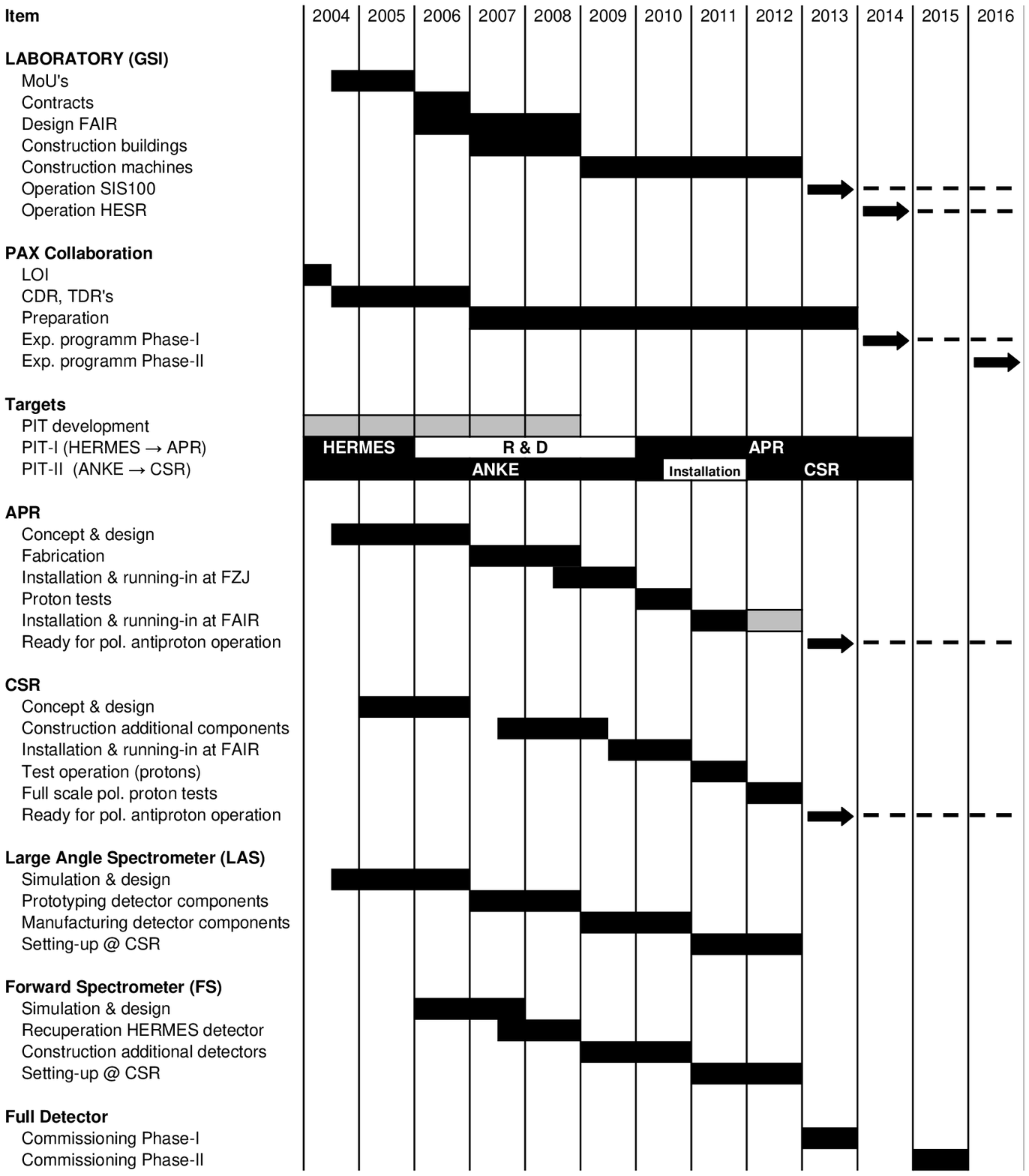}}
\parbox{14cm}{\caption{\label{fig:time-schedule}\small Time Schedule
    for the PAX experiment.}}
\end{figure}

The timelines of the PAX experiment depend on the FAIR schedule, the
availability of recuperated equipment, like components of the HERMES
experiment, and on manpower and funding. A first scenario is
presented in Fig.~\ref{fig:time-schedule}. It is based on the
assumption that the Phase--I polarized antiproton facility is
available in 2013, ready to accept antiprotons from the new antiproton
source.

A central part of PAX is the antiproton polarizer ring (APR) which is
required to be operational in 2010 for proton tests.  Because of the
availability of a polarized proton injector at J\"ulich, we prefer to
set up the APR at COSY for proton tests as soon as possible.
Installation and running--in at FAIR is scheduled for 2011/12.
Furthermore, we assume that the CSR is fully operational at FAIR in
2011, and that the full CSR antiproton facility including the Phase--I
PAX detector is available for proton tests in 2012, including
acceleration of polarized beams.

The PAX detector is located permanently in the CSR straight section.
It consists of the Large--Angle Spectrometer (LAS) and an optional
Forward Spectrometer (FS). The FS is largely based on components
recuperated from the HERMES detector.  Commissioning of the full
Phase--I detector is scheduled for 2013.

Two polarized internal gas targets (PIT) exist which can be utilized
after improvements and modifications.  PIT--I is the present HERMES
target which is available in 2006, after the double--spin program is
terminated.  This target has been in operation since 1996. After a
suitable upgrade, it can be installed at the APR in 2009 for proton
tests. In addition, a development program at INFN Ferrara is conducted
in order to study possible improvements by using superconductive
sextupole magnets for a higher source intensity.  PIT--II is the
present ANKE target which will be available only after 2009 or later.
It is foreseen for installation into the PAX detector at CSR.

\subsection{Cost Estimates}
The cost estimate for the fixed energy APR, listed in
Table~\ref{tab:cost-estimate1}, is based on a ring built from
permanent magnet material, which is a very economic solution, since
there are no power supplies needed. For the same reason, the cost of
operation is substantially reduced compared to a ring built from
electromagnets.

The cost estimate for the PAX detector is divided into four main
categories, which are briefly discussed below.  The resulting
figures, listed in Tab.~\ref{tab:cost-estimate2}, are based on the
1993 figures of the HERMES TDR \cite{hermes-tdr}, and were increased
by 30\% for inflation.

\begin{enumerate}
\item Large--Acceptance Spectrometer: Here the structure of the E835
  detector has been assumed with the addition of a toroidal
  spectrometer magnet, calculated using HERMES figures (Cherenkov and
  Calorimeter), and the price of the HERMES recoil detector for the
  tracking part. 
  
\item Forward Spectrometer: Here the cost is extrapolated from the one
  estimated for the Large--Acceptance Spectrometer, with the
  additional cost for hadron identification (from HERMES RICH).

\item The polarized target is needed for the fixed--target mode only.
  Here parts of the HERMES and/or ANKE targets can be recuperated,
  which should result in a reduction of the order of 40\%.
  
\item Infrastructure: These costs are also based on HERMES figures for
  platform and support structures, cabling, cooling water lines, gas
  supply lines and a gas house, cold gases supply lines, electronic
  trailer with air conditioning etc.
\end{enumerate}

\begin{table}[htb]
\centering
\renewcommand{\arraystretch}{1.3}
  \begin{tabular}{|p{10cm}|r|}
\hline
Polarized target (HERMES)                      &   1.0 MEU \\\hline
Machine (permanent magnet version \& vacuum)   &   6.0 MEU \\\hline
Electron cooler                                &   1.5 MEU \\\hline
Injection/beam lines                           &   0.5 MEU \\\hline
Snake                                          &   0.2 MEU \\\hline
Polarimeter                                    &   0.2 MEU \\\hline\hline
Total                                          &   9.4 MEU\\\hline
  \end{tabular}
\parbox{14cm}{\caption{\label{tab:cost-estimate1} \small Cost estimate for 
the Antiproton Polarizer Ring (APR)}}
\end{table}

\begin{table}[htb]
\centering
\renewcommand{\arraystretch}{1.3}
  \begin{tabular}{|p{10cm}|r|}
\hline
Toroid magnet                                                                        &   3.0 MEU        \\\hline
Polarized target (required only for fixed--target mode)                        &   1.0 MEU        \\\hline
Large--Acceptance Spectrometer (LAS)                                                 &   12.0 MEU       \\\hline
Forward Spectrometer (FS) (optional)                                                 &   (5.0) MEU      \\\hline
Infrastructure (cabling, cooling, shielding)                                         &    3.0  MEU      \\\hline\hline
Total, without (with) FS                                                             &  19.0 (24.0) MEU \\\hline
  \end{tabular}
\parbox{14cm}{\caption{\label{tab:cost-estimate2} \small Cost estimate for 
the PAX detector.}}
\end{table}

\clearpage
\pagestyle{myheadings} \markboth{Technical Proposal for
  ${\cal PAX}$}{Bibliography}

\cleardoublepage
\part{Appendix A} 
\pagestyle{myheadings} \markboth{Appendix A to the Technical Proposal
for ${\cal PAX}$}{Polarization--Transfer Technique}
\begin{appendix}
\section{Polarization Transfer Technique
\& Applications}
This mini--review
addresses the polarization--transfer technique and its high--energy
  applications.
\subsection{Breit Hamiltonian
and Antiproton Polarizer}
The spin-filtering by electromagnetic antiproton-electron interaction
is at the core of the proposed PAX experiment\footnote{A historical
introduction to the subject, a review of the FILTEX experiment carried
out in 1992 at the Test Storage Ring at MPI Heidelberg, detailed
evaluations of the attainable polarization based on the
Horowitz--Meyer calculations, the accelerator set--up to achieve the
maximal polarization of stored antiprotons and further references are
found in Secs. 7 and 8 of part~\ref{sec:APR}.}  and a brief review on
the QED foundations of this technique is in order.  In the
spin--filtering method one depends on the $e\bar{p}$ interaction at
nonrelativistic kinetic energy of electrons in the proton rest frame
$$ T_e = {m_e\over m_p+m_e}T_{\bar{p}}\,,
$$ where $T_{\bar{p}}$ is the kinetic energy of antiprotons in the
Antiproton Polarizer Ring (APR)\footnote{The optimization for the
polarization figure of merit suggests the preferred energy
$T_{\bar{p}}\approx 50$ MeV (see part~\ref{sec:APR}).}.  This
interaction is described by the celebrated Breit Hamiltonian. In its
application to the electron--proton interaction, ever since its
derivation 75 years ago, in 1929, \cite{Breit}, the Breit Hamiltonian
(improved for the anomalous magnetic moment of nucleons) has been the
fundamental tool of atomic physics. It is found in any textbook on QED
and atomic physics \cite{AkhPomer,LandauLifshitzBreit} and need not be
reproduced here. From the point of view of spin--dependence, it
contains the spin--orbit, spin--tensor and hyperfine spin--spin
interactions.

Now, recall that in the scattering of spin--0 particles off
spin--${1\over 2}$ particles the target--spin asymmetry (analyzing
power) and normal polarization of the recoil spin--${1\over 2}$
particle equal each other: $P_y=A_y$.  In the more complex case of
$$ {1\over 2}+{1\over 2}\to {1\over 2}+{1\over 2}
$$ scattering, the number of polarization observables is much higher
\cite{NNreview}.  Evidently, the existence of spin--tensor and
hyperfine interactions would give rise to a beam--target double--spin
asymmetry.  Simultaneously, they entail the correlation between
induced polarizations of scattered particles and, most importantly for
the present discussion, of the polarization transfer from the
polarized beam particle to the recoil target particle. For the Breit
interaction the relationships between these observables are trivial;
the general discussion with application to nucleon--nucleon
interactions is found in the classic review by Bystricky et al.
\cite{NNreview}.

The Horowitz--Meyer derivation \cite{HorowitzMeyer} of the
polarization transfer from electrons in the polarized hydrogen atom to
scattered (anti)protons can be described as based on the Breit
Hamiltonian, although they use directly the relativistic approach and
skip Breit's reduction to the non--relativistic formalism. In terms of
the Breit Hamiltonian, the factor of 2 difference between the transfer
of the transverse and longitudinal polarizations from the electron to
proton stems from the spin--tensor interaction.

\subsection{Polarized Electron--Proton  Elastic Scattering: Theory}
Now we turn to a brief overview of applications to high--energy
electron--nucleon scattering and the independent determination of the
charge and magnetic moment structure of nucleons. Our coverage is far
from being complete as we only wanted to emphasize an enormous advance
in the form factor studies resulting from the electron--to--nucleon
polarization transfer technique.

In 1957 Akhiezer et al. published the first analysis of the scattering
of polarized electrons on polarized protons \cite{Akhiezer1957}.  They
noticed correctly that polarization effects could in principle be
applied to determine $G_E$.  In the quest for methods to determine the
longitudinal polarization of leptons produced in weak interactions,
Bincer in 1957 considered the electromagnetic scattering of
longitudinally polarized leptons on leptons
\cite{Bincer}. Electron--proton scattering for arbitrary
polarizations, with allowance for the anomalous magnetic moments, has
been treated by Scofield in 1959 (\cite{Scofield1959}, see also the
follow--up paper on three and four spin observables
\cite{Scofield1966}).

The modern treatment of the scattering of polarized leptons at high
energy is for the most part based on the 1969 Rev.~Mod.~Phys. \ paper
by N. Dombey \cite{Dombey}. For ultrarelativistic electrons the
dependence on the transverse polarization of the electron vanishes.
This is a consequence of the conservation of the helicity of
ultrarelativistic electrons in high--energy QED, first noticed in 1954
by Yennie, Ravenhall and Wilson (\cite{Yennie1954}, see also the
textbook \cite{LandauSCHC}). This is precisely the phenomenon by which
the transverse polarization effects can only be seen in the
annihilation processes $e^{-\uparrow} \vec{e}^{+\uparrow} \to
\mu^+\mu^-, q\bar{q}$, or in the Drell--Yan reaction $q^{\uparrow}
\bar{q}^{\uparrow} \to e^+e^-$. Dombey shows that when the incident
lepton is longitudinally polarized, the virtual photon is in a pure
polarization state which is a coherent superposition of an
elliptically polarized transverse state and a longitudinal state.

Extracting the new information contained in the interference between
$xy$, i.e., $TT'$, and $yz$, i.e., $LT$ components of the photon
absorption amplitude, requires scattering off polarized targets or,
alternatively, measurements of the recoil polarization.  Dombey made a
strong point, implicitly contained in Ref.~\cite{Akhiezer1957} and
especially in the 1968 paper by Akhiezer and Rekalo
\cite{Akhiezer1968}, that the most interesting experiment is to
measure an interference term of the form $G_E G_M$ and so find $G_E$,
including its sign. Dombey emphasized that the possibility of $G_E$
changing its sign at large $Q^2$ is not excluded. The target must be
polarized perpendicular to the virtual photon in the Breit frame, i.e.
perpendicular to the direction of the recoil nucleon in the laboratory
and in the scattering plane.  In modern language, one must measure the
double--spin longitudinal--transverse asymmetry. The double--spin
longitudinal--longitudinal asymmetry is proportional to $G_M^2$ (see
also Eq. (1) below).  Both the double--spin asymmetry and
electron--to--proton polarization transfer techniques have been
applied subsequently at SLAC, MIT--Bates, MAMI and Jlab.

\subsection{Beam--target Double--spin Asymmetry and $G_E/G_M$}
The first determination of the sign of $G_E/G_M$ by the
double--polarization asymmetry in $\vec{e}\vec{p}$ elastic scattering
at $Q^2=0.765$ GeV$^2$ was performed in 1976 at SLAC
\cite{SLACALT}. The authors comment that the practical usefulness of
the method at higher $Q^2$ is limited by low counting rates. The first
double polarization measurements of the neutron electric form factor
via neutron knockout from a polarized $^3$He target,
$^3\vec{\mathrm{He}}(\vec{e},e'n)$, were carried out by the MAMI
collaboration (\cite{MeyerhoffMAMI}, for the subsequent MAMI
experiments see \cite{RoheMAMI,GlazierMAMI} and references therein).
In several Jlab experiments the neutron form factor has been evaluated
from quasielastic scattering off polarized deuterons,
$\vec{d}(\vec{e},e'n)$ \cite{PasschierJlab,ZhuJlab,WarrenJlab}. An
excellent summary of these data is shown below in Fig.~\ref{fig:GeN},
borrowed from the MAMI publication by Glazier et
al.~\cite{GlazierMAMI}.

\begin{figure}[hbt]
 \begin{center}
 \includegraphics[width=0.8\linewidth]{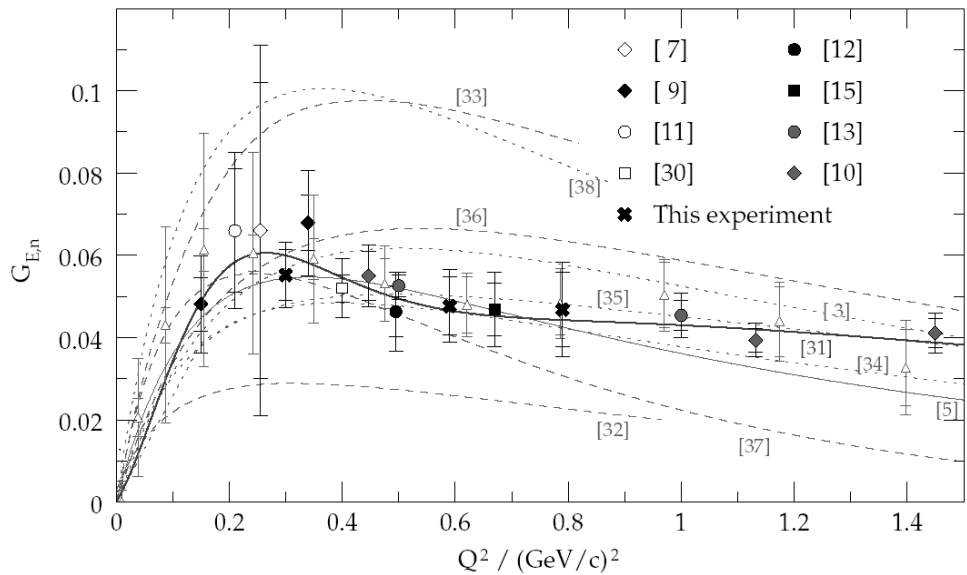}
  \parbox{14cm}{\caption{\label{fig:GeN}\small 
$G_{E,n}$ from double--polarization experiments as compiled
  by Glazier et al. (MAMI) \cite{GlazierMAMI}.  Polarization--transfer
  measurements on the deuteron \cite{Ede94b,Her99,Mad03} are marked
  with diamonds, experiments using polarized Deuterium
  \cite{PasschierJlab,ZhuJlab,WarrenJlab} or $^3\vec{\rm He}$
  \cite{Ber03,Gol01} targets are shown as circles and squares,
  respectively.  "This experiment" refers to the MAMI experiment by
  Glazier et al.  \cite{GlazierMAMI}.  Open triangles refer to the
  analysis \cite{Sch01} of unpolarized data. The thin full curve
  represents the original Galster parameterization \cite{Gal71}, the
  thick line represents the ``pion--cloud'' parameterization
  \cite{Fri03} (see text). The legend of the dashed and dotted lines
  is found in Ref.  \cite{GlazierMAMI}.
}}
\end{center}
\end{figure}
\subsection{Electron Proton Polarization Transfer and  $G_E$/$G_M$: Theory}
The advantages of the electron--to--proton and electron--to--neutron
polarization transfer technique were recognized in 1974 by Akhiezer
and Rekalo \cite{Akhiezer1974} and elaborated in 1981 by Arnold,
Carlsson and Gross \cite{Arnold1981}. The method requires that the
polarization of the recoiling nucleon be measured in a second,
analyzing, scattering. The recoil nucleon is polarized in the
scattering plane and has polarization components either transverse
(sideways), $P_T$, or longitudinal, $P_L$, to its momentum:
\begin{eqnarray}
P_T&=&hD_{LT}= {h\over I_0}\cdot \left(-2\sqrt{\tau (1+\tau)} 
G_M G_E \tan{\theta_e\over 2}\right)\nonumber\\
P_L&=&hD_{LL}= {h\over I_0}\cdot{E+E'\over M_p}
\cdot \sqrt{\tau (1+\tau)} 
(G_M)^2 \tan^2{\theta_e\over 2}\, .
\end{eqnarray}
Here $h$ is the helicity of the incident electron, $E$ and $E'$ are
the laboratory energies of the incident and scattered electron,
respectively, $\theta_e$ is the electron scattering angle and $I_0$ is
the unpolarized cross section (excluding the Mott cross section). In
accordance with general theorems \cite{NNreview}, there is a
one--to--one correspondence between the double--spin beam--target
asymmetries and electron--to--nucleon polarization--transfer.  The
recoil polarimeters give access to the transverse polarization of the
recoil particle; using spin rotation in the special spin--rotator
and/or spectrometer magnets, one can measure simultaneously both $P_T$
and $P_L$ in the same polarimeter. Then the ratio of the two
polarization components would give the ratio of form factors,
\begin{eqnarray}
{G_E \over G_M}=-{P_T \over P_L} \cdot  {E+E'\over 2M_p}\cdot 
\tan{\theta_e\over 2}\, .
\end{eqnarray}
Neither the beam polarization nor the analyzing power needs to be
known. Because the two polarization observables are measured
simultaneously, this technique avoids a major systematic uncertainty
of the Rosenbluth separation. An important virtue of the
polarization--transfer technique is that it is viable at large $Q^2$.

\begin{figure}[hbt]
 \begin{center}
 \includegraphics[width=0.8\linewidth]{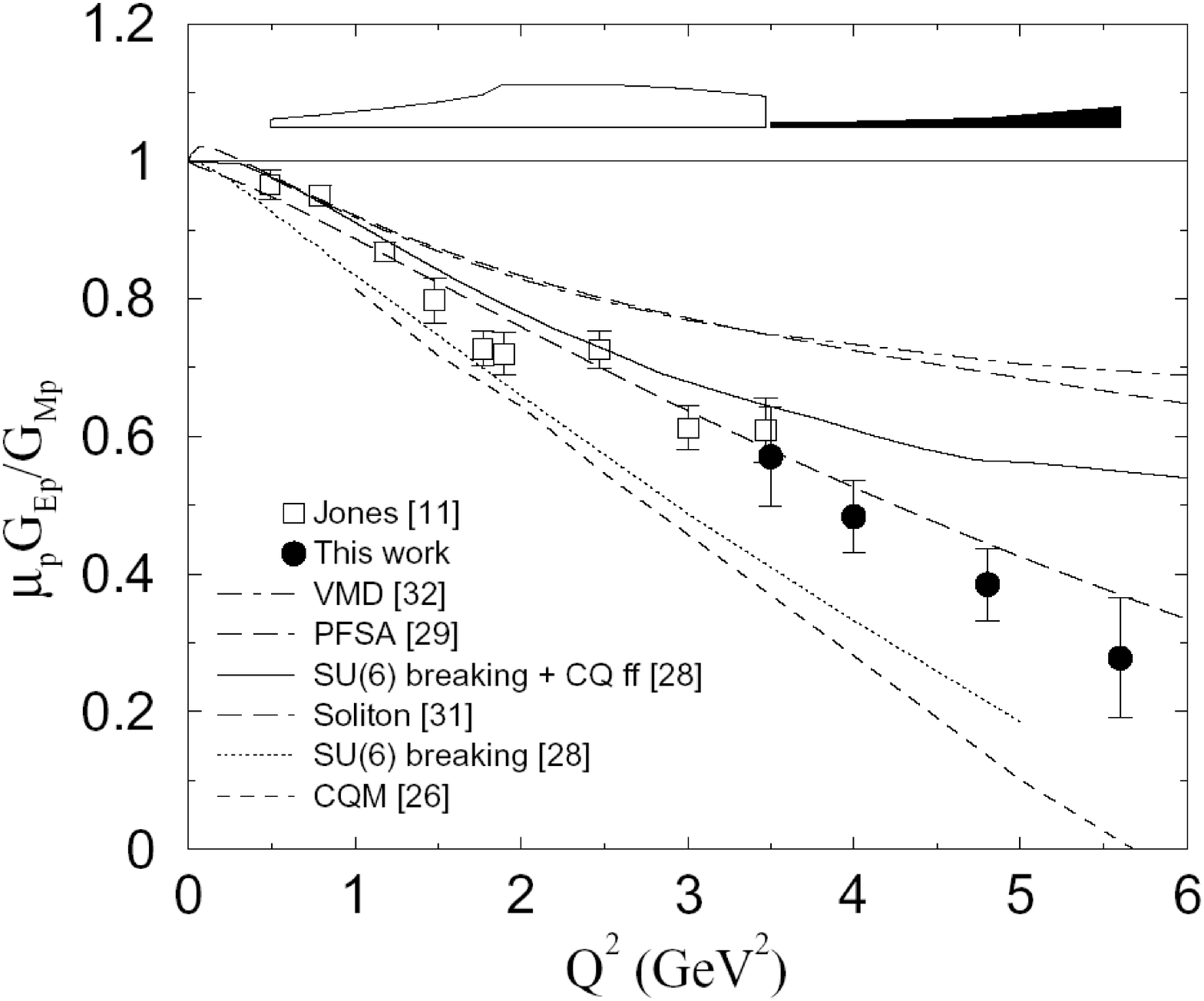}
  \parbox{14cm}{\caption{\label{fig:gegm_new2}\small A compilation of
  the experimental data on the ratio $\mu_p G_{Ep}/G_{Mp}$ from the
  Jefferson lab experiments \protect{\cite{JonesJlab,GayouJlab}}
  compared with theoretical calculations (for the references see
  \protect\cite{GayouJlab}).  Systematic errors are shown as a band at
  the top of the figure.  }}
\end{center}
\end{figure}

\begin{figure}[hbt]
 \begin{center}
 \includegraphics[width=0.7\linewidth]{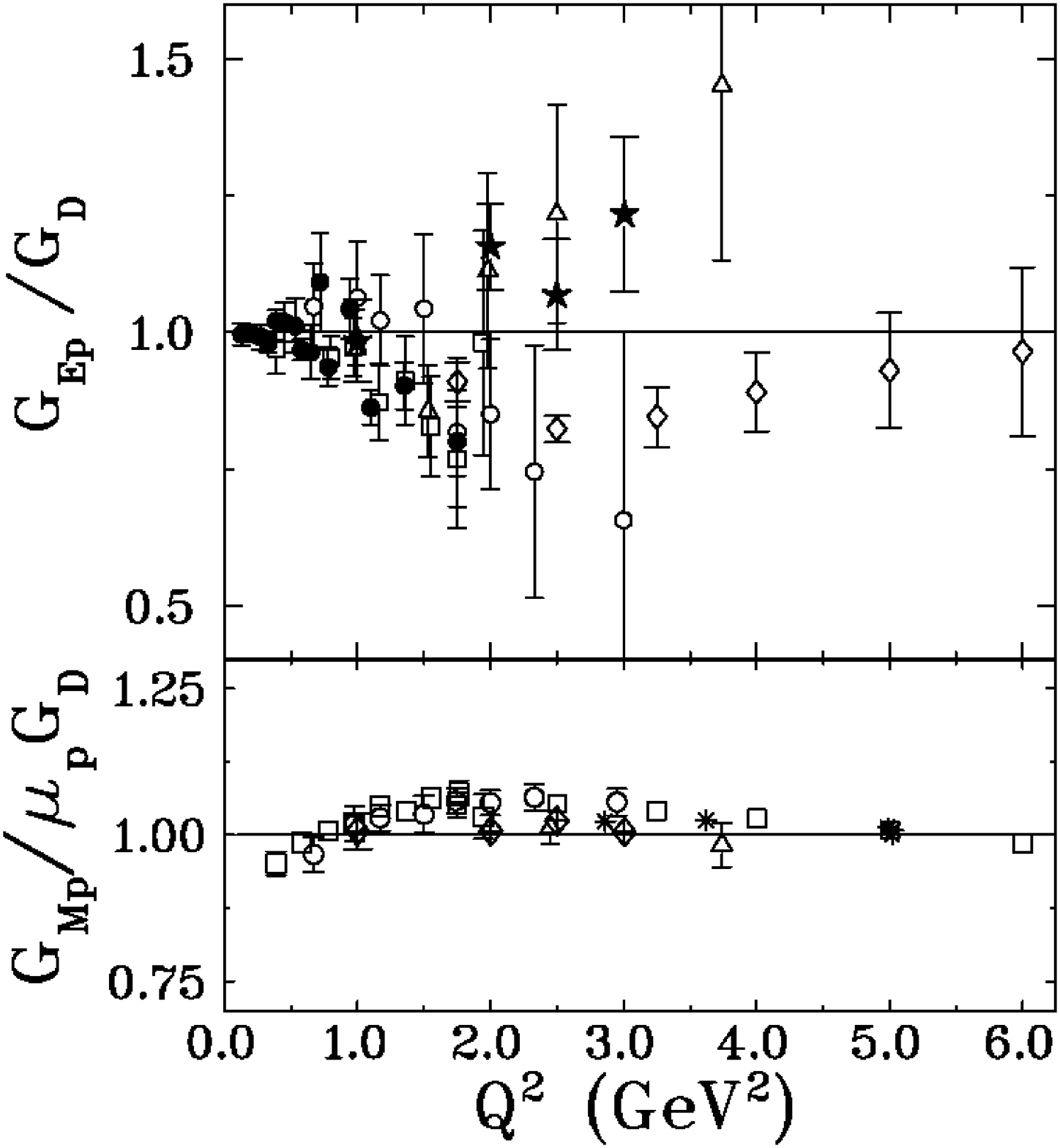}
  \parbox{14cm}{\caption{\label{fig:GeOld}\small World data prior to
1998 for the ratio of form factors to the dipole parameterization
$G_D$: (a) $G_{E}^p/G_D$ and (b) $G_{M}^p/\mu_p G_D$ versus $Q^2$.
Refs. Litt {\it et al.} ($\triangle$) \cite{litt}, Berger {\it et al.}
($\Box$) \cite{berger}, Price {\it et al.} ($\bullet$) \cite{price},
Bartel {\it et al.}  ($\circ$) \cite{bartel}, Walker {\it et al.}
($\star$) \cite{walker}, Andivahis {\it et al.}  ($\Diamond$)
\cite{andivahis} and Sill {\it et al.} ($\ast$) \cite{sill}.  }}
\end{center}
\end{figure}

\subsection{Electron Proton Polarization Transfer: Experiment}
Milbrath et al.\ were the first to apply recoil polarimetry to the
determination of $G_E/G_M$ at low $Q^2$ for protons in the MIT--Bates
experiment \cite{MilbrathBates}. Similar measurements have been
carried out at MAMI \cite{PospischilMAMI}. Particularly noteworthy is
a series of beautiful large--$Q^2$ experiments at Jlab
\cite{JonesJlab,GayouJlab} which resulted in the discovery of a steep
decrease of the ratio $G_E^p/G_M^p$ with $Q^2$ which continues to the
largest $Q^2$ studied (see Fig.~\ref{fig:gegm_new2}, borrowed from
Ref.~\cite{GayouJlab}).  These results must be compared to the world
data prior to 1998 shown in Fig.~\ref{fig:GeOld}, borrowed from
Punjabi et al. \cite{Punjabi}.  The latter paper gives an excellent
introduction to the experimental realization of the
polarization--transfer technique.

The polarization--transfer technique has led to a dramatic improvement
in the determination of $G_E^p/G_M^p$, which inspired the recent
flurry of theoretical interest in the time--like form factors
discussed in Section 4 of the PAX Technical Proposal. A very detailed
discussion of implications of the new data for the theoretical
description of form factors is found in Ref. \cite{Punjabi}.

\subsection{Electron--to--Neutron Polarization Transfer and  $G_E^n/G_M^n$: Experiment}

The polarization--transfer technique proved equally fruitful in
applications to the quasielastic scattering off polarized deuterons
and as a source of information on the neutron form factors.
Arenh\"ovel et al.\ made an important point \cite{Arenhoevel} that in
the application of the technique to quasifree scattering off the
deuteron, the corrections for final state interactions, meson exchange
currents, isobar configurations, and the models of the deuteron
structure, are well under control (for a further theoretical analysis
of quasifree scattering off $^3\mathrm{He}$ within the Faddeev
approach and the comparison of extraction of $G_E^n$ from quasifree
scattering off the $^3\mathrm{He}$ and deuterium targets see
\cite{Gol01}). This point has been confirmed experimentally by a
direct comparison of the determinations of $G_E^p/G_M^p$ from recoil
polarimetry in elastic, $\vec{e}p\to \vec{p}e$, and quasielastic,
$\vec{e}(d,\vec{p}e')$, scattering.  The MIT--Bates experiment by
Milbrath et al.~\cite{MilbrathBates} was the first to do so at low
$Q^2$, and the results from the large--$Q^2$ Jlab experiment are
reported in Ref.~\cite{MilbrathJlab}, where one would find further
references.

Eden et al. in 1994 were the first to apply the polarization--transfer
technique at MIT--Bates \cite{Ede94b}; subsequently it has been used
at MAMI \cite{GlazierMAMI} and Jlab (\cite{Mad03} and references
therein). Madey et al. \cite{Mad03} make a convincing point that the
recoil polarimetry made facilitated measurements at $Q^2$ as large as
1.13, and 1.45 (GeV/c)$^2$, which had never been achieved before in
polarization measurements.  Figure \ref{fig:GeN}, borrowed from Ref.
\cite{GlazierMAMI}, summarizes the experimental data on $G_E^n$ from
double--polarization experiments.

Besides the much smaller error bars compared to those provided by
analysis of the earlier data from unpolarized scattering, the
interpretation of the recoil--polarimetry data is free of model
uncertainties inherent in the interpretation of the unpolarized data.
This particular point prompted the use of polarization transfer in
$^4\mathrm{He}(\vec{e}, e'\vec{p})^3\mathrm{H}$ by a recent Jlab
experiment \cite{StrauchJlab} as a sensitive probe of the long sought
modification of the form factors of a proton bound in a nucleus
(\cite{StrauchJlab} and references therein).

\subsection{Summary}

The electron-to-proton polarization transfer at energies of the
Antiproton Polarizer Ring (APR) is a simple quantum-mechanical problem
based on the Breit Hamiltonian which has been derived in 1929 from
fundamental QED. Ever since its derivation the Breit Hamiltonian has
been at the core of the atomic physics. At relativistic energies
precisely the same QED mechanism describes double-spin asymmetries in
the scattering of longitudinally polarized electrons on polarized
nucleons and/or the polarization transfer from longitudinally
polarized electrons to recoil nucleons.  In high-energy experiments
the interest is not in testing the straightforward QED calculations
per se, rather the polarization-transfer, in conjunction with the
recoil polarimetry, has become a fundamental tool for the
high-precision determinations of the charge and magnetic moment
distributions in protons and neutrons. The unprecedented accuracy
achieved in those experiments uncovered the properties of the form
factors which triggered a lively theoretical dispute on the onset of
hard pQCD asymptotics both in the space like and time like region,
which is behind the experimental program outlined in this document.

\cleardoublepage
\part{Appendix B \label{appendix-B}}
\pagestyle{myheadings} \markboth{Appendix B to the Technical Proposal
  for ${\cal PAX}$}{Polarized Antiproton-Proton Soft Scattering}
\section{Polarized Antiproton-Proton Soft Scattering}
In this Appendix
we comment in more detail on the impact of the spin-dependence of
  antiproton-proton interaction on the interpretation of the
  Coulomb-Nuclear Interference data and on the special interest in
  double-polarized antiproton-proton scattering at very low energies
  in view of the indications for the protonium state.

\subsection{Spin effects in the Interpretation of Coulomb--Nuclear Interference}
The compilation of the experimental data on the ratio of the real to
imaginary part, $\rho$, of the $p\bar{p}$ forward scattering
amplitude, shown in Sec. 6.1 of the PAX Technical Proposal
suggests a substantial departure of the experimental data points from
the dispersion theory calculations (DT). The extraction of $\rho$ from
the experimentally measured differential cross section of elastic
scattering in the Coulomb-Nuclear Interference (CNI) region is usually
done assuming a negligible spin dependence of the elastic scattering
amplitude:
\begin{eqnarray}
{d\sigma\over dt} &=& {\pi \over p^2} \bigl|f_C\exp(i\delta_B) +f_N\bigr|^2
={d\sigma_C\over dt}+{d\sigma_{int}\over dt}+{d\sigma_N\over dt}\,,\\
\label{eq:CNI1}
{d\sigma_C\over dt} &=&{4\pi \alpha_{em}^2 G_E^{4}(t) \over \beta^2t^2}\,,\\
\label{eq:CNI2}
{d\sigma_N\over dt} &=&  {\sigma_{tot} ^2 (1+ \rho^2)
\over 16\pi}\cdot \exp(-B|t|),\\
\label{eq:CNI3}
{d\sigma_{int}\over dt}&=& {\alpha_{em}^2  \sigma_{tot} G_E^{2}(t)\over \beta
|t|}\cdot (\rho \cos \delta_B + \sin \delta_B) \cdot  \exp(-{1\over 2} B|t|) \, .
\label{eq:CNI4}
\end{eqnarray}
Jakob and Kroll \cite{KrollCoulombPhase2} make a point that
\begin{itemize}
\item The optical point at $t=0$,
\begin{equation}
 {d\sigma \over  dt}\bigr|_{t=0} = {1\over 16\pi}\left({1\over 4} \sigma_{s} ^2 (1+ \rho_s^2)
+{3\over 4} \sigma_{t} ^2 (1+ \rho_t^2)\right) 
\label{eq:CNI5}
\end{equation}
will differ from given by formula (3) with the spin averaged
quantities,
\begin{eqnarray}
\sigma_{tot}&=& {1\over 4} \sigma_{s} +{3\over 4} \sigma_{t}\, \\
\label{eq:CNI6}
\rho &=& {1\over \sigma_{tot}} \left( {1\over 4} \sigma_{s}\rho_s +{3\over 4} \sigma_{t}\rho_t\right)\,.
\label{eq:CNI7}
\end{eqnarray}
where, for the sake of illustration, we only consider the simplest
example with
\begin{equation}
f_N= \left( {1\over 4} f_{s} +{3\over 4}f_{t}\right) + (f_t-f_s)(\vec{s}_p\vec{s}_{\bar{p}})\,,
\label{eq:CNI8}
\end{equation}
the discussion of the general case is found in
\cite{KrollCoulombPhase2}.

\item The $t$-dependence of amplitudes $f_{s,t}$ for the spin-singlet
  and spin-triplet states could be different, i.e., taking one and the
  same diffraction slope $B$ in the exponential $t$-dependent factors
  in the strong interaction term (\ref{eq:CNI3}) and the CNI term
  (\ref{eq:CNI3}) is an assumption which must be tested experimentally
 
\item The so-called Bethe phase, $\delta_B$, between the Coulomb and
  the strong-interaction amplitudes can vary from one spin state to
  another.
\end{itemize}

\begin{figure}[hbt]
 \begin{center}
\vspace{0.2cm}
 \includegraphics[width=0.8\linewidth]{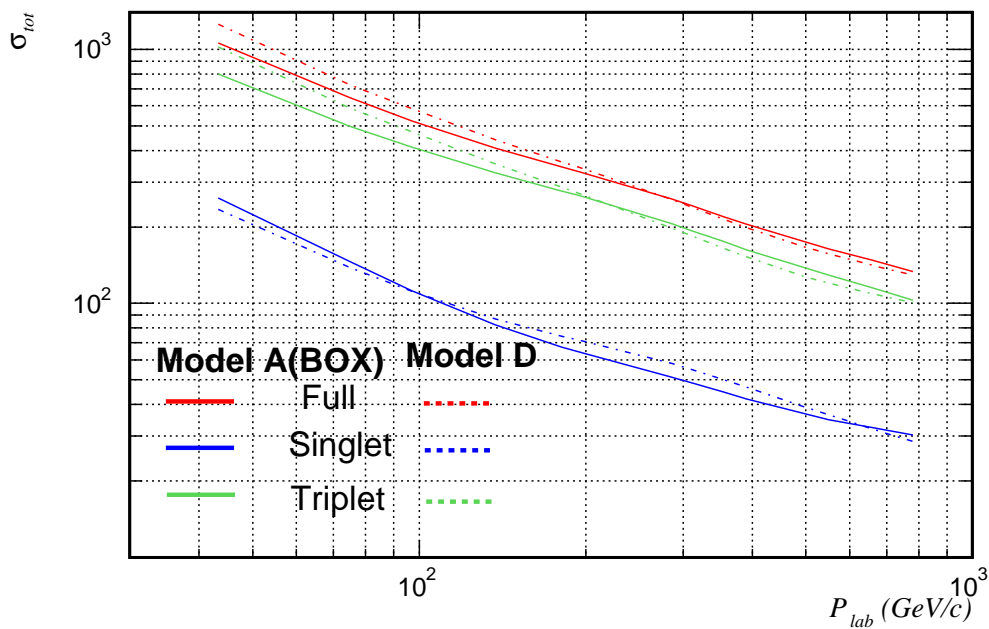}
  \parbox{14cm}{\caption{\label{fig:NbarNsigma}\small Predictions for
  the momentum-dependence of the spin-singlet, spin-triplet and
  spin-averaged antiproton-proton total cross section for the two
  models of $p\bar{p}$ interaction described in the text.  }}
\end{center}
\end{figure}

The spin-dependence of antiproton-proton scattering is an entirely
uncharted territory and in order to get an idea on the expected
effects one must resort to predictions of phenomenological models.
Such models were developed earlier in connection with the experimental
studies of the interactions of unpolarized antiprotons at LEAR and
CPLEAR.  We illustrate the major points on an example of the
spin-singlet and spin-triplet forward scattering amplitudes evaluated
in two $N\bar N$ models developed by the J\"ulich group
\cite{Hippchen,Mull} within the meson-exchange picture. Specifically,
we show here predictions of the models A(BOX) and D introduced in
Refs.~\cite{Hippchen} and \cite{Mull}, respectively.
The elastic part of these $N\bar N$ interaction models is obtained by
a G-parity transform of the full Bonn $NN$ potential \cite{MHE}. In
case of A(BOX) annihilation is accounted for by a phenomenological
\hbox{spin-}, isospin- and energy-independent complex optical
potential of Gaussian form.
Model D utilizes the same elastic part. However, annihilation is now
described in part in terms of microscopic baryon-exchange processes
based on $N$, $\Delta$, $\Lambda$, $\Sigma$, and $Y^*$ exchange and
involving $N\bar N{\to}$ 2 meson decay channels with all possible
combinations of $\pi$, $\eta$, $\rho$, $\omega$, $a_0$, $f_0$, $a_1$,
$f_1$, $a_2$, $f_2$, $K$, $K^*$ - see Ref. \cite{Mull} for details.
For both models the results for the total and the integrated elastic
and charge-exchange cross sections as well as for angular dependent
observables are in good agreement with the available experimental
information \cite{Hippchen,Mull}.

\begin{figure}[hbt]
 \begin{center}
\vspace{0.2cm}
 \includegraphics[width=0.8\linewidth]{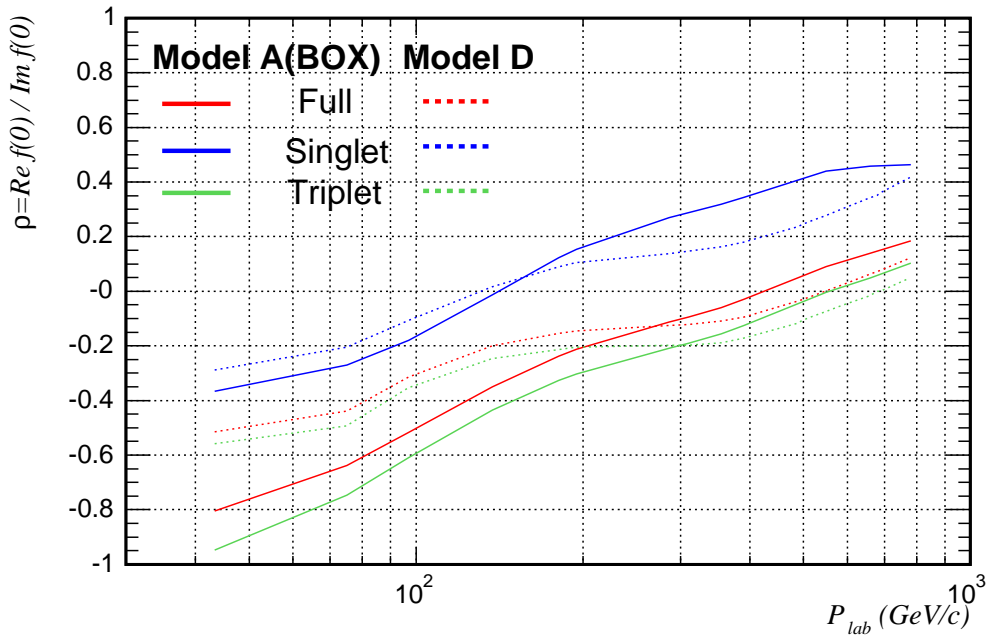}
  \parbox{14cm}{\caption{\label{fig:NbarNRho}\small The
 momentum-dependence of the ratio of real and imaginary parts of the
 spin-singlet, spin-triplet and spin-averaged forward
 antiproton-proton scattering amplitudes as predicted by the two
 models for $p\bar{p}$ interaction described in the text.  }}
\end{center}
\end{figure}

In Fig.~\ref{fig:NbarNsigma} we show the momentum dependence of the
total cross section for the spin-triplet, spin singlet and
spin-averaged interactions. Both models predict a dominance of the
spin-triplet cross section over that in the spin-singlet case,
typically by the factor $\sim 5$. The difference between the
predictions from the two models is marginal.  In
Fig.~\ref{fig:NbarNRho} we show the ratio of the real to imaginary
part for the amplitude of forward elastic scattering in the
spin-triplet and spin-singlet states. To this end one must recall, see
Figure 8 in the PAX TP, that in the considered momentum range the
statistical error bars in the experimental data points for $\rho$ are
of the order of $\Delta \rho \sim 0.05$.  This must be contrasted with
the very large variation of $\rho$, by $\approx 0.4$, from the
spin-triplet to the spin-singlet state.  Furthermore, the departure of
predictions of the models from each other is of the order of 0.1.
These results, in conjunction with the related results for the total
cross section, suggest a substantial spin dependence of low and
intermediate energy antiproton-proton interactions and make a strong
case for the low-$t$ physics with polarized antiprotons scattering on
polarized protons.

Such measurements at Phase-I will cover precisely the most interesting
range of momentum, in which the early experimental data on $\rho$
exhibit a nontrivial momentum dependence and where the DT calculations
and the data from the previous experiments seem to diverge most.  The
further measurements can also be extended to Phase-2 in the fixed
target mode.

\subsection{Separation of Spin--Singlet and Spin--Triplet Scattering at Low Energy}
If polarized antiprotons stored in CSR can be decelerated down to
$\sim 10$ MeV or still lower energies, one can also study the
spin-dependence of the $N\bar N$ interaction at very low energies.
This interaction is of strong relevance for the interpretation of the
$p\bar p$ mass spectrum measured recently by the BES collaboration in
the decay $J/\Psi \to \gamma p\bar p$ \cite{BES}, which shows a strong
enhancement near the $p\bar p$ threshold. The observed enhancement led
to speculations that one has found a signal of an $N\bar N$ bound
state (protonium) or an indication for a so far unobserved narrow
resonance with the quantum numbers $J^{PC}$ = $0^{-+}$ or $0^{++}$
\cite{BES}.  Investigations by Sibirtsev et al. \cite{Sibirtsev}
(among others) have shown that the interpretation of this enhancement
in terms of the final state interaction in the $p\bar p$ system is not
excluded. However, one has to keep in mind that the $p\bar p$ system
in the $J/\Psi \to \gamma p\bar p$ decay is in a spin-singlet state
near threshold and the corresponding $p\bar p$ cross section is very
small compared to the spin-triplet cross section, as mentioned above,
see Figure \ref{fig:NbarNsigma}.  As a consequence the spin-singlet
amplitudes predicted by the models are only poorly constrained by the
presently available $N\bar N$ data and, therefore, solid conclusions
on the origin of the strong enhancement seen by the BES collaboration
in the $p\bar p$ mass spectrum cannot be made at present. A
measurement of the spin-dependence of $N\bar N$ interaction at low
energies would certainly help to clarify this issue.  Challenging
though such a deceleration is, the cross sections one needs to measure
are in the several hundred millibarn range, and the recoil detector
described in Section 16.8 of the PAX TP can still be operated even at
such low energies.

\subsection{The Impact of the Spin--Dependence of Antiproton--Proton 
Scattering on the Polarization Buildup \label{impact}}
\begin{figure}[hbt]
 \begin{center}
 \includegraphics[width=0.8\linewidth]{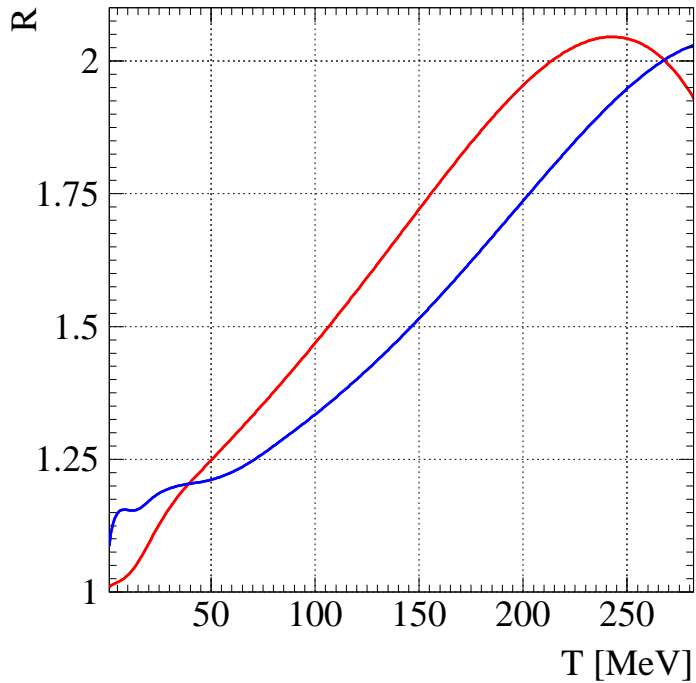}
  \parbox{14cm}{\caption{\label{fig:em_and_had}\small Energy
 dependence of the ratio $R=(\sigma_{e||} + \sigma_{\bar{p}p}) /
 \sigma_{e||}$. The predictions of the two models A(Box) (red line)
 and D (blue) indicate a sizeable increase of the polarizing cross
 section in the energy range of interest.}}
\end{center}
\end{figure}

The above presented model calculations suggest a non--vanishing
proton--to--antiproton polarization transfer. The suggested approach
to the polarization buildup, as outlined in Sec. 8.2 of the PAX
Technical Proposal, is based on the injection into the hydrogen gas
target of two hyperfine states such that the nuclear polarization of
antiprotons is close to zero. The rate of polarization buildup is
determined by the reliably known electron--to--proton polarization
transfer.  The method of the direct experimental determination of the
proton--to--antiproton polarization transfer is described in Sec. 6.2
of the PAX TP. It might happen that the proton--to--antiproton
polarization transfer is so strong that one can profit from the
constructive interference of the electron and proton contributions,
operating the polarized target at a lower density with one hyperfine
state only.  The current theory of spin effects in antiproton--proton
interactions can not be trusted enough to serve as a basis for the
decision about the polarization buildup mode employed for PAX. In
Fig.~\ref{fig:em_and_had}, we give for the case longitudinal spin
transfer the ratio $R=(\sigma_{e||} + \sigma_{\bar{p}p}) /
\sigma_{e||}$, where $\sigma_{e||}$ is the $ep$ spin transfer cross
section, depicted in Fig.~10. The spin--dependent hadronic part
$\sigma_{\bar{p}p}$ was taken from the two model predictions, A(Box)
(red line) and D (blue), described earlier. The behavior should be
taken as an indication that a sizeable improvement in the
spin--dependent cross section, responsible for the polarization
buildup of antiprotons, could be achieved.  During spin--filtering
using both electromagnetic and hadronic contributions, $\sigma_{e||} +
\sigma_{\bar{p}p}$, the polarized target has to be operated by
injection of a single hyperfine state into the storage cell, whereby
the target density is reduced by a factor of two compared to
spin--filtering with a purely electron--polarized gas target, for
which two hyperfine states can be injected. Thus, in order to benefit
from the constructive interference of electron and proton
contributions, the value of $R$ must be larger than two.

\subsection{Summary}
The spin--dependence of the antiproton--proton interactions is an
entirely uncharted territory. The available theoretical models were
developed more than a decade ago. They give a good description of the
wealth of the experimental data, predominantly on the two--body final
states, from CPLEAR; however, they have never been tested against the
experimental data on double--spin observables. The models indicate a
fairly strong spin effects, for instance, a very strong suppression of
the spin--singlet cross section.


\cleardoublepage
\part{Appendix C}
\pagestyle{myheadings} \markboth{Appendix C to the Technical Proposal
  for ${\cal PAX}$}{Electromagnetic Form Factors in the Time-like
  Region}
\section{Resonance Structures and Phase Motion of the Electromagnetic Form Factors 
in the Time-like Region}
The experimental investigation of the time-like electromagnetic form
factor of the proton has caused considerable interest in resonances in
the vicinity of the proton-antiproton threshold. The $\bar{p}p
\rightarrow e^+e^-$ data show considerable enhancements above the
threshold, see Figure 5 in Section 4 of the PAX Technical Proposal,
which can be explained by subthreshold mesons, see
e.g. Refs.~\cite{Williams95,Hammer96,hammer03, dub03,iach04}. The point
made in the PAX TP is that such structures seen in the modulus of the
form factor imply, by virtue of the analyticity arguments, a
nontrivial variation of the phase of form factors as a function of
$q^2$, which can be measured in single- and double-polarized $\bar{p}p
\rightarrow e^+e^-$ at Phase-I of PAX experiment. In this Appendix
we discuss to more detail the impact of recent electron-positron
collider results on proton-antiproton physics.

\begin{figure}[hbt]
 \begin{center}
 \includegraphics[width=0.49\linewidth]{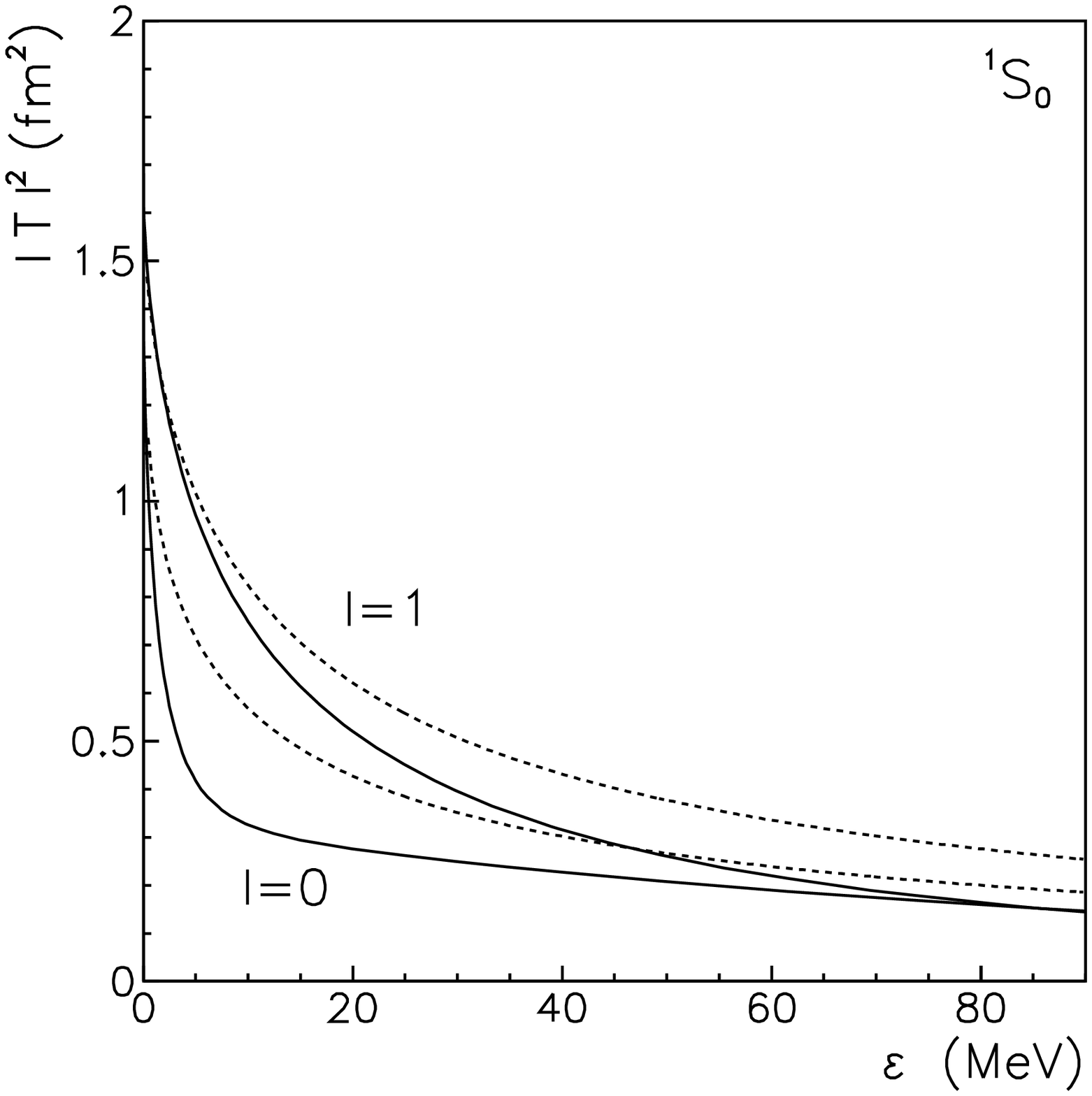}
 \includegraphics[width=0.49\linewidth]{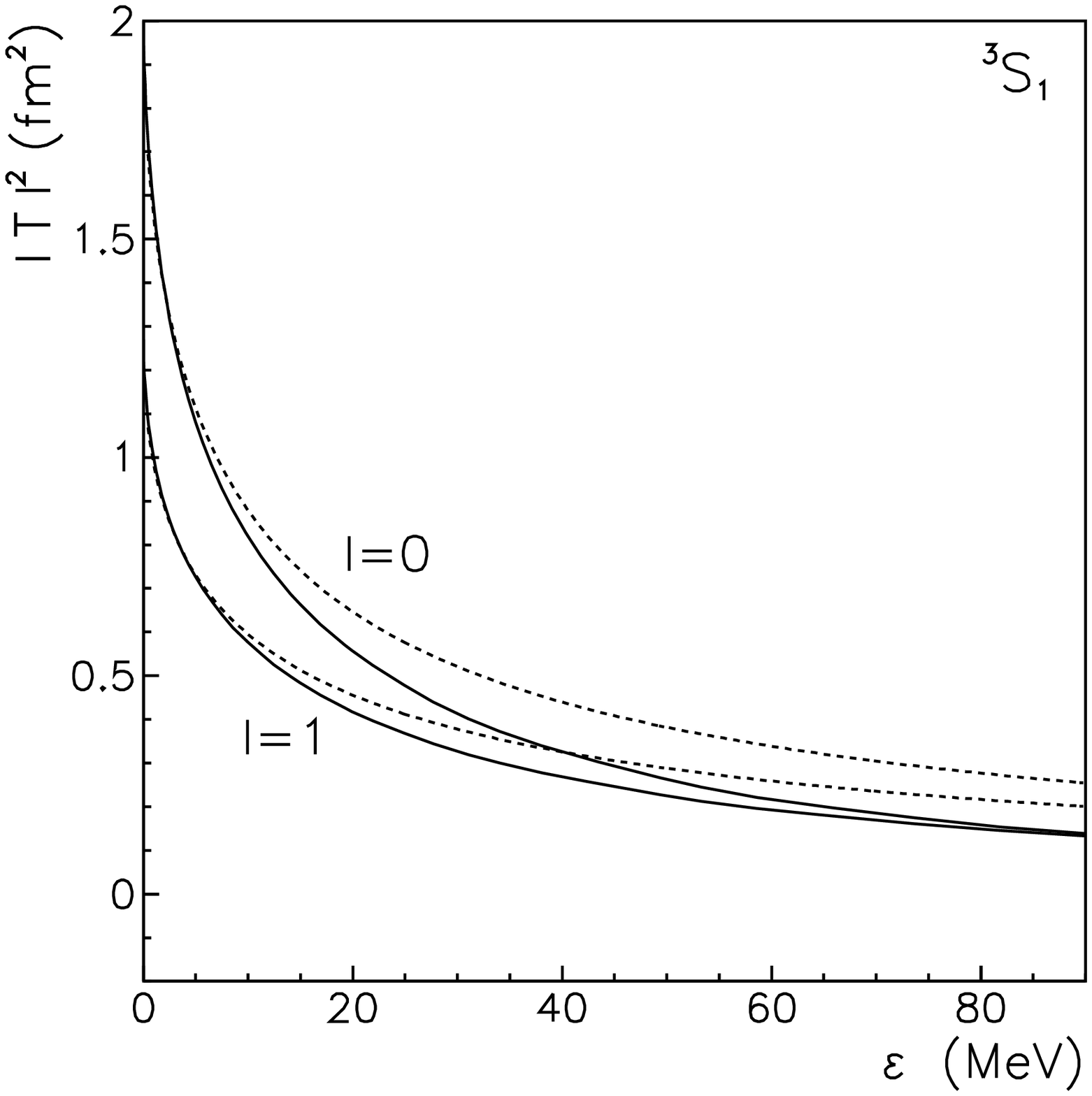}
  \parbox{14cm}{\caption{\label{pbar6}\small The $p{\bar p}$
scattering amplitudes for the $^1S_0$ and $^3S_1$ partial waves as a
function of the center-of-mass energy.  The solid lines show the
results of the J\"ulich model while the dashed lines indicate the
scattering length approximation.  }}
\end{center}
\end{figure}

\subsection{Resonance Structures and Nuclear Baryonium States in
$N\bar{N}$ Interactions}
In the past, antiproton-proton reactions have made extremely important
contributions to the knowledge about meson-like hadronic resonances.
The Low-Energy Antiproton Ring LEAR at CERN has operated with low
momentum antiprotons (momenta between 60 MeV/c and 140 MeV/c) in the
time period between 1983 and 1996.  A rich harvest of light mesons has
been collected~\cite{amsler98}. Our knowledge of scalar mesons has
been substantially increased by the firm establishment of the
$a_0(1450), f_0(1370)$, and $f_0(1500)$ and their decay modes. Two
isoscalar $2^{-+}$ mesons, $\eta_2(1645)$ and $\eta_2(1870)$, and the
$0^{-+} \eta(1410)$ have been observed. Moreover, an exotic
$\bar{\rho}(1450)$ meson with quantum numbers $1^{-+}$ has been
identified by its $\pi \eta$ decay. There is a general consensus that
we have seen non-$\bar{q}q$ mesons~\cite{Amsler}.

The experimental quest for {\it nuclear baryonia}, i.e. quasibound
$\bar{N}N$ states or resonances, on the other hand, never matched
theoretical expectations, mainly because annihilation broadens any
possible resonance structure.  Some moderately broad baryonia
candidates can survive, though ~\cite{lacombe84}.  More details and
references to early works can be found in a recent review by Klempt et
al.~\cite{klempt02}.

\begin{figure}[hbt]
 \begin{center}
 \includegraphics[width=0.49\linewidth]{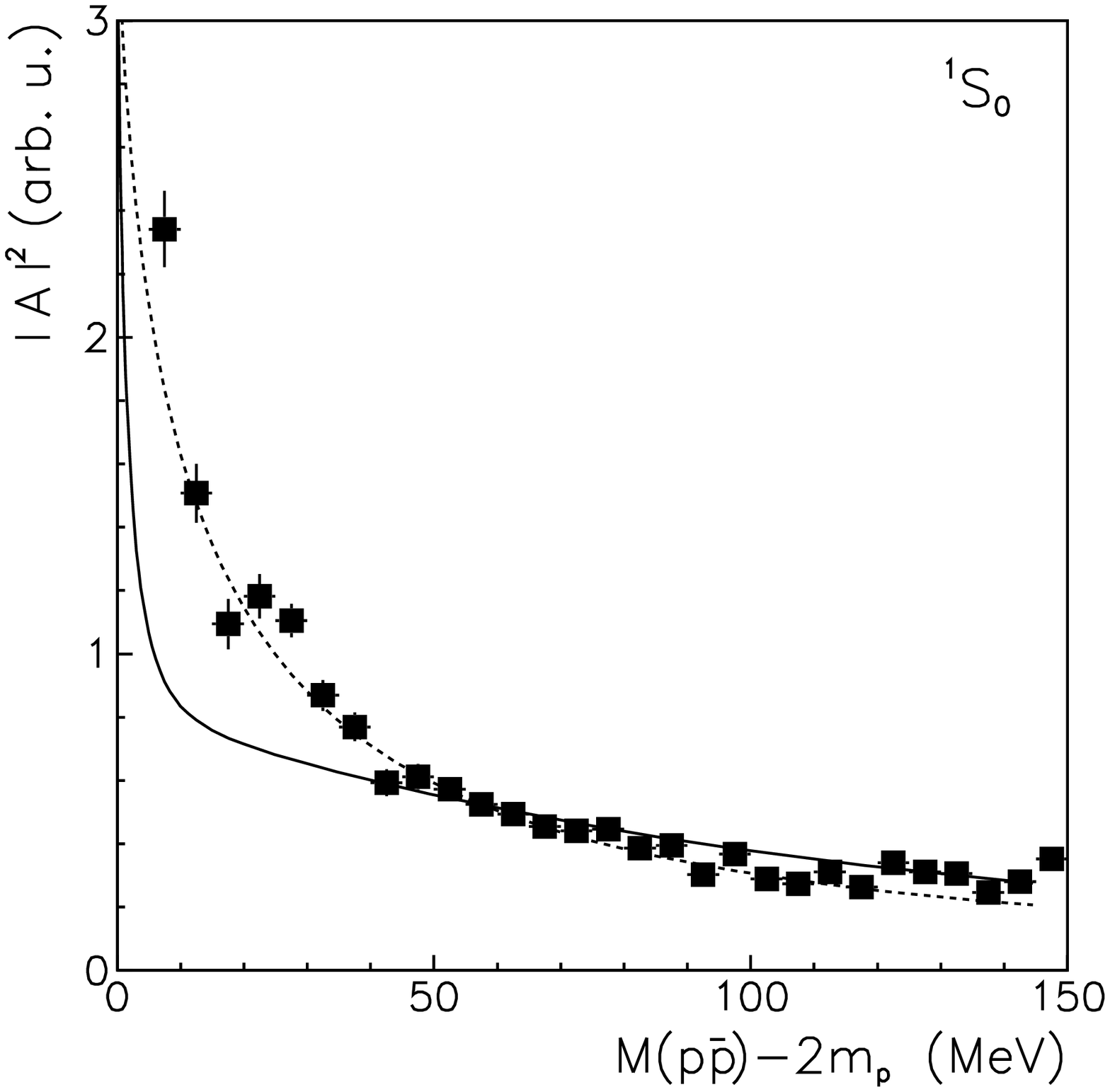}
 \includegraphics[width=0.49\linewidth]{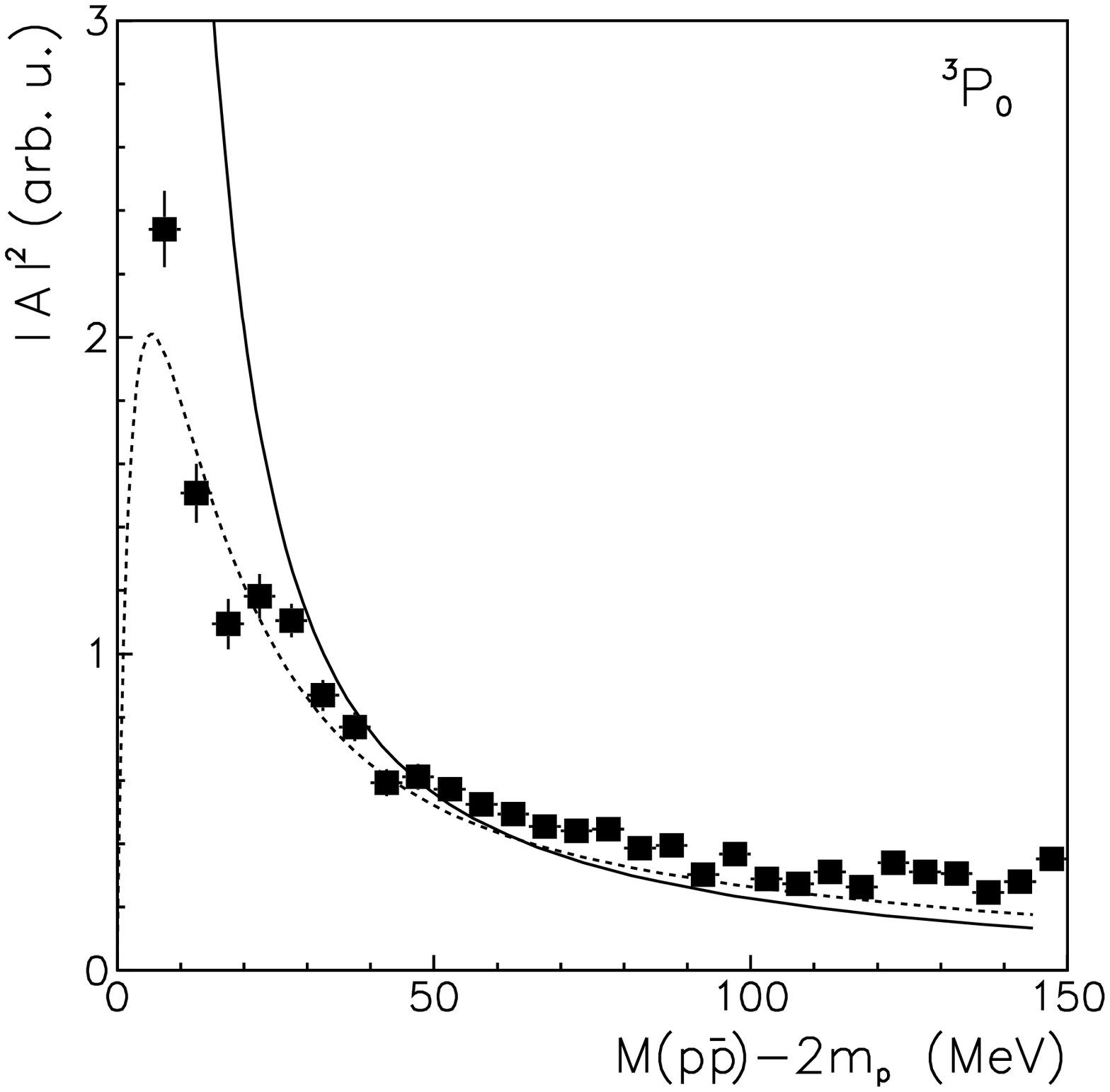}
  \parbox{14cm}{\caption{\label{pbar4}\small Invariant
$J/\Psi{\to}\gamma p{\bar p}$ amplitude $|A|^2$ as a function of the
$p{\bar p}$ mass. The squares represent the experimental values of
$|A|^2$ extracted from the BES data.  The curves are the scattering
amplitude squared ($|T|^2$) predicted by the $N\bar N$ model A(OBE)
for the $^1S_0$ and $^3P_0$ partial waves and the $I{=}0$ (solid) and
$I{=}1$ (dashed) channels, respectively. Note that the latter results
have been normalized to $|A|^2$ at $M(p\bar p){-}2m_p{=}$50 MeV.  }}
\end{center}
\end{figure}

The new generation of electron-positron colliders has started to
contribute to this discussion.  BELLE observes threshold enhancements
in B-decay, $ {B^+} \rightarrow p \bar{p} K^+ $ and $ \bar{B^0}
\rightarrow p \bar{p} D^{0}$, see Refs.~\cite{belle_02a,belle_02b}.
BES finds an even more pronounced threshold enhancement in the
reaction $ J/\psi \rightarrow p \bar{p} \gamma$, which the
collaboration has interpreted as evidence for a $0^{-+}$-resonance
with a mass of 1859 MeV/c$^2$, see Refs. ~\cite{bes_03a,bes_03b}.  The
resonance claim has been challenged recently because a cusp effect
might explain the experimental findings as well~\cite{bugg_04}. The
J\"ulich theory group has performed calculations based on the J\"ulich
meson-exchange potential for the $\bar{p}p$-reaction and finds that
final state interactions are important but do not suffice to explain
the BES data~\cite{sib05}.  This finding leaves room for a possible
resonance interpretation of the enhancement seen by BES. A recent
preprint by Loiseau and Wycech is even more specific and claims that a
new version of the Paris potential is compatible with a resonance in
the $^{11}S_0$ partial wave~\cite{loiseau}.

\subsection{$N\bar{N}$ Interaction and the Interpretation 
of the BES Results}

The difficulties in interpreting the BES data are illustrated in the
following figures.  The scattering amplitudes for the $^1S_0$ partial
wave are presented in Fig.~\ref{pbar6}. The solid lines are the result
for the full amplitude while the dashed lines are based on the
scattering length approximation.  Note that the scattering lengths
predicted by the J\"ulich $N\bar N$ model are
$a_0{=}(-0.18{-}i1.18)$~fm and $a_1{=}(1.13{-}i0.61)$~fm for the
isospin $I{=}0$ and $I{=}1$ channels, respectively. It is evident that
the scattering length approximation does not reproduce the energy
dependence of the scattering amplitude that well. For the $I{=}1$
channel the difference at an excess energy of 50 MeV amounts as much
as 50 \%.  The difference is even more pronounced for the $I{=}0$
channel, where we already observe large deviations from the full
result at rather low energies. This strong failure of the scattering
length approximation is due to the much smaller scattering length
predicted by the J\"ulich model for the $I{=}0$ partial wave.

Results for the $^3S_1$ partial wave are shown in the right panel of
Fig.~\ref{pbar6}.  Here the scattering lengths predicted by the $N\bar
N$ model are $a_0{=}(1.16{-}i0.82)$~fm and $a_1{=}(0.75{-}i0.84)$~fm
for the $I{=}0$ and $I{=}1$ channels, respectively. This partial wave
cannot contribute to the reaction $J/\Psi{\to}\gamma p{\bar p}$.

One has to realize that the main uncertainty in estimating $p\bar p$
FSI effects does not come from the scattering length approximation but
from our poor knowledge of the $p\bar p$ $^1S_0$ amplitudes near
threshold and of the $J/\Psi{\to}\gamma p{\bar p}$ reaction mechanism.
The scattering lengths employed in the literature are spin-averaged
values. Since the $^3S_1$ partial wave contributes with a weighting
factor 3 to the $p\bar p$ cross sections (and there are no
experimental data on the spin-dependent observables at low energies
that would allow one to disentangle the spin-dependence) it is obvious
that their value should correspond predominantly to the $^3S_1$
amplitude.  Thus, it is questionable whether it should be used for
analyzing the BES data at all because the contribution of the $^3S_1$
partial wave to the decay $J/\Psi{\to}\gamma p{\bar p}$ is forbidden
by charge-conjugation invariance.

\begin{figure}[hbt]
 \begin{center}
 \includegraphics[width=0.8\linewidth]{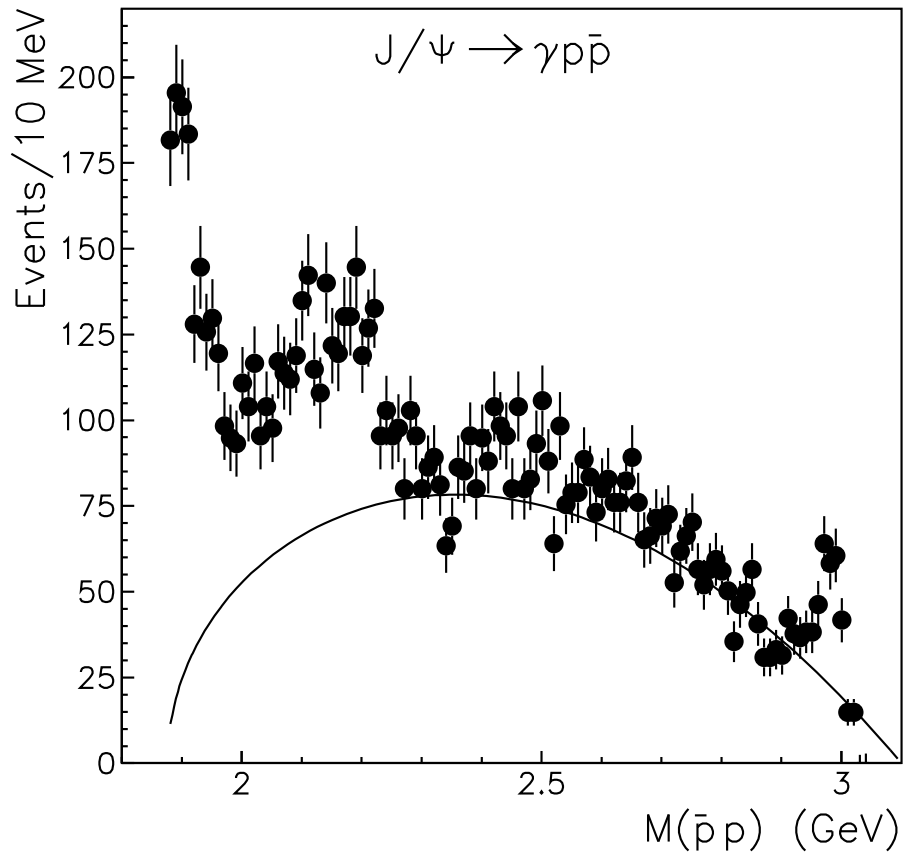}
  \parbox{14cm}{\caption{\label{pbar5}\small The $p{\bar p}$ mass
spectrum from the decay $J/\Psi{\to}\gamma p{\bar p}$.  The circles
show experimental results of the BES Collaboration~\cite{bes_03a},
while the solid line is the spectrum obtained assuming a constant
reaction amplitude.  }}
\end{center}
\end{figure}

The solid lines in Fig. \ref{pbar4} show the $p{\bar p}$ invariant
scattering amplitudes squared for the $^1S_0$ and $^3P_0$ partial
waves and the $I{=}0$ and $I{=}1$ channels.  We consider the isospin
channels separately because the actual isospin mixture in the final
$p{\bar p}$ system depends on the reaction mechanism and is not known.
The reaction $J/\Psi{\to}\gamma p{\bar p}$ can in principle have any
isospin combination in the final $p\bar p$ state.  Note that all
squared $p\bar p$ scattering amplitudes $|T|^2$ were normalized to the
BES data at the invariant mass $M(p{\bar p}){-}2m_p${=}50 MeV by
multiplying them with a suitable constant.
The results indicate that the mass dependence of the BES data can
indeed be described with FSI effects induced by the $^1S_0$ scattering
amplitude in the $I{=}1$ isospin channel. The $I{=}0$ channel leads to
a stronger energy dependence which is not in agreement with the BES
data. We can also exclude dominant FSI effects from the $^3P_0$
partial waves.  Here the different threshold behavior due to the
$P$-wave nature cannot be brought in line with the data points very
close to threshold.

One should note that the BES-data show a resonance structure in the
$\bar{p}p$-invariant mass spectrum near 3 GeV/c$^2$ which corresponds
to the $\eta_c$ meson, see Fig. \ref{pbar5}.  Another broad
enhancement is seen near 2.3 GeV which might correspond to the
$f_0(2200)$ meson.

\subsection{Summary: Baryonium and $N\bar{N}$ Resonances in
 $p\bar{p} \to e^+e^-$}

In the $p\bar{p} \to e^+e^-$ reaction the annihilation proceeds from
the $S$-wave and $D$-wave spin-triplet, $J=1$ states. The charge and
magnetic form factors receive different contributions from the $S$-
and $D$-annihilation \cite{DubnickaNuovoCimento}.  The above discussed
BES data, in conjunction with the already available experimental
evidence for the near-threshold structure in the time-like form
factor, suggest that similar resonance behavior is quite likely also
in the spin-triplet $S$- and $D$-waves.  The phase motion of partial
waves is a well known indicator for resonances; in the case of
$p\bar{p} \to e^+e^-$ this phase motion is directly related to the
relative phase of the time-like charge and magnetic form factors. The
latter is measurable via the single- and double-spin asymmetries as
discussed in Section 4 of the PAX TP. The Phase-I PAX experiment can
explore the broad kinematical range from near-threshold to moderately
high values of $q^2$ and is ideally suited to look for the phase
variations associated with the expected resonance structures.


\cleardoublepage
\part{Appendix D}
\pagestyle{myheadings} \markboth{Appendix D to the Technical Proposal
  for ${\cal PAX}$}{Drell--Yan Cross Sections and Spin Asymmetries in
  the PAX Kinematic Regime}
\section{Comments on Drell-Yan Cross Sections
and Spin Asymmetries in the PAX Kinematic Regime}
The Drell-Yan (DY) event rates and spin asymmetries reported in
Section 2 of the PAX Technical Proposal are based on the leading-order
(LO) formulas for the Drell-Yan cross section. The kinematical range
of DY masses $M$ and of the principal scaling variable $\tau = M^2/s$
accessible in the asymmetrical collider mode at Phase-II are typical
of the high-energy regime studied in the previous high-energy
fixed-target experiments. Specifically, the masses $M$ above the
$J/\Psi,\Psi'$ resonances will readily be accessible.  Here the
situation with the higher order pQCD corrections to the LO formalism,
as described in terms of the so-called $K$-factors, is well
established. On the other hand, in the fixed-target mode at Phase-II,
with Drell-Yan masses $M$ of $2-5$~GeV and relatively low
center-of-mass energies of $\sqrt{s}\approx 5.5-6.7$~GeV, one is not
in the ``classic'' regime discussed so far. A detailed theoretical
understanding of Drell-Yan physics in this regime is crucial, as the
interpretation of the experimentally observed $A_{TT}$ in terms of
transversity relies exactly on the applicability of parton model ideas
and factorization relations. Here we comment briefly on the origin of
the $K$-factors and on ongoing work on the assessment of higher orders
in perturbation theory as well as of non-perturbative corrections to
the cross sections and spin asymmetries in this new kinematical
domain.

\subsection{Factorization and Perturbation Theory for the Drell-Yan Process}

At high energies and large dimuon invariant mass $M$ the Drell-Yan
cross section factorizes into convolutions of parton densities and
perturbative partonic hard-scattering cross sections. Schematically,
\begin{equation}
M^4\,\frac{d\sigma}{dM^2}=\sum_{a,b}\, f_a\otimes f_b\otimes
\frac{M^4d\hat{\sigma}_{ab}}{dM^2}\;+\; {\cal O}\left(
\frac{\lambda}{M}\right)^p \; .
\label{eq:DY1}
\end{equation} 
For brevity, we have considered here the unpolarized cross section,
and we have also integrated over the rapidity of the lepton pair. The
partonic cross sections, $\hat{\sigma}_{ab}$, for the reactions $ab\to
\gamma^{\ast}X$ may be calculated in QCD perturbation theory. Their
expansion in terms of the strong coupling constant $\alpha_s(M)$ reads
\begin{equation}
d\hat{\sigma}_{ab}=d\hat{\sigma}_{ab}^{(0)}+\frac{\alpha_s(M)}{\pi}\,
d\hat{\sigma}_{ab}^{(1)}+\left( \frac{\alpha_s(M)}{\pi}\right)^2
d\hat{\sigma}_{ab}^{(2)}+\ldots \; ,
\label{eq:DY2}
\end{equation}
corresponding to lowest order (LO), next-to-leading order (NLO), and
so forth.  In the unpolarized Drell-Yan case, even the complete NNLO
corrections are known~\cite{Hamberg} (see this reference for an
account of the more than a decade long story of the theoretical
derivation of these corrections. For a good summary of the
experimental data, see Ref.~\cite{Stirling}). The corrections are
often presented in terms of the so-called $K$-factor, the ratio of the
higher-order cross section to the LO one. Roughly speaking, at typical
fixed-target energies, the perturbative corrections to the DY cross
section evaluated with the parton densities determined from the deep
inelastic lepton scattering (DIS) data, increase the predicted LO DY
cross section by about 50\% or even more~\cite{rijken}. Although this
is a fairly large correction, its origin is well understood and under
control theoretically, as we will discuss in the next paragraph.
Taking into account the perturbative corrections is important for
using DY data for precision determinations of the antiquark densities
in the proton and, at colliders, for precision predictions of the
$W,Z$-boson production cross sections.

Because of the intricate interplay of the virtual and real-emission
perturbative corrections, the kinematical dependence of the $K$-factor
may be very important. This will be the case in particular for the
Phase-II experiments in the fixed-target mode.  The variable
$\tau=M^2/s$ is typically quite large for the corresponding
kinematics, $0.2 \lesssim \tau\lesssim 0.7$.  This is a region where
higher-order corrections to the partonic cross sections are
particularly important. Specifically, for a given $M$, $z=\tau/x_a
x_b=1$ sets a threshold for the partonic reaction to proceed and, as
$z$ approaches unity, very little phase space for real gluon radiation
remains in the partonic process, since most of the initial partonic
energy is used to produce the virtual photon. Virtual and
real-emission diagrams then become strongly imbalanced, and the
infrared cancellations leave behind large logarithmic higher-order
corrections to the partonic cross sections, the so-called threshold
logarithms.  At the $k$-th order in perturbation theory, the leading
logarithms are of the form $\as^k\ln^{2k-1}(1-z)/(1-z)$. For
sufficiently large $z$, the perturbative calculation at fixed order in
$\alpha_s$ will not be useful anymore, since the double logarithms
will compensate the smallness of $\alpha_s(M)$ even if $M$ is of the
order of a few GeV.  If $\tau$ is itself close to unity, as is the
case for the Phase-II fixed-target kinematics, the region of large
$z\lesssim 1$ completely dominates, and it is crucial that the terms
$\as^k\ln^{2k-1}(1-z)/(1-z)$ be resummed to all orders in
$\alpha_s$. Such a ``threshold resummation'' is a well established
technique in QCD. In fact, it was developed first for the Drell-Yan
process a long time ago \cite{dyresum}.  It turns out that the
soft-gluon effects exponentiate, not in $z$-space directly, but in
Mellin moment space.

The NLO corrections for the {\it transversely polarized} Drell-Yan
cross section have been calculated in~\cite{DYNLO,Ratcliffe:2004we}.
The evaluation of double-transverse spin asymmetry for the closely
related direct photon production $p^\uparrow p^\uparrow \to \gamma X$
at RHIC energies can be found in \cite{photon}.  They are technically
somewhat harder to obtain than the corresponding corrections in the
unpolarized or longitudinally polarized cases, because the transverse
spin vectors lead to a non-trivial $\cos(2\phi)$-dependence on the
azimuthal angle of one of the Drell-Yan leptons, so that one cannot
integrate over its full phase space.  One way of dealing with this is
by using a projection method~\cite{photon}.

Close to partonic threshold, the transversely polarized cross section
is subject to the same large logarithmic corrections as described for
the unpolarized one above. The crucial point is that these corrections
are {\it spin-independent}, which means that the spin asymmetry
$A_{TT}$ is expected to be very robust with respect to higher-order
corrections. The underlying spin-independence of soft-gluon emission
is associated with the nature of the quark-gluon vertex and is also
responsible for similar cancellations of the corrections in the case
of the double-longitudinal spin asymmetry $A_{LL}$ observed
in~\cite{Ratcliffe:1983yj}.

\subsection{Phenomenological Studies in the PAX Kinematic Regimes}
The definition of the $K$-factor for the unpolarized DY process is
quite straightforward, because the pQCD expansion (\ref{eq:DY2}) is
uniquely defined (within a given choice of the factorization scheme)
in terms of parton densities determined from DIS. In this sense, the
transversity distribution $h_1(x,Q^2)$ is a special case since it
cannot be measured independently in inclusive DIS.  Nevertheless, once
a factorization scheme (such as the customary
$\overline{\mathrm{MS}}$-scheme) is adopted, the calculation of
higher-order effects is completely specified.  Of course, we currently
have no knowledge about transversity, so that in order to make
estimates of the expected $A_{TT}$ model assumptions need to be made,
for instance, imposing \cite{VogelsangTrieste} the saturation of the
Soffer inequality \cite{soffer} at some initial scale. An alternative
is to start with equal helicity and transversity distributions at a
low scale~\cite{Ratcliffe:2004we}. These are of course just simple
assumptions -- measurements of $A_{TT}$ in the PAX experiment at GSI
FAIR will hopefully give us the true picture!

\begin{figure}[t!]
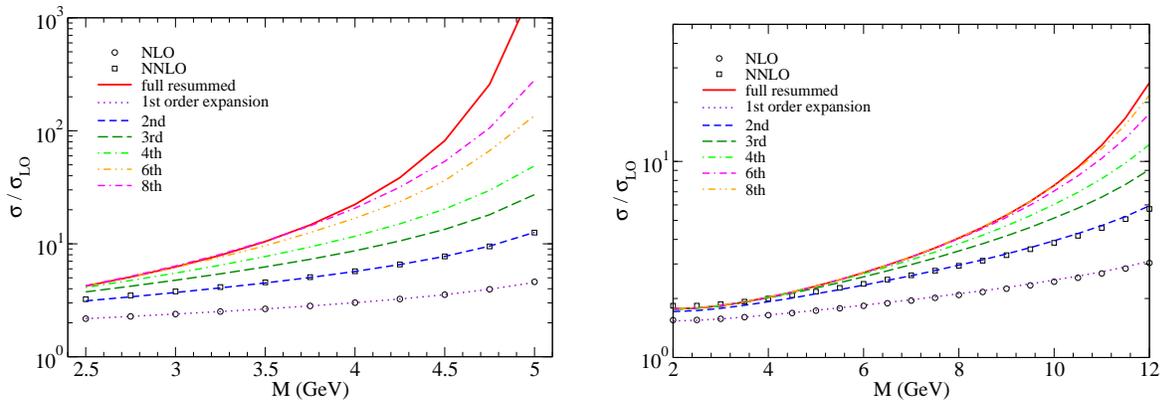

\begin{center}
\hspace*{-5mm}
\epsfig{figure=kfac-fixed.eps,width=0.45\textwidth}
\hspace*{5mm}
\epsfig{figure=kfac-s210.eps,width=0.45\textwidth}
\parbox{14cm}{\caption{\small ``$K$-factors'' relative to LO for the
Drell-Yan cross section in fixed-target $\bar{p}p$ collisions at
$S=30$~GeV$^2$ (left) and for an asymmetric collider mode with
$S=210$~GeV$^2$ (right), as functions of lepton pair invariant mass
$M$. The symbols denote the results for the exact NLO and NNLO
calculations, the curves are for the NLL resummed case and various
fixed-order expansions.  Taken
from~\cite{WernerNLO}. \label{figappd1}}}
\end{center}
\vspace*{0.cm}
\end{figure}

\begin{figure}[t!]
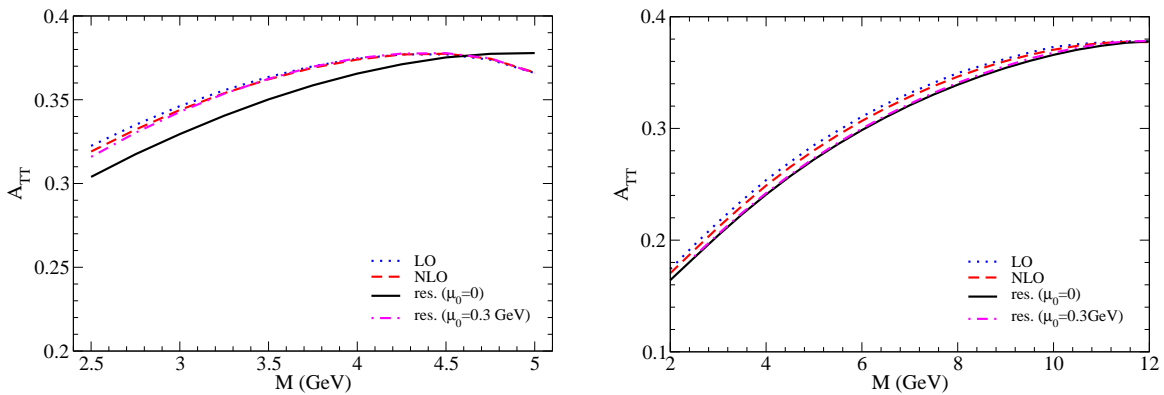

\begin{center}
\vspace*{0.7cm}
\hspace*{-5mm}
\epsfig{figure=asym-s30.eps,width=0.45\textwidth}
\hspace*{5mm}
\epsfig{figure=asym-s210.eps,width=0.45\textwidth}
\parbox{14cm}{\caption{\small Corresponding spin asymmetries
$A_{TT}(\phi=0)$ at LO, NLO and for the NLL resummed
case. \label{figappd2}}}
\end{center}
\end{figure}

Figure~\ref{figappd1} shows results of~\cite{WernerNLO} for the $K$
factors for the unpolarized Drell-Yan cross section at $S=30$~GeV$^2$
(left) and $S=210$~GeV$^2$ (right), at NLO, NNLO, and for the
next-to-leading logarithmic (NLL) resummed case, along with various
higher-order expansions of the resummed result.  As can be seen, the
corrections are very large, in particular in the lower-energy
case. Figure~\ref{figappd2} shows the corresponding spin asymmetries
$A_{TT}$. Here, saturation of Soffer's inequality has been assumed in
order to model the transversity densities.  $A_{TT}$ indeed turns out
to be extremely robust and remarkably insensitive to higher-order
corrections.  Perturbative corrections thus appear to make the cross
sections larger independently of spin. They would therefore make
easier the study of spin asymmetries, and ultimately transversity
distributions. Additional work on the perturbative higher-order
corrections is ongoing~\cite{EnzoNLO}.


\subsection{Resummations and Nonperturbative Power Corrections}

The measured spin asymmetry $A_{TT}$ can only be interpreted in terms
of the transversity densities if the power corrections in
(\ref{eq:DY1}) can either be shown to be small in the accessible
kinematic domain, and/or if they are sufficiently well
understood. There is a close relation between resummation and the
nonperturbative power corrections. It has been shown that perturbative
resummation suggests~\cite{cspv} the form of nonperturbative,
power-suppressed, dynamics. There is reason to
believe~\cite{WernerNLO} that the large enhancement predicted by
perturbation theory at $S=30$~GeV$^2$ (see Fig.~\ref{figappd1}) is
only partly physical. As has been shown there, the very large
corrections arise from a region where the resummed expression becomes
sensitive to the behavior of the strong coupling at small
scales. Further ongoing work focuses on the treatment of this
``far-infrared" limit of resummed perturbation theory and its
consequences.

\subsection{Conclusions}

Current work~\cite{WernerNLO,EnzoNLO} on the Drell-Yan cross section
in the PAX kinematic regimes addresses NLO corrections, higher-order
resummations, and also the study of nonperturbative power corrections.
Large perturbative corrections to the Drell-Yan cross sections have
been found which, however, cancel to a very large extent in the
double-transverse spin asymmetry $A_{TT}$. Further studies aim to give
an idea of the scale of non-perturbative power corrections.  Studies
performed so far suggest~\cite{WernerNLO} that the Drell-Yan process
is theoretically better understood for a GSI FAIR $\bar{p}p$ collider
option than for the fixed-target case.

We finally emphasize that Drell-Yan measurements at PAX would allow us
to enter uncharted territory in QCD: never before have precise
Drell-Yan measurements been performed in this kinematic regime. Not
only do we hope to learn about $A_{TT}$ and transversity. Also,
measurements of the unpolarized cross section alone would shed light
on the relationship between fixed orders, perturbative resummation and
nonperturbative dynamics in hadronic scattering and thus enhance our
understanding of QCD dynamics near the transition between the
perturbative and nonperturbative regimes.


\cleardoublepage
\part{Appendix E}
\pagestyle{myheadings} \markboth{Appendix E to the Technical Proposal
  for ${\cal PAX}$}{Beam Dynamics Simulations}
\section{Beam Dynamics Simulations for the PAX  using the BETACOOL code}
\subsection{Introduction}
The simulations described in this appendix were carried out for the
proton--antiproton collider mode of the PAX experiment using the CSR
and the HESR
The simulations made use of the RMS beam dynamics algorithm of the
BETACOOL code \cite{betacool}. The physical model of this algorithm is
based on the following general assumptions:

\begin{enumerate}
\item the ion beam has a Gaussian distribution over all degrees of
freedom, and this is not changed during the simulation;

\item the algorithm is considered to provide a solution of the
equations for the RMS values of the beam phase space volume in three
degrees of freedom, i.e. at this stage no tracking of individual
particles is performed;

\item the maxima of all distribution functions coincide with the
equilibrium orbit, all instability factors (linear and nonlinear
resonances, space--charge effects, beam--beam tune shift, etc.) are
not taken into account during the simulation.
\end{enumerate}

The following effects are included in the simulation:

\begin{itemize}
\item Electron cooling (EC),
\item Intrabeam scattering (IBS),
\item Scattering on the residual gas (RG), and 
\item Particle losses (PL) from the hadronic interaction at the
interaction point (IP).
\end{itemize}

The EC is taken into account using the Parkhomchuk model
\cite{parkhomchuk} of the friction force. The IBS growth rates are
calculated with the Martini model \cite{martini} using ring lattice
functions imported from the output file of the MAD
program~\cite{lattice-functions}. Interactions at the IP are used for
the simulation of the luminosity and beam--beam parameters. The
following PL effects were used during simulation: electron capture in
EC for the proton beam, losses due to interactions at the IP with a
total $\bar{p}p$ cross section of 40~mbarn, scattering on the residual
gas (electron capture, single scattering, nuclear reactions).

The goal is to provide in the collider mode a luminosity in excess of
$10^{30}$~cm$^{-2}$s$^{-1}$. In order to avoid the hourglass effect,
the bunch lengths have to be about 30~cm, which is equal to the
beta--functions at the IP. The parameters of the RF system are
responsible to ensure a bunch length of 30~cm in equilibrium. The
simulation for the collider mode was carried out for the highest
energies achievable in each of the rings (HESR: 15~GeV/c, and CSR:
3.65~GeV/c). A list of initial parameters used in the simulation is
given in Table~\ref{tab:beam-dynamics}.

The initial emittances of the proton and the antiproton beam in the
collider mode were chosen such that the diameter of the ion beam is
smaller than the diameter of the electron beam in the cooler section
(Table~\ref{tab:beam-dynamics}). After equilibrium is reached, both
the proton and the antiproton beam have approximately the same radius
and bunch length at the collision point. The cooling rates in
equilibrium are equal to the IBS growth rates.

The electron cooler for the HESR has the same design parameters as
those required for experiments with a dense internal hydrogen pellet
target of PANDA \cite{PANDA}. The electron cooler for the CSR requires
a strong cooling force in order to provide cooling of the short proton
bunches. The required length of the cooling section for the CSR is
about 10 m, i.e. about 3 times longer than the 2~MV cooler to be built
to provide a test bed for high energy electron cooling at COSY
\cite{ecoolerforcosy}.
\begin{table}[h!]
\begin{center}
\renewcommand{\arraystretch}{1.11}
\begin{tabular}{|p{7cm}|c|c|}
\hline 
\bf Initial Parameters                             & \bf CSR                      & \bf HESR          \\\hline\hline
Particles                                          & proton                       & antiproton        \\
Momentum [GeV/c]                                   & 3.65                         & 15                \\
Relativistic factor $\gamma$                       & 4.04                         & 16.1              \\          
RF Harmonic Number                                 & 10                           & 30                \\
RF Voltage [kV]                                    & 200                          & 200               \\
Number of particles per bunch                      & $10^{11}$                    & $10^{10}$         \\
Number of bunches                                  & 10                           & 30                \\
Beta function at IP  [m]                           & 0.3                          & 1                 \\
Cross section at IP [mbarn]                        & 40                           & 40                \\
Transverse emittance [mm mrad]                     & 1                            & 0.13              \\
Momentum spread $\Delta P/P$                       & $10^{-3}$                    & $5\times 10^{-4}$         \\\hline\hline
\multicolumn{3}{|c|}{\bf Electron Cooler} \\\hline\hline
Cooler length [m]                                  & 10                           & 30                \\
Magnetic field [kG]                                & 2                            & 5                 \\
Beam radius [cm]                                   & 0.5                          & 0.5               \\
Beam current [A]                                   & 3                            & 1                 \\
Horizontal beta function [m]                       & 14                           & 100               \\
Vertical beta function   [m]                       & 14                           & 100               \\\hline\hline             
\multicolumn{3}{|c|}{\bf Equilibrium Parameters}                                \\\hline\hline                        
Beam parameter                                     & $3\times 10^{-3}$            & $6\times 10^{-3}$ \\
Transverse emittance [mm mrad]                     & 0.42                         & 0.032             \\
Momentum spread $\Delta P/P$                       & $2.5 \times 10^{-4}$         & $1.9 \times 10^{-4}$         \\
Bunch length [cm]                                  & 27                           & 22                           \\
Transverse cooling/heating rate [$s^{-1}$]         & 0.059                        & 0.012                        \\
Longitudinal cooling/heating rate [$s^{-1}$]       & 0.102                        & 0.014                        \\
Cooling time [s]                                   & $\approx 100$                & $\approx 1500$               \\
Peak luminosity [cm$^{-2}$s$^{-1}$]                  & \multicolumn{2}{c|}{$1.6\times 10^{30}$}                     \\\hline\hline
\multicolumn{3}{|c|}{\bf Particle losses}                                \\\hline\hline                  
Interaction point [s$^{-1}$]                       & $6.5\times 10^{-8}$          & $2.2 \times 10^{-7}$         \\
Electron Cooler     [s$^{-1}$]                     & $6.1\times 10^{-6}$          & $1.2 \times 10^{-7}$ $^*$    \\
Rest gas        ($10^{-10}$ mbar) [$s^{-1}$]       & $6.8 \times 10^{-8}$         & $1.3\times 10^{-7}$          \\
Total beam life time [h]                           & $\approx 45$                 & $\approx 800$                \\\hline\hline
\end{tabular}
\parbox{14cm}{\caption{\label{tab:beam-dynamics}\small Initial
parameters used in the simulation, parameters of the electron cooler,
equilibrium parameters, and particle losses. (The particle losses in
the electron cooler of the HESR ($^*$) are calculated for protons, for
antiprotons, this loss mechanism is absent.)  }}
\end{center}
\end{table} 

\subsection{The Cooling Process}
\begin{figure}[hbt]
 \begin{center}
 \includegraphics[width=0.35\linewidth]{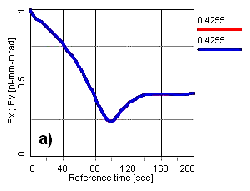}
 \includegraphics[width=0.35\linewidth]{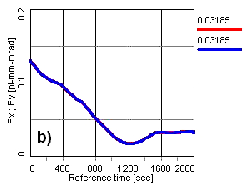}
 \includegraphics[width=0.35\linewidth]{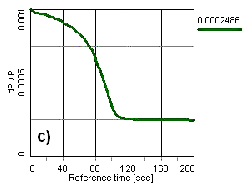}
 \includegraphics[width=0.35\linewidth]{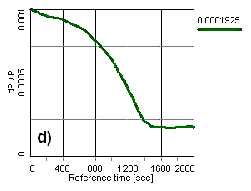}
 \includegraphics[width=0.35\linewidth]{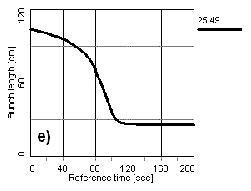}
 \includegraphics[width=0.35\linewidth]{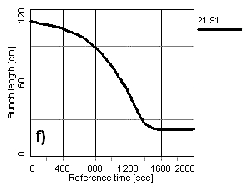}
 \includegraphics[width=0.35\linewidth]{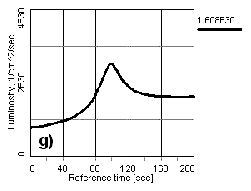}
 \includegraphics[width=0.35\linewidth]{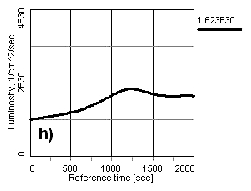}
  \parbox{14cm}{\caption{\label{fig:beamdyn1}\small Results of the RMS
beam dynamics calculation for CSR and HESR in the collider mode. The
panels on the left side are for the CSR, those on the right for the
HESR. From top to bottom, the panels denote the beam emittance,
momentum spread, bunch length, and luminosity. The final numbers of
each parameter are shown on the top right next to each panel.}}
\end{center}
\end{figure}

The behavior of the RMS beam parameters (emittance, momentum spread,
bunch length, and luminosity) during the cooling process is presented
in Fig.~\ref{fig:beamdyn1}. The panels on the left side correspond to
the CSR, those on the right to the HESR. The initial values are listed
in Table~\ref{tab:beam-dynamics}. In the CSR, it takes about 200~s
until the beam parameters reach equilibrium, while the in the HESR, a
cooling time around 1500~s is required. After cooling, all parameters
reach a constant value and do not change for a long time. Particle
loss rates are a few order of magnitudes smaller than the cooling time
and were not taken into account during the simulation.

\begin{figure}[hbt]
 \begin{center}
 \includegraphics[width=0.49\linewidth]{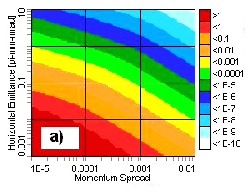}
 \includegraphics[width=0.49\linewidth]{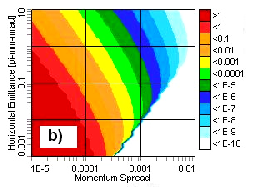}
 \includegraphics[width=0.49\linewidth]{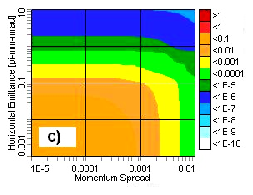}
 \includegraphics[width=0.49\linewidth]{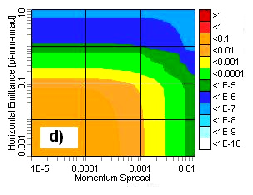}
 \includegraphics[width=0.49\linewidth]{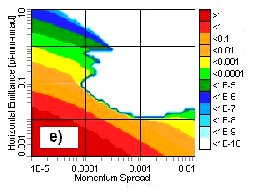}
 \includegraphics[width=0.49\linewidth]{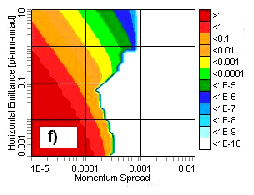}
  \parbox{14cm}{\caption{\label{fig:beamdyn2}\small 3D diagrams of
growth rates [s$^{-1}$] for the HESR as function of the momentum
spread, in accordance with the RMS beam dynamics results, shown in
Fig.~\ref{fig:beamdyn1}. The panels on the left side apply to
transverse, those on the right to longitudinal components. From top to
bottom, the panels denote the IBS growth rates (a and b), cooling
rates of the EC (c and d), and the combined effect of heating and
cooling rates (e and f).}}
\end{center}
\end{figure}

\begin{figure}[hbt]
 \begin{center}
 \includegraphics[width=0.8\linewidth]{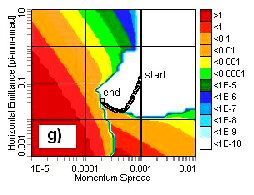}
  \parbox{14cm}{\caption{\label{fig:beamdyn3}\small 3D diagram
(transverse emittance vs momentum spread) of the transverse and
longitudinal components of cooling and heating rates, arrived at by
combining the panels e and f of Fig.~\ref{fig:beamdyn2} for the
HESR. The open circles indicate the evolution of the beam parameters
from a starting point (start) to the equilibrium point (end). The
equilibrium point is unique, i.e. it does not depend on the particular
choice of the starting point.}}
\end{center}
\end{figure}

The dip of the emittance and, respectively, of the luminosity as
function of time can be explained with the help of three--dimensional
diagrams, shown in Fig.~\ref{fig:beamdyn2}, where the transverse
emittance is shown as a function of the momentum spread. The vertical
emittance is assumed to be equal to the horizontal one.
The cooling rates for the EC (Fig.~\ref{fig:beamdyn2}, panels c and d)
are calculated in accordance with the Parkhomchuk formula of the
cooling force \cite{parkhomchuk}. The transverse and longitudinal
components of the cooling rates show approximately the same behavior.
The combined effect of cooling and heating rates is presented in
Fig.~\ref{fig:beamdyn2} (e and f). The boundaries between colored and
white areas indicate the equilibrium between IBS and EC for the
transverse and longitudinal components. The regions of the equilibrium
can be found if one combines all four panels (a, b, c, and d) of
Fig.~\ref{fig:beamdyn2}, which results in Fig.~\ref{fig:beamdyn3}. The
final position of the equilibrium point does not depend on the initial
coordinate.

In Fig.~\ref{fig:beamdyn3} the dependence of the transverse emittance
on the momentum spread during the cooling process for the RMS dynamics
of Fig.~\ref{fig:beamdyn1} (b and d) is shown.
Initially, the electron cooling force achieves equilibrium with the
transverse component of IBS. During this process, the emittance and
the momentum spread are decreased. Subsequently, the cooling process
continues and the beam parameters change in accordance with the
equilibrium boundary of the transverse component. The momentum spread
continues to decrease but the transverse emittance begins to increase.
When the cooling force also reaches equilibrium with the longitudinal
component of the IBS, the beam parameters converge to an equilibrium
point, which does not depend on the initial parameters. The RMS
dynamics is rather different and the cooling time can show large
changes. This indicates that the initial parameters of the ion beam do
not influence the equilibrium point but have a large effect on the
cooling time.

\subsection{Cycling of APR, CSR and HESR}
\begin{figure}[hbt]
 \begin{center}
 \includegraphics[angle=90,width=0.60\linewidth]{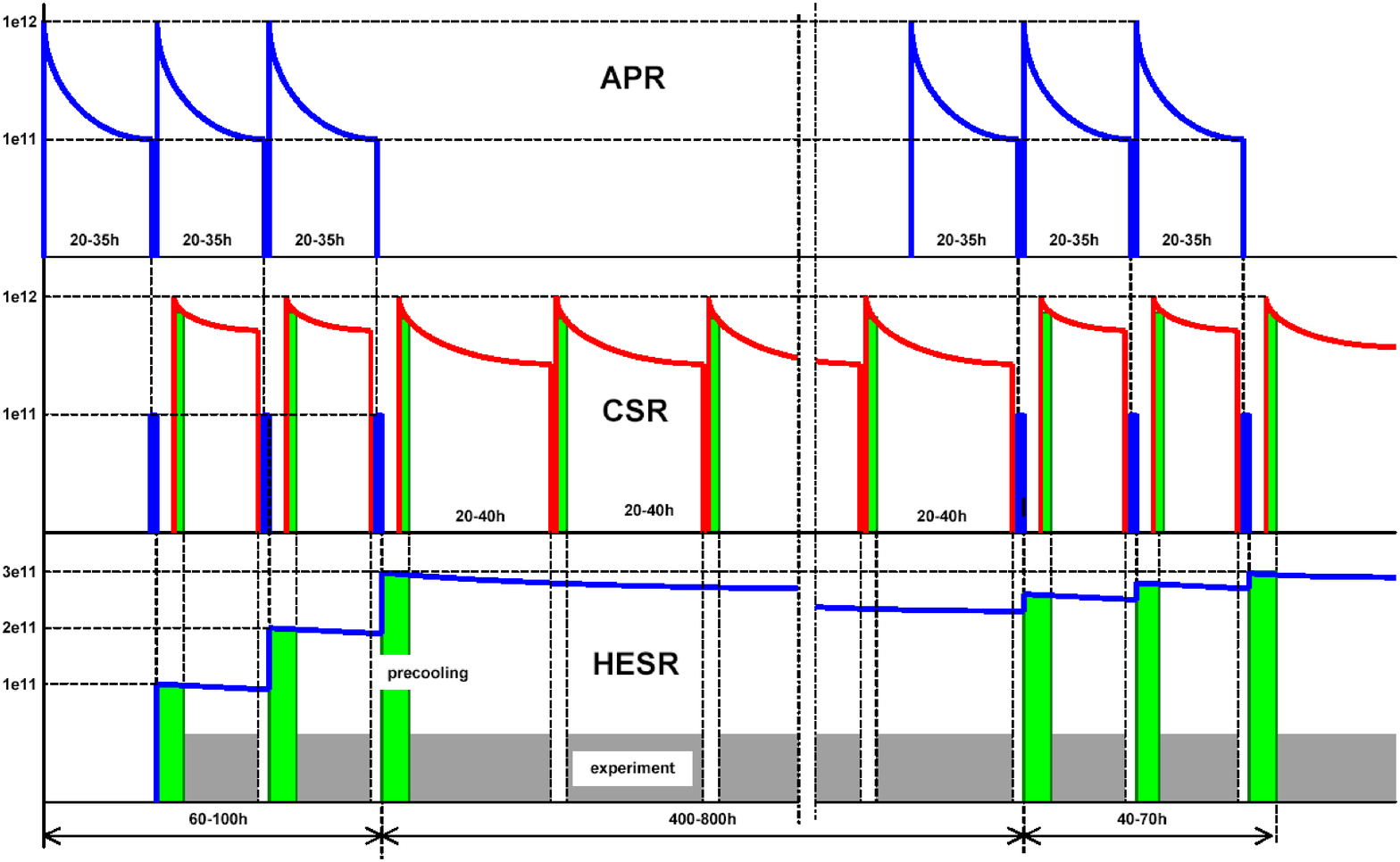}
  \parbox{14cm}{\caption{\label{fig:beamdyn4}\small In the cycling
scheme presented here, blue lines correspond to antiproton beam, red
lines to proton beam . After the cooling process (green areas) in the
CSR and in the HESR, the collider experiments can be started (black
fill areas). The electron coolers stay in operation in both storage
rings, CSR and HESR, to suppress intrabeam scattering during the
experiments. Here we assume a conservative approach to the filling of
the HESR. New fills from the APR are used to replace the antiprotons
in the HESR with the largest dwell time. }}
\end{center}
\end{figure}

The three rings, APR, CSR and HESR have to be operated together to
provide the maximum luminosity for the collider experiments at the PAX
experiment. A scheme showing how the three rings are cycled
altogether, is shown in Fig.~\ref{fig:beamdyn4}. Some details of the
ring operation do not require much time, i.e. they are fast and do not
affect the integrated luminosity. These are: injection, acceleration,
bunching, and the cooling time. The following effects define the time
table of experiment:

\begin{enumerate}
\item Production rate of antiprotons,
\item Polarization buildup time of antiprotons in the APR,
\item Space charge limit of particles at the injection energy, and
\item Beam lifetime in all storage rings.
\end{enumerate}

The production rate of antiprotons is assumed to amount to about
$10^7$~s$^{-1}$ ($=3.6 \times 10^{10}$~h$^{-1}$). The polarization
buildup time in the APR is defined by the lifetime of the antiproton
beam in the interaction with the hydrogen target (see Sec. 8.2.2). In
order to achieve the maximum polarization of about 0.4, a ring
acceptance angle of $\psi_\mathrm{acc} = 50$~mrad is required, which
corresponds to a beam lifetime of $\tau_\mathrm{APR} = 17$~h. The
number of particles which can be injected into the APR at each
injection is
\begin{eqnarray}
N_{\bar{p}} = 2 \times R \times \tau_{\mathrm{APR}} = 1.2 \times 10^{12}\;.
\end{eqnarray}
This value is close to the space charge limit for the APR at the
injection energy. After spin--filtering for two beam lifetimes, the
number of antiprotons decreases by about one order of magnitude,
i.e. to about $N_{\bar{p}} = 10^{11}$.

Subsequently, antiprotons are injected into the CSR. After
acceleration and bunching, the antiproton beam has 10 bunches with
$10^{10}$ particles per bunch. Then antiprotons are injected into the
HESR.  The new fill should then be added to the antiprotons already
circulating in the HESR, thus the antiproton beam should be
decelerated to the injection energy.  The antiprotons are then
injected into the HESR, accelerated up to the experimental energy, and
the cooling process can start.

At the same time, a proton beam with intensity $N_{\bar{p}} = 10^{12}$
is injected into the CSR, which also corresponds to the space charge
limit at injection energy. After acceleration and bunching of the
proton beam, the electron cooler is switched on. After both beams are
cooled down, colliding beam experiments can be started.  The proton
beam lifetime in the CSR is mainly defined by electron capture in the
cooler section (Table~\ref{tab:beam-dynamics}). But this effect is
absent for antiprotons in the HESR and the beam lifetime is defined by
the total cross section at the IP and the residual gas pressure. The
vacuum pressure at HESR should not exceed about $10^{-10}$~mbar.

The beam lifetime in the CSR is about one order of magnitude shorter
less than in the HESR. The CSR should therefore be refilled with
protons a couple of times, while antiprotons are circulated in the
HESR. After the first injection of polarized antiprotons into the
HESR, the experiments in the collider mode can be started. But the
peak luminosity will be achieved only after injection of three cycles
from the APR.

The cycling of the different rings for the PAX experiment is mainly
defined by the lifetime of antiprotons in the HESR. After each cycle a
new portion of antiproton beam replaces particles which were
circulating in the ring before. In this scheme, the average luminosity
is not so different from the peak luminosity. Because of the very long
beam lifetime of the antiproton beam in the HESR ($\approx$ 800~h), in
the collider mode antiprotons do not have to be delivered continously
to the APR.

\subsection{Possible Improvements}
One way to increase the luminosity for PAX in the collider mode could
consist of raising the injection energy in the CSR to avoid running
into the space charge limit. But the lifetime of the antiproton beam
is linearly proportional to the density of the proton beam. An
increased proton beam intensity simply leads to a decrease of the beam
lifetime of the antiproton beam.

Another option would be to increase the number of antiprotons in the
HESR, which would lead to larger IBS growth rates and require longer
cooling times. The time required to fill the HESR with antiprotons is
linearly proportional to the particle number due to the fixed
production rate of antiprotons.  The estimate for the peak luminosity
of ${\cal L}=1.6\times 10^{30}$~cm$^{-2}$s$^{-1}$ (listed in
Table~\ref{tab:beam-dynamics}) is based on conservative assumptions
about the number of antiprotons accumulated in the HESR. The number of
antiprotons in the HESR can be increased by the transfer of more than
just three shots from the APR, as depicted in Fig.~\ref{fig:beamdyn4}.
It should, however, be noted that the preparation (polarization
buildup) of a single shot in the APR takes more than one day, thus the
accumulation of ten shots in the HESR would take more than one week!
In addition, any increase in the number of particles leads to larger
instabilities due to space charge. In the present simulations, neither
these effects nor instabilities due to resonances have been taken into
account.
\subsection{Summary}
The present simulations show that in fact a high luminosity of ${\cal
L}\approx 1.5\times10^{30}$~cm$^{-2}$s$^{-1}$ in the collider mode can
be achieved, however, to that end, a strong cooling force should be
applied in both the CSR and the HESR. The parameters of the proton and
antiproton beam are defined by the equilibrium between electron
cooling and intrabeam scattering with values of heating growth rates
of about 0.01--0.1 s$^{-1}$. The main particle loss mechanism in the
CSR is electron capture in the cooler section. Particle losses in the
HESR are mainly caused by scattering on the rest gas and by the
hadronic interaction at the interaction point. The role of
space--charge effects for the stability of the proton and antiproton
beam should be further studied for the collider mode.


\cleardoublepage
\part{Appendix F}
\pagestyle{myheadings} \markboth{Appendix F to the Technical Proposal
for ${\cal PAX}$}{Detector simulations}
\section{Detector Simulation for PAX}
\subsection{PAX detector concept}
The primary goal of this appendix to the PAX Technical Proposal is to
 show that the most challenging and outstanding measurement of the PAX
 experimental program, the direct measurement of the $h_1^q$
 transversity distribution, is feasible. Moreover, the appendix is
 intended to show that other studies, like the measurement of the
 phases of the electromagnetic form factors of the proton and the spin
 correlations in the elastic proton-antiproton scattering, are much
 less demanding tasks, due to the high reaction rates involved.

\begin{figure}[h]
  \centering \includegraphics*[width=0.79\linewidth]{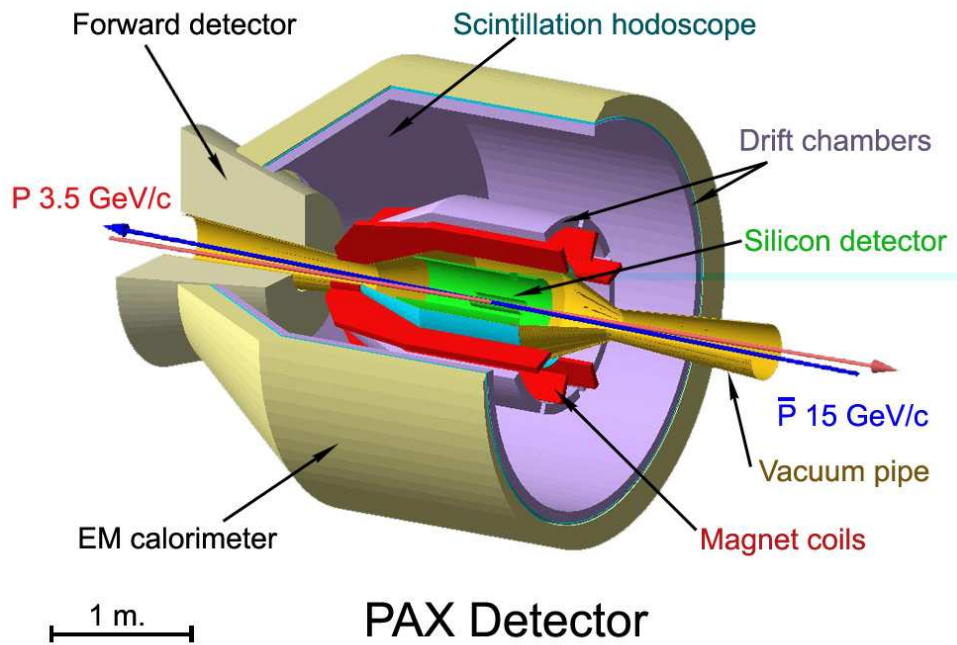}
  \parbox{14cm}{\caption{\small \label{artistic2} This conceptual
   design of the PAX detector is employed to estimate the performance
   of the detector and to show the feasibility of the transversity
   measurement in the asymmetric antiproton--proton collider mode at
   PAX.  The artists view is produced by {\sc Geant}.}}
\end{figure}

An extensive program of studies has been started to investigate
different options for the PAX detector configuration, aiming at an
optimization of the achievable performance. We concentrate here on the
detector design proposed in the PAX Technical Proposal
(Fig.~\ref{artistic2}), which is well--suited to provide large
invariant-mass $e^+ e^-$ pair detection, from both Drell-Yan reactions
and $\overline{p}p$ annihilations.  In addition, such a detector is
capable to efficiently detect secondaries in two body reactions, like
elastic scattering events, where the over--constraint kinematics
simplifies the event reconstruction and reduces the requirements for
the particle identification.  Alternative detector scenarios,
e.g. based on $\mu^+\mu^-$ Drell--Yan pair detection, instrumented in
the forward detector region or an extended hadron particle
identification, will be studied at a later stage. The present detector
is based on driving principles, outlined below. \newline The detector
should:

\begin{itemize}
\item provide a large angular acceptance. Good azimuthal coverage and
symmetry are needed to be sensitive to the dependence of the
observables on the angle between production plane and target spin
orientation. Several benchmark observables require an acceptance
optmized for large polar angles, i.e. the $A_{TT}$ asymmetry in
Drell-Yan reactions (Eq.~\ref{att}) and the single ${\cal A}_y$ and
double ${\cal A}_{yy}$, ${\cal A}_{zx}$ spin asymmetries in
$\overline{p}p\rightarrow e^+e^-$ annihilations (Eq.~\ref{py}) are
weighted by trigonometric functions of the scattering angle, while the
transverse spin effects on elastic scattering concentrate at large
transverse momenta~\cite{ZGS}.
\item be sensitive to electron pairs. Several detection tools allow
one to efficiently identify electrons without an adverse effect on the
momentum resolution. The overwhelming hadronic background requires
excellent lepton identification. High momentum resolution is needed to
be sensitive to $h_1$ dependence on Bjorken $x$; in addition it opens
the interesting possibility to extend the measured range down to 2 GeV
dilepton mass, thereby enlarging the Bjorken $x$ coverage of the $h_1$
measurement and facilitating the study of spin effects in the
resonance production versus continuum region. A high resolution device
with excellent particle identification constitutes a flexible and
complete facility which can cope with new physics goals that may
emerge in the upcoming years.
\item use a toroid magnet. The spectrometer should provide high
momentum resolution and measure the charge of secondaries. This is
crucial in order to identify the wrong--charge control sample and
subtract the combinatorial background. The spectrometer magnet should
not affect the transverse spin orientation of the beam and provide an
environment to ensure the operation of the \Cer detector. The toroid
has almost negligible fringe-fields outside its active volume, both
internally along the beam line and externally inside the tracking
volume.
\end{itemize}

\subsection{Phase--I: Electromagnetic form factors of the proton and hard elastic scattering}
This section presents the signal estimates for the benchmark
measurements in {Phase--I} of PAX physics program. Here a (polarized)
antiproton beam with momentum up to 3.6 GeV/c scatters off a polarized
internal gas target in the CSR ring.  The signal estimates show that
the reaction rates are large enough to not put stringent requirements
on the experimental set--up. The luminosity in fixed target mode is
calculated as ${\cal L}=N_{\overline{p}} \cdot f \cdot d_t$ where
$N_{\overline{p}}=10^{11}$ is the number of antiprotons stored in CSR,
$f=L_{\rm CSR}/\beta_{\overline{p}} c \sim 1$ MHz is the antiproton
revolution frequency depending on the antiproton velocity
($\beta_{\overline{p}}c$) and on the length of the CSR ring ($L_{\rm
CSR}$), and $d_t=\Ord{14}$ $\rm cm^{-2}$ is the areal density of the
target.

\subsubsection{Electromagnetic form factors of the proton}
Using the $p\overline{p}\rightarrow e^+e^-$ cross-section measured by
PS170~\cite{PS170_2} it is possible to estimate the running time
required to get a precise measurement of the relative phases of the
time--like electric and magnetic form--factors of the proton.  For
single spin asymmetries (SSA) and double spin asymmetries (DSA), the
statistical error scales as
$$\Delta A_{\rm SSA}= \frac{1}{Q} \frac{1}{\sqrt{N_{\rm SSA}}}
\hspace*{2cm} \Delta A_{\rm DSA}= \frac{1}{QP} \frac{1}{\sqrt{N_{\rm
DSA}}}$$ where $Q=0.8$ is the proton target and $P=0.3$ is the
expected antiproton beam polarization. $N_{\rm SSA}$ ($N_{\rm DSA}$)
is the number of collected events in the single (double) polarized
mode. The following table lists the running time required to reduce
the error down to $\Delta A=0.05$ for a few typical beam momenta,
accessible in the CSR:

\vspace*{0.5cm}
\begin{tabular}{|c|c|c|c|c|c|}
\hline
Beam & c.m. energy & $\sigma^{p\overline{p}\rightarrow e^+e^-}$ & ${\cal L}$ & Running time & Running time \\
momentum & $\langle s \rangle$ & PS170 & & DSA & SSA \\ 
(MeV/c) & ($\rm GeV^2$) & (nbarn) & ($\rm cm^{-2} s^{-1}$) & (days) & (days) \\ \hline
549 & 3.76 & 7.3 & $7.8\ord{30}$ & 2.9 & 0.3\\
900 & 4.18 & 3.7 & $1.1\ord{31}$ & 4.7 & 0.5 \\
3600 & 8.75 & 0.044 & $1.5\ord{31}$ & 132 & 13 \\ \hline
\end{tabular}

\vspace*{0.5cm} \noindent Here a 50 \% acceptance for
$\sigma^{p\overline{p}\rightarrow e^+e^-}$ events is estimated basing
on the conceptual PAX detector described in the previous section.
Most of the measurements can be performed in a relatively short time,
from less than 1 day up to few weeks.  Only the most challenging
measurement of double polarized asymmetries at the largest momenta
requires a few months of data--taking. Note that the CSR ring be
operated with polarized antiproton beam down to 200 MeV/c.  Additional
studies are foreseen to relax this limit and work even closer near
threshold.

\subsection{Hard elastic scattering}
An estimate can be performed for the rate of hard $\overline{p}p$
elastic scatterings at the maximum transverse momentum achievable in
the CSR. At the higher CSR antiproton beam energy of 3.6 GeV/c and at
the largest scattering angles within the PAX detector acceptance
(around $120^\circ$), the achievable momentum transfer is $t_{\rm
PAX}=3.9$~$\rm GeV^2/c^2$. This is a benchmark experimental condition
since there the cross--section is smallest, where one expects to
observe the largest transverse spin effects. As a starting point we
take the cross section measured by E838 at momentum transfer $t_{\rm
E838}=5$~$\rm GeV^2/c^2$~\cite{E838}:
$$\left. \frac{d\sigma}{dt}\right|_{\rm E838} = 10^{-4} \hspace{0.3cm}
\frac{\rm mb}{\rm GeV}.$$ In a reasonable approximation the cross
section scales with a tenth--power of the transverse momentum $t$,
thus the E838 value can be rescaled to the PAX kinematics through
$$\left.  \frac{d\sigma}{dt}\right|_{\rm PAX}=\left. \frac{d\sigma}{dt}\right|_{\rm E838} \cdot 
\left[ \frac{t_{\rm E838}}{t_{\rm PAX}} \right]^{10}\sim 6\ord{-4} 
\hspace{0.3cm} \frac{\rm mb}{\rm GeV}.$$ Given the above estimated
luminosity ${\cal L}=1.5\ord{31}$ $\rm cm^{-2}s^{-1}$, the event rate
in a $\Delta t=0.1$~$\rm GeV^2/c^2$ interval, centered around the
selected working point, is of the order of 1 Hz. The most challenging
double--polarized measurement therefore requires only a few hours of
data--taking to reach a precision of
$$\Delta A_{\rm DSA} = \frac{1}{QP} \frac{1}{\sqrt{N_{\rm DSA}}} = 0.05.$$  

\subsection{Phase--II: The transversity measurement}

The requirements to be fulfilled by the PAX detector have already been
discussed in Sec.~15. Here we emphasize that the detector should be
capable to cope with the overwhelming, $\sim 10^7$ times larger
background than the Drell--Yan signal, and still should allow for
reconstruction of the Drell--Yan kinematics with high resolution. A
high resolution of the dilepton invariant--mass (and Feynman--$x_F$)
allows one to efficiently isolate the resonance region from the
continuum. Moreover, a precise determination of the Bjorken--$x$ of
the proton and antiproton is important in order to maximize the
sensitivity for the $x$--dependence of the $h_1^q(x)$ distribution
function.

For these studies, the asymmetric--collider option, described in
Sec.~7, was adopted as the most promising scenario, where a 15~GeV/c
polarized antiproton beam from the HESR collides with a 3.5~GeV/c
polarized proton beam from the CSR to produce $e^+ e^-$ Drell-Yan
events.  In the following, we will label the Drell-Yan $e^+e^-$
candidate as {\it right--sign dilepton pairs}, whereas the background
dilepton pairs, $e^+e^+$ and $e^-e^-$, will be labeled as {\it
wrong--sign dilepton pair}.

\subsubsection{Detector Simulation}
\begin{description}
\item {\bf Software} \\ The proton-antiproton collisions are generated
with the {\sc Pythia} package implemented in {\sc Root}. Particles
generated from the interaction in antiproton--proton collisions are
traced and particular detector responses are generated by the {\sc
Geant 4} package. Whenever not explicitly stated, all physical
processes regarding particles passing through the detector material,
are accounted for in the simulation.

\item {\bf Data Samples} \\ The interaction point was fixed at the
origin of the coordinate system in order to maximize the detector
acceptance in particular for the background events.  Two data samples
are generated:

\begin{description}
\item {\it Drell-Yan sample}: it contains $10^5$ pure Drell--Yan
events into electron--positron pairs. This sample is used to test the
acceptance and resolution of the PAX detector.
\item {\it Background sample}: it consists of $2\ord{8}$ minimum--bias
background events where Drell--Yan events and proton--antiproton
elastic scattering events were excluded.  This sample corresponds to
about 40 minutes of data--taking with a cross-section of 40 mb and a
luminosity of $2\ord{30}$ $\rm cm^{-2}s^{-1}$. The corresponding
$e^+e^-$ Drell--Yan yield is of the order of 10 events.
\end{description}

Although the leptonic decays of charm resonances ($\JPsi$ and $\Psi$)
provide an alternative access to transversity~\cite{abdn2}, the
simulation of these decays is not implemented yet.

\item{\bf Detector} \\ The detector setup employed in the simulation
is shown in Fig.~\ref{artistic2}: it consists of a somewhat simplified
version of the one described in part~\ref{partIV}, composed of a
barrel section covering the $60^\circ$ to $120^\circ$ interval of
polar angles, complemented by a detector part of conical shape,
covering the smaller polar angles from $20^\circ$ to $60^\circ$.
The right--handed coordinate system is defined as follows: the
$z$--axis is pointing along the antiproton beam of 15~GeV/c momentum,
$x$ points sideways, and the $y$--axis points upward. The origin of
the coordinate system is located at the interaction point (IP).\\

{\it Silicon detector} (Si0, Si1): two layers of double--sided silicon
strip detectors are placed close to the IP inside the vacuum of the
beam pipe. They are utilized to measure the part of the track before
the magnet, to reconstruct the vertex, and to veto neutral particles
(especially gammas). The first layer has a thickness of 300~$\rm \mu
m$ and a distance of 5~cm from the beam axes.  The second layer has a
thickness of 300~$\rm \mu m$ and a distance of 22~cm from the beam
axes. The resolution is assumed to be 20~$\rm \mu m$ in both the
longitudinal ($z$) and transverse ($R\phi$)
coordinates~\cite{VTXbabar}.

{\it Vacuum chamber} (VC): the beam pipe connects with a vacuum
chamber of 30 cm radius in correspondence with the IP. The windows of
the vacuum chamber are made from a thin stainless--steel foil of
0.1~mm thickness.

{\it Hodoscopes} (H0, H1): two scintillation hodoscopes provide fast
signals for triggering and time--of--flight information of
low--momentum particles. The first scintillator, with a thickness of
4~mm, is placed just behind the vacuum window. The second one, with a
thickness of 10~mm, is placed in front of the electromagnetic
calorimeter.

{\it Drift chambers} (DC0, DC1): two drift chambers are placed at a
distance of 65 and 135~cm from the beam axis
to measure the track segments behind the magnet and to provide a
momentum resolution of the order of 1~\%. The assumed spatial
resolution is 200 $\rm \mu m$ for both coordinates.

{\it \Cer detector} (CER): the detector is inserted into the
free--space of the tracking arm of the drift chambers to provide an at
least 60~cm thick radiator along the particle path. Since no
particle--identification (PID) procedure is implemented into the
simulation yet, the detector response is not generated.  Nevertheless
the detector material is accounted for during the particle tracking.

{\it Calorimeter} (EC): the electromagnetic calorimeter is a
homogeneous detector with full azimuthal coverage and a length along
the particle path extending up to 16 $X_0$, able to contain 5~GeV
showers. In the present simulation it is assumed to consist of 14 cm
long radiation--hard $\rm Pb W0_4$ scintillator crystals.  The
response for electromagnetic particles ($e^\pm$ and $\gamma$) is
parameterized by $\sigma_E/E=3 \%/\sqrt{E} \oplus 0.5 \%$ where
$\oplus$ stands for summing in quadrature the two
contributions~\cite{ECperform}.  All other particles are fully tracked
inside the calorimeter material.  Since the assumed resolution is
achievable by other types of commonly used detectors, like lead glass
or ionization calorimeters~\cite{ECE760,ECNOMAD,ECNA48}, the validity
of the result is not limited to the scintillating material.

  
{\it Spectrometer Magnet}: the toroidal magnetic field for the
momentum analysis of charged particles is generated by eight
superconducting coils arranged symmetrically around the beam
pipe. Each coil occupies a non--instrumented azimuthal sector of
$5^\circ$. The acceptable solid angle is reduced by less than 11~\%
and ensures an azimuthal acceptance in excess of 80~\% for Drell--Yan
dilepton events. Although a coil design similar to the one of {\sc
Atlas} can be anticipated~\cite{SCatlas} and detailed studies on the
realistic field map are in progress, the magnet details have not been
taken into account yet in the simulation in order to save computing
time. A homogeneous toroidal field is assumed with a total bending
power of 0.4 Tm.
Deep electromagnetic showers generated by the coil materials may
prevent efficient event reconstruction within a time interval
comparable with DC drift--time. Preliminary simulations have shown
that the number of these showers is small enough to ensure a good
detector performance.

\item {\bf Event Reconstruction} \\ Track segments are constructed
from the hits in the silicon layers (Si0 and Si1, inner segments
before the magnet) and in the DC0 and DC1 (outer segments after
passing the magnetic field).  The impact point on the EC is assumed to
correspond to the hit position in the DC2, located just in front of
the calorimeter. Pattern recognition is simplified in directions
perpendicular to the bending plane: only pairs of hits within an
azimuthal $2^\circ$ interval are combined to form a track
segment. Tracks from the inner segments are accepted if the minimal
distance with the beam axis is smaller than 5~mm. In addition, inner
and outer track segments should not differ by more than $2^\circ$ in
azimuthal angle.

The outer segment should point to an electromagnetic cluster in the EC
with an energy deposit of more than 300~MeV. This threshold is chosen
to be above the one for minimum--ionizing particles (MIPs).

The Drell-Yan sample is simulated with the magnetic field turned on in
order to test the momentum resolution of the detector. The large
background sample is traced inside the detector without magnetic field
and without drift chamber digitization to improve the speed of the
simulation. In this case, the track is reconstructed using only an
inner segment pointing toward an EC cluster, 300~MeV above threshold.

\end{description}

\subsubsection{Background Evaluation}
The major sources of background to the Drell--Yan process are the
combinatorial background from Dalitz--decays ($\pi^0$, $\eta$) and
gamma conversions, which can be studied and subtracted using the
wrong--sign candidates, and the decays of charmed--mesons, which can
be studied and reduced by reconstructing the secondary vertex of the
decays.

\begin{description}
\item Light meson decay: electron prongs from mesons decaying in
flight are rejected by requiring that the track points when extended
backwards towards the IP d not appear in the bending plane. Fast
Dalitz decays (mainly $\pi^0$) appear to originate directly from the
interaction point. Dileptons from the decay of one single light meson,
can be identified by their low invariant mass. Only multiple decays
may generate a dilepton with invariant mass larger than 2~$\rm
GeV/c^2$. Although an additional electromagnetic particle in the event
can be used to identify the parent $\pi^0$ and to reject the
candidate, this kind of request is not implemented yet into the
simulation: the result can be considered as conservative. Due to its
combinatorial origin, this background can be studied and subtracted at
large invariant mass by investigating wrong--sign candidates (control
sample).

\item Gamma conversion: gamma conversions are vetoed requiring a
charged hit in the first silicon layer, belonging to an electron
candidate track. To reject conversions taking place before or inside
the first tracking layer, one may in addition require silicon hits
with a twice as large energy deposit or the second prong of the lepton
pair to be reconstructed by the DC. In the present simulation, only the
veto from the first silicon layer is implemented, providing a
conservative result. The background from gamma conversions has a
combinatorial origin like the one from meson decay. Hence, the
residual background can be finally subtracted using wrong--sign
control sample.

\item Decay of charmed mesons: due to their associated production,
large mass and short lifetime, charmed--mesons tend to produce
dangerous right--sign candidates at high invariant mass. In the
fixed--target mode, the center--of--mass energy is too low to generate
a significant contamination from charmed mesons. At the higher
collider energies, charm background may become a serious issue. This
kind of background can be studied and eventually reduced by
reconstructing the secondary vertex of the decay with the silicon
detector. As an example, we note that the {\sc Babar} vertex detector
was already successfully employed to study $D^0-\overline{D}^0$
mixing, by measuring the vertex distribution of semileptonic decays of
$D$ mesons ~\cite{D0babar}.  Charm background was accounted for in the
present simulation, but no criteria were adopted yet to reduce it.

\item Misidentifications: The PAX detector is designed to provide
redundant high--level information about the particle type. This
includes a signal above threshold in the \Cer detector, a close to one
$E/p$ ratio (between the energy $E$ deposited in the EC and the
momentum $p$ measured in the spectrometer), a compact lateral profile
of the EC cluster. In addition $dE/dx$ measurements in the tracking
system and time--of--flight information can be eventually employed. In
the present simulation, the background by particle misidentification
is assumed to be negligible and neglected.
\end{description}

\begin{figure}[htb]
\begin{center}
  \includegraphics[width=0.65\linewidth]{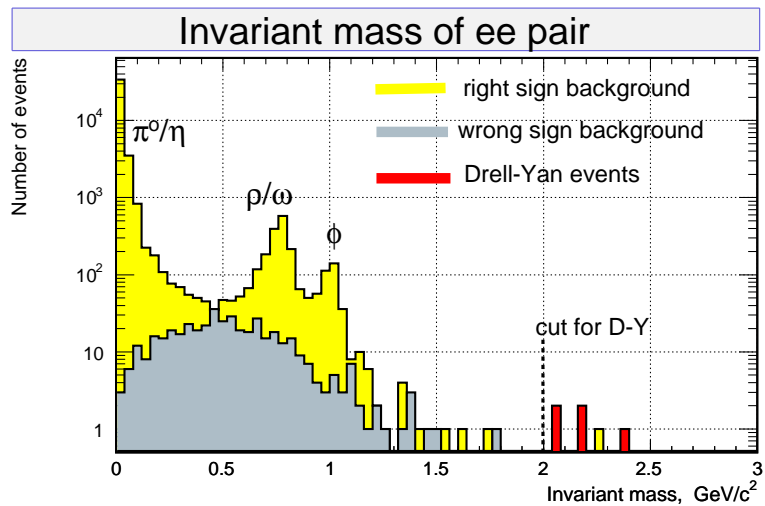}
  \parbox{14cm}{\caption{ \label{PAXback}\small Background estimate
      for the PAX collider mode: right--sign background (yellow
      histogram) is dominated by single meson decay at low invariant
      mass. At high invariant mass the major component has
      combinatorial origin and can be subtracted by investigating the
      wrong--sign background (gray histogram). Here electron tracks
      with energy greater than 300 MeV should originate close to the
      beam axis; gamma conversions are vetoed by requiring a charged
      hit in the first silicon layer.  
      PID and tracking system is assumed.  One background event is
      found above the threshold imposed to Drell-Yan candidates,
      $M_{ee}=2$ $\rm GeV/c^2$. The generated statistics is $2\ord{8}$
      $\overline{p}p$ inelastic interactions (40 minutes of data
      taking at a luminosity of $2\ord{30}$ $\rm cm^{-2}s^{-1}$).  The
      corresponding yield of $e^+ e^-$ Drell--Yan events inside the
      PAX detector acceptance is of the order of several events (red
      histogram).  Although limited in statistics, this result
      supports the expectation of a signal over background ratio close
      to one {\it before} background subtraction. This is in agreement
      with what was anticipated in the TP basing on a simplified
      simulation without detailed detector description.}}
\end{center}
\end{figure}

The dielectron invariant mass distribution for the background sample
($2\ord{8}$ minimum-bias $p\overline{p}$ interactions) is shown in
Fig.~\ref{PAXback}.
At low invariant mass, the right--sign candidates are dominated by the
not--dangerous contribution from single meson decays~\cite{CERES}.
This is indicated by the peaks in the distribution due to the high
mass--resolution. The combinatorial background becomes the largest
background source at dilepton masses larger than 1 $\rm GeV/c^2$.
There it can be studied and subtracted by means of the wrong--sign
control sample.  Within the available statistics of $2\ord{8}$
$p\overline{p}$ collisions, the excess of right--sign candidates at
large invariant masses due to the charm background is hardly
visible. One background event is found with invariant mass above 2
$\rm GeV/c^2$. The corresponding Drell--Yan dielectron signal, inside
the PAX detector acceptance with a mass greater than 2 $\rm GeV/c^2$,
is expected to be of the order of several events.  Note that the
signal over background ratio is of the order of one {\it before}
combinatorial background subtraction.  Although limited in statistics,
this study supports the view that the background for the $e^+e^-$
Drell-Yan measurement is well under control.

\subsubsection{Detector Performance}
\begin{figure}[htb]
\begin{center}
  \includegraphics[height=6.7cm]{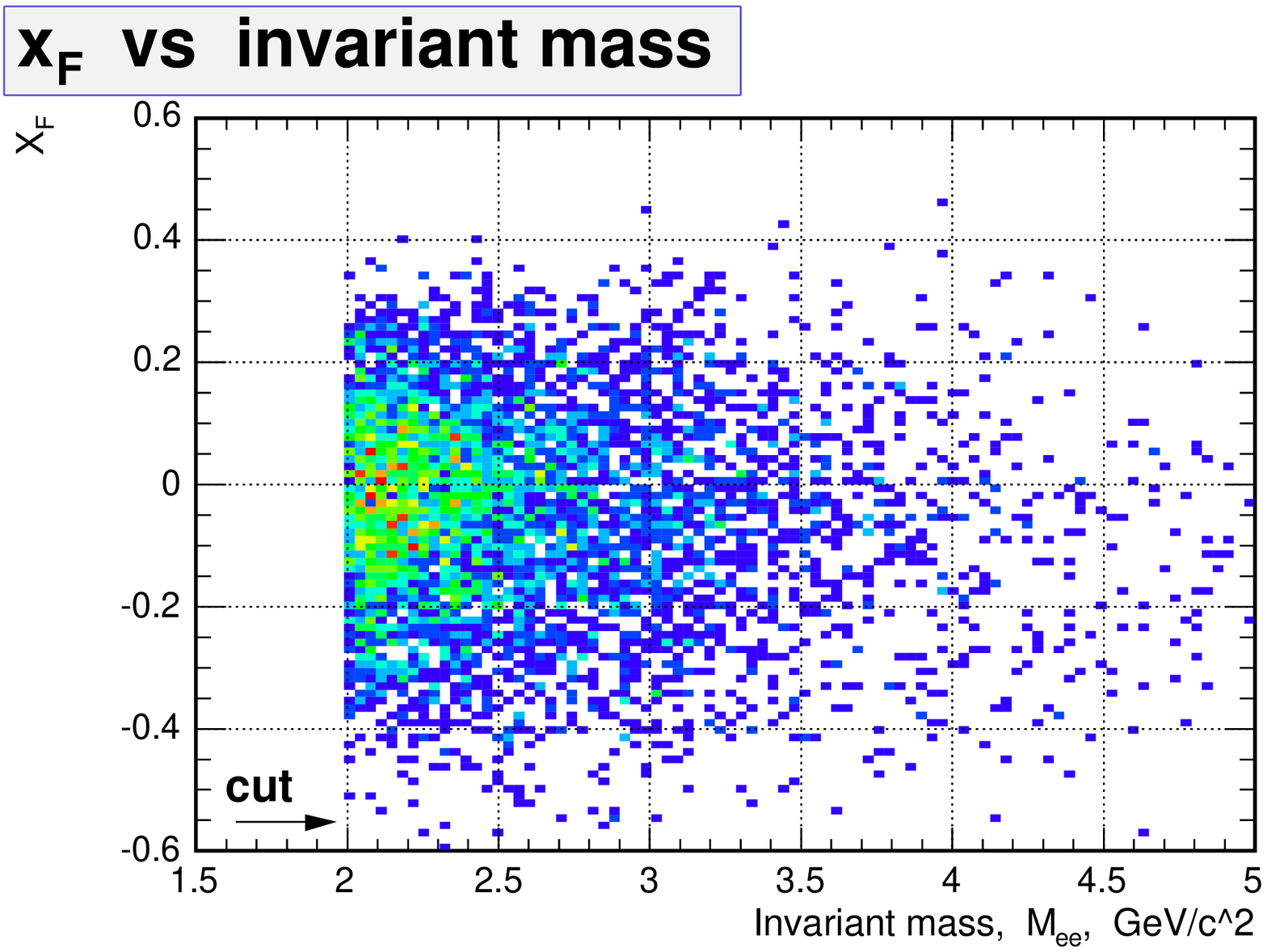}
  \includegraphics[height=6.7cm]{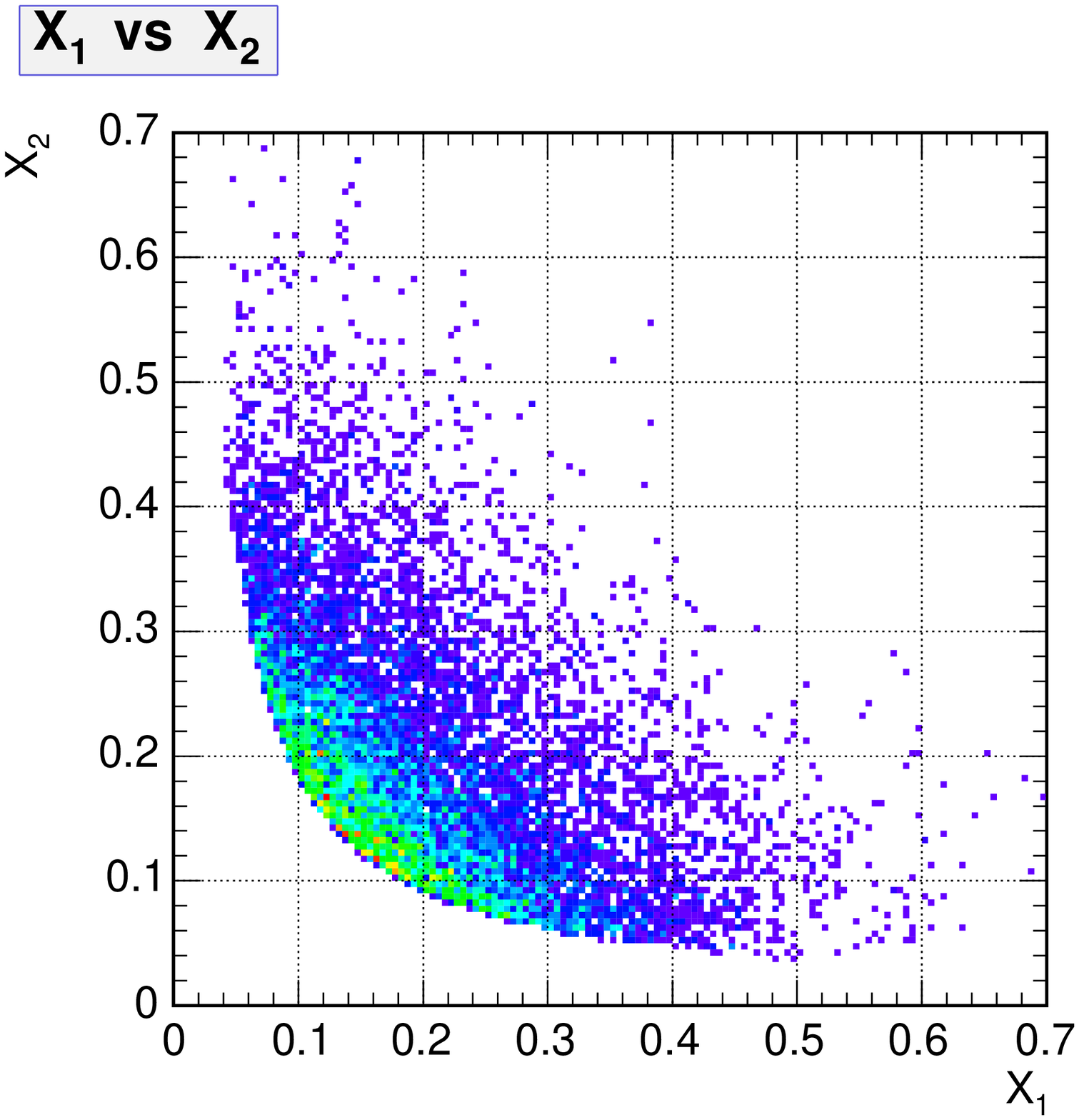}
  \parbox{14cm}{\caption{ \label{PAXkine}\small Kinematic distribution
   of the Drell--Yan events with $M_{ee}>2$ $\rm GeV/c^2$: Feynman
   $x_F=x_1-x_2$ versus invariant mass $M_{ee}=\sqrt{x_1 x_2 s}$ (left
   panel) and Bjorken $x_2$ versus $x_1$ (right).}}
\end{center}
\end{figure}

\begin{figure}[hbt]
\begin{center}
  \includegraphics[width=0.45\linewidth]{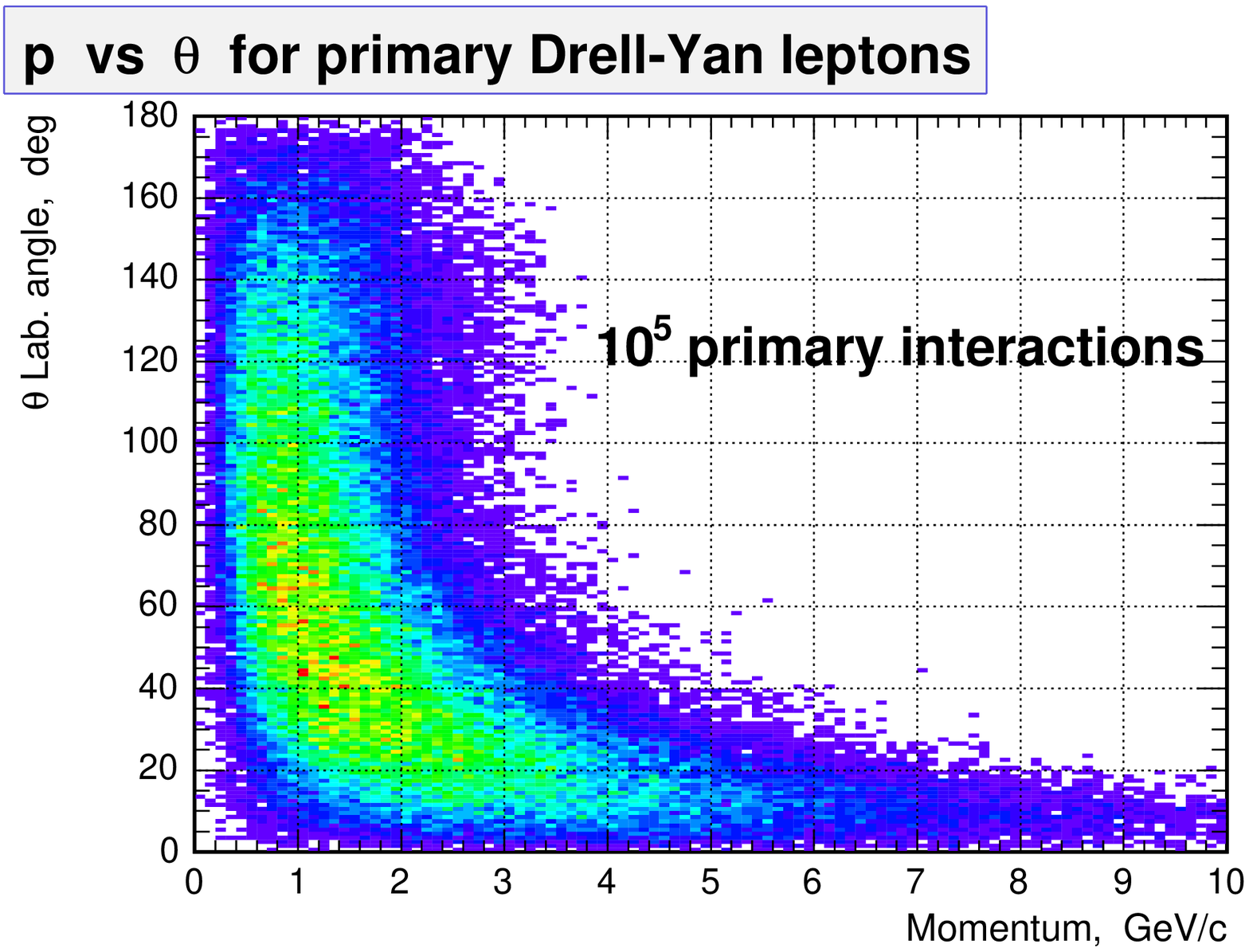}
  \includegraphics[width=0.45\linewidth]{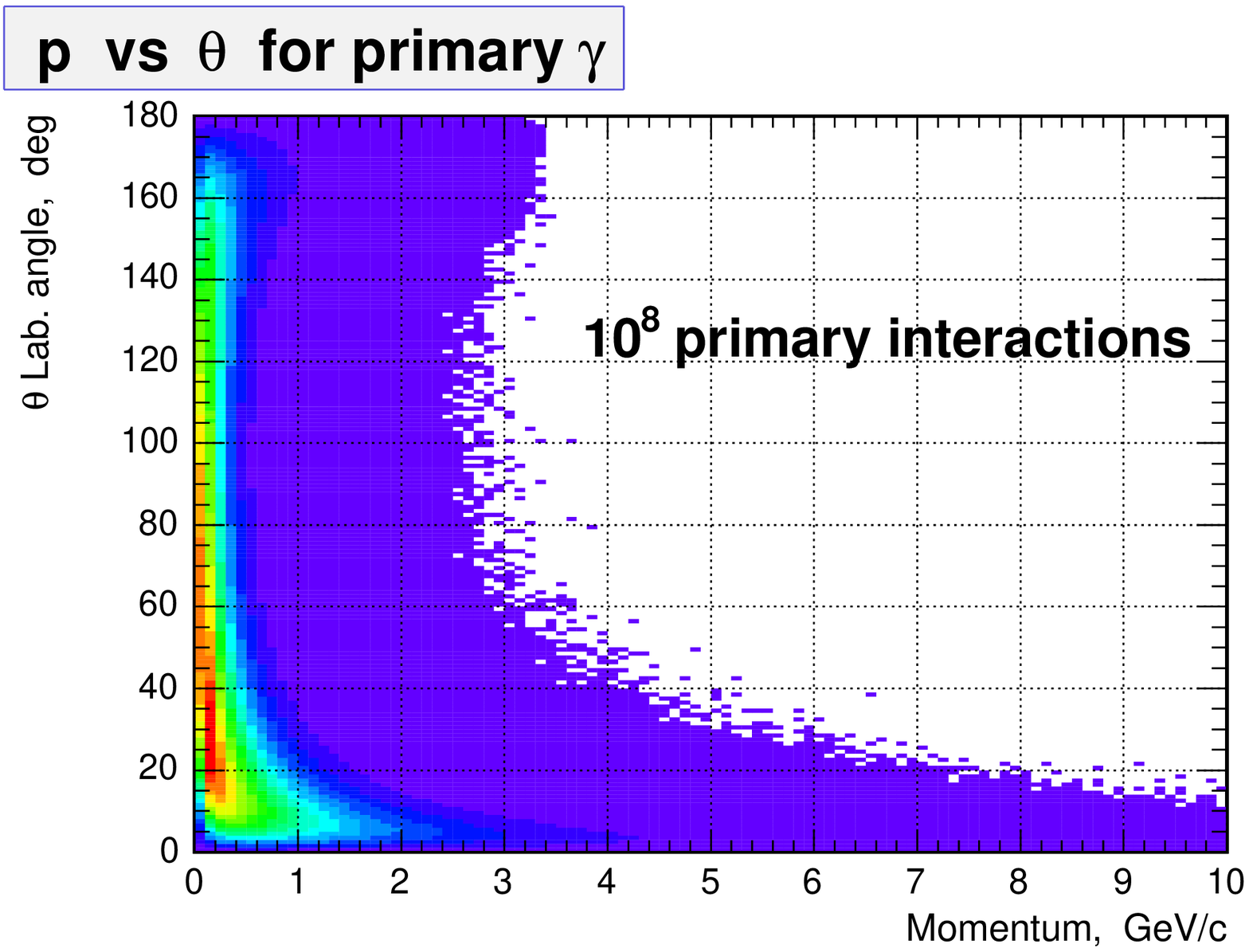}
  \includegraphics[width=0.45\linewidth]{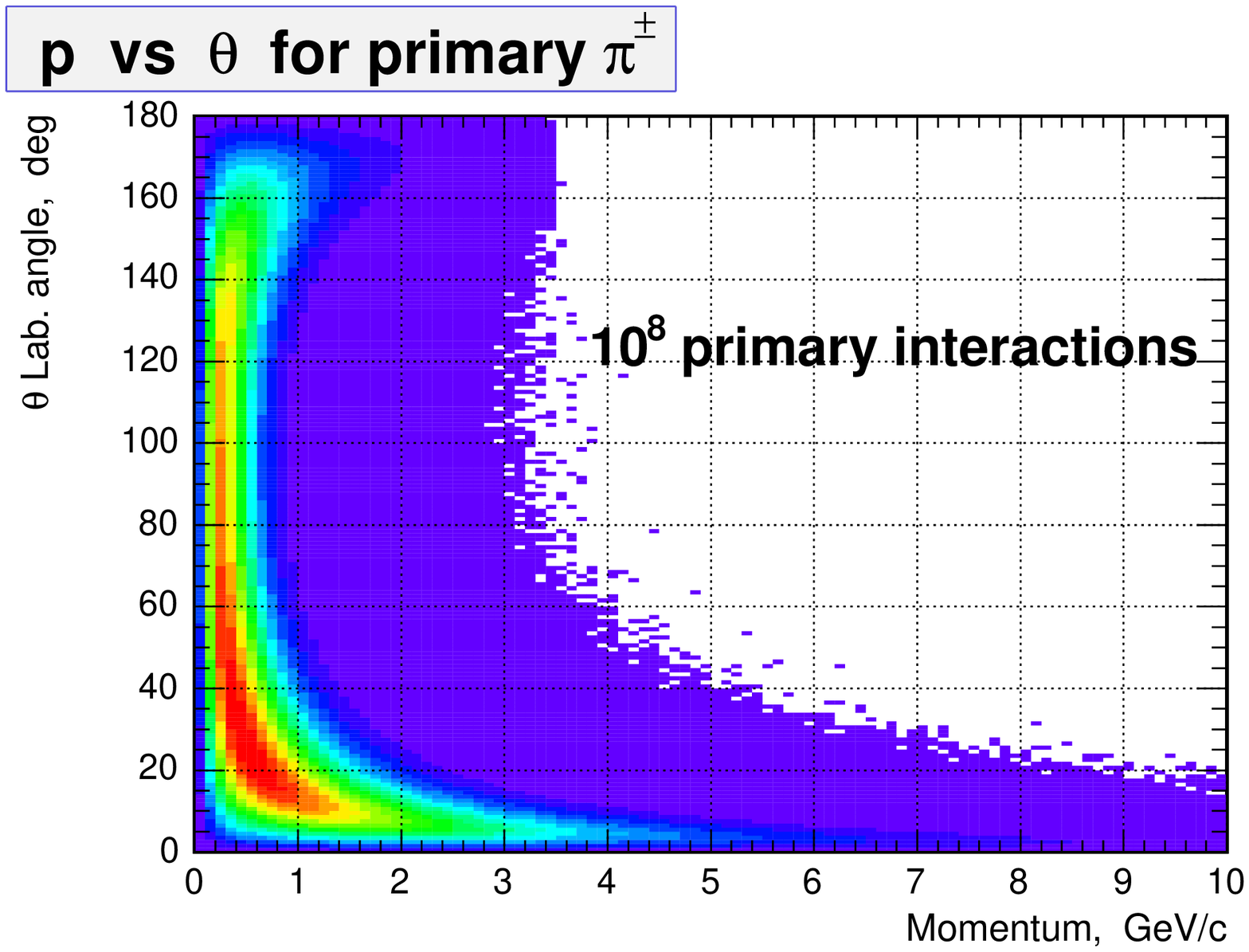}
  \parbox{14cm}{\caption{ \label{PAXphase}\small Phase space of
   Drell--Yan and background events: polar angle $\theta$ versus
   momentum $p$ of electrons (top left panel ) gamma (top right) and
   pions (bottom). Background particles concentrate at low energy and 
   in the forward direction.}}
\end{center}
\end{figure}

For every Drell--Yan event, the Bjorken $x_1$ of the proton and $x_2$
of the antiproton can be extracted from the measured invariant mass
($x_1 x_2 s=M^2=Q^2$) and from the longitudinal momentum
($x_1-x_2=x_F=2 p_L/\sqrt{s}$) of the lepton pair, where $s$ is the
center--of--mass energy, as shown in Fig.~\ref{PAXkine}.
In the $u$ dominance hypothesis, the $\Att(x_1,x_2)$ asymmetry is
related to the convolution of the transversity distributions
$h_1^u(x_1)\cdot h_1^u(x_2)$ of the proton through the relation
\begin{eqnarray}
\Att(x_1,x_2)=\hat{a}_{TT} \;
\frac{h_1^u(x_1)}{u(x_1)}\frac{h_1^u(x_2)}{u(x_2)} \hspace*{2cm}
\hat{a}_{TT}=\frac{\sin^2 \theta}{1+\cos^2 \theta} \cos (2\phi)\;.
\end{eqnarray}
where $u(x)$ is the up-quark unpolarized distribution and
$\hat{a}_{TT}$ is the asymmetry of the elementary process
$q\overline{q}\rightarrow e^+ e^-$.  The transverse asymmetry vanishes
at small polar angles $\theta$. Moreover, the background from generic
$p\overline{p}$ interactions concentrates in forward direction
(Fig.~\ref{PAXphase}).
In the present simulation the PAX detector is not instrumented in the
forward region, where the sensitivity to the $h_1^u$ is small.  The
active area of the detector is assumed to cover polar angles between
$20^\circ$ and $120^\circ$. Although this reduces the acceptance for
Drell--Yan events by about 50~\%, it maximizes the sensitivity to the
$h_1^u$ signal. The coils of the toroid cover only 11~\% of the
azimuthal acceptance. They are located outside of the $\phi=n \cdot
\pi/2$ positions where the transverse asymmetry reaches its maximum.

\begin{figure}[hbt]
\begin{center}
  \includegraphics[width=0.49\linewidth]{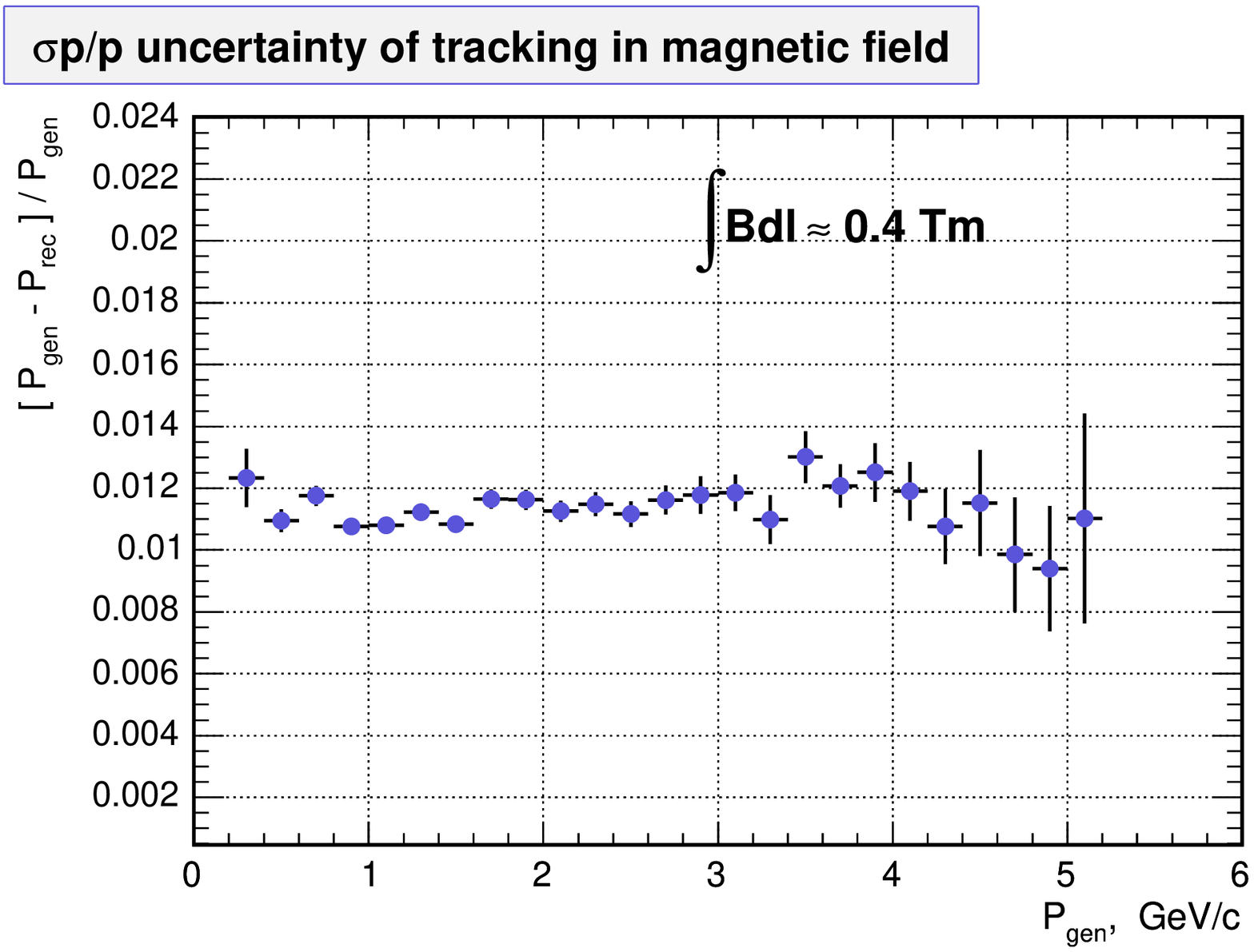}
  \includegraphics[width=0.49\linewidth]{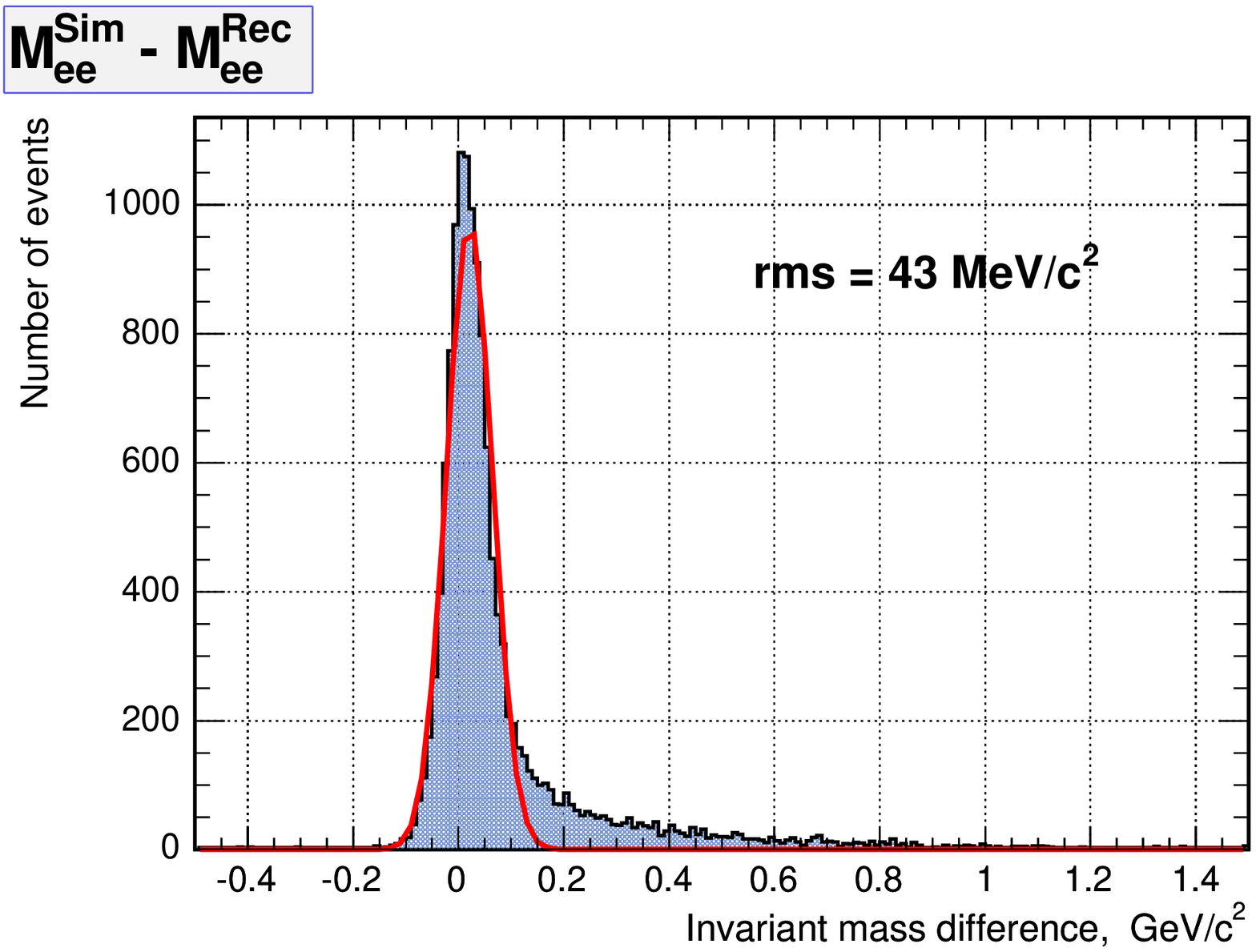}
  \parbox{14cm}{\caption{ \label{PAXreso}\small A 1 \% resolution in
   the particle momentum (left panel) and a better than 2 \%
   resolution in the dilepton invariant mass (right panel) are
   achievable by PAX spectrometer with a 0.4~Tm integrated field.}}
\end{center}
\end{figure}

\begin{figure}[hbt]
\begin{center}
  \includegraphics[width=0.55\linewidth]{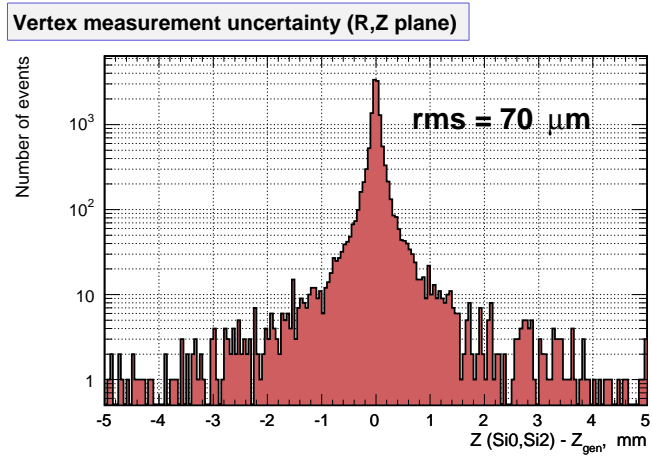}
  \parbox{14cm}{\caption{ \label{PAXvtx}\small Resolution in the
   vertex position achievable with the PAX silicon detector.  As
   demonstrated by {\sc Babar}~\cite{D0babar}, this resolution is
   sufficient to study secondary vertices of charm--meson decays.}}
\end{center}
\end{figure}

A resolution of the order of 1~\% in the lepton momentum and below
2~\% in the dilepton invariant mass is achievable with the PAX
spectrometer employing a magnetic bending power of 0.4~Tm, as shown in
Fig.~\ref{PAXreso}.
Such a resolution is required to efficiently distinguish the resonant
contribution from the continuum contribution and to precisely extract
the dependence of the $h_1^u$ distribution on the relevant kinematic
variables (the Bjorken $x$ and the square of the four-momentum
transfer $Q^2$). The silicon detector provides a $\sim 70$ $\rm \mu m$
resolution on the vertex position (Fig.~\ref{PAXvtx}).  This value is
in agreement with the design performance of the similar {\sc Babar}
silicon detector~\cite{VTXbabar}.

The Drell--Yan process is the reaction with the highest demand on
luminosity among the ones proposed to be studied by PAX. The
experimental uncertainty for double--spin asymmetries depends on the
number of observed events $N$ as well as on the degrees of
polarization of the two beams. A value of $P_p \gsim 0.80$ can be
assumed for the proton beam polarization, whereas values of
$P_{\bar{p}}~\approx~0.30$ are anticipated for the antiproton beam
polarization~\cite{ap2}. The statistical error of the transverse
asymmetry $\Att$ is then roughly given by $(P_p
P_{\bar{p}}\sqrt{N})^{-1} = 4/\sqrt{N}$.  At a luminosity ${\cal
L}=2\ord{30}$ $\rm cm^{-2}s^{-1}$, the expected signal rate is several
hundreds of events per day inside the PAX detector acceptance.  During
one year of data--taking with the above assumed luminosity, the
statistical error reduces to about 0.015.  This uncertainty must be
compared to the value of the measurable asymmetry which is ten times
larger, of the order of $\Att \sim \langle \hat{a}_{TT} \rangle \cdot
0.3 \sim 0.15$. Here an indicative $A_{TT}/\hat{a}_{TT}=0.3$ value is
assumed being supported by theoretical predictions~\cite{abdn2,gms2},
whereas $\langle \hat{a}_{TT}\rangle \sim 0.5$ is the average value
inside the PAX detector acceptance. It should be noted that an
extensive study is foreseen to optimize the spin--filtering process:
any beam polarization acquired in addition in the antiproton beam
leads to a linear reduction of the experimental uncertainty.

\begin{figure}[hbt]
\begin{center}
  \includegraphics[width=0.6\linewidth]{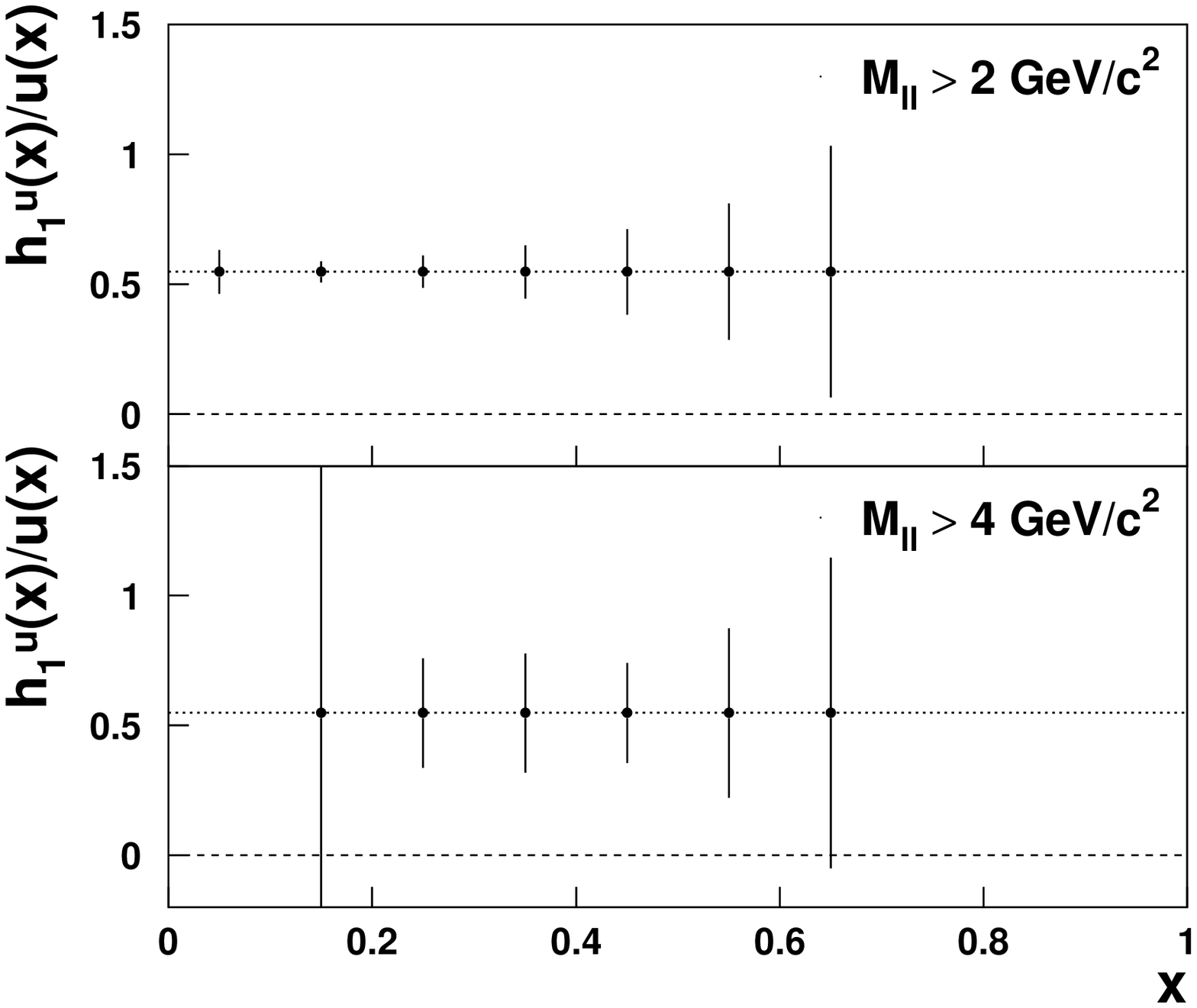}
  \parbox{14cm}{\caption{ \label{h1resu2}\small Expected precision of
      the $h_1^u(x)$ measurement for one year of data taking in the
      collider mode at PAX. A luminosity of $2\ord{30}$ $\rm
      cm^{-2}s^{-1}$ and a polar angle acceptance between $20^\circ$
      and $120^\circ$ were assumed. An indicative
      $A_{TT}/\hat{a}_{TT}=0.3$ value (supported by theoretical
      predictions~\cite{abdn2,gms2}) is used as input to the
      simulation. The data points are plotted along the corresponding
      value of the $h_1^u(x)/u(x)$ ratio. The precision achievable
      within the full $Q^2>4$ $\rm GeV^2$ kinematic range is of the
      order of 10 \% (top panel). The bottom panel shows the precision
      achievable in the restricted $Q^2>16$ $\rm GeV^2$ range. By
      tuning the beam energies, it should be possible to explore with
      high precision different $x$--intervals (see
      text).}}
\end{center}
\end{figure}

The achievable precision of the ratio of the transverse $h_1^u(x)$ to
the well--known unpolarized $u(x)$ distributions of the proton, in
different intervals of Bjorken--$x$ and after one year of data--taking
is shown in Fig.~\ref{h1resu2}.


The $h_1^u(x)$ transverse distribution can be measured in a wide $x$
range, from $x=0.7$ down to $x=0.05$, covering the most interesting
valence region and extending to low values of $x$, where the
theoretical predictions show the largest deviations.  It should be
noted that in principle the beam energies can be tuned to best explore
different $x$ intervals. Indeed the highest sensitivity is achievable
for $x\sim1/\sqrt{p_p p_{\overline{p}}}$ (the center--of--mass energy
$s\sim 4 p_p p_{\overline{p}}$, and the Drell-Yan cross section peaks
at low invariant masses of $M=\sqrt{x_1 x_2 s} \sim 2$ $\rm GeV/c^2$).
These numbers entail only the non--resonant contribution to the
Drell--Yan process: the exploitation of the \JPsi\ resonance region
will lead to a considerable enhancement of the number of events in the
$M^2$~=~9--16~GeV$^2$ range.

\subsection{Summary}
Extensive studies have been started to investigate different options
for the PAX detector configuration, aiming at an optimization of the
achievable performance.

\subsubsection{Phase-I}
In the Phase--I of the PAX physics program, where a (polarized)
antiproton beam with momentum up to 3.6 GeV/c scatters off a polarized
internal gaseous target, a luminosity of $1.5\ord{31}$ $\rm
cm^{-2}s^{-1}$ can be safely achieved.  Estimations of the expected
signals show that

\begin{figure}[hbt]
  \centering \includegraphics*[width=0.82\linewidth]{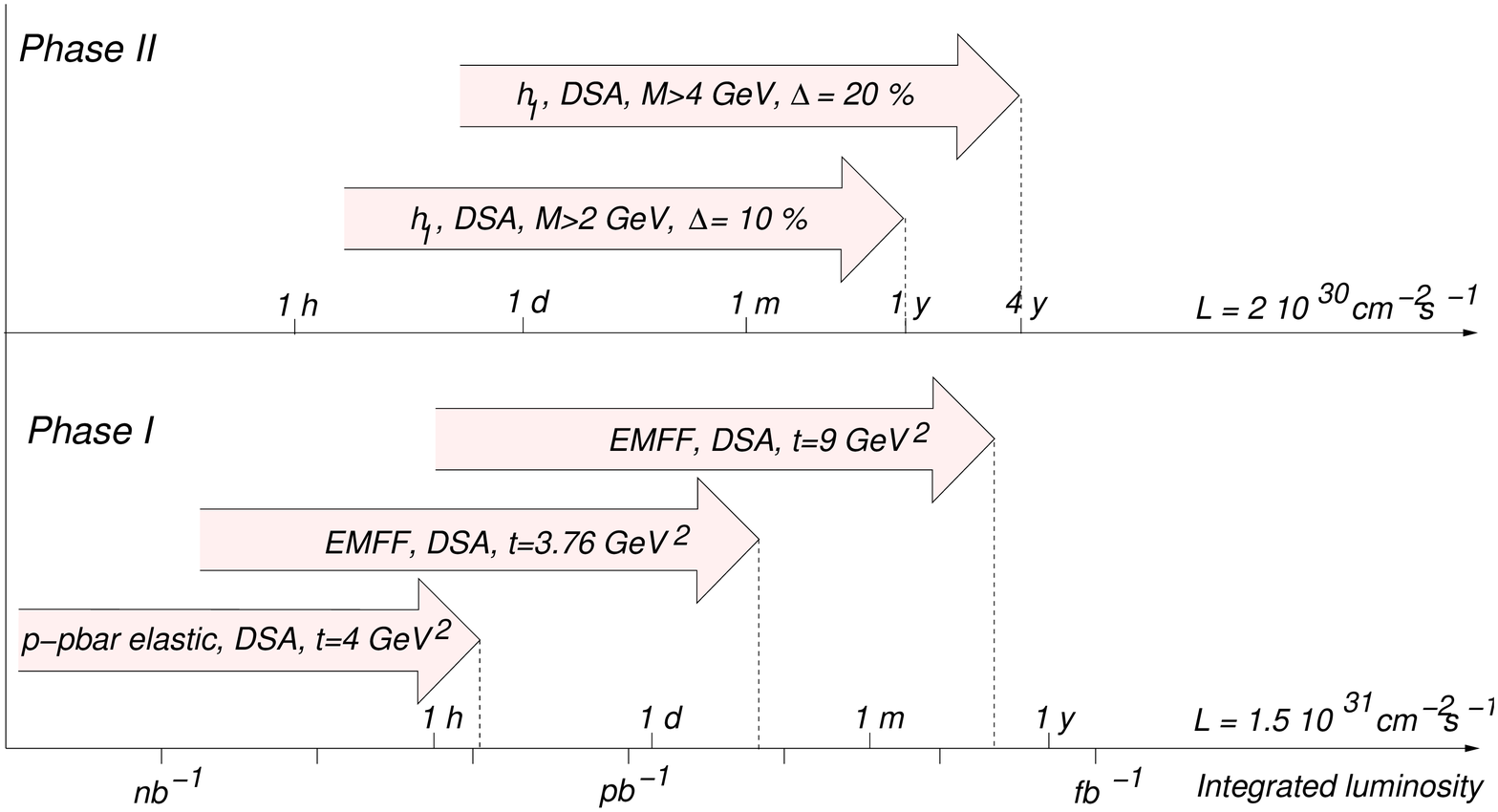}
  \parbox{14cm}{\caption{\small \label{Figrates} Integrated
luminosities required to precise measure double--spin asymmetries
(DSA) for the benchmark physics cases of the PAX program. The
corresponding running time (h=hour, d=day, m=month, y=year) is
indicated, as a function of the expected luminosity in Phase--I (fixed
target, ${\cal L}=1.5\ord{31}$ $\rm cm^{-2}s^{-1}$) and in Phase--II
(collider, ${\cal L}=2\ord{30}$ $\rm cm^{-2}s^{-1}$). An antiproton
beam polarization of $P_{\overline{p}}=0.3$ and a proton polarization
of $P_p=0.8$ are assumed.  In Phase--I, polarized hard elastic
scattering observables can be measured in a few hours of data--taking
with an absolute error of $\Delta=0.05$, whereas the different
measurements of the proton electromagnetic form factors entail from a
few days up to a few months of data--taking to reach the same
precision. In Phase--II, one year of running is enough to achieve a
relative error of 10 \% for the transversity distribution, if dilepton
invariant masses larger than 2 GeV are accepted. In a few years, the
$h_1$ measurement can be refined in the same mass range and verified
with a relative error of 20 \% at a higher mass scale, larger than 4
GeV and above the charm resonances.}}
\end{figure}

\begin{itemize}
\item the benchmark measurements of Phase--I, like the relative phases
of electric and magnetic form factors of the proton or transverse spin
effects in hard elastic scatterings, have no limiting reaction rates
(see Fig.~\ref{Figrates}).
\end{itemize}

\subsubsection{Phase--II}
The primary goal was to prove that the most challenging and
outstanding measurement of the PAX experimental program, the direct
measurement of the $h_1^q$ transversity distribution, is feasible. The
asymmetric--collider option, described in Sec.~7 of the PAX Technical
Proposal, was adopted as the most promising scenario, where a 15~GeV/c
polarized antiproton beam from the HESR collides with a 3.5~GeV/c
polarized proton beam from the CSR to produce $e^+ e^-$ Drell-Yan
events. A detailed simulation of the performances of the detector
proposed in the PAX Technical Proposal, which is designed to provide
$e^+ e^-$ Drell--Yan pair detection, has been performed. Although
preliminary, the results are very encouraging:

\begin{itemize}
\item the background is under control and can be studied with control
samples: in particular the signal over background ratio is expected to
be of the order of one but the major background component, being of
combinatorial origin, can be subtracted by means of the measured
wrong-sign candidates;
\item a vertex detector similar to already working devices can provide
a resolution in the vertex position better than 100~$\rm \mu m$: this
is useful to control background from charm--meson decays;
\item conventional tracking detectors can provide a resolution better
than 2 \% in dilepton invariant mass: this is sufficient to
efficiently distinguish resonance from continuum contributions and to
investigate the $h_1^q(x)$ dependence. As a consequence, the
requirements on the calorimeter performance can be relaxed and
solutions cheaper than scintillating devices can be investigated;
\item during one year of data--taking, the most interesting valence
region can be explored and the $h_1^q$ transverse distribution can be
measured with a precision better than 10~\%.
\end{itemize}

A conservative estimate of the luminosity achievable in the
accelerator scheme presented in the PAX Technical Proposal, (and
described in Appendix E) provides a value around $2\ord{30}$ $\rm
cm^{-2}s^{-1}$. With this luminosity, a few years of data-taking would
be sufficient to obtain a definite measurement of the last leading
piece of the partonic description of the nucleon
(Fig.~\ref{Figrates}).


\end{appendix}

\begin{thebibliography}{99}
\bibitem{pol-conservation}Preservation of beam polarization is
  routinely accomplished at COSY--J\"ulich \cite{lehrach}, and at
  higher energies at the AGS of BNL \cite{khiari,huang,bai}.
\end{thebibliography}
\end{document}